\documentclass[12pt]{report}
\usepackage[utf8]{inputenc}
\usepackage{amsmath}
\usepackage{adforn}
\usepackage{amssymb}
\usepackage{braket}
\usepackage{tikz}
\usepackage{color,soul}
\usepackage{simplewick}
\usepackage{xcolor}
\usepackage{graphicx}
\usepackage{calc}
\usepackage{comment}
\usepackage[framemethod=TikZ]{mdframed}
\mdfdefinestyle{MyFrame}{
    linecolor=blue,
    outerlinewidth=2pt,
    roundcorner=20pt,
    innertopmargin=\baselineskip,
    innerbottommargin=\baselineskip,
    innerrightmargin=20pt,
    innerleftmargin=20pt,
    backgroundcolor=white}
\usepackage{geometry}
 \usepackage{setspace}
\newlength{\depthofsumsign}
\usetikzlibrary{positioning}
\usetikzlibrary{decorations.markings}
\DeclareMathOperator{\Tr}{Tr}
\def\blank{\medskip\hrule\medskip}
\allowdisplaybreaks
\newcommand{\nsum}[1][1.4]{% only for \displaystyle
    \mathop{%
        \raisebox
            {-#1\depthofsumsign+1\depthofsumsign}
            {\scalebox
                {#1}
                {$\displaystyle\sum$}%
            }
    }
}
\newlength{\depthoflbracket}
\newlength{\depthofrbracket}
\newcommand{\?}{\stackrel{?}{=}}
\tikzset{->-/.style={decoration={
  markings,
  mark=at position #1 with {\arrow{>}}},postaction={decorate}}}

\usepackage{afterpage}

\setul{0.5ex}{0.3ex}
    \definecolor{Red}{rgb}{1,0.0,0.0}
    \setulcolor{Red}

\usepackage[english]{babel}
\usepackage[backend=biber]{biblatex}
\nocite{*}

\usepackage[colorlinks=true,linkcolor=blue,citecolor=blue]{hyperref}
\newcommand{\btz}{\begin{tikzpicture}}

\newcommand{\etz}{\end{tikzpicture}}

\newcommand{\size}[2]{\scalebox{#1}{#2}} % #2 is the tikz picture

\newcommand{\posn}[3]{\hspace*{#1}\raisebox{#2}{#3}}

\newcommand{\nd}[3]{\draw #1 node[scale=#2] {#3} }

\newcommand{\pt}[2]{
\filldraw[black] #1 circle (2pt) node[scale=2,anchor=north]{#2};
}

\newcommand{\sld}[2]{
\draw #1--#2;
}

\newcommand{\eql}[2]{\draw #1 node[scale=#2]{=};
}

\begin{document}
\setul{0.5ex}{0.3ex}
    \definecolor{Red}{rgb}{1,0.0,0.0}
    \setulcolor{Red}
\begin{titlepage}

    \begin{center}
        \vspace*{1cm}
            
        \LARGE
        \textbf{Diagrammatics for Wightman Correlators in General States: The Cases of the Simple Harmonic and Anharmonic Oscillators}
            
        \Large
            
        \vspace{2 cm}
            
        \textbf{Shridhar Vinayak}
            
        \vspace{1.5 cm}
            
        \textit{A dissertation submitted for the partial fulfilment of BS-MS dual degree in Physics}
        
        \vspace{1.5 cm}
            
        \includegraphics[width=0.35\textwidth]{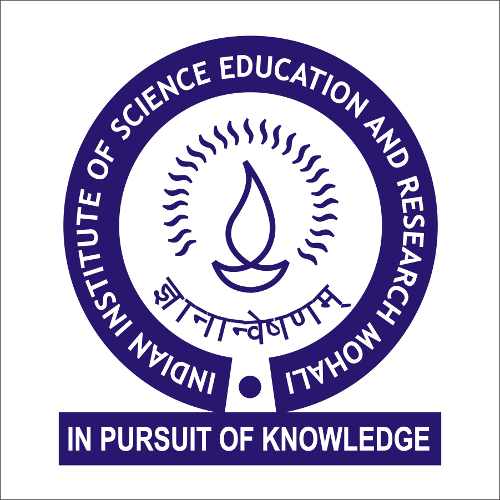}
        \\[20pt]
        \Large
        \textbf{Department of Physical Sciences}\\
        \textbf{Indian Institute of Science Education and Research Mohali}\\
        \textbf{April 2021}

    \end{center}
\end{titlepage}
\newpage
\begin{center}
\LARGE\underline{\textbf{Acknowledgements}}
\end{center}\vspace{20pt}
I am genuinely of the view that this section is the toughest to write. Physics has the advantage of mathematics being its language. As a result, the gist of most physical ideas can be conveyed precisely in a few sets of equations. But alas! The same does not hold true for writing acknowledgements. Even if one writes pages and pages of it, one remains far from expressing what one truly feels. But I will try my bit.

I begin by thanking my supervisor, Dr. Loganayagam R. He has completely changed the way I look at not only research, but life in general. He taught me how important it is for one to first think of genuine and solid motivations before attacking a problem. He made me realise that a good theoretical physicist must also have the ability to look at a problem from the point of view of an experimentalist. It is not always good to think in very abstract terms. More fruitful may be the approach wherein one thinks in terms of simple examples and simple experiments which one may devise to test the idea in mind. Through his guidance, I also learnt how to write the English accompanying Physics well, wherein one needs to strike a fine balance between brevity and clarity. He has taught me that it is not important how productive one's efforts turn out to be, but how much effort actually went in. And one must never sit empty and wait for one's flow to return; one must keep working through the wait.\iffalse I would also like to genuinely thank him for always giving me time to cope up with my recurring personal and academic problems.\fi

I also take this opportunity to thank Dr. Kinjalk Lochan for giving me many opportunities to present the work I had been doing to his research group. I would also like to thank him for his critical remarks. \iffalse and for being very accommodating regarding deadlines. A sincere thanks goes to the Dean (Academics), IISER Mohali Prof. Jasjeet Bagla too in this regard.\fi

I then thank my home institute IISER Mohali for  the wonderful and well-rounded education it has imparted. I hope this thesis does it justice from my side. I also sincerely express gratitude towards \textbf{ICTS-TIFR Bengaluru} for supporting me through their \textbf{S.N Bhatt Memorial Excellence Fellowship Programme} and their \textbf{Long-Term Visiting Students' Programme}. I would also like to thank Chandramouli Chowdhury, Vinay Kumar and Ritwik Mukherjee for helping me in submitting this thesis to the arXiv.

\iffalse To my father, who is solely responsible for building my foundations in Physics through the many hours he dedicated for teaching me the subject throughout my school life. He has been the best teacher I have had. 

To my aunt and uncle Nirja Jha and Anil Jha for having been extremely strong pillars of support throughout. To my other aunt Sujata Sanghamitra for her so very selfless and warm affection and for her constant surprise visits to Delhi. To Dr. Shubhra Sharma for all the guidance, care and support she has given. To my cousins Niranjan and Nirupama who had become like my new hostel-mates during the lockdown. To my other cousin Tanvi for bringing splashes of life everytime we talked or met. To my grandparents Ashok and Reena Pandey for giving me the strong roots of my childhood on which I am growing. To my younger brother Tanmay who should have been the older one and to Maximus for his ever-growing unconditional affection.

To my friends Tinku, Gaurav, Yash, Kabeer, Abhijit, Anubhav and Hayman for making my college days the ones I would always cherish.

And to my mother, Dr. Shailaja Thakur, because of whom I am what I am and because of whom I will be what I will be.\fi
\newpage
\begin{center}
\LARGE \underline{\textbf{Abstract}}\\[40pt]
\end{center}
The machinery of computing vacuum expectation values of a time-ordered sequence of position operators of the simple harmonic oscillator is already well established. It rests on a Wick theorem, which enables one to decompose such a quantity in terms of products of \emph{pairwise contractions}, which are vacuum expectation values of a time-ordered sequence of position operators taken two at a time. This result naturally leads to a diagrammatic approach of computing such correlators, and is already well known in the form of Feynman diagrams. We generalise this setup to encompass expectation values of a general ordered sequence of position operators (Wightman sequences) in general density matrices of the simple harmonic oscillator. A Wick theorem is first developed for this situation and consequently a diagrammatics is laid down. Wightman correlators in general density matrices of the \emph{anharmonic} oscillator are also analysed and a diagrammatic formalism is developed for them too.
\newpage
\tableofcontents
\newpage
\chapter{Introduction}\label{intro}
\section{Background}
Quantum field theory, or abbreviated QFT, has emerged as being one of the most successful theories developed by mankind. In addition to being theoretically very deep, it has also fared exceptionally well on experimental grounds. The most glorified example of such a victory is how the QFT result of the anomalous magnetic dipole moment of the electron matches with its experimentally determined value to more than 10 significant digits! Courtesy to this theory, the magnetic moment of the electron holds the accolade of being the most accurately verified prediction in the history of Physics.

QFT has a lot of predictive power. As already mentioned, the magnetic moment of the electron is one such observable one can calculate using this theory. However, there are a bunch of many others which are computed using the techniques of QFT, and indeed have been directly verified in laboratories. Scattering cross-sections of a variety of high-energy processes and decay rates of particles are some examples among many.

But how does one carry out such calculations? 

To begin with, one has to realise that in order to model processes like scattering and decay as actually seen in Nature using QFT, the quantum fields under consideration must be \emph{interacting}. A typical example of such an interacting QFT is $\phi^4$-theory, based on the Lagrangian:
\begin{equation}
    \mathcal{L}=\frac{1}{2}\partial_{\mu}\phi\;\partial^{\mu}\phi-\frac{1}{2}m^2\phi^2-\frac{\lambda\phi^4}{4!},
\end{equation}
where $\phi\equiv\phi(t,\textbf{x})$ is a scalar field and the third term, namely:
\begin{equation}
    \frac{\lambda\phi^4}{4!},
\end{equation}
introduces interactions into the system. The parameter $\lambda$ is termed the \emph{coupling}. It is precisely this interaction term which allows the particles corresponding to the scalar field to interact among themselves. If interactions were absent, all the scattering processes described by the theory would have been trivial; the particles corresponding to the field would have passed through each other unaffected, a phenomenon rarely observed in Nature. On the other hand, decay processes do not even make sense without introducing interactions.

Once interactions have been introduced into the theory, the doors of perturbation theory are knocked in most of the cases. As a part of this protocol, one assumes the QFT being studied to be \emph{weakly coupled}. And then, one expands the observable(s) under consideration in a power series in the coupling. Ignoring the mathematical subtleties involved, one is then, in principle, in a position to predict the value of the observable(s) being analysed to all orders in the coupling.

A class of observables which make frequent appearances in QFT calculations despite being of physical significance in their own right are \emph{time-ordered field correlators in the vacuum}. They are also called the \emph{Green's functions} of the theory. They assume the form:
\begin{equation}\label{eq:1}
    \bra{\Omega}T\phi_1\phi_2\dots\phi_n\ket{\Omega},
\end{equation}
where $\phi_i=\phi(x_i)\equiv\phi(t_i,\textbf{x}_i)$ are Heisenberg-picture field operators, $T$ is the time-ordering symbol and $\ket{\Omega}$ is the vacuum state of the interacting field theory.

The quantities \eqref{eq:1} are of utmost physical significance. As a simple example demonstrating how physical they are, consider the object:
\begin{equation}
    \bra{\Omega}T\phi(x)\phi(y)\ket{\Omega}.
\end{equation}
It represents the probability amplitude for a particle corresponding to the field $\phi$ to propagate from the spacetime point $y$ to the spacetime point $x$. This is indeed very physical.

The objects \eqref{eq:1} play a central role in the LSZ reduction formula \cite{lsz}. To be precise, the residues at the poles of the quantity \eqref{eq:1} are scattering amplitudes. The modulus squared of these amplitudes gives one the scattering cross section of the high-energy process being considered. This result further supports the strong physicality that these objects imbibe.

To analyse these objects, one first switches to the interaction picture, and then makes use of the Gell-Mann and Low theorem to get the well-known formula \cite{pes}:
\begin{align}\label{eq:2}
    \bra{\Omega}T\phi(x_1)\phi(x_2)&\dots\phi(x_n)\ket{\Omega}\notag\\&=\lim_{T\to\infty(1-i\epsilon)}\frac{\bra{0}T\bigl\{\phi_I(x_1)\phi_I(x_2)\dots\phi_I(x_n)e^{-i\int_{-T}^{T}dt\;H_{I}(t)}\bigr\}\ket{0}}{\bra{0}\bigl\{Te^{-i\int_{-T}^{T}dt\;H_{I}(t)}\bigr\}\ket{0}}.
\end{align}
Here, $\phi_I$ represents the interaction-picture field, which evolves in the same way as the \emph{free} field, and $\ket{0}$ is the vacuum state of the \emph{free} theory. $H_I$ is the interaction Hamiltonian in the interaction picture, which is given by:
\begin{equation}
    H_I=\int d^3\textbf{x}\;\frac{\lambda\phi_I^4}{4!}.
\end{equation}
In fact, the true interaction picture of a QFT must also take into account the concepts of counterterms and renormalization, but such subtleties are ignored for the moment.

It is then that perturbation theory is invoked. The exponentials occurring in \eqref{eq:2} are expanded in powers of the coupling. During this process, one encounters correlators of the form:
\begin{equation}\label{eq:3}
\bra{0}T\phi_I^{4}(z_1)\phi_I^{4}(z_2)\dots\phi_I^{4}(z_k)\phi_I(x_1)\phi_I(x_2)\dots\phi_I(x_n)\ket{0}.
\end{equation}
Complicated as these objects may seem, it turns out to be easy to compute them owing to a result known as Wick's theorem \cite{wck}. 

Wick's theorem allows one to express such time-ordered products of field operators in terms of \emph{normal-ordered} products and analytic functions called \emph{Feynman propagators}. Feynman propagators, denoted $D_F(x-y)$, are the objects:
\begin{equation}\label{eq:4}
    D_F(x-y)=\bra{0}T\phi_I(x)\phi_I(y)\ket{0}.
\end{equation}
An additional simplification which occurs in the consideration of the quantities \eqref{eq:1} arises from the fact that they are completely expressed in terms of time-ordered correlators of the free field operators in the vacuum of the free theory $\ket{0}$. Since the expectations of normal-ordered products of the free field operators vanish in this state, the correlators \eqref{eq:1} depend on only the Feynman propagators \eqref{eq:4} owing to Wick's theorem.

Wick's theorem and the fact that one looks at time-ordered correlators in a special state, namely the vacuum state, thus imply that these correlators only depend on Feynman propagators. This realisation leads to a very \emph{visual} way of looking at such correlators. The language of \emph{Feynman diagrams} thus emerges. The starting point for this formalism is the association:
\iffalse\begin{figure}\posn{80pt}{-80pt}{\size{1}{
\btz %Fig 1.1
\nd{(1,0)}{2}{$D_F(x-y)$};
\eql{(3.5,0)}{2};
\pt{(5,0)}{$x$};
\sld{(5,0)}{(7,0)};
\pt{(7,0)}{$y$};
\etz}}
\caption{Diagrammatic representation of the Feynman propagator.}
\end{figure}\fi
\begin{figure}[h]
    \centering
  \includegraphics[scale=0.1]{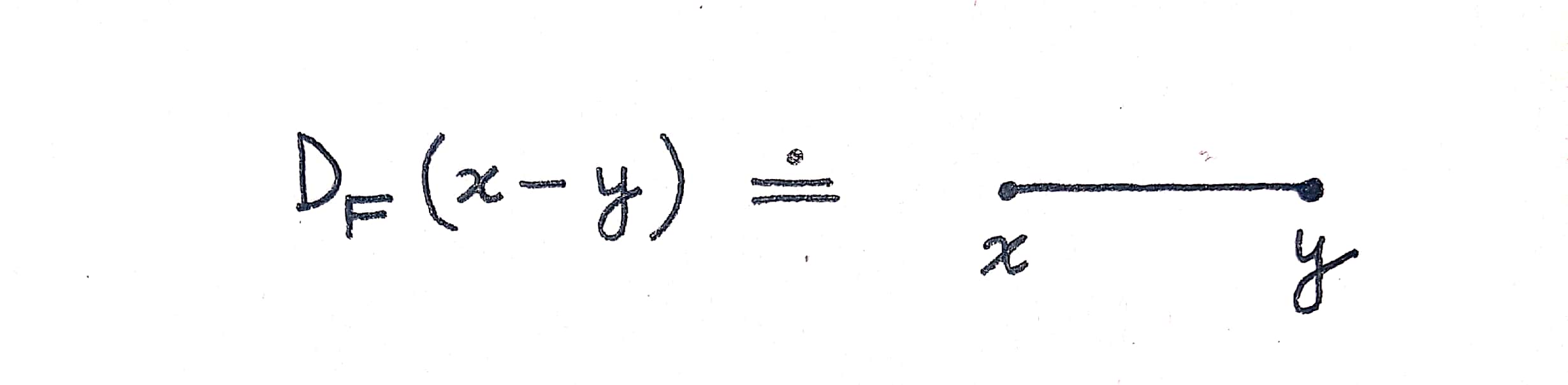}
  \caption{Diagrammatic representation for the Feynman propagator.}
\end{figure}
\\
Further analysis brings in another component to the diagrams called \emph{vertices}, which stand for the analytic factor:
\begin{figure}[h]
    \centering
  \includegraphics[scale=0.1]{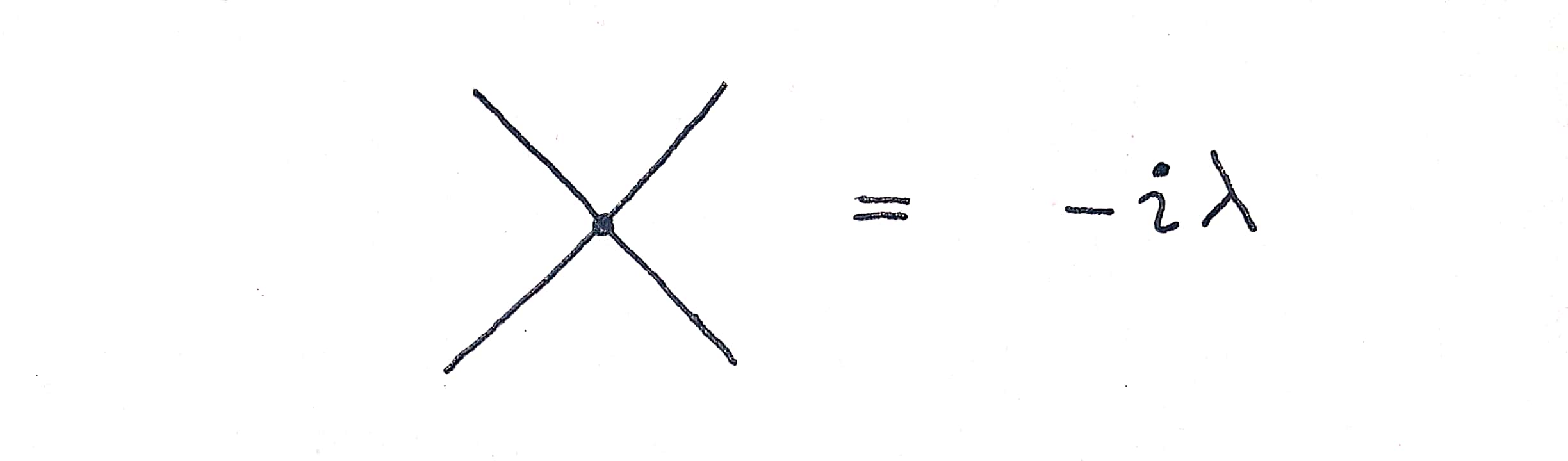}
  \caption{Diagrammatic representation of the interaction vertex.}
\end{figure}
\newpage
An example of a result stated in the new parlance of Feynman diagrams is as follows:
\begin{figure}[h]
    \centering
  \includegraphics[scale=0.16]{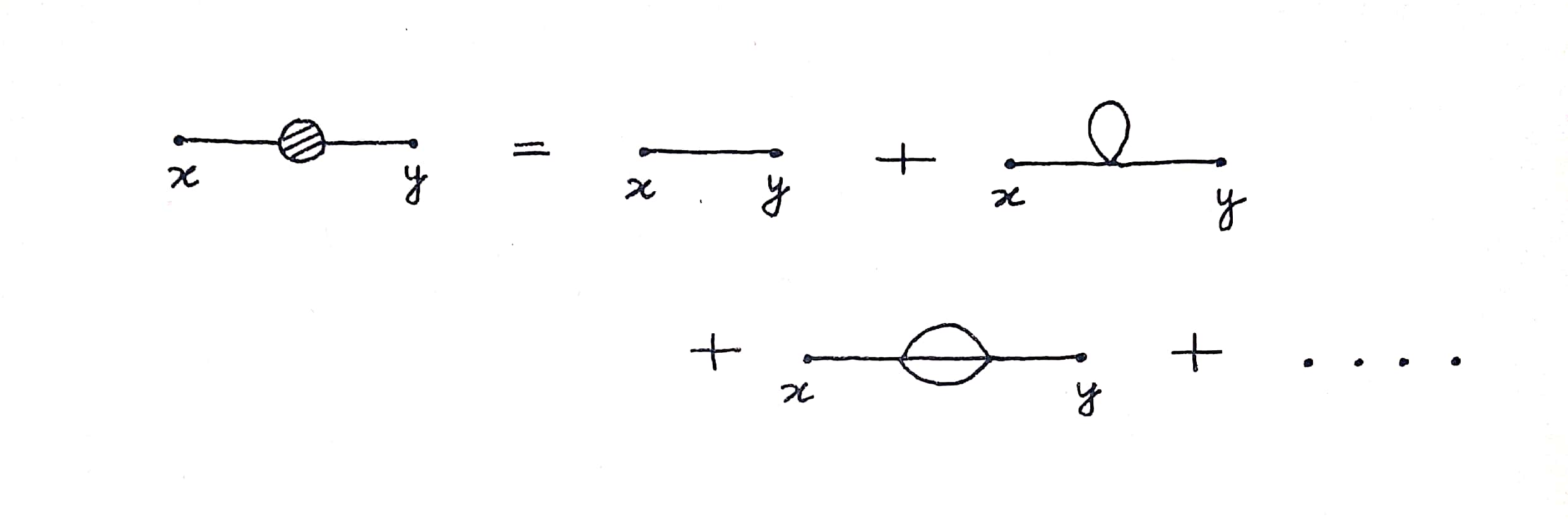}
  \caption{Diagrammatic expansion of $\bra{\Omega}T\phi(x)\phi(y)\ket{\Omega}$.}
\end{figure}
\\
\\
From an analytic point of view, the LHS of the above equation stands for the Green's function:
\begin{equation}\label{eq:5}
    \bra{\Omega}T\phi(x)\phi(y)\ket{\Omega},
\end{equation}
and the RHS is a perturbation expansion for it in terms of the coupling $\lambda$. A clear advantage of this formalism is that one can literally \emph{see} the contributions to the LHS at different orders of the coupling.
Since a vertex brings in a factor of $\lambda$ into the analytic expression, a diagram on the RHS with $n$ vertices is a $\lambda^n$ contribution to the 2-point Green's function \eqref{eq:5} of the interacting theory. 

Not only are the contributions at all orders of the coupling visible as a result of Feynman diagrams, they are also easily calculable. After assigning a simple set of rules called the \emph{Feynman rules} of the theory, one can simply read off the analytic expressions which these diagrams encode. 

In the context of all these advantages, a diagrammatic formalism to compute observables of a QFT seems rather beautiful and inviting.

Though not mentioned as yet, but the whole formalism outlined above also holds true for the analysis of continuous order parameters in thermal states. The analogues of time-ordered correlators in this situation are Euclidean correlators of the order parameter in thermal states. These are nothing but thermal averages of products of the order parameter at different spatial positions. That is, they take on the form:
\begin{equation}
    G^{(n)}_{\beta}(\textbf{x}_1,\textbf{x}_2,\dots,\textbf{x}_n)=\langle\phi(\textbf{x}_1)\phi(\textbf{x}_2)\dots\phi(\textbf{x}_n)\rangle_{\beta},
\end{equation}
where $\phi(\textbf{x})$ is the continuous order parameter and $\beta$ the inverse temperature. $\langle\dots\rangle_{\beta}$ denotes a thermal ensemble average. $G^{(n)}_{\beta}$ is termed the n-point Euclidean correlator of the order parameter $\phi(\textbf{x})$.

These correlators too obey a Wick's theorem and are consequently expressible in the language of Feynman diagrams \cite{klein}.
\section{Generalisations and Novel Directions}
Every theoretical formalism, however beautiful it may seem, always has shortcomings of its own. And these are revealed once one attacks the assumptions it makes. So is the case with the diagrammatic formalism for time-ordered correlators in the vacuum state which was discussed in the previous section. We will now relax the two assumptions made there one by one, and discuss whether there is any physical significance of doing so. If there is, it would serve as a genuine motivation for studying these new objects thus uncovered.

The first assumption made in the previous section was that the system resided in the interacting vacuum $\ket{\Omega}$. This is certainly very reasonable, since the vacuum state is the state of lowest energy. But this is not always the case! Systems in non-equilibrium occupy states different from the interacting vacuum. Such systems are abundant in Nature and have also been studied from the lens of theoretical physics. Examples of setups wherein one encounters such situations include nuclear heavy-ion collisions \cite{dan,bot}, ultracold gases \cite{caz} and optimal control theory \cite{brif}. As the vacuum state, or equivalently, the state of lowest energy is not occupied by these systems, the traditional mechanism of Feynman diagrams does not extend to such cases. Wick's theorem still holds, but due to the fact that the state is not the vacuum, the expectations of normal-ordered products of the degrees of freedom do not vanish. Thus, for example, correlators do not admit simple expressions which involve just propagators and vertices. Some state-dependent diagrammatic elements must also be introduced. The same holds true in the context of statistical field theory, wherein one analyses continuous order parameters. If such a system is in a state which is not thermal, then the whole formalism discussed in the previous section breaks down. And such situations are readily seen in Nature. The transient behaviour of a thermalising system is one such example. 

This presents us with one of our motivations: Can a diagrammatic formalism be developed for systems occupying \emph{general} states?

Such a question has indeed been addressed previously. However, these earlier pursuits have been nestled within the Schwinger-Keldysh formalism \cite{kam}. They have focused on developing a Wick's theorem and consequent diagrammatics for \emph{contour-ordered} correlators \cite{kam} in general initial states \cite{leeu1,leeu2,garny,wagner,craig,hall}.

Let us now attack the second assumption made in the previous section. There, the central objects of analysis were correlators which were \emph{time-ordered}. True, there is a natural bias for time-ordered correlators in QFT, for the mere reason that experiments are bound to flow forward in time. But do \emph{general-ordered} correlators, namely \emph{Wightman functions}, have any physical significance?

Recent developments have shown that they indeed do. What makes them very interesting to study is that they seem to house physical aspects of the system which are not encoded by its time-ordered correlators. For example, in the case of a quantum Brownian particle, Wightman functions encode the noise \cite{feyn,kam,breu}. They have also been shown to be intricately related to chaos \cite{malda} and entanglement \cite{von} in quantum systems.

A possible counter-argument to studying Wightman correlators may be provided by an experimentalist. It has turned out to be quite tough to measure these correlators in laboratories. Many proposals have been suggested, but it is still unclear how such measurements can be carried out. So an experimentalist would assert that if it is so tough to measure these quantities, why study them? 

However, it is still important to explore these objects from a theoretical point of view. For any experiment which is designed to measure some physical quantity rests on solid theoretical foundations. Such a study not only suggests methods through which physical quantities may be measured directly, but also reveals other physical quantities which the object in consideration may affect. These relationships may then be used to make measurements indirectly. Take the example of entropy. It is very tough to measure it directly. But due to theoretical explorations in the form of thermodynamics, it can be related to the heat transferred to a system, which, in turn, is easily measurable.

On the other hand, there are some quantities in physics which are not directly measurable. The entropy and the quantum mechanical wavefunction are some such entities. But does this inability to measure them make them useless? Absolutely not. Introducing objects like these has been extremely important for the development of physics. Though tough to measure directly, they may eventually be connected to experiments. So may be the case with Wightman correlators.

The initial step in studies of such objects is to develop tools to calculate them. Such tools are indispensable for any theoretical endeavour which aims to explore the physics which they encode. Moreover, they also aid in designing prospective experiments to measure them. This thesis plans to develop such tools for Wightman correlators of the simple harmonic and anharmonic oscillators.
\newpage
\section{Outline of the Thesis}
We start by giving an introduction to the simple harmonic oscillator. The traditional Wick's theorem which is used in computing time ordered correlators in the vacuum state is presented in \ref{Wick}. The consequent diagrammatic formalism which emerges, namely the technique of Feynman diagrams, is presented in \ref{fd}. We then shift our attention towards developing a generalised Wick's theorem, which would aid us in calculating Wightman correlators in general states of the simple harmonic oscillator. This exploration leads us to novel objects called cumulants. They are introduced in \ref{cumu}. A generalised Wick's theorem, which is based on this idea of cumulants, is presented in \ref{genwick}. The following section \ref{cumucal} is then dedicated on how to actually compute the cumulants in a given general state of the simple harmonic oscillator. Finally, a diagrammatic formalism to compute Wightman correlators in general states of the simple harmonic oscillator is laid down in \ref{diagform}. The following chapter \ref{anharm} then delves into developing a similar diagrammatic formalism to compute Wightman correlators in general states of the anharmonic oscillator. The section \ref{feynrules} summarises the entire procedure on how one should go about calculating Wightman correlators in general states of the anharmonic oscillator diagramatically.    
\\[70pt]
\begin{center}
\scalebox{2.5}{\adforn{18}}
\end{center}
\chapter{The Simple Harmonic Oscillator}\label{free}
\section{Overview}
The quantum simple harmonic oscillator is a very effective toy model in theoretical physics. One, it is easily solvable and thus, the spectrum of its Hamiltonian is well determined. And two, the analyses for this system naturally extend to those of more realistic systems like quantum fields. Driven by this motivation, we intend to begin our endeavour with this system in mind.

What do we intend to do in this chapter?

We first introduce the simple harmonic oscillator formally. Established formalisms of Wick's theorem and Feynman diagrams for time-ordered correlators in the ground state of this system are briefly given a glimpse of. The objects we would like to analyse in this thesis, namely Wightman correlators in general states, are then brought into the picture.

We begin our investigations by making some guesses at how a similar Wick's theorem, the conventional form of which holds for time-ordered correlators in the ground state, can be developed for Wightman correlators in general states. This leads to the idea of \emph{cumulants}. It turns out that the Wick's theorem for such correlators is best stated in terms of cumulants.

After this, cumulants are explored in detail. Mathematical mechanisms are laid down which would allow the reader to calculate these cumulants for any general state of the oscillator.

A natural diagrammatic formalism to represent Wightman correlators of the oscillator in general states is seen to emerge once one introduces cumulants into the discussion. Such a formalism is developed towards the end of this chapter, followed by some examples. 
\section{The System}
The system we work on is the simple harmonic oscillator in one dimension.

This system has the Hamiltonian:
\begin{equation}
        H=\frac{p^2}{2}+\frac{1}{2}\omega^2 x^2 \qquad (m=1).
\end{equation}
We take the mass of the oscillator to be unity in all our considerations.

The spectrum of this Hamiltonian is well determined:
\begin{equation}\label{eq:6}
        H\ket{n}=\Bigl(n+\frac{1}{2}\Bigr)\hbar\omega\ket{n}\quad;\quad n=0,1,2,\dots.
\end{equation}
Thus, the states $\{\ket{n}\;:\;n=0,1,2,\dots\}$ are the energy eigenstates of this system with energy eigenvalues $\{E_n=(n+\frac{1}{2})\hbar\omega\}$.

Theoretical considerations of this system become very elegant when expressed in terms of a pair of operators called the \emph{creation} and \emph{annihilation} operators. They are denoted as $a$ and $a^{\dagger}$ respectively. Evidently, they are Hermitian conjugates of each other. Their explicit forms in terms of the position and momentum operators of the oscillator are given as:
\begin{gather}
    a=\sqrt{\frac{\omega}{2\hbar}}\Bigl(x+\frac{ip}{\omega}\Bigr),\\
    a^{\dagger}=\sqrt{\frac{\omega}{2\hbar}}\Bigl(x-\frac{ip}{\omega}\Bigr).
\end{gather}
On the other hand, the Heisenberg picture position and momentum operators of the oscillator are then given in terms of the above operators as:
\begin{gather}\label{eq:9}
        x(t_i)=\sqrt{\frac{\hbar}{2\omega}}(ae^{-i\omega t_i}+a^{\dagger}e^{i\omega t_i}),\\
        p(t_i)=-i\sqrt{\frac{\omega\hbar}{2}}(ae^{-i\omega t_i}-a^{\dagger}e^{i\omega t_i}).
\end{gather}
The action of the creation and annihilation operators on the energy eigenstates of the system are given by:
\begin{gather}
        a^{\dagger}\ket{n}=\sqrt{n+1}\ket{n+1},\\
        a\ket{n}=\sqrt{n}\ket{n-1}\quad;\quad a\ket{0}=0.
\end{gather}
They are also defined to obey the commutation relation:
\begin{gather}
    [a,a^{\dagger}]=\mathbb{I}.
\end{gather}
\section{Some Special States of the Oscillator}\label{splstates}
In this section, we introduce some special states of the oscillator, which would be referred to frequently in the following discussions. These are the vacuum state, the excited states, the coherent state and the thermal state.
\subsection{The Vacuum State}
As already introduced in \eqref{eq:6}, the states $\{\ket{n}\;:\;n=0,1,2,\dots\}$ are the energy eigenstates of the oscillator. 

Among these states, the state $\ket{0}$ is called the vacuum state of the oscillator. It is so called because it is the energy eigenstate of lowest energy.
\subsection{Excited States}
Again, as already introduced in \eqref{eq:6}, the states $\{\ket{n}\;:\;n=0,1,2,\dots\}$ are the energy eigenstates of the oscillator.

Among these, the states $\{\ket{n}\;:\;n=1,2,3,\dots\}$ are called the excited states of the oscillator. That is, all the energy eigenstates apart from the vacuum are termed excited states of the oscillator.
\subsection{Coherent States}
A coherent state $\ket{\phi}$ of a harmonic oscillator is defined to be an eigenstate of the annihilation operator:
\begin{equation}
a\ket{\phi}=\phi\ket{\phi}.
\end{equation}
To find $\ket{\phi}$, we first expand it in the $\ket{n}$ basis:
\begin{equation}
\ket{\phi}=\sum_{n=0}^{\infty}c_{n}\ket{n}.
\end{equation}
Putting this into the eigenvalue equation, and comparing the coefficients of the states on both the sides gives us:
\begin{gather}
c_{1}=\phi c_{0},\\
\sqrt{2}c_{2}=\phi c_{1}\implies c_{2}=\frac{\phi^{2}}{\sqrt{2}}c_{0},\\
\sqrt{3}c_{3}=\phi c_{2}\implies c_{3}=\frac{\phi^{3}}{\sqrt{6}}c_{0},\\
\vdots\\
c_{n}=\frac{\phi^{n}}{\sqrt{n!}}c_{0}.
\end{gather}
$c_{0}$ is determined by imposing normalisation on the states $\ket{\phi}$:
\begin{equation}
\braket{\phi|\phi}=\sum_{n=0}^{\infty}|c_n|^2=1.
\end{equation}
Putting $c_n$ in terms of $c_0$, and then assuming $c_0$ to be real for simplicity, one gets:
\begin{align}
c_0=e^{-\frac{|\phi|^2}{2}}.
\end{align}
This gives us:
\begin{align}
c_{n}&=\frac{{\phi}^{n}}{\sqrt{n!}}e^{-\frac{{|\phi|}^{2}}{2}}.
\end{align}
And thus, the coherent states are given as:
\begin{align}
    \ket{\phi}&=\sum_{n=0}^{\infty}\frac{\phi^n}{\sqrt{n!}}\;e^{-\frac{|\phi|^2}{2}}\;\ket{n}.
\end{align}
\subsection{Thermal State}
There are two types of states one talks about in the context of quantum mechanics. They are \emph{pure states} and \emph{mixed states}. A pure state is one which can be represented by a Dirac ket. As examples, the vacuum state $\ket{0}$, the excited states $\ket{n}$ and coherent states $\ket{\phi}$ are all pure states. However, another situation which may arise is one in which the system under study may occupy a set of pure states with certain probabilities. In such a case, the system is said to reside in a mixed state. Let us make this notion a bit more precise.

Suppose a system occupies the set of states $\{\ket{\psi^{(i)}}\}$, with the probability of it being in the pure state $\ket{\psi^{(i)}}$ being $p_i$. This is then an example of a mixed state. Mixed states are best represented through objects called \emph{density matrices}. The density matrix $\rho$ corresponding to the situation we have just highlighted is defined to be:
\begin{equation}
    \rho\equiv\sum_{i}p_i\ket{\psi^{(i)}}\bra{\psi^{(i)}}.
\end{equation}
Representing the state of a quantum mechanical system through density matrices is a more general approach. It clearly encompasses the possibility of the system being in a pure state. This happens when one among the probabilities $\{p_i\}$ is unity, while all the others vanish.

All physical density matrices must be normalised. A density matrix $\rho$ is said to be normalised if:
\begin{equation}
    \Tr\rho=1.
\end{equation}

The expectation value of an observable $\mathcal{O}$ in a density matrix $\rho$, represented as $\langle\mathcal{O}\rangle_{\rho}$, is given by the expression:
\begin{equation}
    \langle\mathcal{O}\rangle_{\rho}=\Tr[\rho\mathcal{O}].
\end{equation}
With this brief introduction to the concept of density matrices, we are ready to state what we mean by a thermal state of a quantum mechanical system.

A thermal state of a quantum mechanical system is a state with the density matrix:
\begin{equation}
    \rho_{\beta}=\frac{e^{-\beta H}}{Z}.
\end{equation}
Here, $H$ is the Hamiltonian of the system and $\beta$ is the system's inverse temperature. $Z$ is a quantity termed the \emph{partition function} of the system, given by the expression:
\begin{equation}
    Z=\Tr[e^{-\beta H}].
\end{equation}
The introduction of the partition function into the definition of the thermal density matrix ensures its proper normalisation.

Since we have realised that density matrices present a more general way through which we can represent the state of a quantum mechanical system, we will work with only these objects hereon.
\section{Time-Ordered Correlators of the Oscillator}
Time-ordered correlators of the simple harmonic oscillator are objects of the form:
\begin{equation}\label{eq:7}
    \langle Tx(t_1)x(t_2)\dots x(t_n)\rangle_{\rho}=\Tr[\rho Tx(t_1)x(t_2)\dots x(t_n)],
\end{equation}
where $\rho$ is a general density matrix of the oscillator and $x(t_i)$ is the Heisenberg-picture position operator of the oscillator at the time $t_i$. $T$ is the time-ordering symbol, which orders the string of Heisenberg operators following it in such a way that the later time operators are placed lefter.

A certain subset of the objects \eqref{eq:7} have been extensively analysed in the literature. These are time-ordered correlators \emph{in the vacuum state}, that is:
\begin{equation}\label{eq:8}
    \bra{0}Tx(t_1)x(t_2)\dots x(t_n)\ket{0}.
\end{equation}
Objects of the type \eqref{eq:8} are easily computed using a result called Wick's theorem \cite{wck}. This result naturally leads to a diagrammatic formalism for computing such correlators; the language of Feynman diagrams. Both these aspects are briefly reviewed in the following sub-sections.
\subsection{Wick's Theorem}\label{Wick}
Wick's theorem allows one to express a time-ordered string of Heisenberg operators in terms of \emph{normal-ordered} strings.

A string of creation and annihilation operators is said to be normal-ordered if all the creation operators are placed to the left of all the annihilation operators.

The formal statement of Wick's theorem is:
\begin{equation}\label{eq:10}
    Tx(t_1)x(t_2)\dots x(t_n)=N\bigl\{x(t_1)x(t_2)\dots x(t_n)+\text{all possible contractions}\bigr\}.
\end{equation}
Here, $N$ stands for the normal-ordering symbol. It normal-orders the string of operators which follow it. Note that it is possible to normal-order a string of position operators because each position operator can be expressed in terms of the creation and annihilation operators owing to \eqref{eq:9}.

A \emph{contraction} is defined as:
\begin{equation}
    \contraction{}{x}{(t_i)}{x}x(t_i)x(t_j)\equiv\bra{0}Tx(t_i)x(t_j)\ket{0}.
\end{equation}
So, for example, \eqref{eq:10} dictates:
\begin{align}
    Tx(t_1)x(t_2)x(t_3)x(t_4)&=N\bigl\{x(t_1)x(t_2)x(t_3)x(t_4)+\contraction{}{x}{(t_1)}{x}x(t_1)x(t_2)x(t_3)x(t_4)\notag\\&\qquad+\contraction{}{x}{(t_1)x(t_2)}{x}x(t_1)x(t_2)x(t_3)x(t_4)+\contraction{}{x}{(t_1)x(t_2)x(t_3)}{x}x(t_1)x(t_2)x(t_3)x(t_4)\notag\\&\qquad+\contraction{x(t_1)}{x}{(t_2)}{x}x(t_1)x(t_2)x(t_3)x(t_4)+\contraction{x(t_1)}{x}{(t_2)x(t_3)}{x}x(t_1)x(t_2)x(t_3)x(t_4)\notag\\&\qquad+\contraction{x(t_1)x(t_2)}{x}{(t_3)}{x}x(t_1)x(t_2)x(t_3)x(t_4)+\contraction{}{x}{(t_1)}{x}x(t_1)x(t_2)\contraction{}{x}{(t_3)}{x}x(t_3)x(t_4)\notag\\&\qquad+\contraction{}{x}{(t_1)x(t_2)}{x}\contraction[1.5ex]{x(t_1)}{x}{(t_2)x(t_3)}{x}x(t_1)x(t_2)x(t_3)x(t_4)+\contraction{}{x}{(t_1)x(t_2)x(t_3)}{x}\contraction[1.5ex]{x(t_1)}{x}{(t_2)}{x}x(t_1)x(t_2)x(t_3)x(t_4)\bigr\}.
\end{align}
Since contractions are scalars, the normal-ordering instruction is blind to them, leading to :
\begin{align}\label{norm}
    Tx(t_1)x(t_2)x(t_3)x(t_4)&=N\bigl\{x(t_1)x(t_2)x(t_3)x(t_4)\bigr\}+\contraction{}{x}{(t_1)}{x}x(t_1)x(t_2)N\bigl\{x(t_3)x(t_4)\bigr\}\notag\\&\qquad+\contraction{}{x}{(t_1)}{x}x(t_1)x(t_3)N\bigl\{x(t_2)x(t_4)\bigr\}+\contraction{}{x}{(t_1)}{x}x(t_1)x(t_4)N\bigl\{x(t_2)x(t_3)\bigr\}\notag\\&\qquad+\contraction{}{x}{(t_2)}{x}x(t_2)x(t_3)N\bigl\{x(t_1)x(t_4)\bigr\}+\contraction{}{x}{(t_2)}{x}x(t_2)x(t_4)N\bigl\{x(t_1)x(t_3)\bigr\}\notag\\&\qquad+\contraction{}{x}{(t_3)}{x}x(t_3)x(t_4)N\bigl\{x(t_1)x(t_2)\bigr\}+\contraction{}{x}{(t_1)}{x}x(t_1)x(t_2)\contraction{}{x}{(t_3)}{x}x(t_3)x(t_4)\notag\\&\qquad+\contraction{}{x}{(t_1)x(t_2)}{x}\contraction[1.5ex]{x(t_1)}{x}{(t_2)x(t_3)}{x}x(t_1)x(t_2)x(t_3)x(t_4)+\contraction{}{x}{(t_1)x(t_2)x(t_3)}{x}\contraction[1.5ex]{x(t_1)}{x}{(t_2)}{x}x(t_1)x(t_2)x(t_3)x(t_4).
\end{align}
Let us now take the expectation of \eqref{norm} over the vacuum state. Again, since contractions are scalars, they come out as plain numbers when we take their expectations over states.

Since $a\ket{0}=0=\bra{0}a^{\dagger}$, \emph{expectations of normal-ordered strings in the vacuum state vanish}. This leads to:
\begin{align}
   \bra{0}Tx(t_1)x(t_2)x(t_3)x(t_4)\ket{0}&=\contraction{}{x}{(t_1)}{x}x(t_1)x(t_2)\contraction{}{x}{(t_3)}{x}x(t_3)x(t_4)+\contraction{}{x}{(t_1)x(t_2)}{x}\contraction[1.5ex]{x(t_1)}{x}{(t_2)x(t_3)}{x}x(t_1)x(t_2)x(t_3)x(t_4)\notag\\&\qquad+\contraction{}{x}{(t_1)x(t_2)x(t_3)}{x}\contraction[1.5ex]{x(t_1)}{x}{(t_2)}{x}x(t_1)x(t_2)x(t_3)x(t_4),
\end{align}
which can be generalised as:
\begin{equation}\label{eq:11}
    \bra{0}Tx(t_1)x(t_2)\dots x(t_n)\ket{0}=\sum\Bigl\{\text{All possible \emph{full} contractions}\Bigr\},
\end{equation}
where a \emph{full} contraction stands for a term in which every position operator is part of a contraction.

A direct result which follows from \ref{eq:11} is that:
\begin{equation}
    \bra{0}T\Bigl\{\text{Odd Number of Position Operators}\Bigr\}\ket{0}=0,
\end{equation}
since there are no possible full contractions possible in this case.
\subsection{Feynman Diagrams}\label{fd}
The relation \eqref{eq:11} motivates a natural diagrammatic formalism to compute time-ordered correlators of the oscillator in the vacuum state.

The starting point of such a formalism is the association:
\begin{figure}[h]
    \centering
  \includegraphics[scale=0.12]{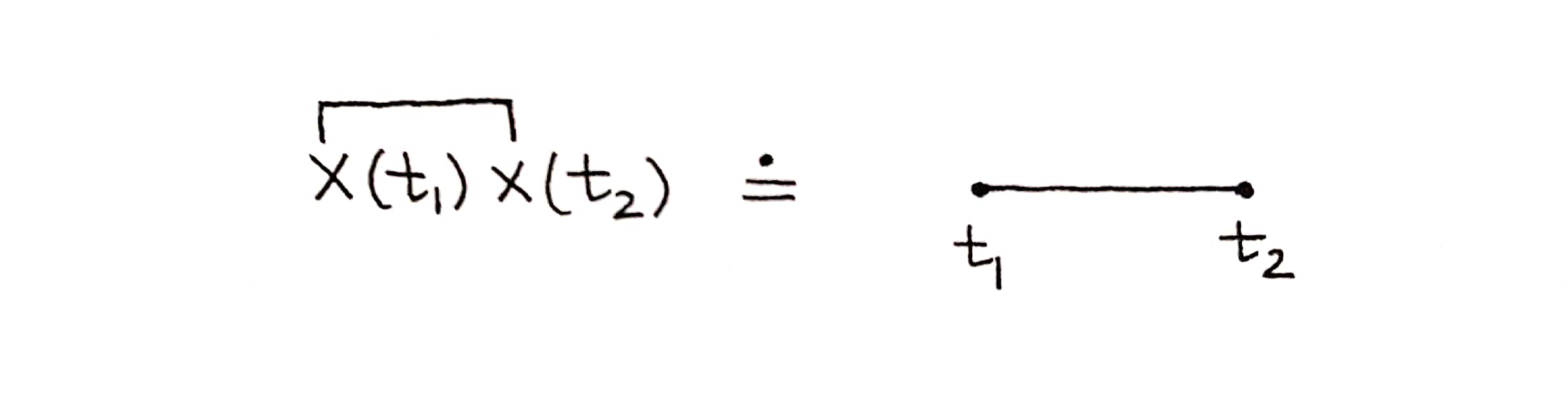}
  \caption{Diagrammatic representation of a Wick contraction.}
\end{figure}
\\
This quantity is termed the \emph{Feynman propagator}, and is denoted $D_F(t_1-t_2)$:
\begin{figure}[h]
    \centering
  \includegraphics[scale=0.1]{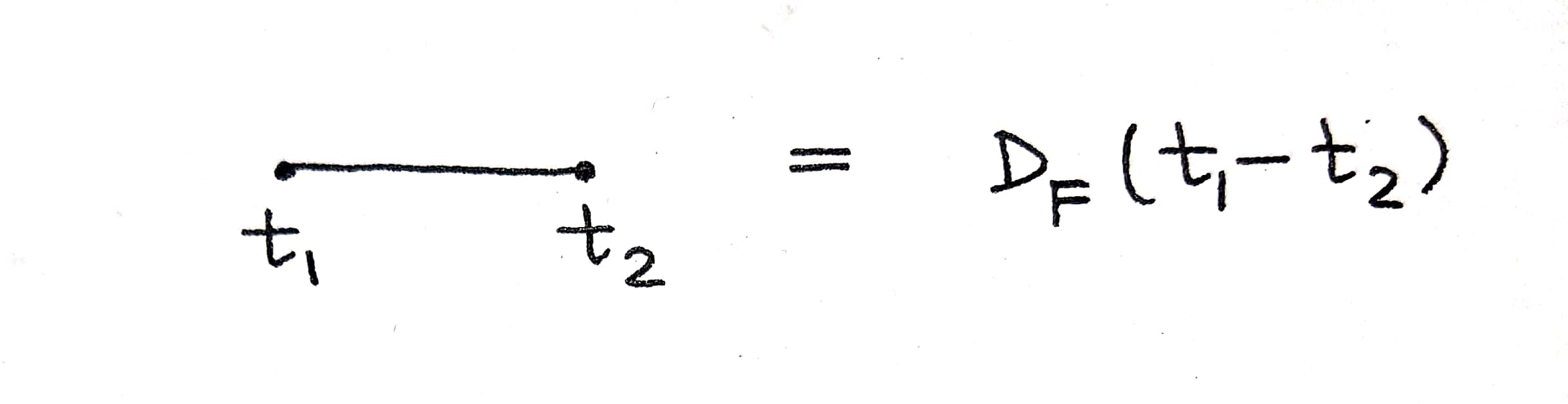}
  \caption{Diagrammatic Representation for the Feynman Propagator.}
\end{figure}
\\
\\
The Feynman propagator can be explicitly computed to be:
\begin{equation}
    D_F(t_1-t_2)=\lim_{\epsilon\to 0}\int_{-\infty}^{\infty}\frac{dE}{2\pi}\frac{ie^{-iE(t_1-t_2)}}{E^2-\omega^2+i\epsilon}.
\end{equation}
Diagrammatically, the result \eqref{eq:11} translates to:
\begin{equation}
    \bra{0}Tx(t_1)x(t_2)\dots x(t_n)\ket{0}=\nsum\Bigl\{\parbox{25em}{All possible diagrams in which each point is connected exactly to one other point through a Feynman propagator}\Bigr\}.
\end{equation}
For example:
\begin{figure}[h]
    \centering
  \includegraphics[scale=0.19]{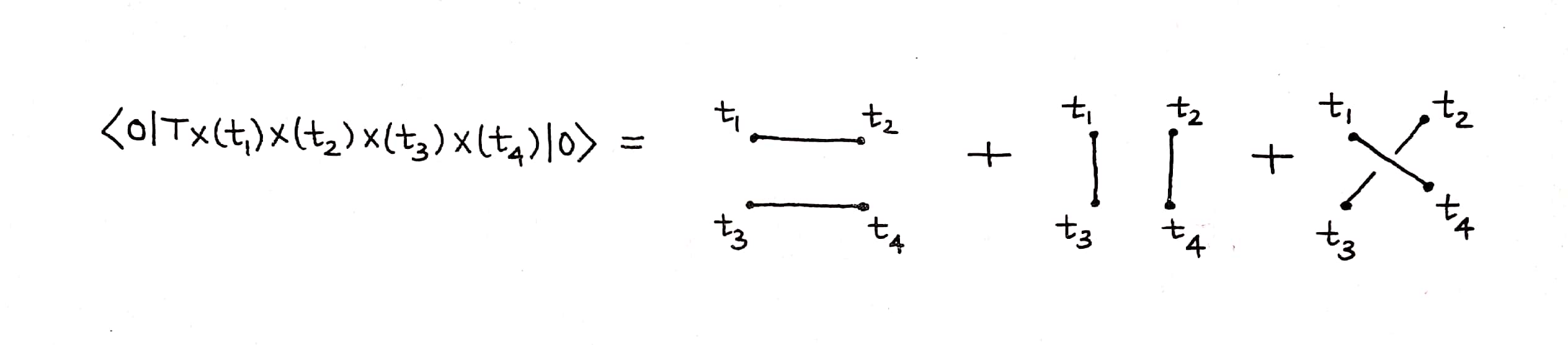}
  \caption{Diagrammatic computation of the 4-point function.}
\end{figure}
\\
\\
This diagrammatic formalism can be extended to weakly interacting systems like the anharmonic oscillator too. A perturbative analysis of such a system brings in another component to the diagrams, namely \emph{vertices}. For more details on this approach, one can refer to \cite{abb}.
\section{Towards a Generalised Wick's Theorem}
We now introduce the central objects of our following analyses. We would be focusing on \emph{Wightman correlators in general states}, objects which take the form:
\begin{equation}\label{eq:12}
    \langle x(t_1)x(t_2)\dots x(t_n)\rangle_{\rho}=\Tr[\rho x(t_1)x(t_2)\dots x(t_n)],
\end{equation}
where $\rho$ denotes a general density matrix of the oscillator. From now on, we will drop the $\rho$ from the subscript of the angular brackets in \eqref{eq:12} to make the notation less cumbersome:
\begin{equation}
    \langle\dots\rangle\equiv\langle\dots\rangle_{\rho}.
\end{equation}
Angular brackets without any subscript are to be understood as expectations taken over a general density matrix.   

Motivated by the Wick's Theorem for time-ordered correlators in the vacuum state, which was presented in \ref{Wick}, we aim to come up with a generalised Wick's theorem which would hold for the objects \eqref{eq:12}.

In this direction, to begin with, we make some naive guesses as to what form this generalised Wick theorem may take. This process eventually leads us to the idea of \emph{cumulants}, and it turns out that the generalised Wick's theorem is best stated in terms of cumulants.

Let us first elaborate on the guesses which can be made about the form of the generalised Wick's theorem.
\subsection{First Guess}\label{fg}
As our first guess, let us naively impose \eqref{eq:11} itself as being our generalised Wick's theorem. That is:
\begin{equation}\label{eq:14}
    \langle x(t_1)x(t_2)\dots x(t_n)\rangle\?\sum\Bigl\{\text{All possible \emph{full} contractions}\Bigr\}.
\end{equation}
The analytic expression corresponding to the contractions defined here may be different from those of the contractions defined in \ref{Wick}, but this is a detail which should be focused on after \eqref{eq:14} is found out to be correct, if at all.

So is \eqref{eq:14} correct in the first place? 

A simple argument shows that it is not. For if it is, then all \emph{odd-point} Wightman correlators in a general state must vanish. But this is not the case! Consider the simplest example, say that of the one-point function in a coherent state. It evaluates to be:
\begin{equation}
    \bra{\phi}x(t)\ket{\phi}=\sqrt{\frac{\hbar}{2\omega}}\;(\phi e^{-i\omega t}+\phi^{*}e^{i\omega t}),
\end{equation}
which, in general, does not vanish. Thus, our guess \eqref{eq:14} is wrong.
\subsection{Second Guess}
The lesson to be learnt from the first guess \ref{fg} is that odd-point correlations are, in general, non-zero and thus, must be carefully considered. A more educated guess can now be made about the generalised Wick's theorem, which takes into account the point made above.

We now guess that \emph{the Wightman correlator $\langle x(t_1)x(t_2)\dots x(t_n)\rangle$ is a sum of products of all its constituting lower-point Wightman correlators}.

For example, we would guess:
\begin{align}
    \langle x(t_1)x(t_2)x(t_3)\rangle&\?\langle x(t_1)\rangle\cdot\langle x(t_2)\rangle\cdot\langle x(t_3)\rangle\notag\\&\quad+\langle x(t_1)\rangle\cdot\langle x(t_2)x(t_3)\rangle\notag\\&\quad+\langle x(t_2)\rangle\cdot\langle x(t_1)x(t_3)\rangle\notag\\&\quad+\langle x(t_3)\rangle\cdot\langle x(t_1)x(t_2)\rangle.
\end{align}
But this guess too has a problem. To understand this shortcoming, we would first need to acknowledge a result which we now state.

The result is that Wick's theorem is found to hold true in its traditional form for Wightman correlators in the vacuum state and in the thermal state. That is, for example:
\begin{align}\label{eq:15}
    \bra{0}x(t_1)x(t_2)x(t_3)x(t_4)\ket{0}&=\bra{0}x(t_1)x(t_2)\ket{0}\cdot\bra{0}x(t_3)x(t_4)\ket{0}\notag\\&\quad+\bra{0}x(t_1)x(t_3)\ket{0}\cdot\bra{0}x(t_2)x(t_4)\ket{0}\notag\\&\quad+\bra{0}x(t_1)x(t_4)\ket{0}\cdot\bra{0}x(t_2)x(t_3)\ket{0},
\end{align}
and similarly:
\begin{align}\label{eq:15a}
    \langle x(t_1)x(t_2)x(t_3)x(t_4)\rangle_{\beta}&=\langle x(t_1)x(t_2)\rangle_{\beta}\cdot\langle x(t_3)x(t_4)\rangle_{\beta}\notag\\&\quad+\langle x(t_1)x(t_3)\rangle_{\beta}\cdot\langle x(t_2)x(t_4)\rangle_{\beta}\notag\\&\quad+\langle x(t_1)x(t_4)\rangle_{\beta}\cdot\langle x(t_2)x(t_3)\rangle_{\beta},
\end{align}
where $\langle\dots\rangle_{\beta}$ denotes an expectation value in a thermal state with inverse temperature $\beta$.

In the light of this result, let us reconsider this second guess of ours for the case of a 6-point Wightman correlator in a general density matrix.
\iffalse\\[25pt]
\begin{center}
\textcolor{blue}{\textbf{(Continued on next page)}}
\end{center}
\newpage\fi
Going by our guess:
\begin{align}\label{eq:16}
    \langle x(t_1)x(t_2)x(t_3)x(t_4)x(t_5)x(t_6)\rangle&\?\bigl\{\langle x(t_1)x(t_2)\rangle\cdot\langle x(t_3)x(t_4)\rangle\notag\\&\qquad\cdot\langle x(t_5)x(t_6)\rangle+\parbox{6em}{Permutations}\bigr\}\notag\\[8pt]&\quad\;+\bigl\{\langle x(t_1)x(t_2)\rangle\cdot\langle x(t_3)x(t_4)x(t_5)x(t_6)\rangle\notag\\&\qquad\quad+\parbox{6em}{Permutations}\bigr\}\notag\\&\hspace{100pt}\vdots\notag\\[8pt]&\quad\;+\langle x(t_1)\rangle\cdot\langle x(t_2)\rangle\cdot\langle x(t_3)\rangle\cdot\langle x(t_4)\rangle\cdot\langle x(t_5)\rangle\cdot\langle x(t_6)\rangle.
\end{align}
Let the general state being considered now be the vacuum state $\ket{0}$. 

For this choice of the state, only the first two sets of terms listed on the RHS of \eqref{eq:16} survive, since odd-point position correlations vanish in the vacuum state. 

But due to the fact that Wick's theorem holds in its traditional form for Wightman correlators in the vacuum, the first set of terms on the RHS of \eqref{eq:16} are enough to meet the equality. The second set of terms violate it. Thus, our second guess is also not true.
\subsection{The Idea of Cumulants}\label{cumu}
Let us summarise the lessons we have learnt from our two guesses about the form of a generalised Wick's theorem.

One, that odd-point Wightman correlators must be considered seriously, since they \emph{do not} vanish in a general state. 

Two, that simply decomposing a Wightman correlator in a general state as a sum of products of its constituent lower-point Wightman correlators does not work.

It turns out that \emph{cumulants} are the best objects to introduce at this juncture. Not only do they counter the problems faced by our two guesses, but also take on nice values in the states where the traditional Wick's theorem for Wightman correlators holds!

The notion of cumulants has indeed been explored in previous investigations of contour-ordered correlators in general states within the Schwinger-Keldysh formalism \cite{fauser1,fauser2} and time-ordered correlators in general states \cite{brouder}.

Within the purview of our present analysis of Wightman correlators in general states, how are these objects defined? In a general state, the \emph{first cumulant}, denoted $C_1$, is defined as:
\begin{equation}\label{eq:17}
    C_1(t_1)\equiv\langle x(t_1)\rangle.
\end{equation}
As an example, consider the general state being considered to be a coherent state $\ket{\phi}$. In this case, the first cumulant $C_1$ turns out to be:
\begin{align}\label{coh1}
    C^{\phi}_1(t_1)&=\bra{\phi}x(t_1)\ket{\phi}\notag\\
    &=\sqrt{\frac{1}{2\omega}}\;(\phi e^{-i\omega t_1}+\phi^{*}e^{i\omega t_1}).
\end{align}
The \emph{second cumulant}, denoted $C_2$, is defined as:
\begin{align}\label{eq:18}
    C_2(t_1,t_2)\equiv&\langle x(t_1)x(t_2)\rangle\notag\\&\;-C_1(t_1)\cdot C_1(t_2).
\end{align}
\newpage
Again, for the case of the general state being a coherent state $\ket{\phi}$, the second cumulant $C_2$ explicitly turns out to be:
\begin{align}\label{coh2}
C^{\phi}_2(t_1,t_2)&=\bra{\phi}x(t_1)x(t_2)\ket{\phi}\notag\\
&\quad -\bra{\phi}x(t_1)\ket{\phi}\cdot\bra{\phi}x(t_2)\ket{\phi}\notag\\
&=\frac{\hbar}{2\omega}e^{-i\omega(t_1-t_2)}.
\end{align}
Similarly, the \emph{third cumulant}, denoted $C_3$, is defined as:
\begin{align}\label{eq:19}
    C_3(t_1,t_2,t_3)\equiv & \langle x(t_1)x(t_2)x(t_3)\rangle\notag\\&\;-C_1(t_1)\cdot C_2(t_2,t_3)-C_1(t_2)\cdot C_2(t_1,t_3)\notag\\&\;-C_1(t_3)\cdot C_2(t_1,t_2)\notag\\&\;-C_1(t_1)\cdot C_1(t_2)\cdot C_1(t_3),
\end{align}
and so on. For the sake of completeness, let us also evaluate the third cumulant $C_3$ for the case of the general state being a coherent state. It turns out that:
\begin{align}
    C^{\phi}_3(t_1,t_2,t_3)&= \bra{\phi}x(t_1)x(t_2)x(t_3)\ket{\phi}\notag\\&\quad-C^{\phi}_1(t_1)\cdot C^{\phi}_2(t_2,t_3)-C^{\phi}_1(t_2)\cdot C^{\phi}_2(t_1,t_3)\notag\\&\quad-C^{\phi}_1(t_3)\cdot C^{\phi}_2(t_1,t_2)\notag\\&\quad-C^{\phi}_1(t_1)\cdot C^{\phi}_1(t_2)\cdot C^{\phi}_1(t_3)\notag\\
    &=0,
\end{align}
after putting in the values of $C^{\phi}_1$ and $C^{\phi}_2$ as evaluated in \eqref{coh1} and \eqref{coh2}.

How do cumulants fill up the gaps that were noticed in our initial guesses?

Firstly, it is evident that the cumulants take odd-point correlations into account. This is clear, say, from the expression for the first cumulant as given in \eqref{eq:17}.

To appreciate how cumulants rectify the shortcoming with our second guess, let us invert, for example, the relation \eqref{eq:19}:
\begin{align}\label{eq:20}
    \langle x(t_1)x(t_2)x(t_3)\rangle\equiv &\;\;\; C_3(t_1,t_2,t_3)\notag\\&\;+C_1(t_1)\cdot C_2(t_2,t_3)+C_1(t_2)\cdot C_2(t_1,t_3)\notag\\&\;+C_1(t_3)\cdot C_2(t_1,t_2)\notag\\&\;+C_1(t_1)\cdot C_1(t_2)\cdot C_1(t_3).
\end{align}
It is clear from this that a n-point Wightman correlator in a general state is now not expressed as a sum of products of its constituent lower-point Wightman correlators, but instead as a sum of products of $n^{th}$ and lower \emph{cumulants}. In fact, \eqref{eq:20} is precisely an example of the generalised Wick's theorem at work. We now state this theorem formally.
\subsection{The Generalised Wick's Theorem}\label{genwick}
The generalised Wick's theorem is the following simple statement:
\\[13pt]
\fcolorbox{black}{blue!10}{\parbox{35em}{The $n$-point Wightman correlator in a general state is the sum of products of the cumulants $\{C_i,C_j,\dots,C_k\;:i+j+\dots+k=n\:\text{and}\:i,j,\dots,k\geq 0\}$ in that state.}}
\\[13pt]
For example, owing to this theorem, the 3-point Wightman correlator in a general state reads:
\begin{align}
    \langle x(t_1)x(t_2)x(t_3)\rangle\equiv &\;\;\; C_3(t_1,t_2,t_3)\notag\\&\;+C_1(t_1)\cdot C_2(t_2,t_3)+C_1(t_2)\cdot C_2(t_1,t_3)\notag\\&\;+C_1(t_3)\cdot C_2(t_1,t_2)\notag\\&\;+C_1(t_1)\cdot C_1(t_2)\cdot C_1(t_3).
\end{align}
As another example, this theorem dictates that the 4-point Wightman correlator in a general state would be:
\begin{align}
    \langle x(t_1)x(t_2)x(t_3)x(t_4)\rangle\equiv &\;\;\; C_4(t_1,t_2,t_3,t_4)\notag\\&\;+C_1(t_1)\cdot C_3(t_2,t_3,t_4)+C_1(t_2)\cdot C_3(t_1,t_3,t_4)\notag\\&\;+C_1(t_3)\cdot C_3(t_1,t_2,t_4)+C_1(t_4)\cdot C_3(t_1,t_2,t_3)\notag\\&\;
    +C_1(t_1)\cdot C_1(t_2)\cdot C_2(t_3,t_4)+C_1(t_1)\cdot C_1(t_3)\cdot C_2(t_2,t_4)\notag\\&\;+C_1(t_1)\cdot C_1(t_4)\cdot C_2(t_2,t_3)+C_1(t_2)\cdot C_1(t_3)\cdot C_2(t_1,t_4)\notag\\&\;+C_1(t_2)\cdot C_1(t_4)\cdot C_2(t_1,t_3)+C_1(t_3)\cdot C_1(t_4)\cdot C_2(t_1,t_2)\notag\\&\;+C_2(t_1,t_2)\cdot C_2(t_3,t_4)+C_2(t_1,t_3)\cdot C_2(t_2,t_4)+C_2(t_1,t_4)\cdot C_2(t_2,t_3)\notag\\&\;+C_1(t_1)\cdot C_1(t_2)\cdot C_1(t_3)\cdot C_1(t_4).
\end{align}
It is also easy to see that the traditional Wick's theorem holding for the vacuum and thermal states implies that the third and higher cumulants in these states vanish.
\subsection{Cumulants and Connected Wightman Correlators}
The equation \eqref{eq:20} brings to light an important aspect. Looking at it, one realises that the cumulants in a given general state are nothing but \emph{connected} Wightman correlators in the same state. To be precise, the $n^{th}$ cumulant $C_n$ in a general state is the connected n-point Wightman correlator in that state.

To appreciate this association, we must, for now, divert our attention to another physical setup. Consider analysing time-ordered correlators of the anharmonic oscillator in the interacting vacuum state. The anharmonic oscillator being talked about is assumed to have the Hamiltonian:
\begin{equation}
    H=\frac{p^2}{2}+\frac{1}{2}\omega^2 x^2+\frac{gx^3}{3!}+\frac{\lambda x^4}{4!} \qquad (m=1).
\end{equation}
Let us denote the \emph{full} n-point correlator of this system in the interacting vacuum state:
\begin{equation}
    \bra{\Omega}Tx(t_1)x(t_2)\dots x(t_n)\ket{\Omega},
\end{equation}
as $G^{(n)}(t_1,t_2,\dots,t_n)$. Moreover, we would write the \emph{connected} n-point correlator as $G_c^{(n)}(t_1,t_2,\dots,t_n)$. After a traditional generating functional analysis, one gets the result:
\begin{align}
    G^{(3)}(t_1,t_2,t_3)&=G_c^{(3)}(t_1,t_2,t_3)\notag\\&\quad+G_c^{(1)}(t_1)\cdot G_c^{(2)}(t_2,t_3)+G_c^{(1)}(t_2)\cdot G_c^{(2)}(t_1,t_3)\notag\\&\quad+G_c^{(1)}(t_3)\cdot G_c^{(2)}(t_1,t_2)\notag\\&\quad+G_c^{(1)}(t_1)\cdot G_c^{(1)}(t_2)\cdot G_c^{(1)}(t_3).
\end{align}
This equation matches exactly in structure with \eqref{eq:20}! This similarity prompts us to make the association that the cumulants in our consideration are, in fact, connected Wightman correlators.
\section{Cumulants In General States}\label{cumucal}
We arrived at a generalised Wick's theorem in the previous section. The key objects in this theorem turned out to be cumulants. In this section, we shift our focus onto how one can actually calculate these cumulants in a general state of the oscillator.
\subsection{The Coefficients $\{\xi\}$}
We start our analysis in this direction by defining a set of coefficients for general states of the free oscillator, which we denote as $\{\xi\}$. For a general state of the free oscillator, the coefficients $\{\xi\}$ associated to it are defined as:
\begin{equation}\label{defxi}
    \xi_{mn}\equiv\langle(a^{\dagger})^{m}a^{n}\rangle.
\end{equation}
As an example, the $\{\xi\}$ for a coherent state $\ket{\phi}$ of the oscillator read:
\begin{align}
    \xi^{\phi}_{mn}&=\bra{\phi}(a^{\dagger})^{m}a^{n}\ket{\phi}\notag\\
    &=(\phi^*)^{m}\phi^{n}.
\end{align}
As outlined in one of the appendices \ref{appA}, it is possible to express Wightman correlators in general states of the free oscillator in terms of the coefficients $\{\xi\}$ associated to the state. To be specific, a Wightman correlator in a general state of the free oscillator is a linear combination of certain functions of time. Each function of time in this linear combination is weighted by the coefficients $\{\xi\}$ corresponding to the state.

For example, the 3-point Wightman correlator in a general state of the free oscillator reads:
\begin{align}
    \langle x(t_1)x(t_2)x(t_3)\rangle&=\xi_{03}f_{+++}+\xi_{30}f_{---}\notag\\
    &\quad+\xi_{12}(f_{+-+}+f_{++-}+f_{-++})\notag\\&\quad+\xi_{21}(f_{--+}+f_{-+-}+f_{+--})\notag\\ 
    &\quad+\xi_{01}(f_{+-+}+2f_{++-})+\xi_{10}(f_{-+-}+2f_{+--}),
\end{align}
where:
\begin{equation}
    f_{\sigma_1\sigma_2\sigma_3}=\Bigl(\frac{\hbar}{2\omega}\Bigr)^{3/2}\exp{\bigl(-i\omega\;[\sigma_1t_1+\sigma_2t_2+\sigma_3t_3]\;\bigr)}.
\end{equation}
Owing to the definition of the cumulants as laid down in \ref{cumu}, it is sufficient to know the Wightman correlators in a general state to compute the cumulants in that state. 

For example, as listed in \ref{cumu}, the third cumulant $C_3$ has the definition:
\begin{align}
    C_3(t_1,t_2,t_3)\equiv & \langle x(t_1)x(t_2)x(t_3)\rangle\notag\\&\;-C_1(t_1)\cdot C_2(t_2,t_3)-C_1(t_2)\cdot C_2(t_1,t_3)\notag\\&\;-C_1(t_3)\cdot C_2(t_1,t_2)\notag\\&\;-C_1(t_1)\cdot C_1(t_2)\cdot C_1(t_3).
\end{align}
Now since the first cumulant $C_1$ and the second cumulant $C_2$ involved in this definition are themselves obtainable in terms of Wightman correlators (\ref{cumu}), the whole of $C_3$ can be expressed in terms of Wightman correlators.

As we have just seen, Wightman correlators in general states of the free oscillator can be written as expressions involving the coefficients $\{\xi\}$ of the state. Thus, the cumulants in a general state too can be written in terms of the coefficients $\{\xi\}$ corresponding to that state.

For example, the third cumulant $C_3$ in a general state, when written in terms of the $\{\xi\}$ associated to that state, adopts the form:
\begin{align}\label{cu3}
    C_3&=[2 \xi_{01}^3-3 \xi_{02} \xi_{01}+\xi_{03}]f_{+++}+[2 \xi _{10}^3-3 \xi _{20} \xi _{10}+\xi _{30}]f_{---}\notag\\&\quad+[2 \xi _{10} \xi_{01}^2-2 \xi _{11} \xi_{01}-\xi_{02} \xi _{10}+\xi _{12}](f_{-++}+f_{++-}+f_{+-+})\notag\\&\quad+[2 \xi_{01} \xi _{10}^2-2 \xi _{11} \xi _{10}-\xi_{01} \xi _{20}+\xi _{21}](f_{--+}+f_{-+-}+f_{+--}).
\end{align}
After expressing the cumulants in a general state in terms of the coefficients $\{\xi\}$ associated to that state, we are ready to introduce a new set of coefficients associated to the state. These new coefficients are termed the $\{\chi\}$, and they are explained in detail through the following section.
\newpage
\subsection{The Coefficients $\{\chi\}$}
Before introducing the coefficients $\{\chi\}$, it proves to be profitable to first introduce a compact notation for the functions of time which appear in the expressions for the cumulants in a general state.

Analogous to the functions of time $f_{\sigma_1\sigma_2\sigma_3}$ which make up the third cumulant $C_3$ \eqref{c3}, the $n^{th}$ cumulant $C_n$ is observed to be made up of the functions $f_{\sigma_1\sigma_2\dots\sigma_n}$, where:
\begin{equation}
    f_{\sigma_1\sigma_2\dots\sigma_n}=\Bigl(\frac{\hbar}{2\omega}\Bigr)^{n/2}\exp{\bigl(-i\omega\;[\sigma_1t_1+\sigma_2t_2+\dots+\sigma_nt_n]\;\bigr)}.
\end{equation}

With this context, we define a new function $F_{nk}(t_1,t_2,\dots,t_n)$ as:
\begin{equation}\label{fnk}
    F_{nk}\equiv\sum_{\pi_k}f_{\sigma_1\sigma_2\dots\sigma_n},
\end{equation}
where $\pi_k$ denotes a permutation of the list $(\sigma_1\sigma_2\dots\sigma_n)$ such that $k$ of them are (+) and the remaining $n-k$ are (-). The definition \eqref{fnk} then directs us to sum over all the $n\choose k$ such possible permutations.

As examples:
\begin{gather}
    F_{31}=f_{+--}+f_{-+-}+f_{--+},\notag\\
    F_{42}=f_{++--}+f_{+-+-}+f_{+--+}+f_{-++-}+f_{-+-+}+f_{--++}.
\end{gather}
We are now in a position to introduce the coefficients $\{\chi\}$. 

Supported by explicit computations carried out in the appendix \ref{appA}, the coefficients $\{\chi\}$ for a general state of the free oscillator are defined through the equation:
\begin{equation}\label{defchi}
    C_{n}\equiv\sum_{m=0}^{n}\chi_{m(n-m)}F_{n(n-m)}+\delta_{n,2}f_{+-}.
\end{equation}
where $C_n$ stands for the $n^{th}$ cumulant in the general state being considered.

For example, \eqref{defchi} dictates:
\begin{align}
    C_{3}&=\sum_{m=0}^{3}\chi_{m(n-m)}F_{n(n-m)}\notag\\
    &=\chi_{03}F_{33}+\chi_{12}F_{32}+\chi_{21}F_{31}+\chi_{30}F_{30}.
\end{align}
Rather than analyse cumulants in their entirety, it proves to be more useful to study the coefficients $\{\chi\}$ instead. There are two reasons for this.

One, the $\{\chi\}$ are very \emph{nice} objects. We say this because they attain very simple numerical values in quite a few states of the free oscillator. For example, in the vacuum state, all the $\{\chi\}$ apart from $\chi_{00}$ vanish. In the thermal state, all the $\{\chi\}$ apart from $\chi_{00}$ and $\chi_{11}$ vanish. As another example, in coherent states, all the $\{\chi\}$ apart from $\chi_{00}$,\hspace{3pt}$\chi_{01}$ and $\chi_{10}$ vanish. This is explicitly shown in the second appendix \ref{appB}.

Two, it turns out that the $\{\chi\}$ occupy a very crucial role in the diagrammatic representations of Wightman correlators in general states. This is not very surprising once one realises that the cumulants are nothing but \emph{connected} Wightman correlators. And since developing such a diagrammatic formalism is one of our aims, exploring these objects in detail proves to be very inviting.

\subsection{Relations Between the $\{\xi\}$ and $\{\chi\}$}
Having motivated the importance of the coefficients $\{\chi\}$, our next task is to figure out how to explicitly compute them for a general state of the free oscillator. A consideration which proves to be fruitful in this pursuit is to express the coefficients $\{\chi\}$ in terms of the coefficients $\{\xi\}$.

The basic procedure which would be followed in doing so would be the following. We would first compute the $n^{th}$ cumulant $C_n$ in a general state of the oscillator in terms of the coefficients $\{\xi\}$. The fact that one can do so has already been illustrated through the example \eqref{cu3}. Once this is done, we would compare the expression of the cumulant we thus get with its associated expression arising from the definition \eqref{defchi}. Doing this yields us mathematical expressions for the coefficients $\{\chi\}$ in terms of the coefficients $\{\xi\}$.

Let us illustrate this procedure through an example.

As already pointed out through \eqref{cu3}, the third cumulant $C_3$ stated in terms of the coefficients $\{\xi\}$ reads:
\begin{align}\label{c31}
    C_3&=[2 \xi_{01}^3-3 \xi_{02} \xi_{01}+\xi_{03}]f_{+++}+[2 \xi _{10}^3-3 \xi _{20} \xi _{10}+\xi _{30}]f_{---}\notag\\&\quad+[2 \xi _{10} \xi_{01}^2-2 \xi _{11} \xi_{01}-\xi_{02} \xi _{10}+\xi _{12}](f_{-++}+f_{++-}+f_{+-+})\notag\\&\quad+[2 \xi_{01} \xi _{10}^2-2 \xi _{11} \xi _{10}-\xi_{01} \xi _{20}+\xi _{21}](f_{--+}+f_{-+-}+f_{+--}).
\end{align}
Using the new functions $F_{nk}$ defined through \eqref{fnk}, the above equation becomes:
\begin{align}\label{c32}
    C_3&=[2 \xi_{01}^3-3 \xi_{02} \xi_{01}+\xi_{03}]F_{33}+[2 \xi _{10}^3-3 \xi _{20} \xi _{10}+\xi _{30}]F_{30}\notag\\&\quad+[2 \xi _{10} \xi_{01}^2-2 \xi _{11} \xi_{01}-\xi_{02} \xi _{10}+\xi _{12}]F_{32}\notag\\&\quad+[2 \xi_{01} \xi _{10}^2-2 \xi _{11} \xi _{10}-\xi_{01} \xi _{20}+\xi _{21}]F_{31}.
\end{align}
On the other hand, the definition \eqref{defchi} dictates:
\begin{equation}\label{c33}
    C_{3}=\chi_{03}F_{33}+\chi_{30}F_{30}+\chi_{12}F_{32}+\chi_{21}F_{31}.
\end{equation}
Comparing \eqref{c32} with \eqref{c33} yields:
\begin{gather}
    \chi_{03}=2 \xi_{01}^3-3 \xi_{02} \xi_{01}+\xi_{03}\\
    \chi_{12}=2 \xi _{10} \xi_{01}^2-2 \xi _{11} \xi_{01}-\xi_{02} \xi _{10}+\xi _{12}.
\end{gather}
Note that $\xi_{ji}=\xi_{ij}^{*}$. This is clear from the definition \eqref{defxi}. As a consequence, it also follows that $\chi_{ji}=\chi_{ij}^{*}$.

A similar procedure is then to be repeated for all order cumulants to get the complete set of mathematical relationships expressing the coefficients $\{\chi\}$ in terms of the coefficients $\{\xi\}$. This is explicitly carried out till the fourth cumulant in the appendix \ref{appA}.

Though this procedure is foolproof, it is certainly not convenient. The process of computing cumulants of all orders is unending. It would be much better if \emph{all} the relationships between the coefficients $\{\chi\}$ and the coefficients $\{\xi\}$ could be condensed into one single equation. It turns out that this can indeed be done. But before we can state that equation, we would need to introduce new objects called \emph{generating functions} for these sets of coefficients.
\subsection{Generating Function for Cumulants}
We begin this section by defining an object called the \emph{generating function} for the coefficients $\{\xi\}$ in a general state of the free oscillator.

The generating function for the coefficients $\{\xi\}$ in a general state of the free oscillator is defined as:
\begin{equation}\label{gfxi}
Z_{P}(\lambda,\bar{\lambda})\equiv\sum_{m,n=0}^{\infty}\frac{\lambda^m}{m!}\frac{\bar{\lambda}^{n}}{n!}\;\xi_{mn}
\end{equation}
The coefficients $\{\xi\}$ corresponding to the general state are then obtained from the generating function $Z_{P}$ through the formula:
\begin{equation}
    \xi_{mn}=\frac{\partial^{m+n}}{\partial^{m}\lambda\;\partial^{n}\bar{\lambda}}\;Z_{P}(\lambda,\bar{\lambda})\Bigl|_{\lambda,\bar{\lambda}=0}.
\end{equation}
On similar grounds, we can define the generating function for the coefficients $\{\chi\}$ in the general state of the free oscillator as:
\begin{equation}\label{gfchi}
Z_{\chi}(\lambda,\bar{\lambda})\equiv\sum_{m,n=0}^{\infty}\frac{\lambda^m}{m!}\frac{\bar{\lambda}^{n}}{n!}\;\chi_{mn}
\end{equation}
The coefficients $\{\chi\}$ corresponding to the general state are then obtained from the generating function $Z_{\chi}$ through the formula:
\begin{equation}\label{s2}
    \chi_{mn}=\frac{\partial^{m+n}}{\partial^{m}\lambda\;\partial^{n}\bar{\lambda}}\;Z_{\chi}(\lambda,\bar{\lambda})\Bigl|_{\lambda,\bar{\lambda}=0}.
\end{equation}
Apart from $\chi_{00}=\xi_{00}=1$, it turns out that the single equation which contains all the relationships between the $\{\xi\}$ and $\{\chi\}$ in it is:
\begin{equation}\label{xichi}
    \boxed{Z_{\chi}=\ln Z_{P}}.
\end{equation}
But what \emph{is} $Z_{\chi}$ exactly? The aim of this section will be achieved only if we can somehow express the generating function $Z_{\chi}$ explicitly in terms of the general state which the system occupies. Can we do so? It turns out that we indeed can.

As stated in \eqref{xichi}, $Z_{\chi}$ is given by:
\begin{equation}
    Z_{\chi}=\ln Z_{P},
\end{equation}
where, according to \eqref{gfxi}:
\begin{align}
    Z_{P}(\lambda,\bar{\lambda})&\equiv\sum_{m,n=0}^{\infty}\frac{\lambda^m}{m!}\frac{\bar{\lambda}^{n}}{n!}\;\xi_{mn},
\end{align}
and by \eqref{defxi}:
\begin{align}
    \xi_{mn}\equiv\langle(a^{\dagger})^{m}a^{n}\rangle=\Tr[\rho (a^{\dagger})^{m}a^{n}].
\end{align}
So now:
\begin{align}
    Z_{P}(\lambda,\bar{\lambda})&\equiv\sum_{m,n=0}^{\infty}\frac{\lambda^m}{m!}\frac{\bar{\lambda}^{n}}{n!}\;\xi_{mn}\notag\\
    &=\sum_{m,n=0}^{\infty}\frac{\lambda^m}{m!}\frac{\bar{\lambda}^{n}}{n!}\;\Tr[\rho(a^{\dagger})^{m}a^{n}]\notag\\
    &=\Tr\Bigl[\rho\sum_{m,n=0}^{\infty}\frac{\lambda^m}{m!}\frac{\bar{\lambda}^{n}}{n!}(a^{\dagger})^{m}a^{n}\Bigr]\notag\\
    &=\Tr\Bigl[\rho\sum_{m=0}^{\infty}\frac{\lambda^m}{m!}(a^{\dagger})^{m}\sum_{n=0}^{\infty}\frac{\bar{\lambda}^{n}}{n!}a^{n}\Bigr]\notag\\
    &=\Tr[\rho e^{\lambda a^{\dagger}}e^{\bar{\lambda}a}].
\end{align}
It is important to make an observation at this juncture. The generating function $Z_P$ is an already well studied object in the context of the quantum-classical correspondence. To be precise, it is the Fourier transform of the \emph{Sudarshan-Glauber P quasiprobability distribution}. This is the reason for the subscript \emph{P} in the notation for it. 

The quantum-classical correspondence allows one to calculate quantities relevant to a quantum mechanical problem through methods of classical statistical mechanics. A detailed introduction in this regard can be found in \cite{carm}.

$Z_{\chi}$ is thus given by:
\begin{equation}\label{s1}
    Z_{\chi}(\lambda,\bar{\lambda})=\ln\Tr[\rho e^{\lambda a^{\dagger}}e^{\bar{\lambda}a}].
\end{equation}
Using \eqref{s1} and \eqref{s2}, one can, in principle, compute all the $\{\chi\}$ for any general state of the oscillator.

The $\{\chi\}$ for some special states of the oscillator, namely the vacuum, the coherent state and the thermal state are calculated in the appendix \ref{appB} using the procedure thus outlined.
\section{Diagrammatic Formalism}\label{diagform}
The stage is now set for developing a diagrammatic formalism to compute Wightman correlators of the free harmonic oscillator in a general state.

As in the traditional case too, the starting point for building such a diagrammatic formalism is a Wick's theorem. In our case, the generalised Wick's theorem \ref{genwick} serves this purpose, as we will shortly see.

Another useful structure which one should establish before presenting diagrams are those of \emph{contractions}. They would especially aid us while doing a perturbative analysis of the anharmonic oscillator in \ref{anharm}.

In the traditional case, as outlined in \eqref{eq:11}, defining \emph{pairwise} contractions is enough. But in our considerations, it is not. A simple justification of this statement is the fact that in a general state, odd-point Wightman correlators do not vanish. If only pairwise contractions were enough, this would not have been possible.

Thus, for the study of Wightman correlators in general states of the oscillator, we would need to define additional contraction structures apart from just pairwise contractions. The inspiration for these definitions would be the generalised Wick's theorem, as we will now see.
\subsection{Contraction Structures}
So how does the generalised Wick's theorem reveal the new contraction structures that we would have to define for studying Wightman correlators in general states?

Just like the traditional case \eqref{eq:11}, we would like Wightman correlators in general states to be expressed as a sum of all possible contractions. That is, it would be good if:
\begin{equation}\label{contr}
    \langle x(t_1)x(t_2)\dots x(t_n)\rangle=\sum\Bigl\{\text{All possible \emph{full} contractions}\Bigr\},
\end{equation}
where a full contraction pattern means a contraction pattern wherein no position operator is left uncontracted.

Keeping this in mind, let us now look at the generalised Wick's theorem. The task is to somehow make the statement of the generalised Wick's theorem \ref{genwick} and \eqref{contr} agree with each other. After a bit of thought, one realises that this can indeed be done if one makes the following associations:
\begin{gather}
    C_1(t_i)\equiv\text{Contraction of $x(t_i)$ with itself},\notag\\
    C_2(t_i,t_j)\equiv\text{Contraction of $x(t_i)$ with $x(t_j)$},\notag\\
    C_3(t_i,t_j,t_k)\equiv\text{Simultaneous Contraction of $x(t_i)$, $x(t_j)$ and $x(t_k)$},
\end{gather}
and so on.

The above makes it clear that for our considerations, apart from pairwise contractions, we would need to define contractions of a position operator with itself and simultaneous contractions of more than two position operators.

Moreover, the above also dictates the analytic expressions which we must associate with the different contraction structures. To be precise, a simultaneous contraction of $n$ position operators ($n\geq 1$) will be assigned an analytic value of $C_n$, where $C_n$ is the $n^{th}$ cumulant.

We now formalise these notions. We define three types of contraction structures, namely \emph{uni-dentate}, \emph{bi-dentate} and \emph{poly-dentate} contractions and explain them one-by-one.
\subsubsection{\underline{Uni-Dentate Contractions}}
A uni-dentate contraction is defined as a contraction of a position operator with itself.

It would be represented as:
\begin{figure}[h]
    \centering
  \includegraphics[scale=0.12]{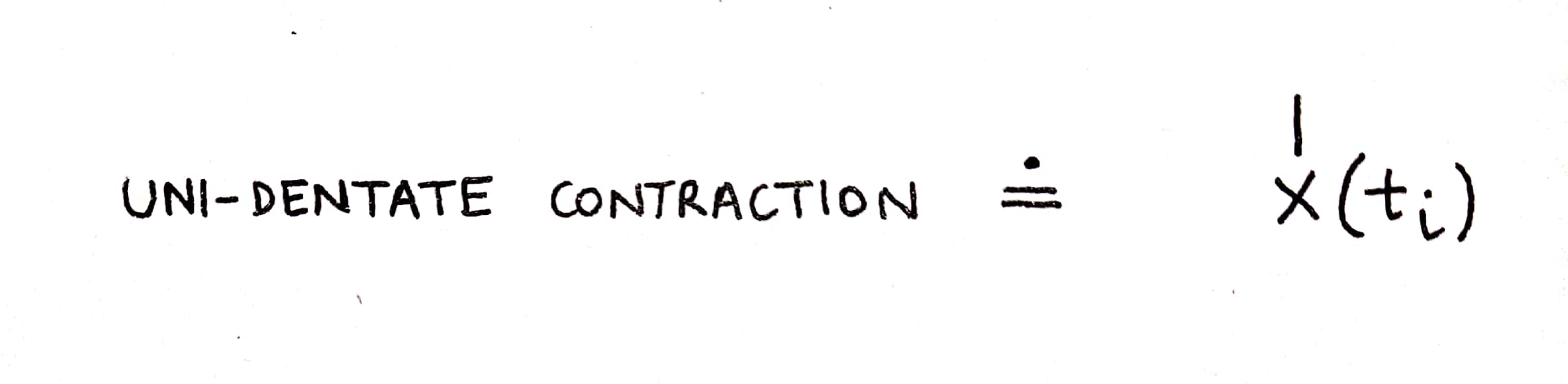}
  \caption{Uni-dentate contractions.}
\end{figure}
\\
The analytic expression corresponding to a uni-dentate contraction is the first cumulant $C_1$. That is:
\begin{figure}[h]
    \centering
  \includegraphics[scale=0.13]{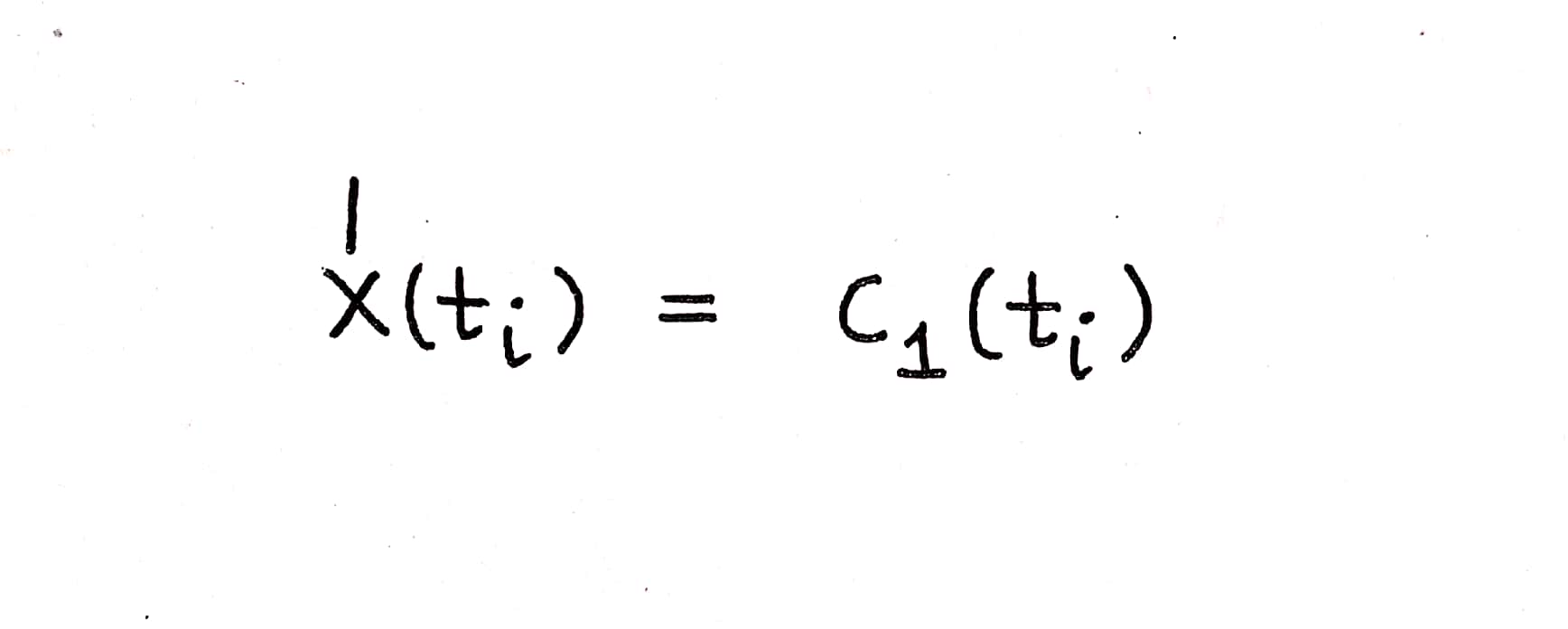}
  \caption{Analytic expression for uni-dentate contractions.}
\end{figure}
\iffalse\\[20pt]
\begin{center}
\textcolor{blue}{\textbf{(Continued on next page)}}
\end{center}\fi
\subsubsection{\underline{Bi-Dentate Contractions}}
A bi-dentate contraction is defined as a contraction of a position operator with another position operator.

It would be represented as:
\begin{figure}[h]
    \centering
  \includegraphics[scale=0.16]{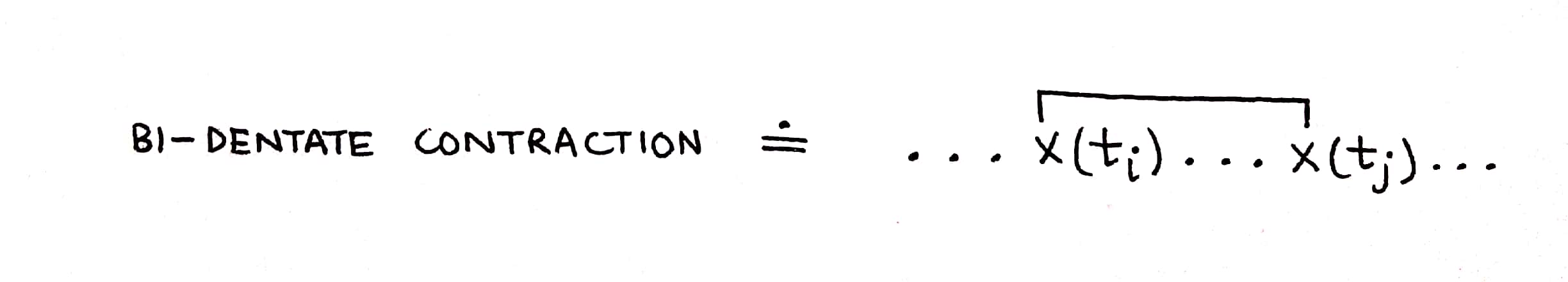}
  \caption{Bi-dentate contractions.}
\end{figure}
\newpage
The analytic expression corresponding to a bi-dentate contraction is the second cumulant $C_2$. That is:
\begin{figure}[h]
    \centering
  \includegraphics[scale=0.13]{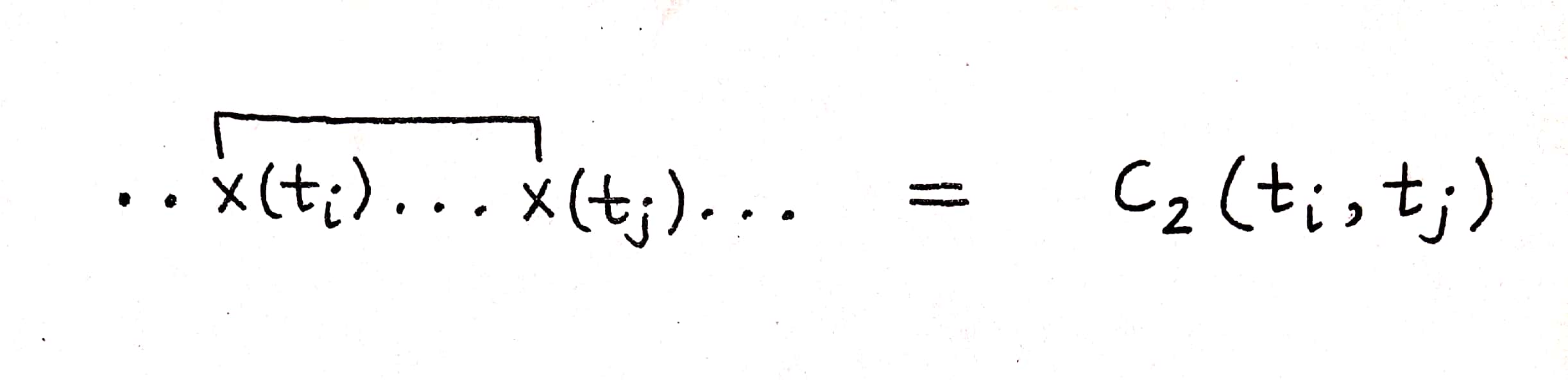}
  \caption{Analytic expression for bi-dentate contractions.}
\end{figure}
\subsubsection{\underline{Poly-Dentate Contractions}}
A poly-dentate contraction is defined as a simultaneous contraction of more than two position operators.

It would be represented as:
\begin{figure}[h]
    \centering
  \includegraphics[scale=0.19]{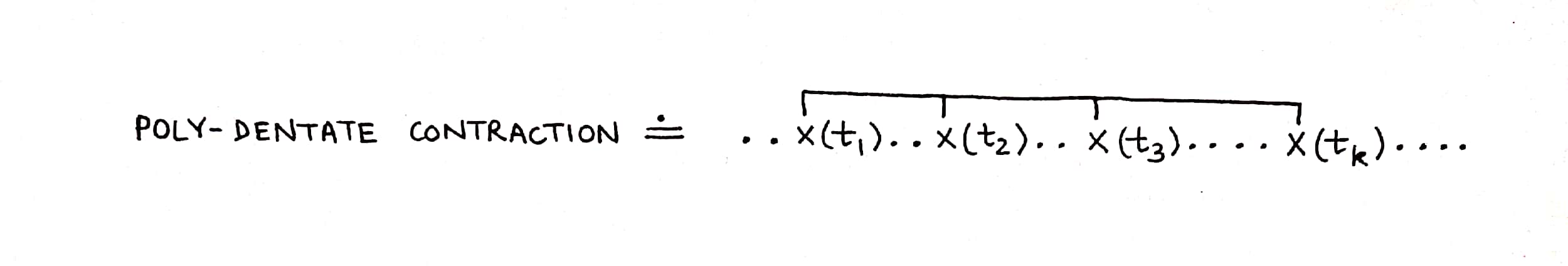}
  \caption{Poly-dentate contractions.}
\end{figure}
\iffalse\\[15pt]
\begin{center}
\textcolor{blue}{\textbf{(Continued on next page)}}
\end{center}\fi
\\
The analytic expression corresponding to a poly-dentate contraction which connects $k$ position operators together ($k>2$) is the $k^{th}$ cumulant $C_k$. That is:
\begin{figure}[h]
    \centering
  \includegraphics[scale=0.19]{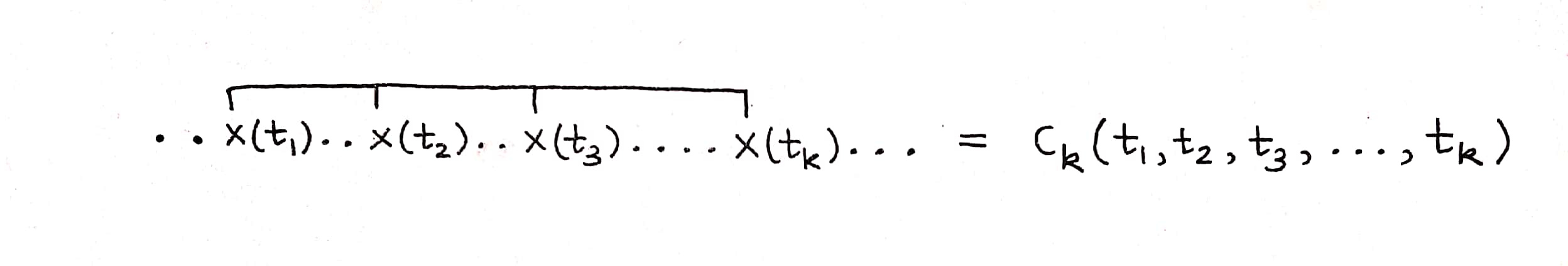}
  \caption{Analytic expression for poly-dentate contractions.}
\end{figure}
\subsection{Generalised Wick's Theorem in Contractions}
For the sake of completeness, let us now state the generalised Wick's theorem \ref{genwick} in terms of the contraction structures just introduced.

In the language of contractions, the generalised Wick's theorem reads:
\begin{equation}
    \langle x(t_1)x(t_2)\dots x(t_n)\rangle=\sum\Bigl\{\text{All possible \emph{full} contractions}\Bigr\},
\end{equation}
where a full contraction pattern means a contraction pattern wherein no position operator is left uncontracted.
\newpage
For example, one computes the 3-point Wightman correlator in a general state as:
\begin{figure}[h]
    \centering
  \includegraphics[scale=0.17]{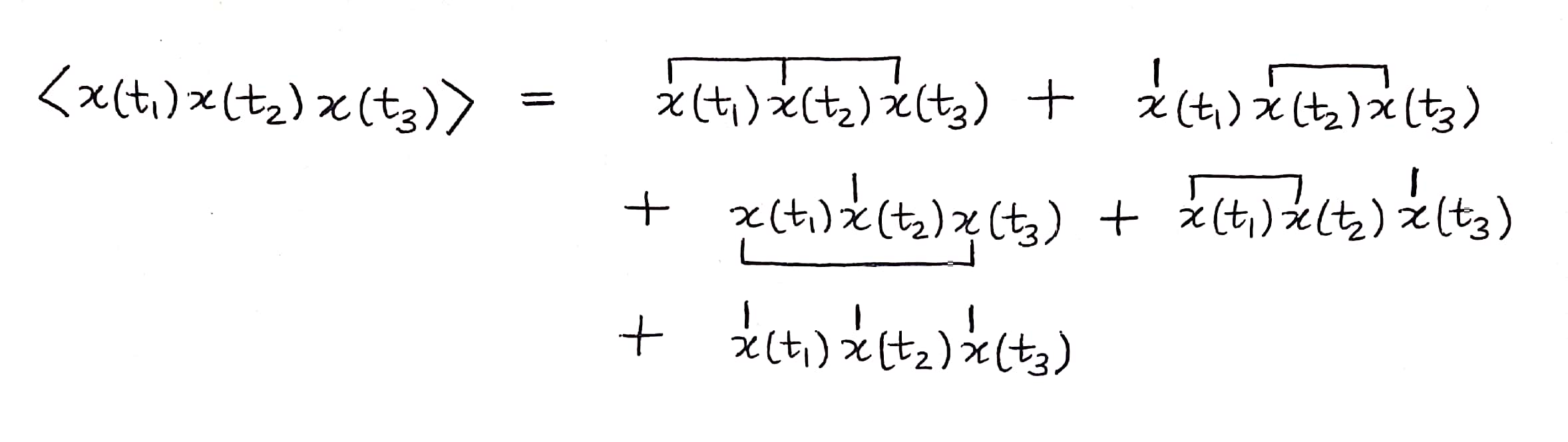}
  \caption{The 3-Point Wightman correlator in a general state.}
\end{figure}
\subsection{The Advent of Diagrams}\label{diagrams}
The only step which remains towards developing a diagrammatic formalism to compute Wightman correlators of the free oscillator in a general state is to associate the different contraction structures introduced in the previous section with diagrams. We will now do so.

To begin with, we first lay down the different diagrammatic components that would play a role in the diagrammatic formalism. Finally, we associate the different contraction patterns to their diagrammatic counterparts.
\subsubsection{Diagrammatic Components}
Here, as the title suggests, we introduce the different components which are involved in the diagrammatic formalism for Wightman correlators of the free oscillator in general states.
\begin{itemize}
    \item\underline{\textbf{External Points}}
    \\The Heisenberg position operator $x(t_i)$ is represented by a point labelled $t_i$:
    \begin{figure}[h]
    \centering
  \includegraphics[scale=0.07]{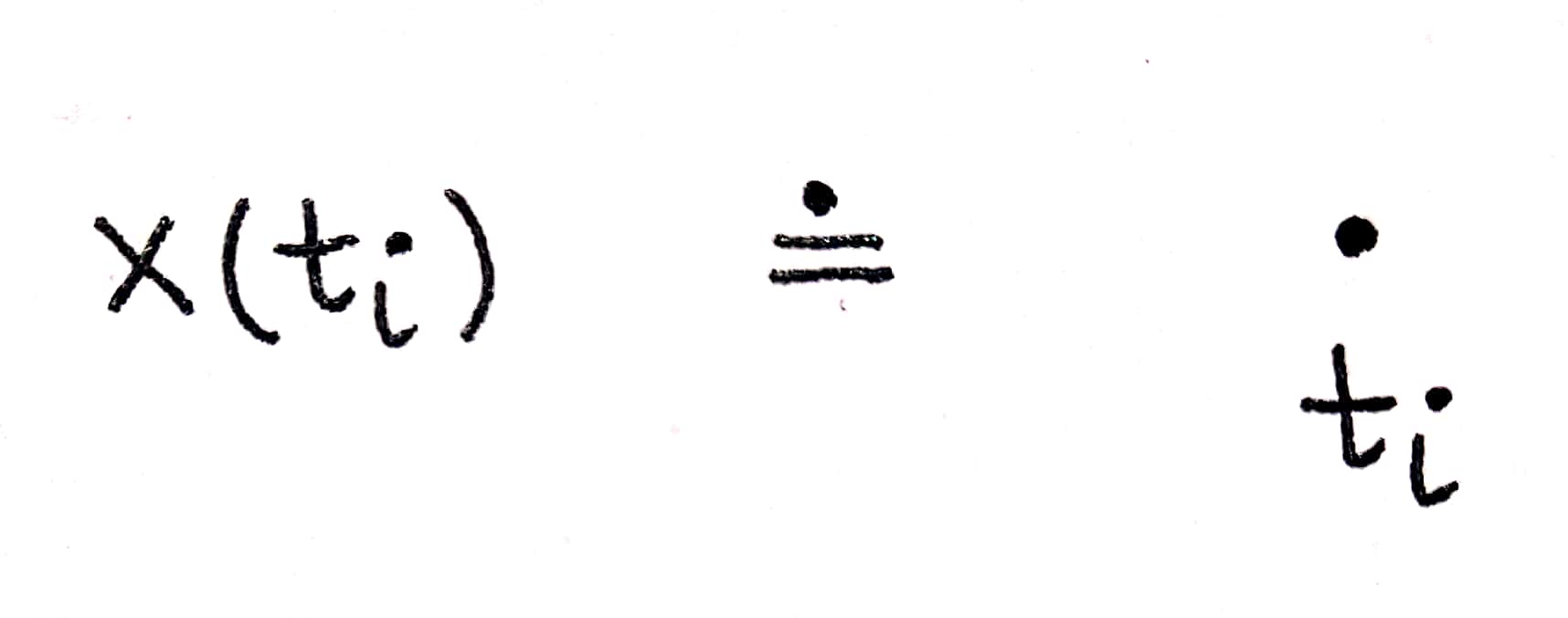}
  \caption{External Points.}
\end{figure}
    \item\underline{\textbf{Solid Lines}}:
    \begin{figure}[h]
    \centering
  \includegraphics[scale=0.09]{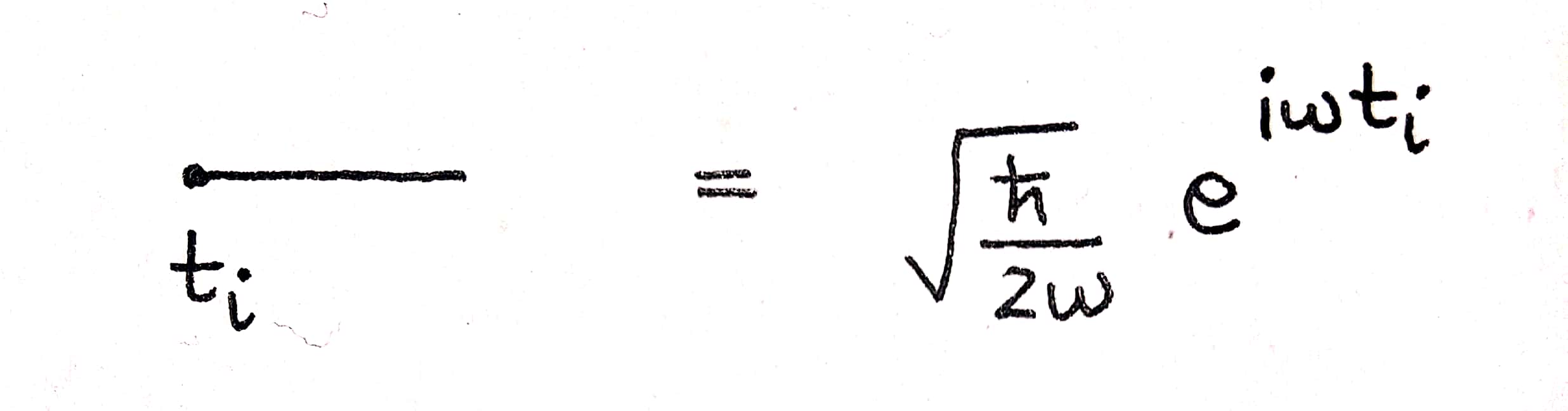}
  \caption{Solid Lines.}
\end{figure}
\newpage
    \item\underline{\textbf{Dotted Lines}}:
    \begin{figure}[h]
    \centering
  \includegraphics[scale=0.09]{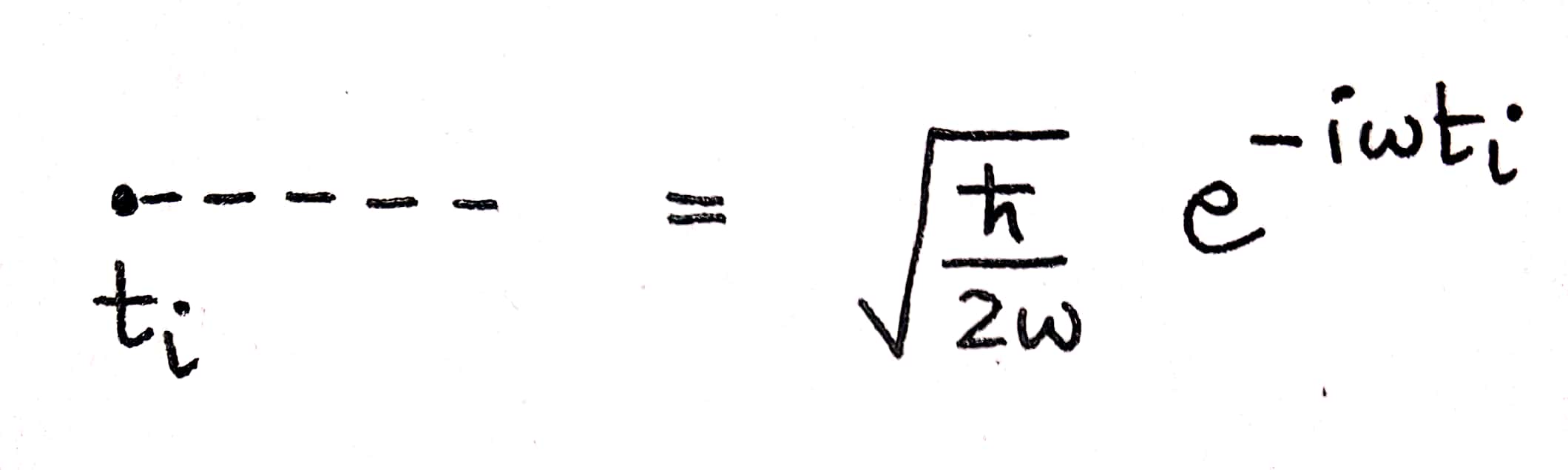}
  \caption{Dotted Lines.}
\end{figure}
    \item\underline{\textbf{Cumulant Blobs}}:
   \begin{figure}[h]
    \centering
  \includegraphics[scale=0.09]{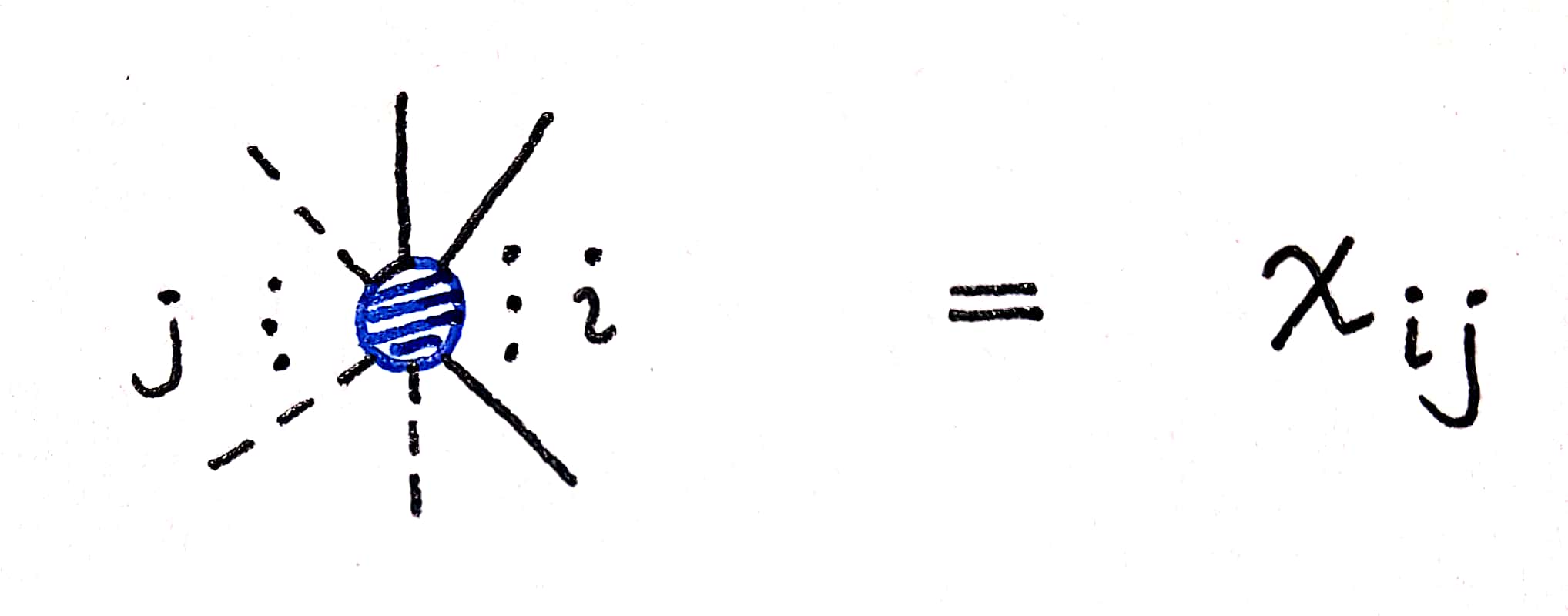}
  \caption{Cumulant Blobs.}
\end{figure}
\\
The $\{\chi_{ij}\}$ mentioned above are obtained through \eqref{s2}. Note that diagramatically, $i$ is the number of solid lines attached to the blob and $j$ is the number of dotted lines attached to the blob.
    \item\underline{\textbf{Symmetrised Lines/Red Propagators}}:
    \begin{figure}[h]
    \centering
  \includegraphics[scale=0.16]{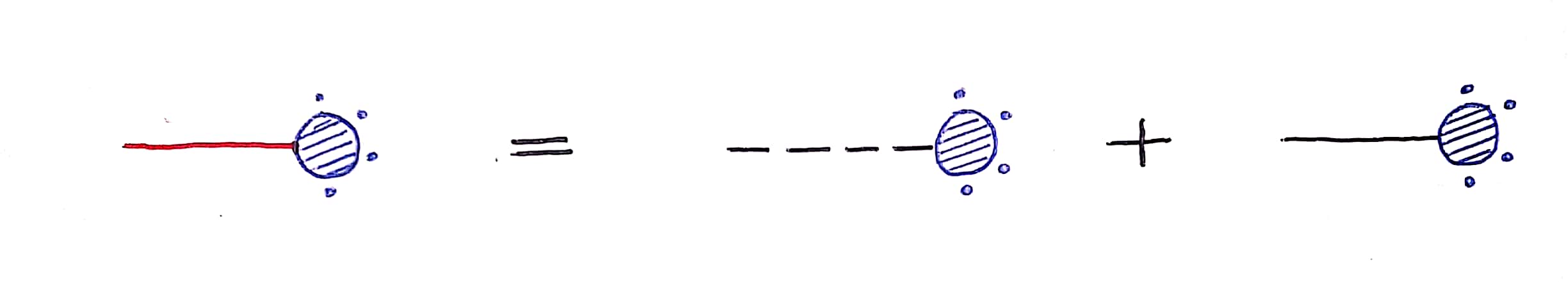}
  \caption{Symmetrised Lines/Red Propagators.}
\end{figure}
\\
As is clear from the above diagram, we will represent symmetrised lines by red coloured lines. From hereon, we would be calling these red lines as \textit{red propagators}. In the above diagram, the dots around the cumulant blob collectively refer to the other lines which may be attached to it.

A red propagator is like a diagrammatic operator which acts on a cumulant blob. Its action is to first attach a dotted line, and then a solid line to the blob and sum over these two configurations.
\iffalse\\[15pt]
\begin{center}
\textcolor{blue}{\textbf{(Continued on next page)}}
\end{center}\fi
\newpage
\item\underline{\textbf{Free Wightman Propagators}}:
    \begin{figure}[h]
    \centering
  \includegraphics[scale=0.1]{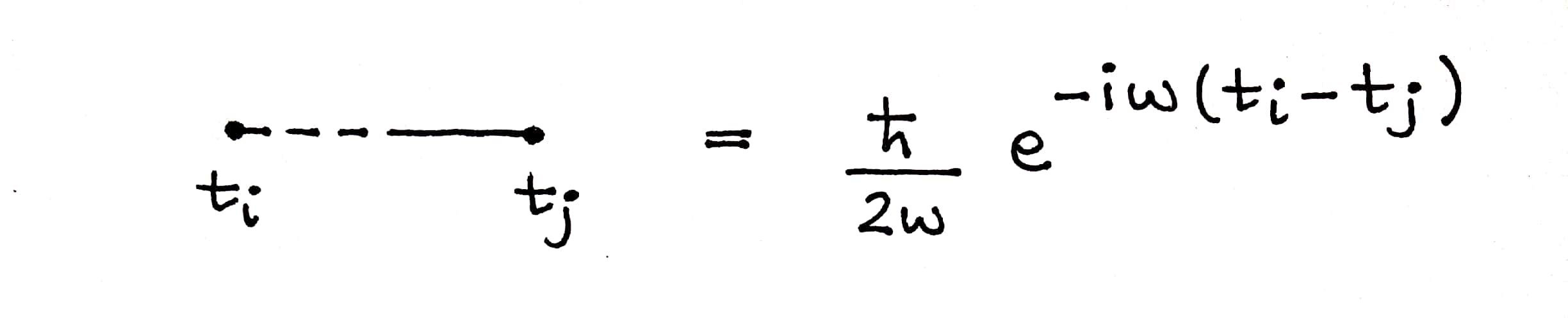}
  \caption{Free Wightman Propagator.}
\end{figure}
\\
This free Wightman propagator will be termed as going \emph{out} from the point $t_i$ \emph{into} the point $t_j$. Observe that the free Wightman propagator is a derived diagrammatic component. It is made up of a solid line and a dotted line. Consequently, its analytic expression can be arrived at through those corresponding to solid and dotted lines.
\end{itemize}
\subsubsection{Association of Contractions with Diagrams}\label{contrdiag}
We now associate the different contraction structures introduced, namely uni-dentate, bi-dentate and poly-dentate contractions, with their diagrammatic counterparts.
\subsubsection{\underline{Uni-Dentate Contractions}}
A uni-dentate contraction will have the following diagrammatic representation:
\begin{figure}[h]
    \centering
  \includegraphics[scale=0.08]{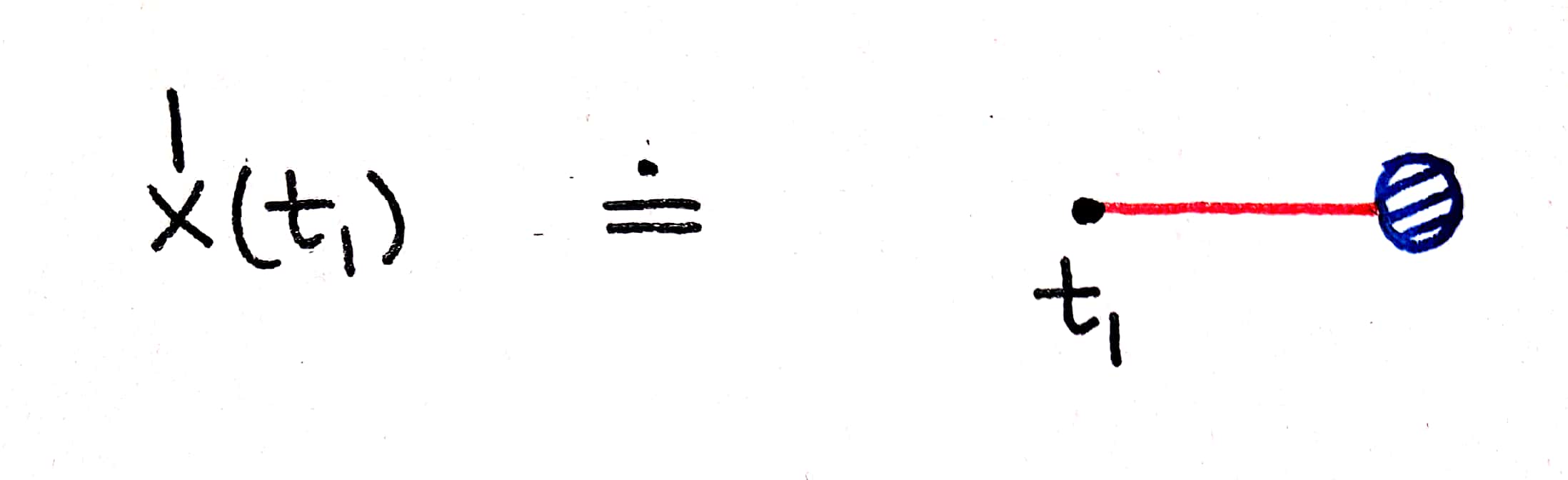}
  \caption{Diagrammatic representation for a uni-dentate contraction.}
  \label{fig:unidentate}
\end{figure}
\\
The LHS of figure \ref{fig:unidentate} is, by definition, equal to the first cumulant $C_1(t_1)$. Using the rules laid down in the previous section, the diagram on the RHS of figure \ref{fig:unidentate} evaluates to:
\begin{equation}
    \sqrt{\frac{\hbar}{2\omega}}\,\chi_{01}\,e^{-i\omega t_1}+\sqrt{\frac{\hbar}{2\omega}}\,\chi_{10}\,e^{i\omega t_1},
\end{equation}
which equals the first cumulant $C_1(t_1)$, as can be evaluated from the guiding definition \eqref{defchi}. Thus, this diagrammatic association is indeed true.
\subsubsection{\underline{Bi-Dentate Contractions}}
A bi-dentate contraction will have the following diagrammatic representation:
\begin{figure}[h]
    \centering
  \includegraphics[scale=0.17]{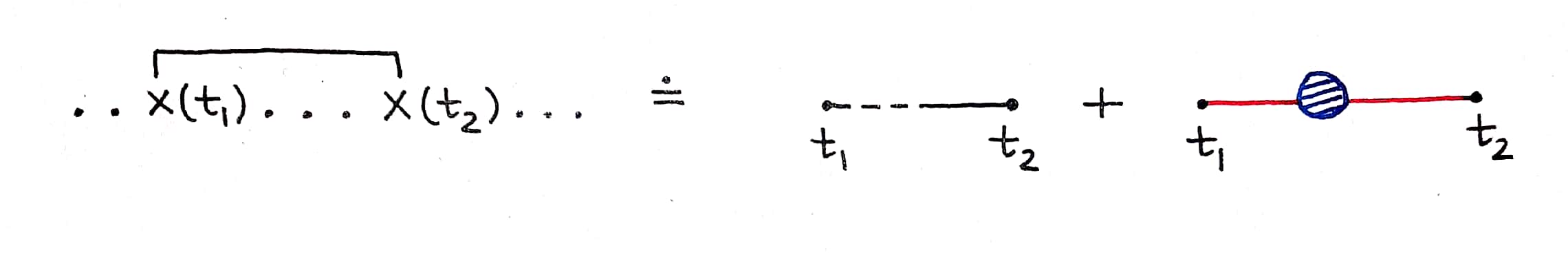}
  \caption{Diagrammatic representation for a bi-dentate contraction.}
  \label{fig:bidentate}
\end{figure}
\\
A very important point to note here is that in the diagrammatic representation of a bi-dentate contraction, the free Wightman propagator always goes \emph{out} from the time placed to the \emph{left} in the given Wightman sequence \emph{into} the time placed to the \emph{right} in the given Wightman sequence.

The LHS of figure \ref{fig:bidentate} is, by definition, equal to the second cumulant $C_2(t_1,t_2)$. Using the rules laid down in the previous section, the diagram on the RHS of figure \ref{fig:bidentate} evaluates to:
\begin{align}
   \Bigl(\frac{\hbar}{2\omega}\Bigr)&\Bigl[\chi_{02}\;e^{-i\omega(t_1+t_2)}+\chi_{20}\;e^{i\omega(t_1+t_2)}\notag\\
    &\;+\chi_{11}\;\bigl\{e^{-i\omega(-t_1+t_2)}+e^{-i\omega(t_1-t_2)}\bigr\}\notag\\
    &\;+e^{-i\omega(t_1-t_2)}\Bigr].
\end{align}
which equals the second cumulant $C_2(t_1,t_2)$, as can be evaluated from the guiding definition \eqref{defchi}. Thus, this diagrammatic association is indeed true.
\subsubsection{\underline{Poly-Dentate Contractions}}
A poly-dentate contraction will have the following diagrammatic representation:
\begin{figure}[h]
    \centering
  \includegraphics[scale=0.1]{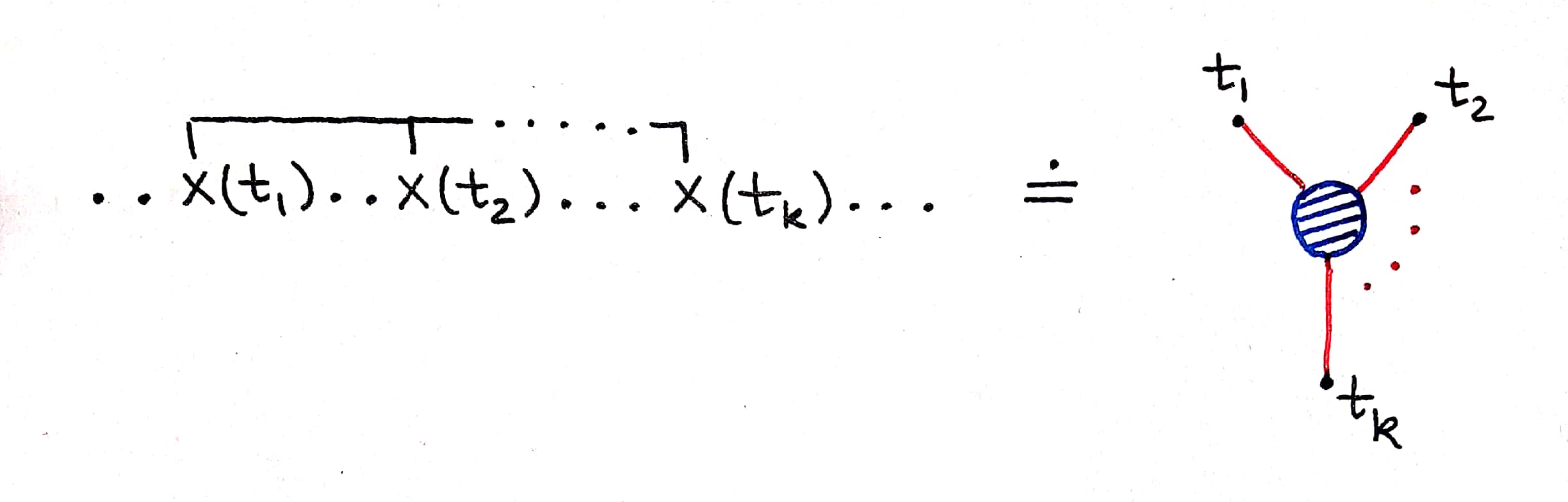}
  \caption{Diagrammatic representation for a poly-dentate contraction.}
  \label{fig:polydentate}
\end{figure}
\\
This definition makes it clear that any number of red propagators may branch out from a cumulant blob. The LHS of figure \ref{fig:polydentate} is, by definition, equal to the $k^{th}$ cumulant $C_k(t_1,t_2,\dots,t_k)$. 
Let us check this diagrammatic association for the simplest case of $k=3$. For this case, the LHS of figure \ref{fig:polydentate} is the third cumulant $C_3(t_1,t_2,t_3)$. The diagram on the RHS of figure \ref{fig:polydentate} would reduce to:
\begin{figure}[h]
    \centering
  \includegraphics[scale=0.05]{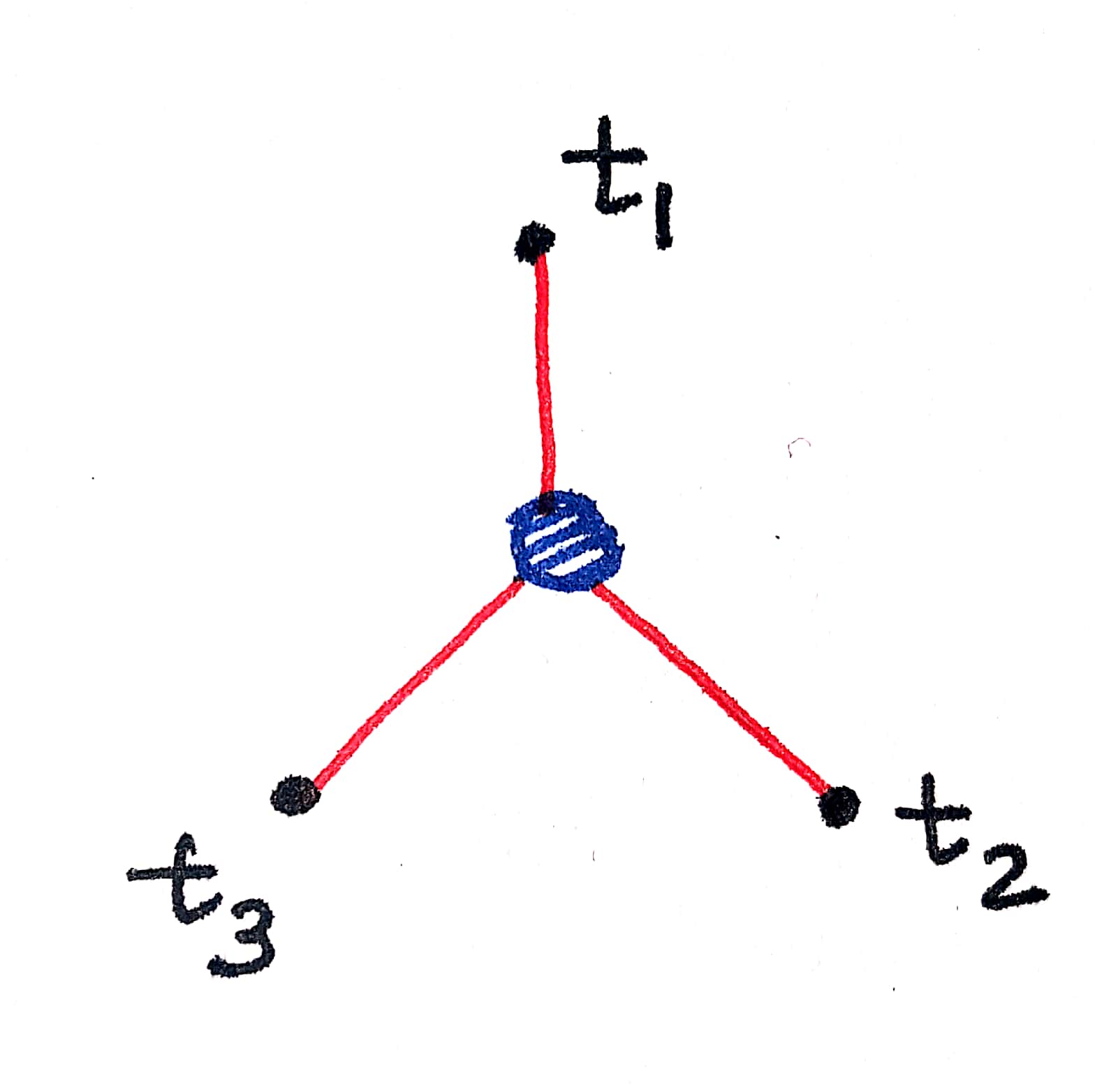}
  \caption{The diagrammatic representation of the third cumulant $C_3$.}
  \label{c3}
\end{figure}
\\
Using the rules laid down in the previous section, this diagram evaluates to:
\begin{align}
    \Bigl(\frac{\hbar}{2\omega}\Bigr)^{3/2}&\Bigl[\chi_{03}\;e^{-i\omega(t_1+t_2+t_3)}+\chi_{30}\;e^{i\omega(t_1+t_2+t_3)}\notag\\
    &\;+\chi_{12}\;\bigl\{e^{-i\omega(t_1-t_2+t_3)}+e^{-i\omega(t_1+t_2-t_3)}+e^{i\omega(t_1-t_2-t_3)}\bigr\}+\text{(on next page)}\notag\\&\;+\chi_{21}\;\bigl\{e^{i\omega(t_1+t_2-t_3)}+e^{i\omega(t_1-t_2+t_3)}+e^{-i\omega(t_1-t_2-t_3)}\bigr\}\Bigr].
\end{align}
which equals the third cumulant $C_3(t_1,t_2,t_3)$, as can be evaluated through the guiding definition \eqref{defchi}.
\subsection{Worked Out Examples}
\begin{figure}[h]
    \centering
  \includegraphics[scale=0.195]{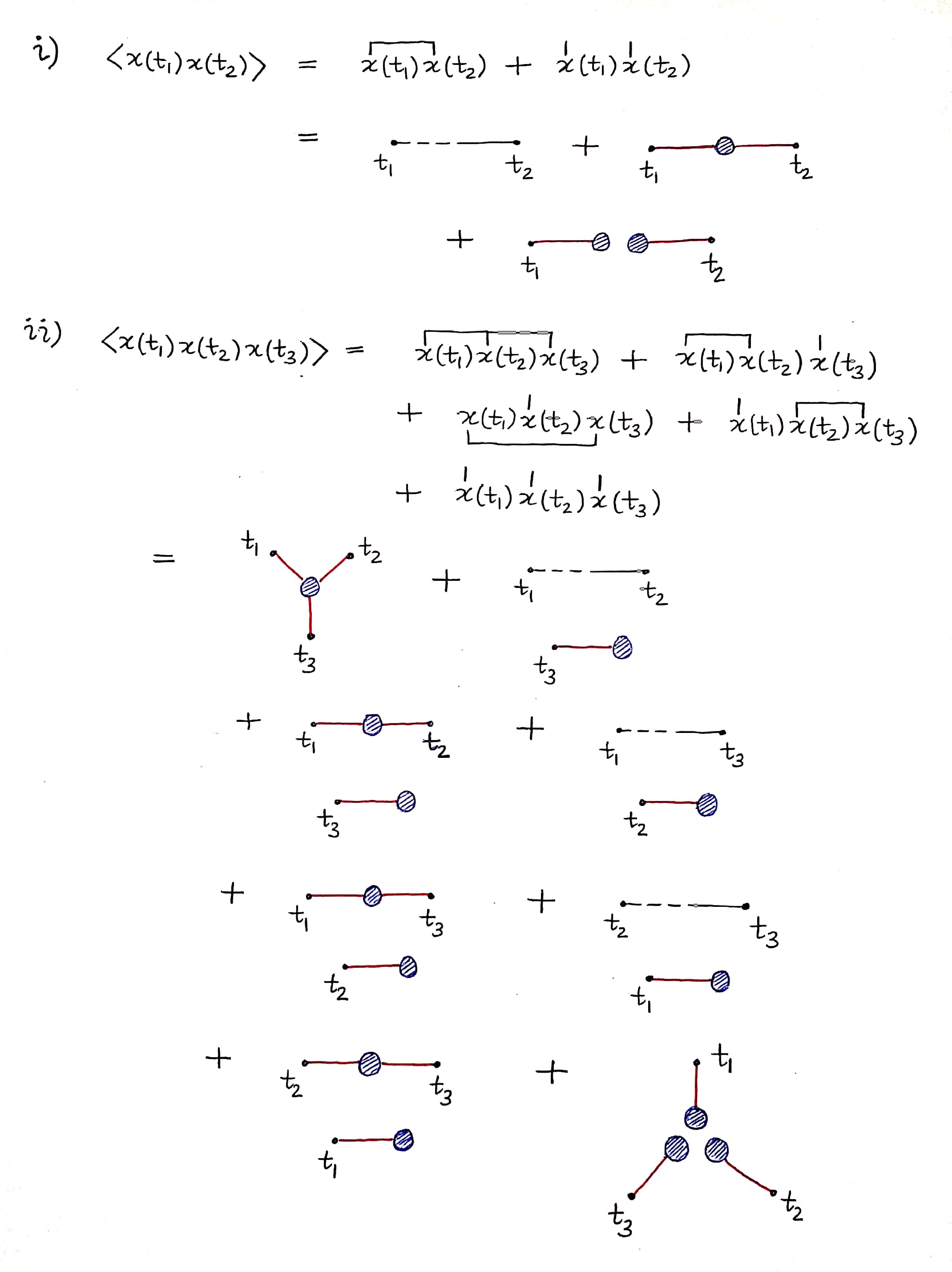}
\end{figure}
\newpage
\subsection{Some Diagrammatic Observations}\label{diagobs}
As a part of this section, we list a few crucial observations concerning the diagrammatic formalism introduced through \ref{diagform}. We do this because they would prove to be useful while studying Wightman correlators in general states of the anharmonic oscillator in \ref{anharm}.
\begin{itemize}
    \item Cumulant blobs are connected to points only through red propagators.
    \item Cumulant blobs are never connected to other cumulant blobs.
    \item Any number of red propagators may originate from a cumulant blob.
    \item A red propagator can never loop back into a cumulant blob.
\end{itemize}
With all this done, let us now delve into our explorations of the simple harmonic oscillator's elder cousin, the anharmonic oscillator.
\vspace{45pt}
\begin{center}
\scalebox{2.5}{\adforn{18}}
\end{center}
\chapter{The Anharmonic Oscillator}\label{anharm}
\section{Motivation}
Before starting this chapter, it would be good to summarise what we have achieved till now.

Motivated by quite a few concrete aspects, we delved into the study of Wightman correlators in general states of the simple harmonic oscillator. We were able to develop a generalised Wick's theorem for these correlators. This theorem then led us to define novel Wick contraction structures which aided us in the computation of these correlators. Finally, these Wick contraction structures gave way to a natural diagrammatic formalism to represent, and consequently calculate, Wightman correlators in general states of the simple harmonic oscillator.

With all this done, why should we study the same in the anharmonic oscillator at all?

This question actually has two questions nested in it. One, why should we study Wightman correlators in general states and two, why should we study them in the anharmonic oscillator? The first has already been answered in the introduction \ref{intro}; the same motivations for the study of Wightman correlators in general states of systems as mentioned there hold for the study of these types of correlators in this new system too. As for the second question, there are two reasons for extending our study to that of the anharmonic oscillator.

One, it is almost impossible to find a simple harmonic oscillator in Nature. Let alone Nature, it is also very tough to setup one in a laboratory. On the other hand, \emph{interacting} oscillators are a relatively more realistic system one could study. The \emph{anharmonic} oscillator is a simplistic model of an interacting oscillator. So if one wants to study Wightman correlators in general states of an interacting oscillator, the anharmonic oscillator serves as an ideal starting point. Methods developed in the study of such correlators of the anharmonic oscillator may be extended to the analyses of these objects in more complicated models of interacting oscillators. 

The second motivation comes keeping quantum fields in mind. As already stated in the introduction \ref{intro}, \emph{interacting} quantum fields are much more prevalent in Nature than free quantum fields. A very strong example in support of this is that the Standard Model of particle physics, the framework which describes all the fundamental forces of Nature apart from gravity, is an interacting quantum field theory. Just as the study of a free harmonic oscillator easily generalises to that of a free quantum field, the study of an interacting oscillator easily generalises to that of an interacting quantum field. So if one wants to analyse Wightman correlators of interacting quantum field theories in general states, it would be best to study them first in an interacting oscillator.

A physical requirement which interacting quantum field theories must fulfill is that they must be \emph{renormalizable}. For self-interacting scalar quantum field theories, this condition restricts the Lagrangian to be only of the following two types:
\begin{equation}\label{phi3}
    \mathcal{L}=\frac{1}{2}\partial_{\mu}\phi\,\partial^{\mu}\phi-\frac{1}{2}m^2\phi^2-\frac{g\phi^3}{3!}-\frac{\lambda\phi^4}{4!},
\end{equation}
or
\begin{equation}\label{phi4}
    \mathcal{L}=\frac{1}{2}\partial_{\mu}\phi\,\partial^{\mu}\phi-\frac{1}{2}m^2\phi^2-\frac{\lambda\phi^4}{4!},
\end{equation}
where $\phi\equiv\phi(t,\textbf{x})$ is a scalar field, $m$ is its mass and $g,\lambda$ are coupling constants.

The corresponding oscillator models which generalise to yield the theories \eqref{phi3} and \eqref{phi4} are ones with the Hamiltonians:
\begin{equation}\label{anharm1}
    H=\frac{p^2}{2}+\frac{1}{2}\omega^2 x^2+\frac{gx^3}{3!}+\frac{\lambda x^4}{4!} \qquad (m=1),
\end{equation}
and
\begin{equation}\label{anharm2}
    H=\frac{p^2}{2}+\frac{1}{2}\omega^2 x^2+\frac{\lambda x^4}{4!} \qquad (m=1),
\end{equation}
respectively.

Both of the models \eqref{anharm1} and \eqref{anharm2} are anharmonic oscillators. Thus, if we aim to finally study Wightman correlators of self-interacting scalar fields in general states, it would be a good idea to first analyse these objects in the anharmonic oscillator. We, however, would only be studying the anharmonic oscillator with the Hamiltonian \eqref{anharm2}. The extension of our investigations made in this direction to the study of the anharmonic oscillator with the Hamiltonian \eqref{anharm1} is trivial.
\section{Overview}\label{anhov}
In this chapter, we first give a formal introduction to the anharmonic oscillator. The objects of analysis, namely Wightman correlators in general states, are then brought into the picture. For studying these objects in the anharmonic oscillator, it proves to be useful to switch to a formalism known as the interaction picture. A brief introduction to this technique is then provided. Finally, without going into any calculational details, it is briefly mentioned how one goes about computing these objects using perturbation theory.   

Three new diagrammatic concepts emerge while studying Wightman correlators of anharmonic oscillators in general states which were not encountered while investigating these objects in the free oscillator. We list them on the next page.
\newpage
The new diagrammatic concepts are:
\begin{itemize}
    \item Interaction vertices
    \item Step function weights for internal propagators
    \item Symmetry factors
\end{itemize}
These three new aspects are then separately studied under different sections. We motivate them all through examples. As a result, a set of rules which must be abided by while carrying out such a diagrammatic computation of a general n-point Wightman correlator in a general state of the anharmonic oscillator are revealed. These rules, the \emph{Feynman rules} for Wightman correlators of the anharmonic oscillator in general states, are then collated under another section. The chapter is put to an end with a section dedicated to a worked out example.
\section{The System}
The anharmonic oscillator on which we would be working in this chapter is characterised by the Hamiltonian:
\begin{equation}\label{anharmham}
    H=\frac{p^2}{2}+\frac{1}{2}\omega^2 x^2+\frac{\lambda x^4}{4!}.
\end{equation}
The mass of the oscillator has been set to unity. $\lambda>0$ is a parameter called the \emph{coupling constant} or the \emph{interaction strength}.

Note that it is the very presence of a non-zero (and positive) coupling constant that makes the oscillator an interacting one.
\section{Objects of Analysis}
The objects of analysis would be the same as those studied in \ref{free}, namely Wightman correlators in general states.

They look like:
\begin{equation}\label{anhwight}
    \langle x_H(t_1)x_H(t_2)\dots x_H(t_n)\rangle=\Tr[\rho x_H(t_1)x_H(t_2)\dots x_H(t_n)],
\end{equation}
where $\rho$ is a general state of the anharmonic oscillator and $x_H(t_i)$ is the Heisenberg position operator of the anharmonic oscillator at the time $t_i$.

The Heisenberg position operators of the anharmonic oscillator evolve according to the equation:
\begin{equation}
    \frac{d^2x_H}{dt^2}+\omega^2x_H+\frac{\lambda x^3_H }{3!}=0.
\end{equation}
The solution of this equation turns out to be:
\begin{equation}
    x_H(t)=U^{\dagger}(t,t_0)x_H(t_0)U(t,t_0)\qquad(t\geq t_0),
\end{equation}
where $U(t,t_0)$ is the time evolution operator given by the expression:
\begin{equation}
    U(t,t_0)=e^{-iH(t-t_0)}.
\end{equation}
Here, $H$ is the total Hamiltonian of the system given by \eqref{anharmham}.
\section{The Interaction Picture}
Before beginning a perturbative analysis for Wightman correlators in general states of an anharmonic oscillator, it proves to be helpful to switch to a new representation of quantum mechanics, namely the interaction picture.

The benefit of doing so is that through this representation, one can express Wightman correlators in general states of the anharmonic oscillator in terms of Wightman correlators in general states of the \emph{free} oscillator. Since the latter has already been explored in the previous chapter \ref{free}, doing so opens the gates for all the techniques developed therewith to be applied in the present analysis of Wightman correlators in general states of the anharmonic oscillator.

We do not review the interaction picture completely in this chapter. We would be directly using those results from it which would aid us in the analysis of Wightman correlators in general states of the anharmonic oscillator. A good introduction to the interaction picture can be found in \cite{tong}.

Invoking the interaction picture, one expresses the Heisenberg position operator of the anharmonic oscillator as:
\begin{equation}
    x_H(t_i)=U^{\dagger}(t_i,t_0)\,x_I(t_i)\,U(t_i,t_0)\qquad (t_i\geq t_0).
\end{equation}
Here, $U(t_i,t_0)$ is the time evolution operator in the interaction picture. $t_0$ is the time when the interaction was switched on, and $x_I(t_i)$ is the interaction picture position operator of the anharmonic oscillator at the time $t_i$. For all practical purposes, it is the same as the Heisenberg position operator of the free oscillator at the time $t_i$. Keeping this in mind, we will, from now on, denote the interaction picture position operator simply as $x(t_i)$ to match it with the notation of the previous chapter \ref{free}. That is:
\begin{equation}\label{nota}
    x_I(t_i)\equiv x(t_i).
\end{equation}
The time evolution operator in the interaction picture has the explicit expression:
\begin{equation}\label{timev}
    U(t_1,t_2)=Te^{-i\int_{t_2}^{t_1}\;dt H_I}\quad(t_1\geq t_2).
\end{equation}
Here, $T$ stands for the time-ordering symbol. $H_I$ is the interacting Hamiltonian in the interaction picture, given by:
\begin{equation}
    H_I=\frac{\lambda{x_I}^4}{4!}=\frac{\lambda{x}^4}{4!}.
\end{equation}
The last equality follows from the notation \eqref{nota} adopted.

Moreover:
\begin{equation}
   U^{\dagger}(t_1,t_2)=U(t_2,t_1)=T^{*}e^{i\int_{t_2}^{t_1}\;dt H_I}\quad(t_1\geq t_2),
\end{equation}
where $T^{*}$ is the anti-time ordering symbol.

Using all the machinery established, the n-point Wightman correlator in a general state of the anharmonic oscillator reads:
\begin{align}\label{anhcor}
    \langle x_H(t_1)x_H(t_2)\dots x_H(t_n)\rangle=\langle&U^{\dagger}(t_1,t_0)\,x(t_1)\,U(t_1,t_0)\notag\\&U^{\dagger}(t_2,t_0)\,x(t_2)\,U(t_2,t_0)\dots\notag\\& U^{\dagger}(t_n,t_0)\,x(t_n)\,U(t_n,t_0)\rangle.
\end{align}
Here, it is important to realise the fact that one cannot combine the time evolution operators sandwiched in between the interaction picture position operators into single time evolution operators. For example, referring to \eqref{anhcor}:
\begin{align}
    U(t_1,t_0)\,U^{\dagger}(t_2,t_0)&=U(t_1,t_0)\,U(t_0,t_2)\notag\\
    &\neq U(t_1,t_2).
\end{align}
This is because \emph{the composition property of time evolution operators holds only for a particular ordering of the times involved}. To be precise, the composition property of time evolution operators dictates:
\begin{equation}\label{comp}
    U(t_1,t_2)\,U(t_2,t_3)=U(t_1,t_3),\quad\text{where}\quad t_1\geq t_2\geq t_3.
\end{equation}
Now since the times in the correlator \eqref{anhwight} obey no particular ordering, we cannot employ the above composition property to simplify \eqref{anhcor} further.
\section{Perturbative Analysis : A Glimpse}\label{pertanh}
In this section, we introduce the formal procedure through which we would be studying Wightman correlators in general states of the anharmonic oscillator from now on. The starting point of all our analyses would be \eqref{anhcor}. How we would be proceeding thereon is explained through the following example.

For the sake of illustration, consider \eqref{anhcor} for the case of a 2-point Wightman correlator in a general state of the anharmonic oscillator:
\begin{equation}\label{twopt}
    \langle x_H(t_1)x_H(t_2)\rangle.
\end{equation}
\newpage
Using the interaction picture, one re-expresses \eqref{twopt} as:
\begin{align}\label{2point}
    \langle x_H(t_1)x_H(t_2)\rangle=\langle&U^{\dagger}(t_1,t_0)\,x(t_1)\,U(t_1,t_0)\notag\\&U^{\dagger}(t_2,t_0)\,x(t_2)\,U(t_2,t_0)\rangle.
\end{align}
From \eqref{timev}, we already know the explicit expressions for the time evolution operators involved in the above equation.

For example, the time evolution operator $U(t_1,t_0)$ involved in \eqref{2point} reads:
\begin{align}\label{u}
    U(t_1,t_0)&=Te^{-i\int_{t_0}^{t_1}\;dt H_I}\notag\\
    &=Te^{-\frac{i\lambda}{4!}\int_{t_0}^{t_1}\;dt\,x^4}.
\end{align}
What we would now want to do is to expand \eqref{u} as a power series in the interaction strength $\lambda$. This is the basic essence of a \emph{perturbative} analysis. But when can such a perturbative analysis be actually carried out?

A simple dimensional analysis exercise reveals that such a condition is met if:
\begin{equation}\label{pertreg}
    \lambda<<\frac{\omega^3}{\hbar}.
\end{equation}
With this ensured, one is allowed to expand \eqref{u} as:
\begin{align}\label{notmot}
    U(t_1,t_0)&=\mathbb{I}+\Bigl(\frac{-i\lambda}{4!}\Bigr)\int_{t_0}^{t_1}dt^{(1)}_{1-}\,x^4(t^{(1)}_{1-})\notag\\&\hspace{20pt}+\Bigl(\frac{-i\lambda}{4!}\Bigr)^{2}\int_{t_0}^{t_1}dt^{(1)}_{1-}\int_{t_0}^{t_1}dt^{(2)}_{1-}\,\{Tx^4(t^{(1)}_{1-})x^4(t^{(2)}_{1-})\}\notag\\&\hspace{20pt}+\mathcal{O}(\lambda^3).
\end{align}
Note that we have introduced a notation here for the internal point which has been generated in the perturbation expansion \eqref{notmot}. This notation would prove to be very important in our consequent analyses. Let us explain it in detail here itself.
\blank
\subsubsection{\underline{Important Notation}}
Consider the 2-point Wightman correlator in a general state of the anharmonic oscillator:
\begin{equation}\label{abba}
    \langle x_H(t_1)x_H(t_2)\rangle.
\end{equation}
As we have just seen, the above expression becomes:
\begin{align}\label{aakhri}
    \langle x_H(t_1)x_H(t_2)\rangle=\langle&U^{\dagger}(t_1,t_0)\,x(t_1)\,U(t_1,t_0)U^{\dagger}(t_2,t_0)\,x(t_2)\,U(t_2,t_0)\rangle,
\end{align}
when one invokes the interaction picture. Now, when one expands the operators $U$ and $U^\dagger$ involved in the above expression in powers of the coupling $\lambda$, internal points are `generated'. 

For example, if one expands the operator $U^\dagger(t_1,t_0)$ in \eqref{aakhri} upto order $\lambda^2$, one gets:
\begin{align}\label{intlabel}
   &\qquad\langle x(t_1)x(t_2)\rangle+\Bigl(\frac{i\lambda}{4!}\Bigr)\int_{t_0}^{t_1}dt\langle\biggl\{x(t)x(t)x(t)x(t)\biggr\}x(t_1)x(t_2)\rangle\notag\\&\quad+\Bigl(\frac{i\lambda}{4!}\Bigr)^2\int_{t_0}^{t_1}dt\int_{t_0}^{t_1}dt'\langle\Bigl\{T^*x(t)x(t)x(t)x(t)x(t')x(t')x(t')x(t')\Bigr\}x(t_1)x(t_2)\rangle.
\end{align}
Clearly, the internal points $t$ and $t'$ have emerged as a result. We now lay down a labelling scheme for these internal points.
\\[5pt]
\fcolorbox{black}{gray!10}{\parbox{35em}{We will call an internal point $t$ generated by the perturbation expansion of the operator $U^\dagger(t_j,t_0)$ as $t_{j+}$.}}
\\[10pt]
For example, in \eqref{intlabel}, the internal point $t$ would be called $t^{(1)}_{1+}$, and the internal point $t'$ would be called $t^{(2)}_{1+}$. The superscripts have been added to distinguish these two integration variables from each other.

On similar grounds:
\\[10pt]
\fcolorbox{black}{gray!10}{\parbox{35em}{We will call an internal point $t$ generated by the perturbation expansion of the operator $U(t_j,t_0)$ as $t_{j-}$.}}
\\[10pt]
To see this, consider some other terms in the perturbation expansion of \eqref{abba}:
\begin{align}
    &\quad\Bigl(\frac{-i\lambda}{4!}\Bigr)\int_{t_0}^{t_1}dt\langle x(t_1)\biggl\{x(t)x(t)x(t)x(t)\biggr\}x(t_2)\rangle\notag\\&+\Bigl(\frac{-i\lambda}{4!}\Bigr)\int_{t_0}^{t_2}dt'\langle x(t_1)x(t_2)\biggl\{x(t')x(t')x(t')x(t')\biggr\}\rangle.
\end{align}
Clearly, the internal point $t$ arises due to the perturbation expansion of $U(t_1,t_0)$. Similarly, the internal point $t'$ arises due to the perturbation expansion of $U(t_2,t_0)$.

Thus, the internal point $t$ would be called $t_{1-}$, and the internal point $t'$ would be called $t_{2-}$.
\blank
An expansion such as \eqref{notmot} is then simultaneously carried out for all the time evolution operators present in \eqref{2point}. As a result, one gets a perturbative expansion in the powers of the coupling $\lambda$ for the 2-point Wightman correlator \eqref{twopt} in a general state of the anharmonic oscillator. It is then trivial to extend such a protocol for studying general n-point Wightman correlators in general states of the anharmonic oscillator.
\newpage
\section{An Observation Regarding Cumulants in Thermal States}
Through \ref{free}, we have seen that objects called cumulants play a very important role in the diagrammatic representation of Wightman correlators in general states of the free oscillator. Now, if one views the anharmonic oscillator as being a small perturbation to the free oscillator, which one is justified in doing in the limit \eqref{pertreg}, cumulants are bound to play an equally important role in the diagrammatic representation of Wightman correlators in general states of the anharmonic oscillator too.

However, there is a subtle point involving the explicit computation of the cumulants associated to the thermal state of an anharmonic oscillator. This aspect deserves some discussion at this juncture.

To explain this point, let us focus on the 2-point Wightman correlator in a general density matrix $\rho$ of the anharmonic oscillator. The perturbation expansion for this object is to be derived from \eqref{2point}. The order $\lambda^0$ term in this expansion reads:
\begin{equation}\label{lamb0}
    \langle x_H(t_1)x_H(t_2)\rangle\bigl|_{\lambda^0}=\Tr[\rho x(t_1)x(t_2)].
\end{equation}
A reader who has gone through chapter \ref{free} would realise that the diagrammatic representation of the RHS of \eqref{lamb0} would involve cumulants of the density matrix $\rho$. These are objects of the form $\{\chi_{ij}\}$, which are generated by the function:
\begin{equation}\label{lamb01}
    Z_{\chi}(\mu,\bar{\mu})=\ln\Tr[\rho e^{\mu a^{\dagger}}e^{\bar{\mu}a}]
\end{equation}
through the formula:
\begin{equation}
\chi_{mn}=\frac{\partial^{m+n}}{\partial^{m}\mu\;\partial^{n}\bar{\mu}}\;Z_{\chi}(\mu,\bar{\mu})\Bigl|_{\mu,\bar{\mu}=0}.
\end{equation}
The point we want to make is that in the anharmonic thermal state, the cumulants themselves depend on the interaction strength $\lambda$. Stated figuratively:
\begin{equation}
    \chi_{\beta}\sim\chi_{\beta}(\lambda),
\end{equation}
where the $\beta$ in the subscript indicates that we are looking at the cumulants of the anharmonic thermal state.

This is easily justified. The thermal density matrix of the anharmonic oscillator reads:
\begin{equation}\label{thermdensity}
    \rho_{th}=\frac{e^{-\beta(\frac{p^2}{2}+\frac{1}{2}\omega^2 x^2+\frac{\lambda x^4}{4!})}}{Z},
\end{equation}
where $\beta$ is the inverse temperature and $Z$ is the partition function given by:
\begin{equation}
    Z=\Tr[e^{-\beta(\frac{p^2}{2}+\frac{1}{2}\omega^2 x^2+\frac{\lambda x^4}{4!})}].
\end{equation}
The thermal density matrix \eqref{thermdensity} clearly depends on the interaction strength $\lambda$. And since it enters the expression for the cumulant generating function \eqref{lamb01} too, the cumulants of the anharmonic thermal state $\{\chi_{\beta}\}$ are also bound to depend on it.

At the diagrammatic level, we may be blind to this subtlety by simply labelling the cumulant blobs (refer to \ref{diagrams}) involved in the diagrammatics by the $\{\chi_{\beta}\}$. But we cannot get away from it. This is because sooner or later, we would have to evaluate the diagrams in their entirety. For this, we would have to explicitly compute the $\{\chi_{\beta}\}$, which would involve the considerations just discussed.

As already mentioned in \ref{anhov}, three new diagrammatic aspects emerge through the analysis of the anharmonic oscillator which were absent in the free oscillator. They are \emph{interaction vertices}, \emph{step-function weights for internal propagators} and \emph{symmetry factors}. Let us now explore them in detail.
\section{Interaction Vertices}
The section \ref{pertanh} has outlined the procedure which one adopts to study Wightman correlators in general states of the anharmonic oscillator in the limit \eqref{pertreg}. Moreover, the earlier chapter \ref{free} has established a diagrammatic formalism to represent Wightman correlators in general states of the free oscillator. The perturbative analysis introduced in \ref{pertanh} expresses the Heisenberg position operators of the anharmonic oscillator in terms of the corresponding operators of the free oscillator. This opens the gates for all the diagrammatic machinery developed in \ref{free} to be put to use in analysing Wightman correlators in general states of the anharmonic oscillator too. When one actually goes through this procedure on paper, three new diagrammatic concepts emerge. They are those of \emph{vertices}, \emph{step function weights for internal propagators} and \emph{symmetry factors}. 

Among these three, the novel aspect of vertices is the one which requires a relatively more detailed discussion, and the present section is dedicated to this very purpose. Since this section would be a bit vast, it is better to first present an overview of it, and then delve into individual explanations.
\subsection{Overview}
We first give the definition of a vertex in \ref{vertexdef}. Following this definition, we give some examples of vertices that make an appearance in the consideration of Wightman correlators in general states of the anharmonic oscillator. Some explicit calculations are also displayed which lead to such vertices. 

The next task is to list down the \emph{vertex factors}, which are mathematical expressions associated to the vertices. Motivated by examples, it is shown in \ref{many} that a vertex in this theory, in general, has more than one vertex factors associated to it.

Once this is realised, a new way to assign vertex factors is developed. The most important step in this regard is \emph{labelling} the vertices of a diagram. Once all vertices of a diagram are labelled, the vertex factors of all these vertices can be read off from these labels. The subsection \ref{labelling} develops a hands-on protocol for labelling the vertices of a diagram. Finally, we end this section with some worked out examples wherein we implement this newly developed method to label the vertices of a couple of diagrams and consequently, read off their vertex factors.
\subsection{What is a Vertex?}\label{vertexdef}
A vertex is defined as a point where four propagators meet. These four propagators can either be free Wightman propagators or red propagators. Refer to \ref{diagrams} for an explanation of these diagrammatic components. 

A convention to be adopted for the sake of making this definition universal is that a \emph{loop} of a free Wightman propagator is to be counted as \emph{two} free Wightman propagators. As a result, this definition dictates that vertices of the form:
\begin{figure}[h]
    \centering
  \includegraphics[scale=0.13]{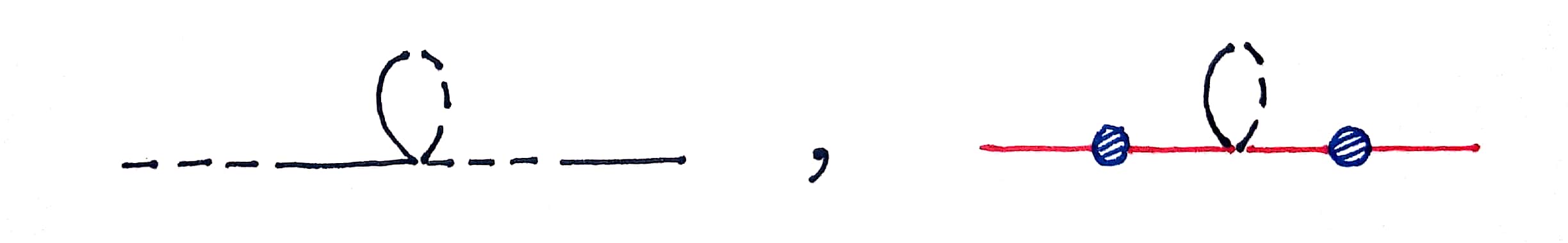}
  \caption{Some vertices present in this theory.}
\end{figure}
\\
would be present in this theory, which they are. If we had not adopted this convention, then the above vertices would have been a meeting point of three free Wightman propagators in the first case, and two red propagators with a third free Wightman propagator in the second case. The definition of vertices would then have ruled both of them out.

Another important point to note at this point is that a cumulant blob (\ref{diagrams}) with four legs is \emph{not} a vertex. As outlined through \ref{diagform}, it is just a diagrammatic representation of the fourth cumulant $C_4$. That is:
\begin{figure}[h]
    \centering
  \includegraphics[scale=0.1]{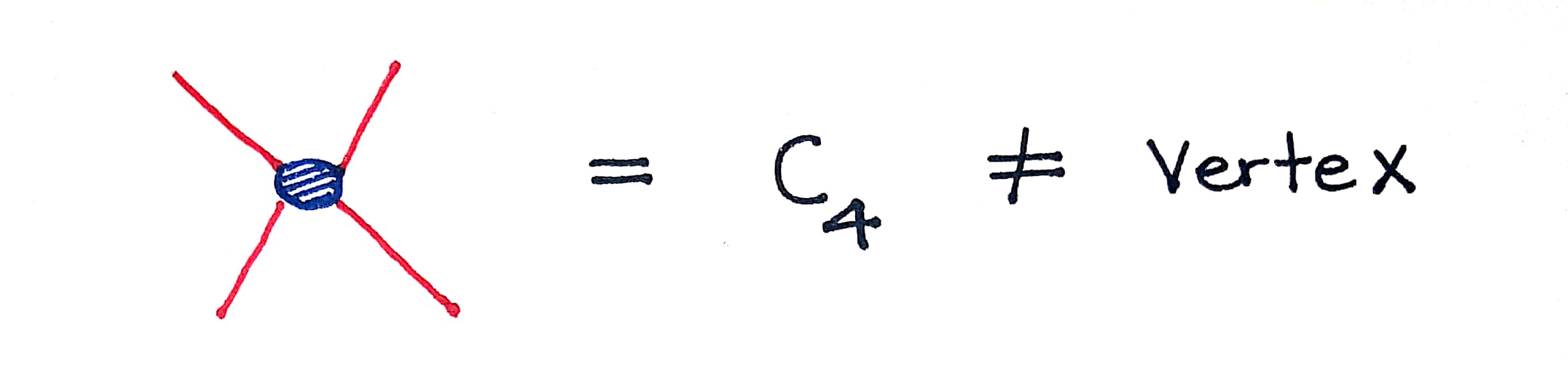}
  \caption{The fourth cumulant is not a vertex.}
\end{figure}
\subsection{Types of Vertices}
With the definition stated in \ref{vertexdef} in place, one would naturally like to try his/her hands at actually drawing the vertices which this theory presents. But certain restrictions have to be abided by while drawing the vertices. They were already observed in (\ref{diagobs}), but let us list them down again on the next page.
\newpage
These restrictions are:
\begin{itemize}
    \item Cumulant blobs are connected to points only through red propagators.
    \item Any number of red propagators may originate from a cumulant blob.
    \item Cumulant blobs are never connected to other cumulant blobs.
    \item A red propagator can never loop back into a cumulant blob.
\end{itemize}
Now, keeping the definition of a vertex and the above restrictions in mind, we are ready to draw the vertices which appear in this theory. 

Some examples of the vertices present in this theory are drawn below:
\begin{figure}[h]
    \centering
  \includegraphics[scale=0.17]{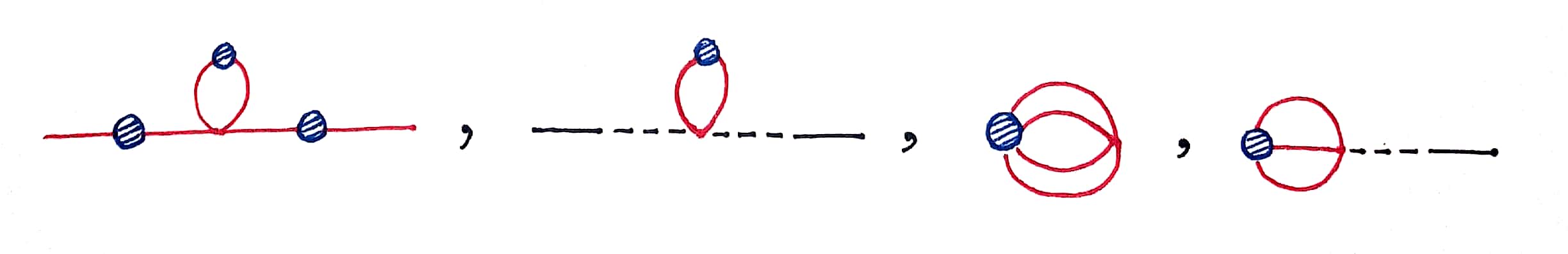}
  \caption{Examples of some interaction vertices.}
  \label{vertices}
\end{figure}
\\
Let us now show how these vertices actually arise from the perturbative analysis as outlined in \ref{pertanh}.

For the sake of illustration, consider the second vertex in figure \ref{vertices}. The claim is that it arises from an order $\lambda$ term in the perturbation expansion for the 2-point Wightman correlator in a general state of the anharmonic oscillator.

\begin{figure}[h]
    \centering
  \includegraphics[scale=0.08]{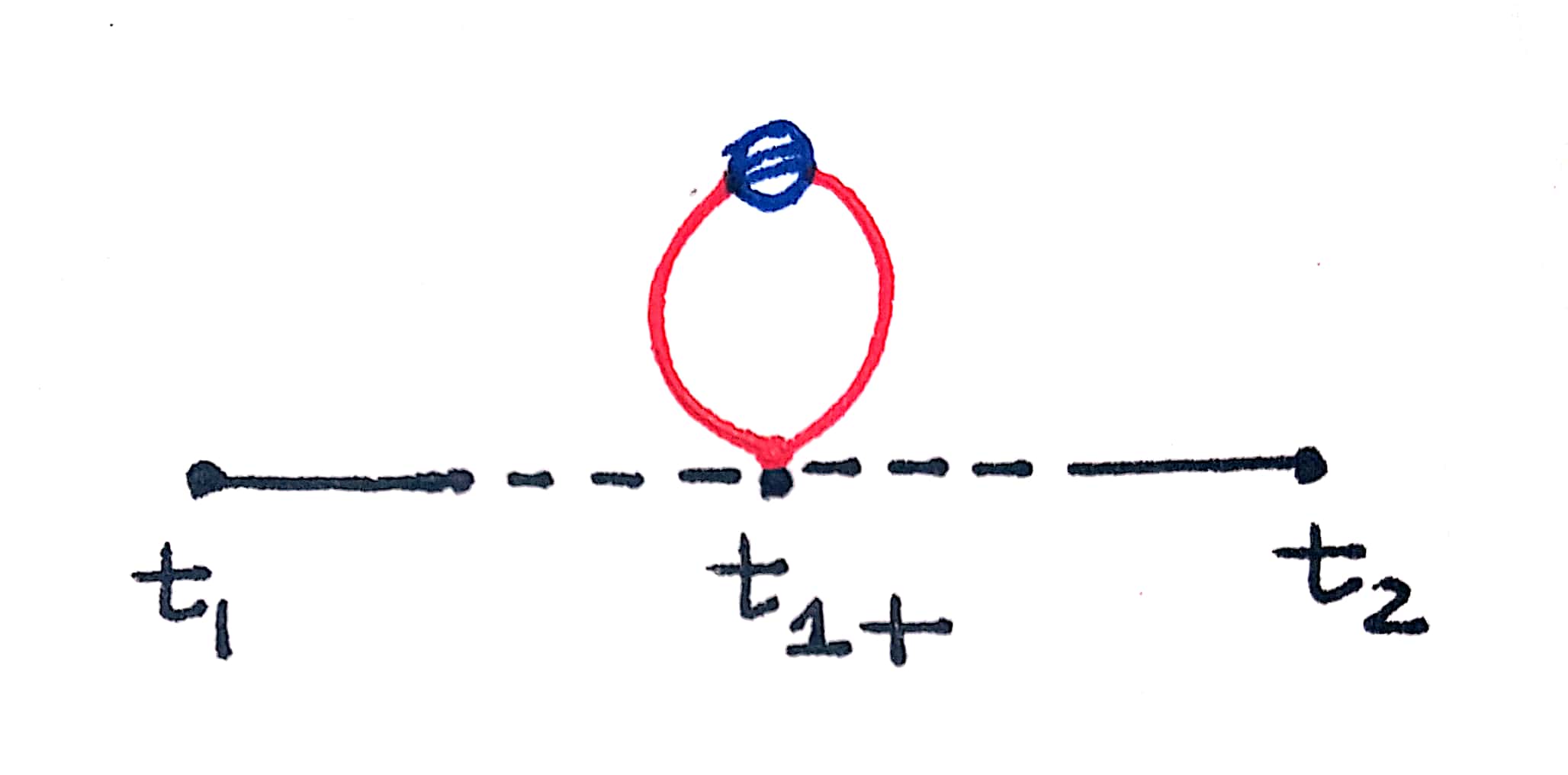}
  \caption{A diagram arising from \eqref{justifydiag}.}
  \label{justifyingdiag}
\end{figure}

The starting point of such a perturbation expansion is \eqref{2point}:
\begin{align}
    \langle x_H(t_1)x_H(t_2)\rangle=\langle&U^{\dagger}(t_1,t_0)\,x(t_1)\,U(t_1,t_0)\notag\\&U^{\dagger}(t_2,t_0)\,x(t_2)\,U(t_2,t_0)\rangle.
\end{align}
Consider expanding only the first anti-time evolution operator $U^{\dagger}(t_1,t_0)$ to the order $\lambda$:
\begin{align}
    \langle x_H(t_1)x_H(t_2)\rangle|_{\lambda}&=\langle x(t_1)x(t_2)\rangle\notag\\&+\Bigl(\frac{i\lambda}{4!}\Bigr)\int_{t_0}^{t_1}dt_{1+}\langle\biggl\{x(t_{1+})x(t_{1+})x(t_{1+})x(t_{1+})\biggr\}x(t_1)x(t_2)\rangle.
\end{align}
If we ignore the numerical factors for the moment, the following contraction pattern of the term listed above:
\begin{equation}\label{justifydiag}
    \contraction{\langle}{x}{(t_{1+})}{x}\contraction{\langle x(t_{1+})x(t_{1+})}{x}{(t_{1+})x(t_{1+})}{x}\contraction[1.5ex]{\langle x(t_{1+})x(t_{1+})x(t_{1+})}{x}{(t_{1+})x(t_1)}{x}\langle x(t_{1+})x(t_{1+})x(t_{1+})x(t_{1+})x(t_1)x(t_2)\rangle,
\end{equation}
gives rise to the diagram \ref{justifyingdiag} shown on the top of the next page.
\\
Similarly, one may justify the emergence of all the other diagrams shown in figure \ref{vertices} from similar perturbative calculations.
\\[50pt]
This section has thus taught us how to draw the different types of vertices which the analysis of Wightman correlators in general states of the anharmonic oscillator reveals. If one actually sits down to draw all the possible vertices present in this theory, one would realise that there are very many of them. In this context, if one adopts the traditional procedure of listing down the vertex factor for each possible vertex, matters would surely become very messy. 

In addition to their large number, there is another complexity associated to the vertices which emerge. It turns out that a vertex, in general, has more than one vertex factors associated to it. This point is made clear through the following section.
\subsection{The Many Vertex Factors of a Vertex}\label{many}
The aim of this section would be to motivate a novel way of assigning vertex factors in the present analysis. Through some examples, we will show that a general vertex in this theory has more than one vertex factors associated to it.

Consider the complete order $\lambda$ term in the perturbative expansion of \eqref{2point}:
\begin{align}\label{orderlamb}
    \langle x_H(t_1)x_H(t_2)\rangle|_{\lambda}&=\langle x(t_1)x(t_2)\rangle\notag\\&\quad+\Bigl(\frac{i\lambda}{4!}\Bigr)\int_{t_0}^{t_1}dt_{1+}\langle\biggl\{x(t_{1+})x(t_{1+})x(t_{1+})x(t_{1+})\biggr\}x(t_1)x(t_2)\rangle\notag\\&\quad+\Bigl(\frac{-i\lambda}{4!}\Bigr)\int_{t_0}^{t_1}dt_{1-}\langle x(t_1)\biggl\{x(t_{1-})x(t_{1-})x(t_{1-})x(t_{1-})\biggr\}x(t_2)\rangle\notag\\&\quad+\Bigl(\frac{i\lambda}{4!}\Bigr)\int_{t_0}^{t_2}dt_{2+}\langle x(t_1)\biggl\{x(t_{2+})x(t_{2+})x(t_{2+})x(t_{2+})\biggr\}x(t_2)\rangle\notag\\&\quad+\Bigl(\frac{-i\lambda}{4!}\Bigr)\int_{t_0}^{t_2}dt_{2-}\langle x(t_1)x(t_2)\biggl\{x(t_{2-})x(t_{2-})x(t_{2-})x(t_{2-})\biggr\}\rangle.
\end{align}
When one draws the diagrams corresponding to all the terms involved in \eqref{orderlamb}, a diagram which consequently emerges is:
\begin{figure}[h]
    \centering
  \includegraphics[scale=0.08]{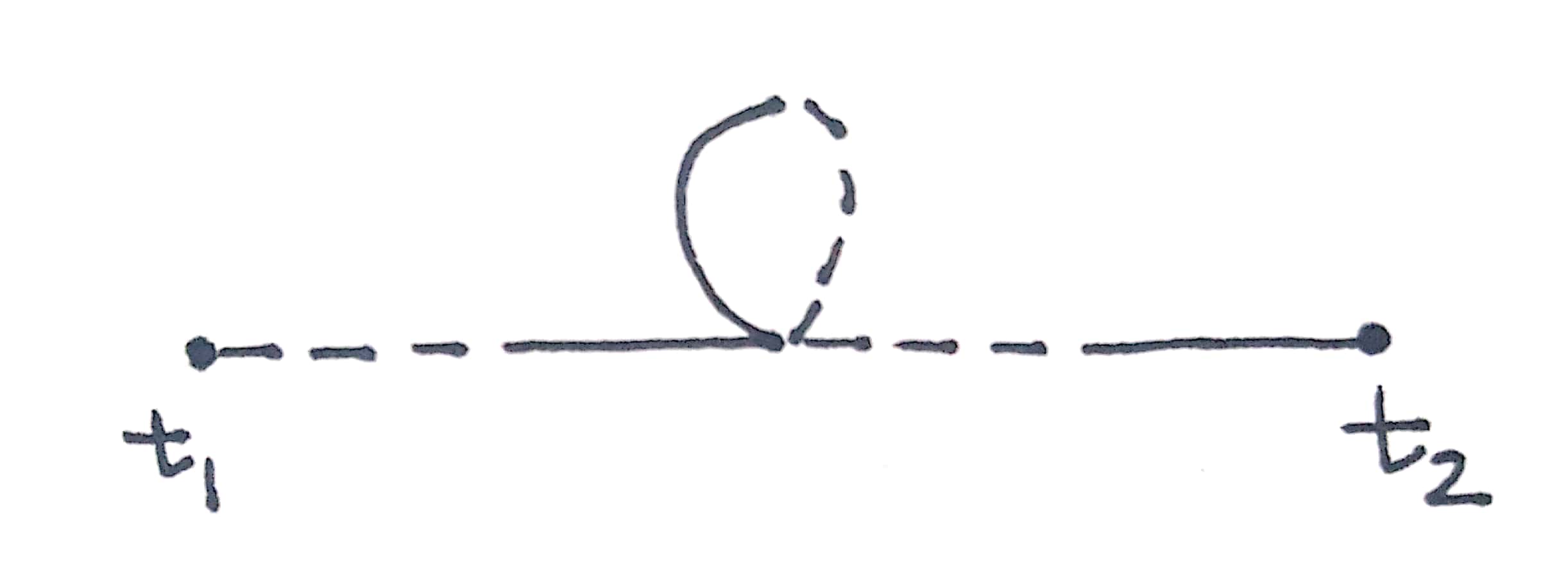}
  \caption{A diagram which emerges from \eqref{orderlamb}.}
  \label{manyvertexfacs1}
\end{figure}
\\
It results from the following two contraction patterns:
\begin{align}\label{diaglamb}
    &\Bigl(\frac{-i\lambda}{4!}\Bigr)\int_{t_0}^{t_1}dt_{1-}\langle\contraction{}{x}{(t_1)}{x}\contraction{x(t_1)x(t_{1-})}{x}{(t_{1-})}{x}\contraction{x(t_1)x(t_{1-})x(t_{1-})x(t_{1-})}{x}{(t_{1-})}{x}x(t_1)x(t_{1-})x(t_{1-})x(t_{1-})x(t_{1-})x(t_2)\rangle\notag\\+&\Bigl(\frac{i\lambda}{4!}\Bigr)\int_{t_0}^{t_2}dt_{2+}\langle\contraction{}{x}{(t_1)}{x}\contraction{x(t_1)x(t_{2+})}{x}{(t_{2+})}{x}\contraction{x(t_1)x(t_{2+})x(t_{2+})x(t_{2+})}{x}{(t_{2+})}{x}x(t_1)x(t_{2+})x(t_{2+})x(t_{2+})x(t_{2+})x(t_2)\rangle.
\end{align}
Let us ignore the combinatoric factors coming from the number of ways in which such contractions as the above can be carried out. Let us also not think about the factor of 4! for the moment. They would all be incorporated into the symmetry factors of the diagrams to be discussed in \ref{symmetry}. With the numerical factors suppressed, \eqref{diaglamb} clearly suggests that the vertex in the diagram \ref{manyvertexfacs1} has \emph{two} vertex factors.

The first vertex factor arises from the first contraction pattern shown in \eqref{diaglamb}, and it is:
\begin{equation}\label{verfaclamb}
    -i\lambda\int_{t_0}^{t_1}dt_{1-}.
\end{equation}
Similarly, the second vertex factor is also revealed through the second contraction pattern in \eqref{diaglamb}. The second vertex factor for the vertex depicted in figure \ref{manyvertexfacs1} is:
\begin{equation}\label{verfaclamb2}
    +i\lambda\int_{t_0}^{t_2}dt_{2+}.
\end{equation}
Summarising:
\begin{figure}[h]
    \centering
  \includegraphics[scale=0.15]{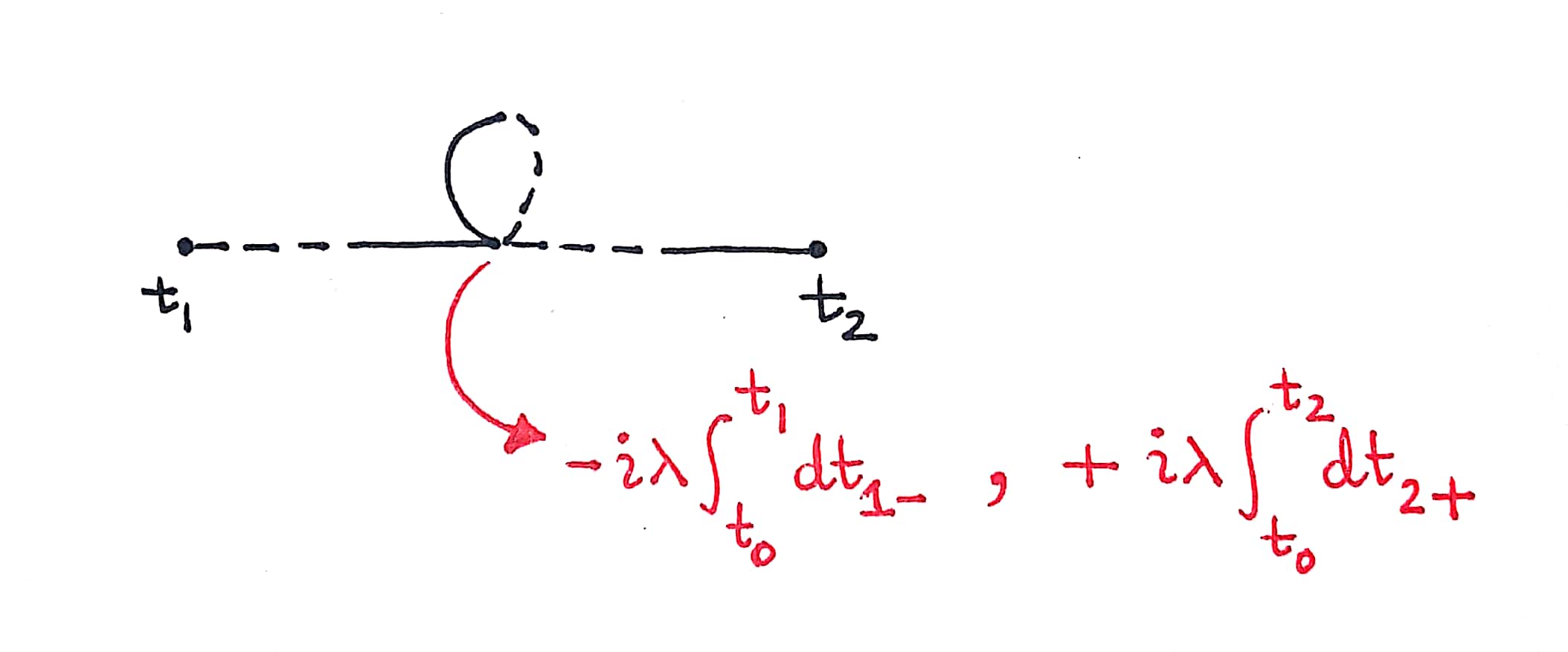}
  \caption{The two vertex factors of diagram \ref{manyvertexfacs1}.}
  \label{manyvertexfacs2}
\end{figure}
\\
Note that both of these vertex factors have a dependence on the other end points of the legs which meet at it. This is reflected in the upper limits $t_1$ and $t_2$ involved in the integrals of \eqref{verfaclamb} and \eqref{verfaclamb2}.

As another example, consider the complete order $\lambda$ term in the perturbative expansion for the 4-point Wightman correlator in a general state of the anharmonic oscillator:
\begin{equation}\label{4ptanh}
    \langle x_H(t_1)x_H(t_2)x_H(t_3)x_H(t_4)\rangle.
\end{equation}
It proves to be impractical to write down the complete order $\lambda$ term of the perturbation expansion of the above object, since it is very long. But considering only three of these terms will fulfill our purpose here. Consider the terms:
\begin{align}\label{terms}
    &\Bigl(\frac{-i\lambda}{4!}\Bigr)\int_{t_0}^{t_2}dt_{2-}\langle x(t_1)x(t_2)\biggl\{x(t_{2-})x(t_{2-})x(t_{2-})x(t_{2-})\biggr\}x(t_3)x(t_4)\rangle\notag\\+&\Bigl(\frac{i\lambda}{4!}\Bigr)\int_{t_0}^{t_3}dt_{3+}\langle x(t_1)x(t_2)\biggl\{x(t_{3+})x(t_{3+})x(t_{3+})x(t_{3+})\biggr\}x(t_3)x(t_4)\rangle\notag\\+&\Bigl(\frac{-i\lambda}{4!}\Bigr)\int_{t_0}^{t_4}dt_{4-}\langle x(t_1)x(t_2)x(t_3)x(t_4)\biggl\{x(t_{4-})x(t_{4-})x(t_{4-})x(t_{4-})\biggr\}\rangle.
\end{align}
By explicit computation, which is not outlined here because of its length, it can be shown that the above three terms indeed form a part of the perturbation expansion of the 4-point Wightman correlator \eqref{4ptanh}.

Now consider the following contraction patterns in evaluating the first and second terms of \eqref{terms}:
\begin{align}\label{terms1}
    &\Bigl(\frac{-i\lambda}{4!}\Bigr)\int_{t_0}^{t_2}dt_{2-}\langle\contraction{}{x}{(t_1)x(t_2)}{x}\contraction[1.5ex]{x(t_1)}{x}{(t_2)x(t_{2-})}{x}\contraction{x(t_1)x(t_2)x(t_{2-})x(t_{2-})}{x}{(t_{2-})x(t_{2-})}{x}\contraction[1.5ex]{x(t_1)x(t_2)x(t_{2-})x(t_{2-})x(t_{2-})}{x}{(t_{2-})x(t_3)}{x}x(t_1)x(t_2)x(t_{2-})x(t_{2-})x(t_{2-})x(t_{2-})x(t_3)x(t_4)\rangle\notag\\+&\Bigl(\frac{i\lambda}{4!}\Bigr)\int_{t_0}^{t_3}dt_{3+}\langle\contraction{}{x}{(t_1)x(t_2)}{x}\contraction[1.5ex]{x(t_1)}{x}{(t_2)x(t_{3+})}{x}\contraction{x(t_1)x(t_2)x(t_{3+})x(t_{3+})}{x}{(t_{3+})x(t_{3+})}{x}\contraction[1.5ex]{x(t_1)x(t_2)x(t_{3+})x(t_{3+})x(t_{3+})}{x}{(t_{3+})x(t_3)}{x}x(t_1)x(t_2)x(t_{3+})x(t_{3+})x(t_{3+})x(t_{3+})x(t_3)x(t_4)\rangle.
\end{align}
Both of the contraction patterns above give rise to the diagram:
\begin{figure}[h]
    \centering
  \includegraphics[scale=0.06]{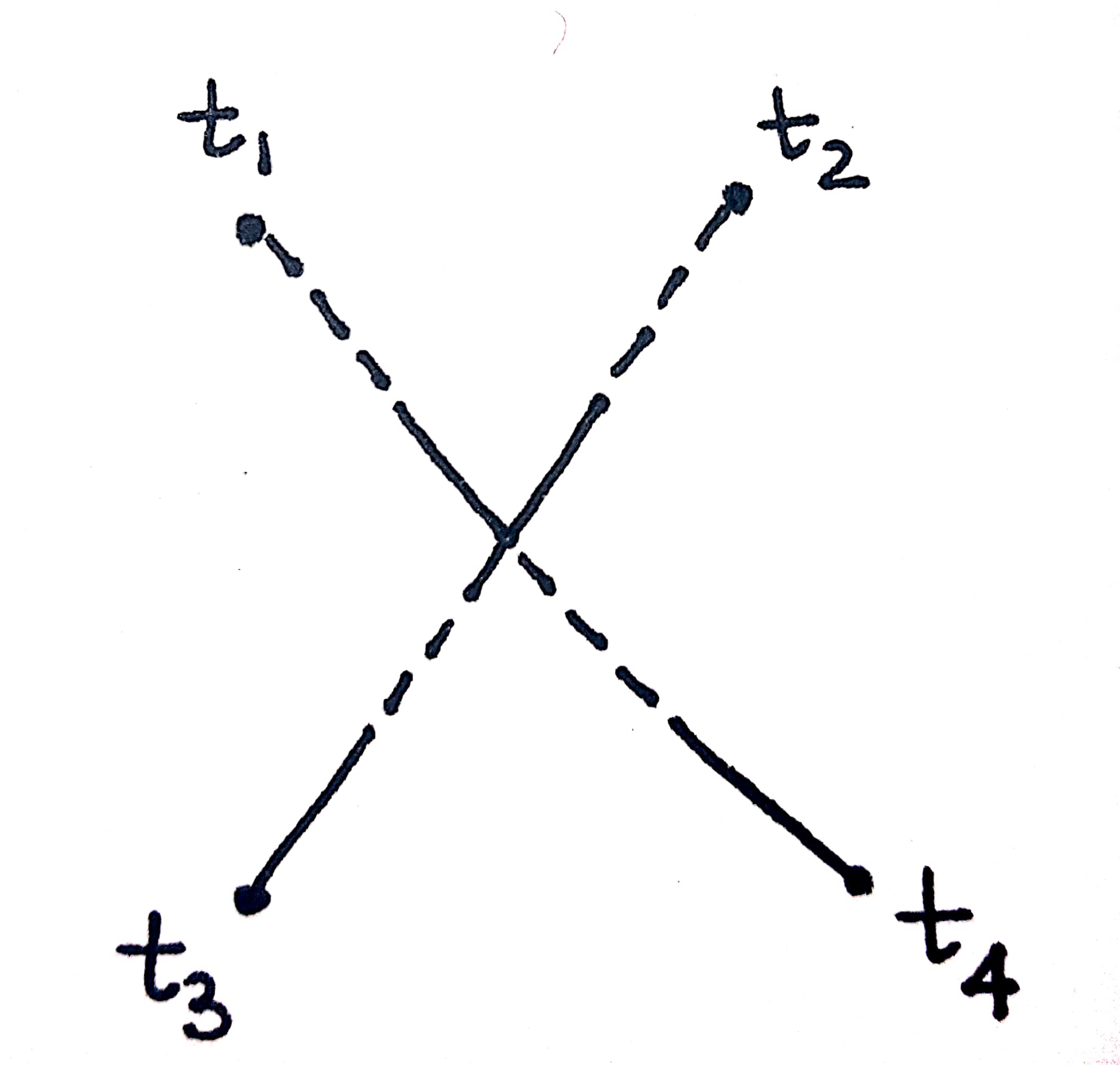}
  \caption{The diagram arising from \eqref{terms1}.}
  \label{manyvertexfacs3}
\end{figure}
\\
It can be checked by explicit calculation that the above diagram is only generated by the contraction patterns \eqref{terms1}.

Again ignoring the combinatoric factors related to the number of ways in which the contractions \eqref{terms1} can be carried out and the factors of 4!, we conclude that the vertex depicted in the figure \ref{manyvertexfacs3} has \emph{two} vertex factors. One of them arises from the first term of \eqref{terms1}, it being:
\begin{equation}
    -i\lambda\int_{t_0}^{t_2}dt_{2-}.
\end{equation}
The other vertex factor associated to the vertex depicted in figure \ref{manyvertexfacs3} is:
\begin{equation}
    +i\lambda\int_{t_0}^{t_3}dt_{3+},
\end{equation}
and it arises from the second term of \eqref{terms1}.

That is:
\begin{figure}[h]
    \centering
  \includegraphics[scale=0.13]{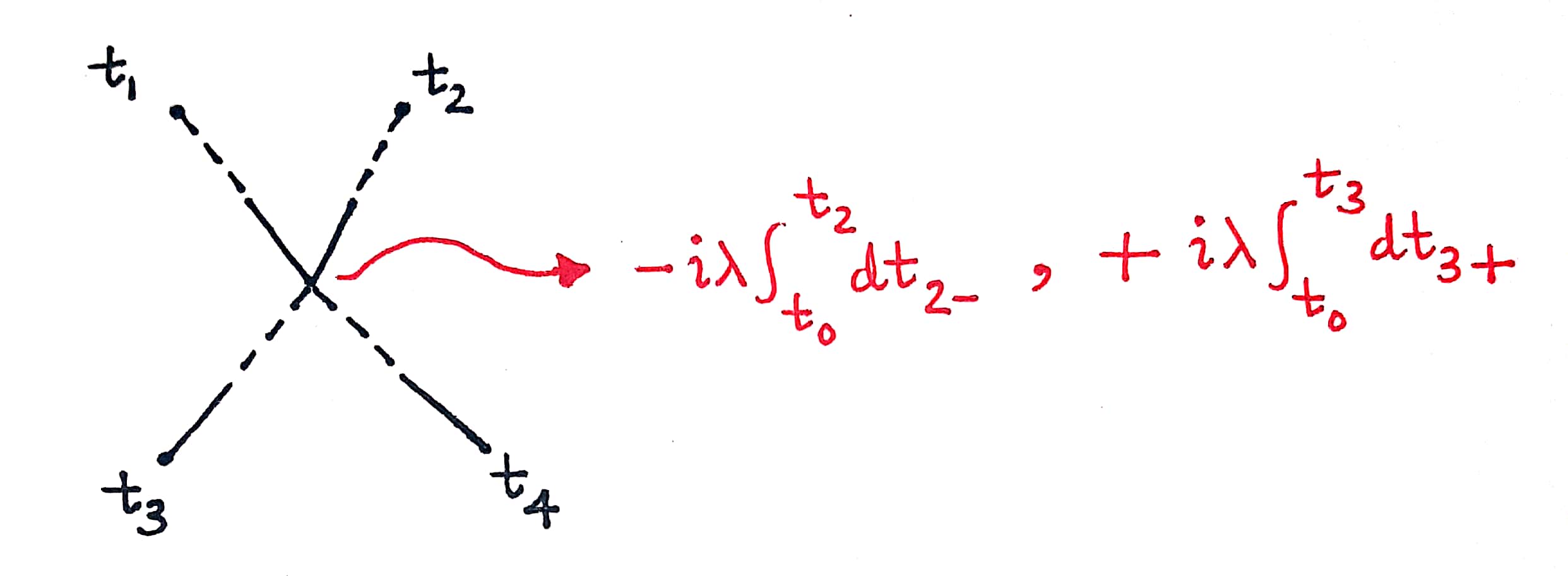}
  \caption{The two vertex factors of the diagram \ref{manyvertexfacs3}.}
  \label{manyvertexfacs4}
\end{figure}
\\
On the other hand, consider the following contraction pattern in evaluating the last term of \eqref{terms}:
\begin{align}\label{terms2}
    \Bigl(\frac{-i\lambda}{4!}\Bigr)\int_{t_0}^{t_4}dt_{4-}\langle\contraction[2ex]{}{x}{(t_1)x(t_2)x(t_3)x(t_4)x(t_{4-})x(t_{4-})x(t_{4-})}{x}\contraction[1.5ex]{x(t_1)}{x}{(t_2)x(t_3)x(t_4)x(t_{4-})x(t_{4-})}{x}\contraction{x(t_1)x(t_2)}{x}{(t_3)x(t_4)x(t_{4-})}{x}\contraction[0.5ex]{x(t_1)x(t_2)x(t_3)}{x}{(t_4)}{x}x(t_1)x(t_2)x(t_3)x(t_4)x(t_{4-})x(t_{4-})x(t_{4-})x(t_{4-})\rangle.
\end{align}
This gives rise to the diagram:
\begin{figure}[h]
    \centering
  \includegraphics[scale=0.07]{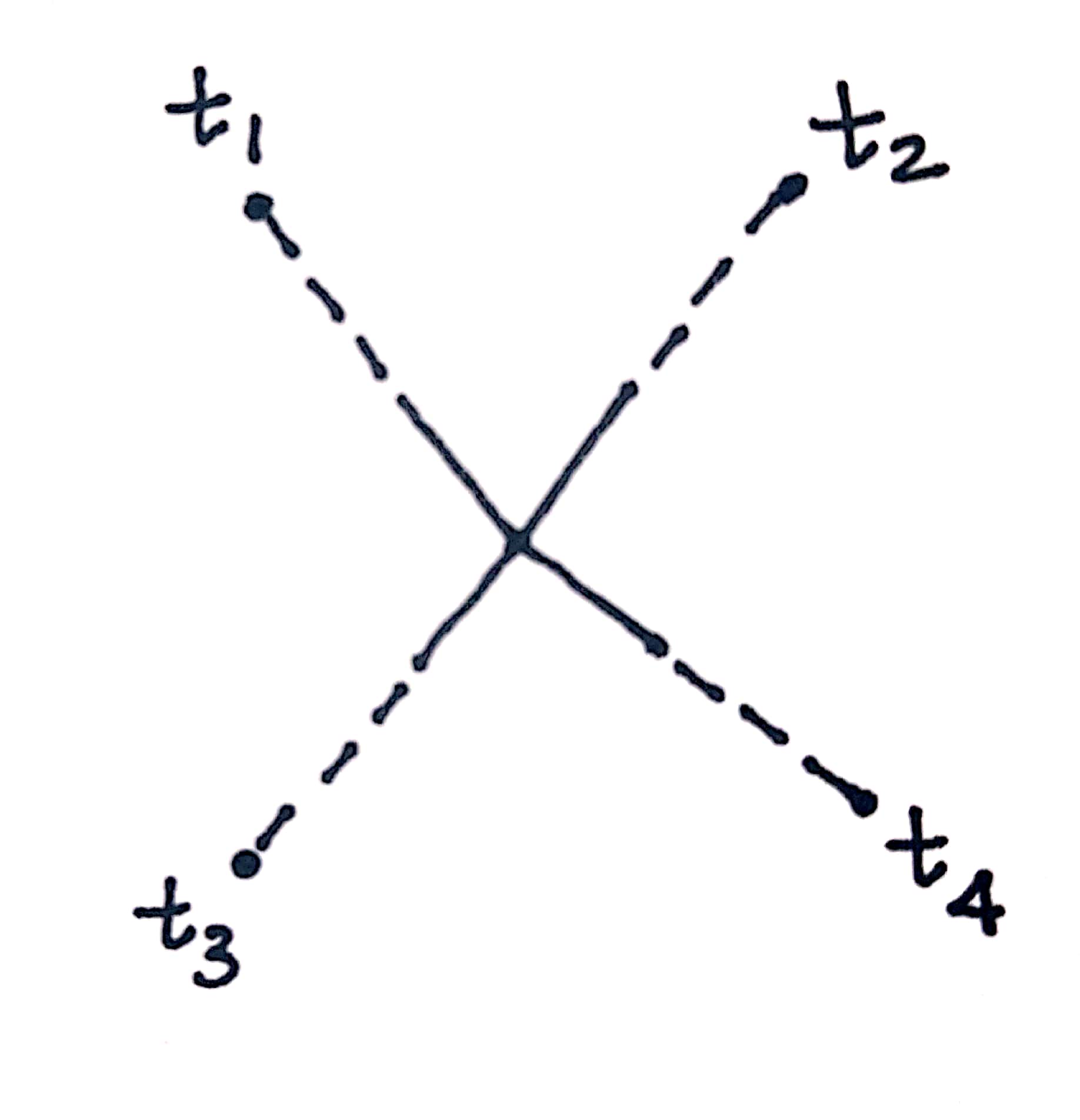}
  \caption{The diagram arising from \eqref{terms2}.}
  \label{manyvertexfacs5}
\end{figure}
\\
It can again be checked by explicit calculation that the above diagram is only generated by the contraction pattern \eqref{terms2}.

Again ignoring the combinatoric factors related to the number of ways in which the contraction \eqref{terms2} can be carried out and the factor of 4!, we conclude that the vertex factor of the diagram \ref{manyvertexfacs5} is:
\begin{equation}\label{verfac}
    -i\lambda\int_{t_0}^{t_4}dt_{4-}.
\end{equation}
\\[15pt]
\begin{center}
\textcolor{blue}{\textbf{(Continued on next page)}}
\end{center}
\newpage
Thus, the vertex depicted in figure \ref{manyvertexfacs5} has only \emph{one} vertex factor, it being \eqref{verfac}. That is:
\begin{figure}[h]
    \centering
  \includegraphics[scale=0.085]{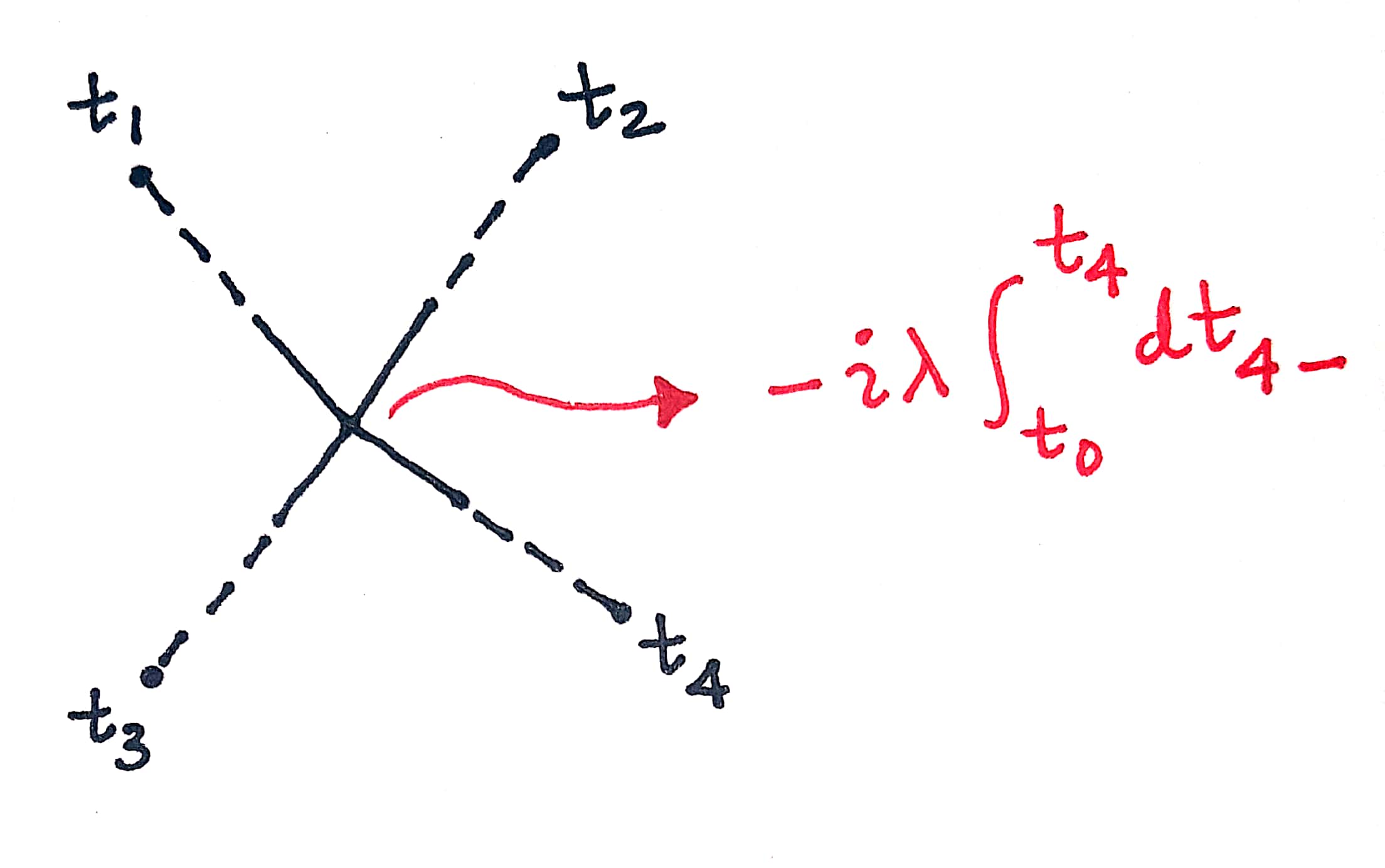}
  \caption{The vertex factor of the diagram \ref{manyvertexfacs5}.}
  \label{manyvertexfacs6}
\end{figure}
\\
All these examples have made it clear that vertices in this theory can have more than one vertex factors associated to them. And, as previously outlined, there are already many types of vertices which the analysis of Wightman correlators in general states of the anharmonic oscillator presents.

In such a situation, if one thinks of going by the usual procedure of listing down the vertex factors of each possible vertex, it would certainly be very impractical. The simplest way of assigning vertex factors in the present consideration proceeds by way of first \emph{labelling} the vertices of a diagram. The vertex factors of the vertices are then simply read off from the labels they have been assigned. 

To begin with, we explain the origin of these labels in the next section. Consequent sections will present the procedure which one must adopt to label the vertices of a given diagram.
\subsection{What Are Vertex Labels?}\label{not}
Simply put, \textit{the labels of a vertex are nothing but the time variables which are integrated over in the corresponding analytic expression for the vertex}. Let us make this clear through an example. 

Consider the case of the 2-point anharmonic Wightman correlator:
\begin{equation}
    \langle x_H(t_1)x_H(t_2)\rangle.
\end{equation}
A diagram which emerges at order $\lambda$ in the perturbation expansion of the above object is:
\begin{figure}[h]
    \centering
  \includegraphics[scale=0.1]{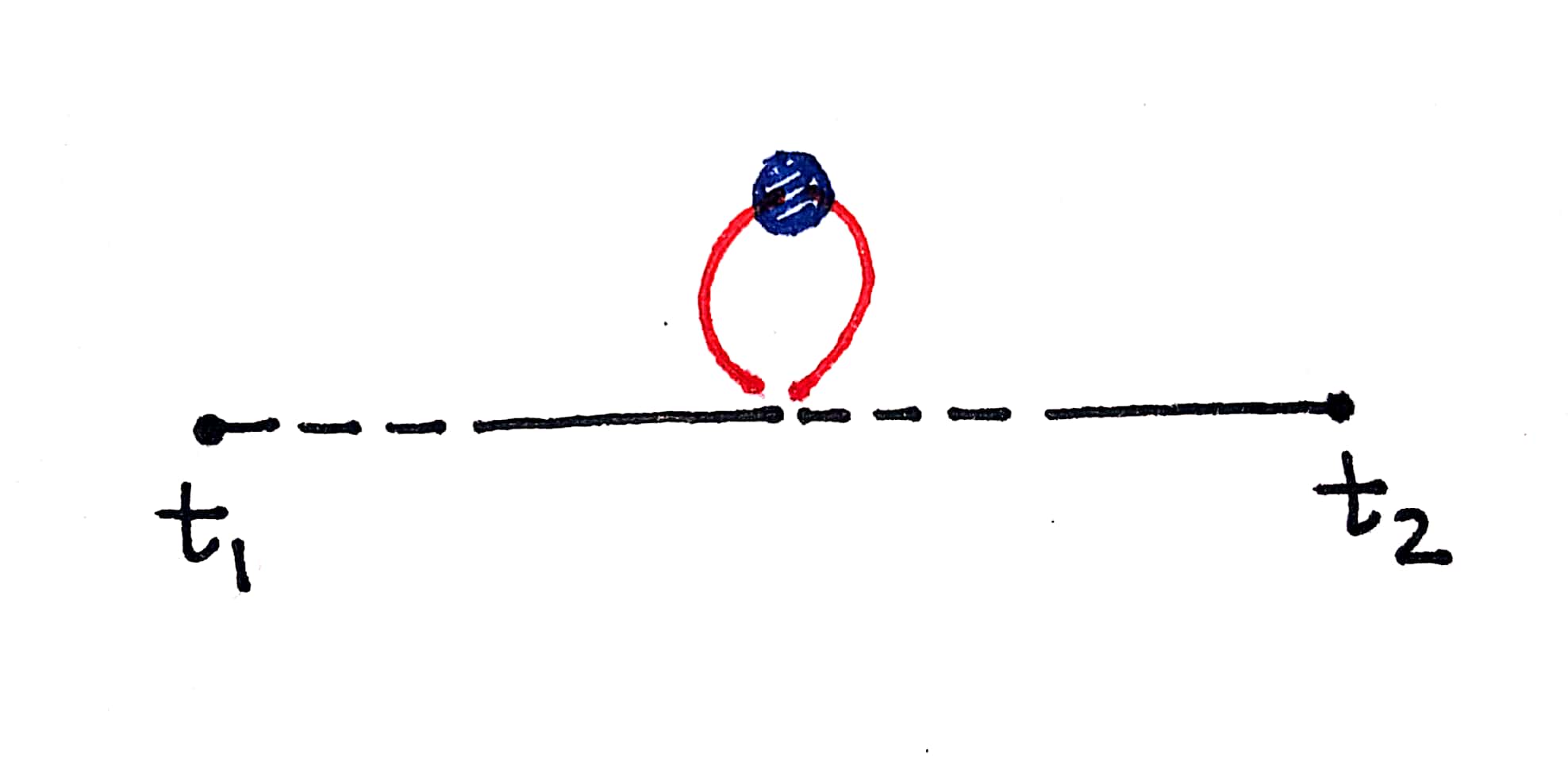}
  \caption{A diagram appearing at order $\lambda$ of the perturbation expansion of $\langle x_H(t_1)x_H(t_2)\rangle$.}
  \label{vertexlabell1}
\end{figure}
\newpage
Let us ask what the labels of the vertex appearing in the above diagram should be. For that, we must go back to the contraction patterns which lead to it. 

As one can check, the diagram \ref{vertexlabell1} is generated only by the following two contraction patterns:
\begin{align}\label{diaglambbb}
    &\Bigl(\frac{-i\lambda}{4!}\Bigr)\int_{t_0}^{t_1}dt_{1-}\langle\contraction{}{x}{(t_1)}{x}\contraction{x(t_1)x(t_{1-})}{x}{(t_{1-})}{x}\contraction{x(t_1)x(t_{1-})x(t_{1-})x(t_{1-})}{x}{(t_{1-})}{x}x(t_1)x(t_{1-})x(t_{1-})x(t_{1-})x(t_{1-})x(t_2)\rangle\notag\\+&\Bigl(\frac{i\lambda}{4!}\Bigr)\int_{t_0}^{t_2}dt_{2+}\langle\contraction{}{x}{(t_1)}{x}\contraction{x(t_1)x(t_{2+})}{x}{(t_{2+})}{x}\contraction{x(t_1)x(t_{2+})x(t_{2+})x(t_{2+})}{x}{(t_{2+})}{x}x(t_1)x(t_{2+})x(t_{2+})x(t_{2+})x(t_{2+})x(t_2)\rangle.
\end{align}
This is because the diagram \ref{vertexlabell1} has a free Wightman propagator going out of the external point $t_1$ and a free Wightman propagator coming into the external point $t_2$. A situation like this can only be achieved when the external point $t_1$ contracts with an internal point which is placed to its right in the correlator and the external point $t_2$ contracts with an internal point which is placed to its left in the correlator. 

With the expression \eqref{diaglambbb} in hand, the labels to the vertex in diagram \ref{vertexlabell1} are readily assigned. In the first contraction pattern of \eqref{diaglambbb}, the time variable getting integrated over is $t_{1-}$. So, $t_{1-}$ is said to be one of the labels of the vertex depicted in diagram \ref{vertexlabell1}. Similarly, in the second contraction pattern of \eqref{diaglambbb}, the time variable getting integrated over is $t_{2+}$. As a result, $t_{2+}$ is said to be the second label of the vertex depicted in \ref{vertexlabell1}.

Thus, the set of labels associated to the vertex in \ref{vertexlabell1} is $\{t_{1-},t_{2+}\}$.

With this done, we can label the diagram \ref{vertexlabell1} as:
\begin{figure}[h]
    \centering
  \includegraphics[scale=0.16]{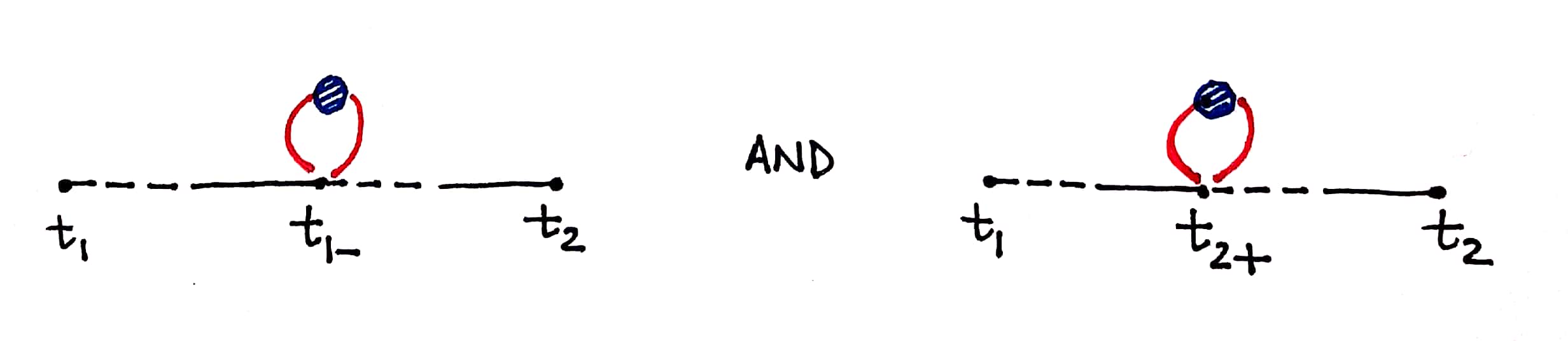}
  \caption{The labels of the vertex in diagram \ref{vertexlabell1}.}
  \label{vertexlabel2}
\end{figure}
\\
Let us go through another example. This example would be a bit detailed, but it is important because it would show us that \emph{labels for vertices cannot be assigned independently}. The label for a vertex, in general, depends on the labels of the other vertices in the diagram.  
\\[15pt]
\begin{center}
\textcolor{blue}{\textbf{(Continued on next page)}}
\end{center}
\newpage
Consider a diagram which emerges at order $\lambda^2$ in the perturbation expansion of the 2-point anharmonic Wightman correlator:
\begin{figure}[h]
    \centering
  \includegraphics[scale=0.12]{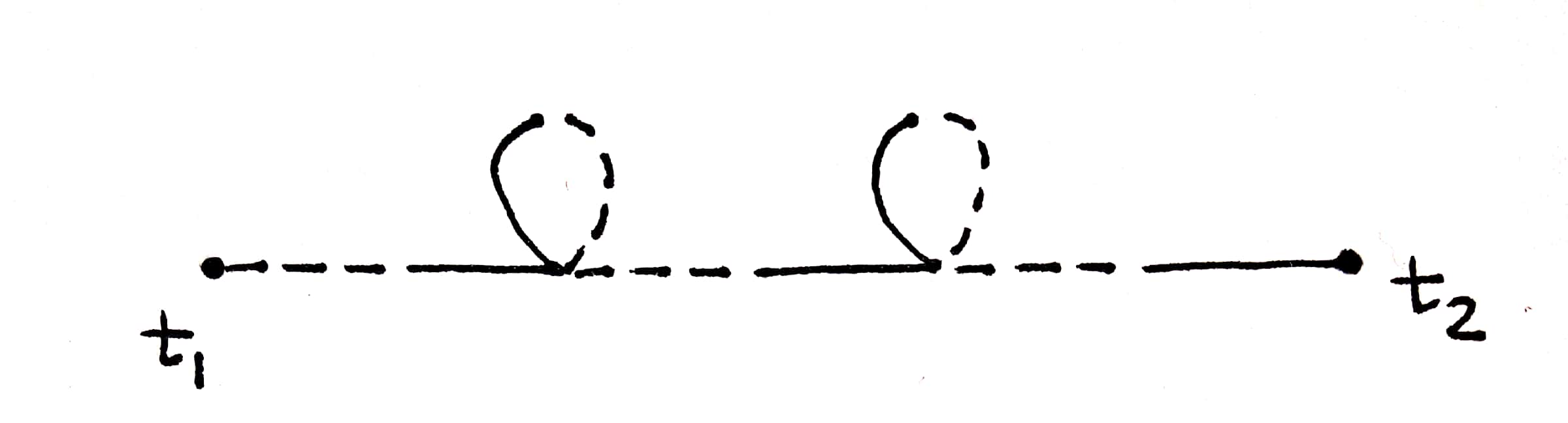}
  \caption{A diagram appearing at order $\lambda^2$ of the perturbation expansion of $\langle x_H(t_1)x_H(t_2)\rangle$.}
  \label{vertexlabell3}
\end{figure}
\\
How do we label the two vertices in the above diagram? As before, to do so, we go into the contraction patterns which lead to the diagram \ref{vertexlabell3}. As one can check, the diagram \ref{vertexlabell3} results only from the terms:
\begin{align}\label{lambdasq}
    &\Bigl(\frac{-i\lambda}{4!}\Bigr)^{2}\int_{t_0}^{t_1}dt^{(1)}_{1-}\int_{t_0}^{t_1}dt^{(2)}_{1-}\langle x(t_1)\biggl\{Tx(t^{(1)}_{1-})x(t^{(1)}_{1-})x(t^{(1)}_{1-})x(t^{(1)}_{1-})\notag\\&\hspace{170pt}x(t^{(2)}_{1-})x(t^{(2)}_{1-})x(t^{(2)}_{1-})x(t^{(2)}_{1-})\biggr\}x(t_2)\rangle\notag\\
    +&\Bigl(\frac{-i\lambda}{4!}\Bigr)\Bigl(\frac{i\lambda}{4!}\Bigr)\int_{t_0}^{t_1}dt_{1-}\int_{t_0}^{t_2}dt_{2+}\langle x(t_1)\biggl\{x(t_{1-})x(t_{1-})x(t_{1-})x(t_{1-})\biggr\}\notag\\&\hspace{170pt}\biggl\{x(t_{2+})x(t_{2+})x(t_{2+})x(t_{2+})\biggr\}x(t_2)\rangle\notag\\+&\Bigl(\frac{i\lambda}{4!}\Bigr)^{2}\int_{t_0}^{t_2}dt^{(1)}_{2+}\int_{t_0}^{t_2}dt^{(2)}_{2+}\langle x(t_1)\biggl\{T^*x(t^{(1)}_{2+})x(t^{(1)}_{2+})x(t^{(1)}_{2+})x(t^{(1)}_{2+})\notag\\&\hspace{170pt}x(t^{(2)}_{2+})x(t^{(2)}_{2+})x(t^{(2)}_{2+})x(t^{(2)}_{2+})\biggr\}x(t_2)\rangle.
\end{align}
The first and the last terms of the above expression involve time ordering and anti-time ordering respectively. For making the further analysis clear, it proves to be fruitful to expand them using step functions of time. The first term of \eqref{lambdasq} can be expanded as:
\begin{align}\label{t}
    &\langle x(t_1)\biggl\{Tx(t^{(1)}_{1-})x(t^{(1)}_{1-})x(t^{(1)}_{1-})x(t^{(1)}_{1-})x(t^{(2)}_{1-})x(t^{(2)}_{1-})x(t^{(2)}_{1-})x(t^{(2)}_{1-})\biggr\}x(t_2)\rangle\notag\\=&\theta(t^{(1)}_{1-}-t^{(2)}_{1-})\langle x(t_1)\biggl\{x(t^{(1)}_{1-})x(t^{(1)}_{1-})x(t^{(1)}_{1-})x(t^{(1)}_{1-})\biggr\}\biggl\{x(t^{(2)}_{1-})x(t^{(2)}_{1-})x(t^{(2)}_{1-})x(t^{(2)}_{1-})\biggr\}x(t_2)\rangle\notag\\+&\theta(t^{(2)}_{1-}-t^{(1)}_{1-})\langle x(t_1)\biggl\{x(t^{(2)}_{1-})x(t^{(2)}_{1-})x(t^{(2)}_{1-})x(t^{(2)}_{1-})\biggr\}\biggl\{x(t^{(1)}_{1-})x(t^{(1)}_{1-})x(t^{(1)}_{1-})x(t^{(1)}_{1-})\biggr\}x(t_2)\rangle.
\end{align}
\newpage
Similarly, the last term of \eqref{lambdasq} can be expanded as:
\begin{align}\label{tstar}
    &\langle x(t_1)\biggl\{T^*x(t^{(1)}_{2+})x(t^{(1)}_{2+})x(t^{(1)}_{2+})x(t^{(1)}_{2+})x(t^{(2)}_{2+})x(t^{(2)}_{2+})x(t^{(2)}_{2+})x(t^{(2)}_{2+})\biggr\}x(t_2)\rangle\notag\\=&\theta(t^{(2)}_{2+}-t^{(1)}_{2+})\langle x(t_1)\biggl\{x(t^{(1)}_{2+})x(t^{(1)}_{2+})x(t^{(1)}_{2+})x(t^{(1)}_{2+})\biggr\}\biggl\{x(t^{(2)}_{2+})x(t^{(2)}_{2+})x(t^{(2)}_{2+})x(t^{(2)}_{2+})\biggr\}x(t_2)\rangle\notag\\+&\theta(t^{(1)}_{2+}-t^{(2)}_{2+})\langle x(t_1)\biggl\{x(t^{(2)}_{2+})x(t^{(2)}_{2+})x(t^{(2)}_{2+})x(t^{(2)}_{2+})\biggr\}\biggl\{x(t^{(1)}_{2+})x(t^{(1)}_{2+})x(t^{(1)}_{2+})x(t^{(1)}_{2+})\biggr\}x(t_2)\rangle.
\end{align}
We will now assume that the vertex nearer the external point $t_1$ in the diagram \ref{vertexlabell3} corresponds to the internal point $t^{(1)}_{1-}$ among $t^{(1)}_{1-}$ and $t^{(2)}_{1-}$. Similarly, among $t^{(1)}_{2+}$ and $t^{(2)}_{2+}$, it is assumed to correspond to $t^{(1)}_{2+}$. Note that these two choices will not affect any of our results. This is because these two sets of times are interchangeable dummy variables. This is clearly seen from the first and last terms of \eqref{lambdasq}. With this choice made, the diagram \ref{vertexlabell3} is a result of the contractions:
\begin{align}\label{big}
    &\Bigl(\frac{-i\lambda}{4!}\Bigr)^{2}\int_{t_0}^{t_1}dt^{(1)}_{1-}\int_{t_0}^{t_1}dt^{(2)}_{1-}\,\theta(t^{(1)}_{1-}-t^{(2)}_{1-})\langle\contraction{}{x}{(t_1)}{x}\contraction[1.5ex]{x(t_1)x(t^{(1)}_{1-})}{x}{(t^{(1)}_{1-})}{x}x(t_1)x(t^{(1)}_{1-})x(t^{(1)}_{1-})x(t^{(1)}_{1-})\notag\\&\hspace{190pt}\contraction[1.5ex]{}{x}{(t^{(1)}_{1-})}{x}\contraction[1.5ex]{x(t^{(1)}_{1-})x(t^{(2)}_{1-})}{x}{(t^{(2)}_{1-})}{x}\contraction[1.5ex]{x(t^{(1)}_{1-})x(t^{(2)}_{1-})x(t^{(2)}_{1-})x(t^{(2)}_{1-})}{x}{(t^{(2)}_{1-})}{x}x(t^{(1)}_{1-})x(t^{(2)}_{1-})x(t^{(2)}_{1-})x(t^{(2)}_{1-})x(t^{(2)}_{1-})x(t_2)\rangle\notag\\+&\Bigl(\frac{-i\lambda}{4!}\Bigr)\Bigl(\frac{i\lambda}{4!}\Bigr)\int_{t_0}^{t_1}dt_{1-}\int_{t_0}^{t_2}dt_{2+}\langle\contraction{}{x}{(t_1)}{x}\contraction{x(t_1)x(t_{1-})}{x}{(t_{1-})}{x}\contraction{x(t_1)x(t_{1-})x(t_{1-})x(t_{1-})}{x}{(t_{1-})}{x}\contraction{x(t_1)x(t_{1-})x(t_{1-})x(t_{1-})x(t_{1-})x(t_{2+})}{x}{(t_{2+})}{x}\contraction{x(t_1)x(t_{1-})x(t_{1-})x(t_{1-})x(t_{1-})x(t_{2+})x(t_{2+})x(t_{2+})}{x}{(t_{2+})}{x}x(t_1)x(t_{1-})x(t_{1-})x(t_{1-})x(t_{1-})x(t_{2+})x(t_{2+})x(t_{2+})x(t_{2+})x(t_2)\rangle\notag\\+&\Bigl(\frac{i\lambda}{4!}\Bigr)^{2}\int_{t_0}^{t_2}dt^{(1)}_{2+}\int_{t_0}^{t_2}dt^{(2)}_{2+}\,\theta(t^{(2)}_{2+}-t^{(1)}_{2+})\langle\contraction{}{x}{(t_1)}{x}\contraction[1.5ex]{x(t_1)x(t^{(1)}_{2+})}{x}{(t^{(1)}_{2+})}{x}x(t_1)x(t^{(1)}_{2+})x(t^{(1)}_{2+})x(t^{(1)}_{2+})\notag\\&\hspace{190pt}\contraction[1.5ex]{}{x}{(t^{(1)}_{2+})}{x}\contraction[1.5ex]{x(t^{(1)}_{2+})x(t^{(2)}_{2+})}{x}{(t^{(2)}_{2+})}{x}\contraction[1.5ex]{x(t^{(1)}_{2+})x(t^{(2)}_{2+})x(t^{(2)}_{2+})x(t^{(2)}_{2+})}{x}{(t^{(2)}_{2+})}{x}x(t^{(1)}_{2+})x(t^{(2)}_{2+})x(t^{(2)}_{2+})x(t^{(2)}_{2+})x(t^{(2)}_{2+})x(t_2)\rangle.
\end{align}
An important point to note here is that the following contraction pattern:
\begin{equation}\label{notcontribute}
    \Bigl(\frac{-i\lambda}{4!}\Bigr)\Bigl(\frac{i\lambda}{4!}\Bigr)\int_{t_0}^{t_1}dt_{1-}\int_{t_0}^{t_2}dt_{2+}\langle\contraction[2ex]{}{x}{(t_1)x(t_{1-})x(t_{1-})x(t_{1-})x(t_{1-})}{x}\bcontraction[2ex]{x(t_1)}{x}{(t_{1-})x(t_{1-})x(t_{1-})x(t_{1-})x(t_{2+})x(t_{2+})x(t_{2+})x(t_{2+})}{x}\contraction{x(t_1)x(t_{1-})}{x}{(t_{1-})}{x}\bcontraction{x(t_1)x(t_{1-})x(t_{1-})x(t_{1-})}{x}{(t_{1-})x(t_{2+})x(t_{2+})x(t_{2+})}{x}\contraction{x(t_1)x(t_{1-})x(t_{1-})x(t_{1-})x(t_{1-})x(t_{2+})}{x}{(t_{2+})}{x}x(t_1)x(t_{1-})x(t_{1-})x(t_{1-})x(t_{1-})x(t_{2+})x(t_{2+})x(t_{2+})x(t_{2+})x(t_2)\rangle,
\end{equation}
does not contribute to the diagram \ref{vertexlabell3}. This is because owing to the definition of a bi-dentate contraction as introduced in \ref{contrdiag}, the above contraction will lead to a situation like:
\begin{figure}[h]
    \centering
  \includegraphics[scale=0.1]{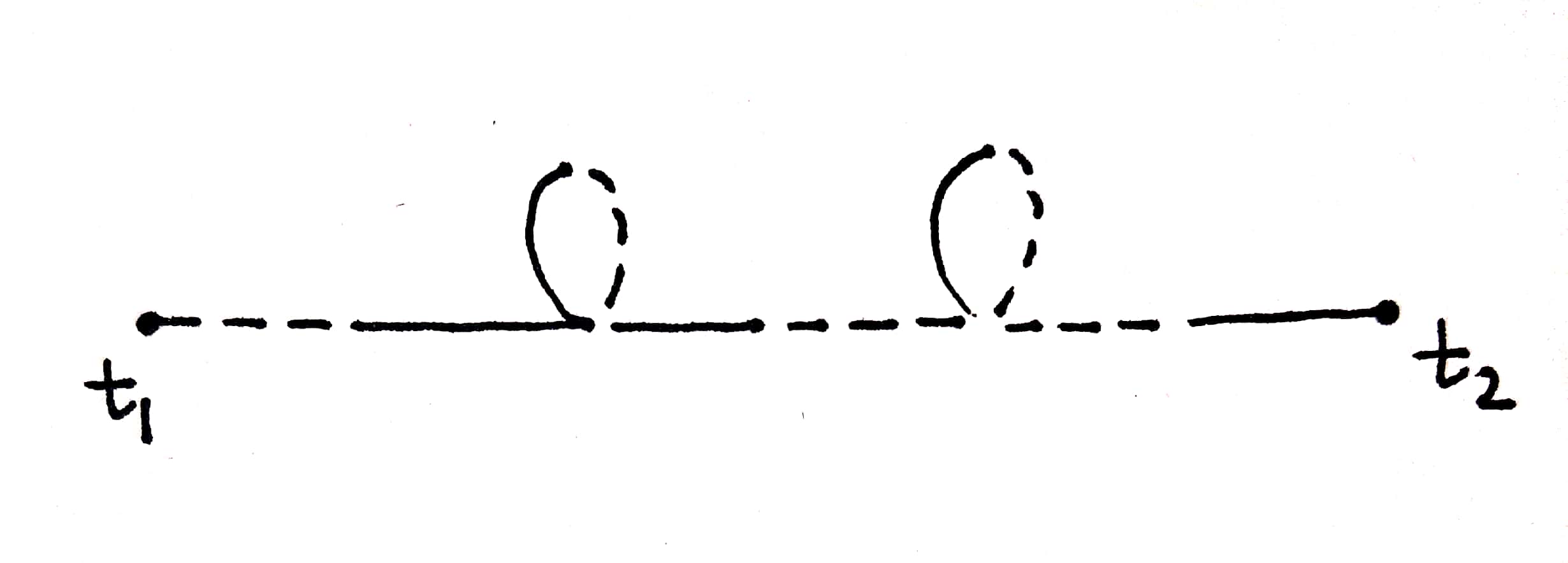}
  \caption{The diagram resulting from \eqref{notcontribute}.}
  \label{vertexlabel4}
\end{figure}
\\
wherein the internal propagator is coming \emph{into} the vertex nearer the external point $t_1$. Clearly, this is not the situation we are considering.

\eqref{big} will now tell us the labels we can assign the two vertices in the diagram \ref{vertexlabell3}. Consider the first term of \eqref{big}. $t^{(1)}_{1-}$ has already been assumed to be the label of the vertex which is nearer $t_1$. Thus, the second vertex automatically gets the label $t^{(2)}_{1-}$. Also note that in this situation, there is an additional factor of $\theta(t^{(1)}_{1-}-t^{(2)}_{1-})$ to be taken care of. This is the basic essence of the step function weights which must be assigned to the internal propagators in a diagram. This aspect will be discussed separately under the section \ref{stepfn}.

Now look at the second term of \eqref{big}. The external time $t_1$ is getting contracted with the internal point $t_{1-}$. This fixes the label of the vertex nearer $t_1$ to be $t_{1-}$. And consequently, the second vertex gets labelled $t_{2+}$.

Finally, let us look at the third term of \eqref{big}. Again, $t^{(1)}_{2+}$ has already been assumed to be the label of the vertex which is nearer $t_1$. Thus, the second vertex automatically gets the label $t^{(2)}_{2+}$. 

Let us introduce a more concise notation at this juncture. Through this example, we have seen labels like $t^{(1)}_{1-}$ and $t^{(2)}_{1-}$. We have also realised that these two are completely equivalent, since they are interchangeable dummy variables. A need to differentiate between them only arises in writing the step function factors which appear in terms like \eqref{t} and \eqref{tstar}. This will be discussed in \ref{stepfn}. As far as labelling vertices is concerned, all labels of the form $t^{(i)}_{j+}$ are equivalent, and would thus be collectively denoted as $t_{j+}$. Similarly, all labels of the form $t^{(i)}_{j-}$ are equivalent, and would thus be collectively denoted as $t_{j-}$.

The task of label assignments to the vertices is thus over. Let us summarise the results in the form of a table:
\begin{table}[ht]
\centering
\begin{tabular}{ |p{3.5cm}|p{3.5cm}|  }
\hline
Label of Vertex 1&Label of Vertex 2\\
\hline
$t_{1-}$ & $t_{1-}$\\
$t_{1-}$ & $t_{2+}$\\
$t_{2+}$ & $t_{2+}$\\
\hline
\end{tabular}
\caption{\label{tab:table}Labelling the vertices of diagram \ref{vertexlabell3}. \emph{Vertex 1} refers to the vertex nearer the external point $t_1$, and \emph{Vertex 2} refers to the other vertex.}
\end{table}
\\
\\
The example just discussed has thus made the point clear that labels for vertices cannot be assigned independently. They depend on the labels which the other vertices have been assigned. This is visible from the table. For example, let us say that the first vertex is assigned the label $t_{2+}$. This corresponds to the last row in the table. Once this is done, one can only assign the label $t_{2+}$ to the second vertex. One cannot assign the label $t_{1-}$ to it, even though it is a completely valid label for it in some other situation.

In this section, we have seen what vertex labels actually are, and how to assign them to vertices from first principles. The procedure we have outlined herein is certainly a very tedious one. It involves going back to all the contraction patterns which give rise to a particular diagram. This process would certainly be very impractical to carry out for bigger diagrams which may have many external points and vertices. This motivates us to develop a more hands-on way to assign labels to vertices in diagrams - a prescription through which one may label vertices by just looking at them.
\subsection{Labelling the Vertices of a Diagram}\label{labelling}
Let us first summarise our situation till now. In \ref{vertexdef}, we first defined what we mean by an interaction vertex in the context of Wightman correlators in general states of the anharmonic oscillator. This definition, though very simple, made us realise that there are indeed many types of vertices present in this study. To add to this, the section \ref{many} revealed that a given vertex has more than one vertex factor associated to it. Both these observations motivated a need for organisation. They made it clear that the traditional route of listing the vertices of a theory along with their vertex factors would prove to be very impractical in this case.

The way out is to first assign labels to all the vertices present in a diagram. The vertex factors can then be simply read off from these labels. In the section \ref{not}, we have made the notion of vertex labels more precise by stating that they are simply the time variables which are integrated over in the corresponding analytic expression for the vertex. In the same section, we also illustrated how one must go about labelling the different vertices present in a diagram through first principles. This method was found to be quite tedious. It would be extremely tough to implement it for diagrams with many vertices and external points. 

In this section, we present a more hands-on way to label the vertices of a diagram. The basic principle behind this method is very simple. The section \ref{vertexdef} has already given us the definition of a vertex that it is a   meeting point of four propagators. We now say that among these propagators which meet at a vertex, a certain type, namely free Wightman propagators coming from any other point apart from that vertex itself, impose some \emph{constraints} on the set of labels which are permissible for that vertex. We then place the natural demand that the final assignment of labels to the vertices in a diagram must be such that all these constraints are fulfilled.

To begin with, we set up some notation in the next section. This would prove to be very useful in our consequent analyses. Finally, in the section following it, we state the constraints which free Wightman propagators coming from another point impose on the set of labels which a particular vertex can acquire.
\subsection*{Setting Up Notation}\label{notation}
We now introduce some notation which would prove to be very useful in our consequent analyses.
\subsubsection{\underline{The Set $L_n$}}
Suppose we are to analyse the n-point anharmonic Wightman correlator:
\begin{equation}\label{wightlab}
    \langle x_H(t_1)x_H(t_2)\dots x_H(t_n)\rangle.
\end{equation}
Then, the set from which we would be picking labels to assign them to vertices which would appear in the diagrammatics of the correlator \eqref{wightlab} would be called the set $L_n$. It is defined as:
\begin{equation}
    L_n\equiv\{t_{1+},t_1,t_{1-},t_{2+},t_2,t_{2-},\dots,t_{n+},t_n,t_{n-}\}.
\end{equation}
For example, this set for the 3-point anharmonic Wightman correlator $\langle x_H(t_1)x_H(t_2)x_H(t_3)\rangle$ reads:
\begin{equation}
    L_3=\{t_{1+},t_1,t_{1-},t_{2+},t_2,t_{2-},t_{3+},t_3,t_{3-}\}.
\end{equation}
It is important to mention why we have included the external points $\{t_1,t_2,\dots,t_n\}$ also into this set despite knowing that they do not fall into the definition of a label as introduced in \ref{not}. We have done so keeping in mind the constraints that free Wightman propagators coming from another point and meeting at a vertex would impose on the set of labels permissible for that vertex. As would be realised in the next section, the mathematical statements of these constraints becomes much simpler if we include the external points too in this set. 
\subsubsection{\underline{The Set $M_n$}}
Given the Wightman correlator $\langle x_H(t_1)x_H(t_2)\dots x_H(t_n)\rangle$, it proves to be convenient to define another set $M_n$ as being the set $L_n$ \emph{without the external points}. That is:
\begin{equation}
    M_n\equiv L_n\setminus\{t_1,t_2,\dots,t_n\}.
\end{equation}
Vertices will always be assigned labels from the set $M_n$. This is trivial to realise if one knows what vertex labels actually are. The section \ref{not} has explained this in detail.

Continuing with the example of the 3-point Wightman correlator, we have:
\begin{equation}
    M_3=\{t_{1+},t_{1-},t_{2+},t_{2-},t_{3+},t_{3-}\}.
\end{equation}
\subsubsection{\underline{The Ordering Structure `$\rightarrow$' on $L_n$}}
Given the set $L_n$, we define an ordering structure `$\rightarrow$' on it such that:
\begin{equation}
    t_{1+}\rightarrow t_1\rightarrow t_{1-}\rightarrow t_{2+}\rightarrow t_2\rightarrow t_{2-}\rightarrow\dots\rightarrow t_{n+}\rightarrow t_n\rightarrow t_{n-}.
\end{equation}
It is also defined to be transitive. This means that:
\begin{equation}
    t_i\rightarrow t_j\quad\text{and}\quad t_j\rightarrow t_k\implies t_i\rightarrow t_k,
\end{equation}
where $t_i,t_j,t_k\in L_n$.

As an example, on the set $L_3$, we would have:
\begin{equation}
    t_{1+}\rightarrow t_1\rightarrow t_{1-}\rightarrow t_{2+}\rightarrow t_2\rightarrow t_{2-}\rightarrow t_{3+}\rightarrow t_3\rightarrow t_{3-}.
\end{equation}
\newpage
And owing to the defined transitivity of this ordering structure, we would also have relations like:
\begin{align}
    t_{1+}&\rightarrow t_{1-},\notag\\
    t_{1+}&\rightarrow t_{2+},\notag\\
    t_{1+}&\rightarrow t_2,\notag\\
    t_{1+}&\rightarrow t_{3+},
\end{align}
and so on.

With all this notation established, we are now ready to state the constraints which free Wightman propagators coming from another point impose on the set of permissible labels for a vertex at which they happen to meet.
\subsection*{Constraints on Vertex Labels}
As already stated in \ref{vertexdef}, a vertex is the meeting point of four propagators. Each free Wightman propagator meeting the vertex and which comes from any other point apart from that vertex itself imposes a certain constraint on its permissible labels. Now, each such free Wightman propagator can either be coming from an external point or from another vertex itself. We will call these two types of free Wightman propagators as \emph{external} and \emph{internal} free Wightman propagators respectively. They impose constraints which are slightly different from each other. We now proceed to state these constraints explicitly. We also follow them up with their appropriate justifications. All through, we will be assuming that we are working with the n-point anharmonic Wightman correlator in a general state:
\begin{equation}
    \langle x_H(t_1)x_H(t_2)\dots x_H(t_n)\rangle.
\end{equation}
\subsubsection{\underline{External Free Wightman Propagator}}
A free Wightman propagator coming \emph{into} a vertex from an external point imposes the constraint:
\iffalse\\[15pt]
\begin{center}
\textcolor{blue}{\textbf{(Continued on next page)}}
\end{center}
\newpage\fi
\begin{figure}[h]
    \centering
  \includegraphics[scale=0.12]{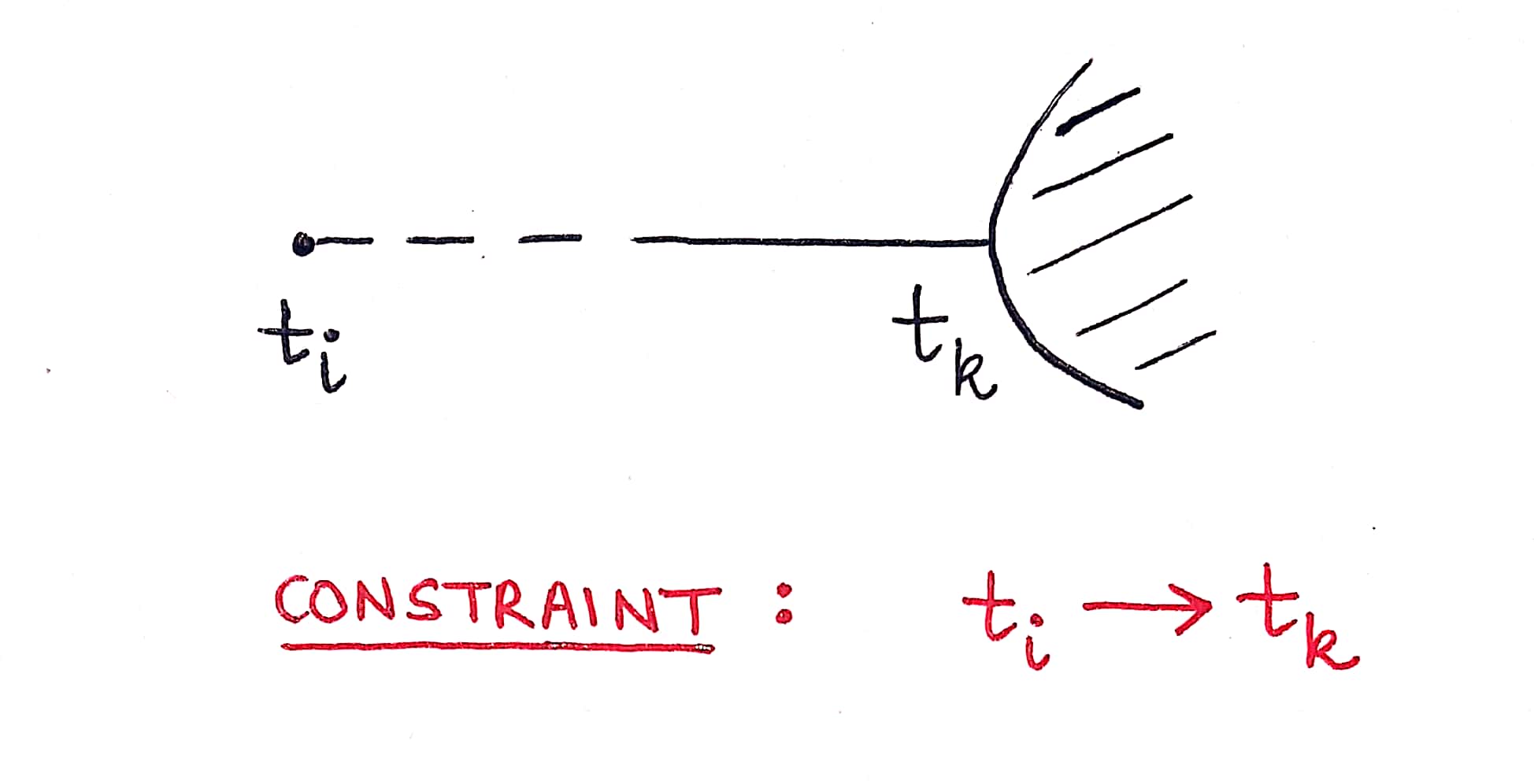}
  \caption{Constraint imposed by an external free Wightman propagator coming into a vertex.}
  \label{constraint1}
\end{figure}
\newpage
The shaded region represents a vertex. $t_i$ is an external point and $t_k\in M_n$ refers to a label assigned to the vertex. The notion of the set $M_n$ and the ordering structure `$\rightarrow$' used here has already been introduced in \ref{notation}.
\subsubsection{\ul{Justification}}
Suppose we are analysing the two point Wightman correlator in a general state of the anharmonic oscillator, which is given by:
\begin{equation}\label{anhcorr}
\langle x_H(t_1)x_H(t_2)\rangle.
\end{equation}
We will now justify the constraint which a free Wightman propagator coming out from the external point $t_1$ into a vertex imposes on the allowed labels for that vertex. That is:
\begin{figure}[h]
    \centering
  \includegraphics[scale=0.14]{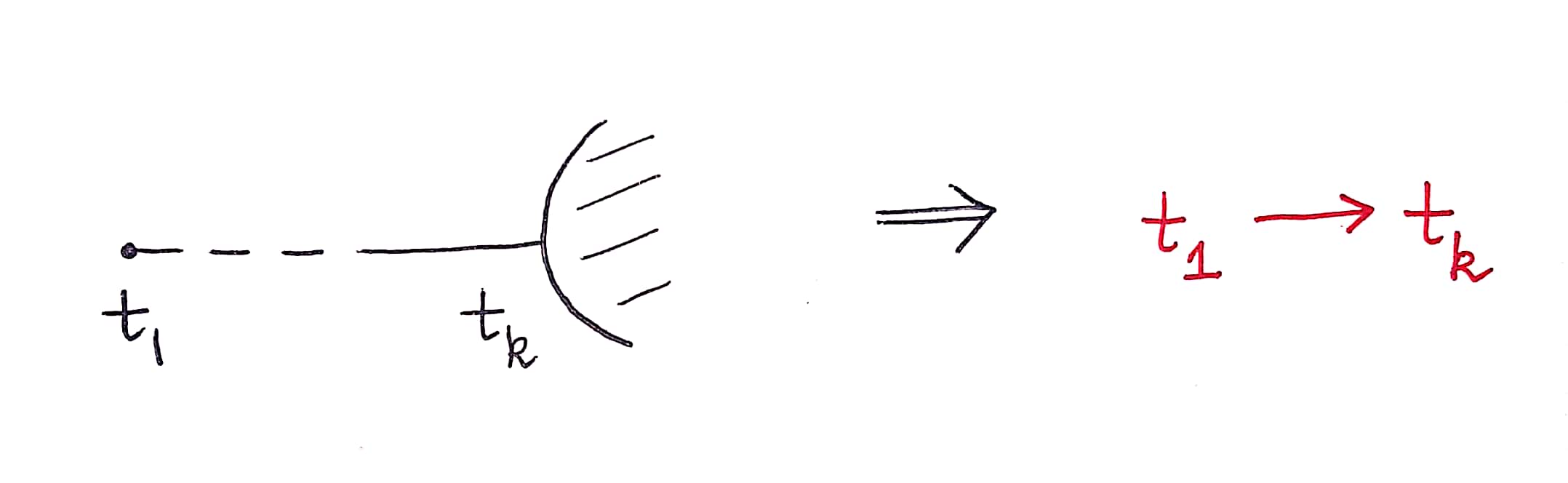}
  \caption{Justifying the constraint imposed by external free Wightman propagators.}
  \label{constraint1just}
\end{figure}
\\
Since only one vertex is involved in the rule we have to justify, it proves to be sufficient to focus on the terms in the perturbation expansion of \eqref{anhcorr} which are only first order in the coupling $\lambda$. Doing this, one gets:
\begin{align}\label{just1}
    \langle x_H(t_1)x_H(t_2)\rangle|_{\lambda}&=\Bigl(\frac{i\lambda}{4!}\Bigr)\int_{t_0}^{t_1}dt_{1+}\langle x(t_{1+})x(t_{1+})x(t_{1+})x(t_{1+})x(t_1)x(t_2)\rangle\notag\\&\quad+\Bigl(\frac{-i\lambda}{4!}\Bigr)\int_{t_0}^{t_1}dt_{1-}\langle x(t_1)x(t_{1-})x(t_{1-})x(t_{1-})x(t_{1-})x(t_2)\rangle\notag\\&\quad+\Bigl(\frac{i\lambda}{4!}\Bigr)\int_{t_0}^{t_2}dt_{2+}\langle x(t_1)x(t_{2+})x(t_{2+})x(t_{2+})x(t_{2+})x(t_2)\rangle\notag\\&\quad+\Bigl(\frac{-i\lambda}{4!}\Bigr)\int_{t_0}^{t_2}dt_{2-}\langle x(t_1)x(t_2)x(t_{2-})x(t_{2-})x(t_{2-})x(t_{2-})\rangle.
\end{align}
To justify the rule, we will naturally have to look at vertices of the type:
\begin{figure}[h]
    \centering
  \includegraphics[scale=0.09]{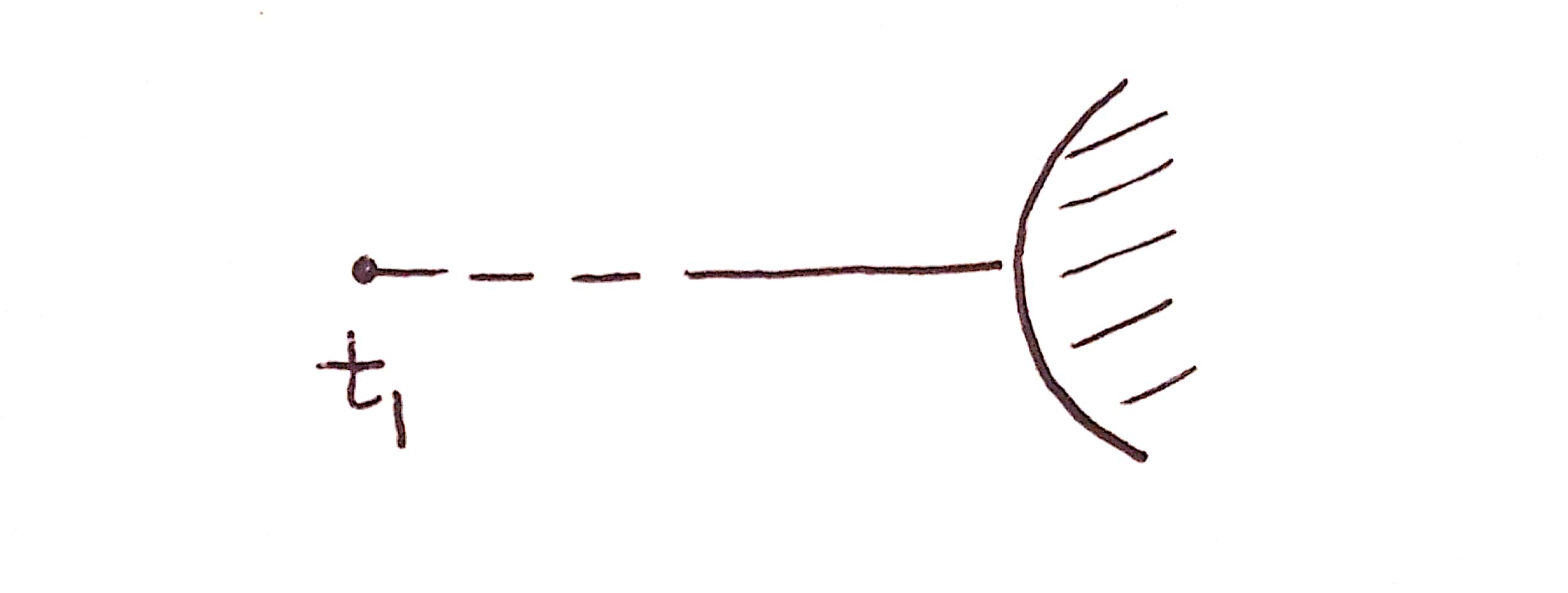}
  \caption{Types of vertices one needs to look at for justifying the constraint depicted in figure \ref{constraint1just}}
  \label{constraint1just2}
\end{figure}
\\
Now, what types of contraction patterns lead to a situation as depicted in figure \ref{constraint1just2} ? If one looks at the diagrammatic representation of a bi-dentate contraction as listed in \ref{contrdiag}, one would realise that a situation as depicted above can only arise if the external point $t_1$ contracts with an internal point which is placed to its right. This means that only the terms:
\begin{align}\label{a}
    &\Bigl(\frac{-i\lambda}{4!}\Bigr)\int_{t_0}^{t_1}dt_{1-}\langle x(t_1)x(t_{1-})x(t_{1-})x(t_{1-})x(t_{1-})x(t_2)\rangle,\notag\\&\Bigl(\frac{i\lambda}{4!}\Bigr)\int_{t_0}^{t_2}dt_{2+}\langle x(t_1)x(t_{2+})x(t_{2+})x(t_{2+})x(t_{2+})x(t_2)\rangle,\notag\\&\Bigl(\frac{-i\lambda}{4!}\Bigr)\int_{t_0}^{t_2}dt_{2-}\langle x(t_1)x(t_2)x(t_{2-})x(t_{2-})x(t_{2-})x(t_{2-})\rangle,
\end{align}
can give rise to the depicted situation. And once one has realised this, it is easy to see that the above terms can lead to the vertex depicted in diagram \ref{constraint1just} only having the labels $t_{1-},t_{2+}$ and $t_{2-}$, which is precisely what the constraint dictates.
\iffalse\\[20pt]
\begin{center}
\textcolor{blue}{\textbf{(Continued on next page)}}
\end{center}
\newpage\fi

A free Wightman propagator going \emph{out} of a vertex \emph{into} an external point imposes the constraint:
\begin{figure}[h]
    \centering
  \includegraphics[scale=0.1]{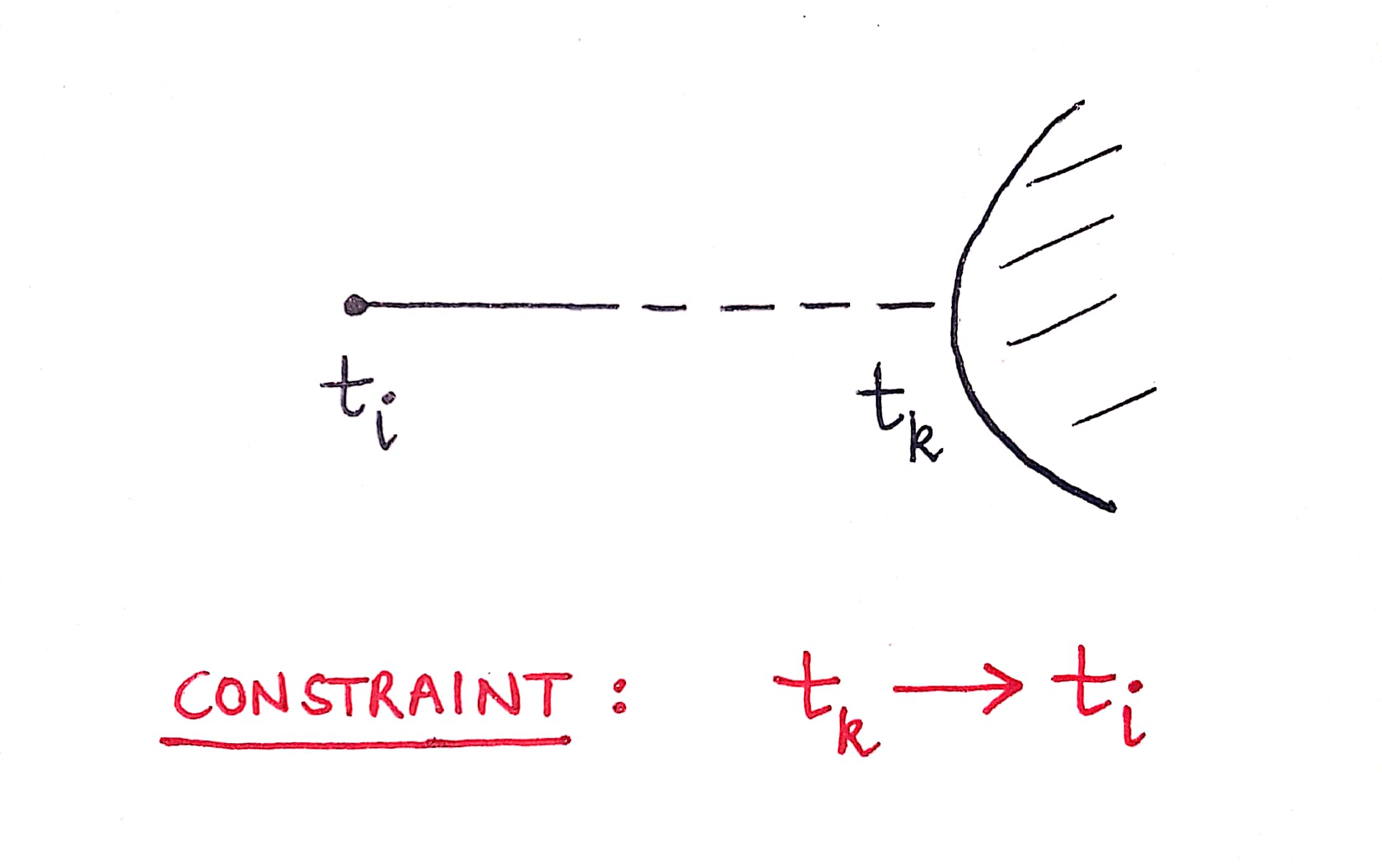}
  \caption{Constraint imposed by an external free Wightman propagator going out of a vertex.}
  \label{constraint2}
\end{figure}
\\
Again, the shaded region represents a vertex. $t_i$ is an external point and $t_k\in M_n$ refers to a label assigned to the vertex. This constraint can be justified in an exactly similar way as for the previous case.

We now state the constraint imposed by an internal free Wightman propagator on the next page.
\newpage
\subsubsection{\underline{Internal Free Wightman Propagator}}
An internal free Wightman propagator imposes the constraint:
\begin{figure}[h]
    \centering
  \includegraphics[scale=0.13]{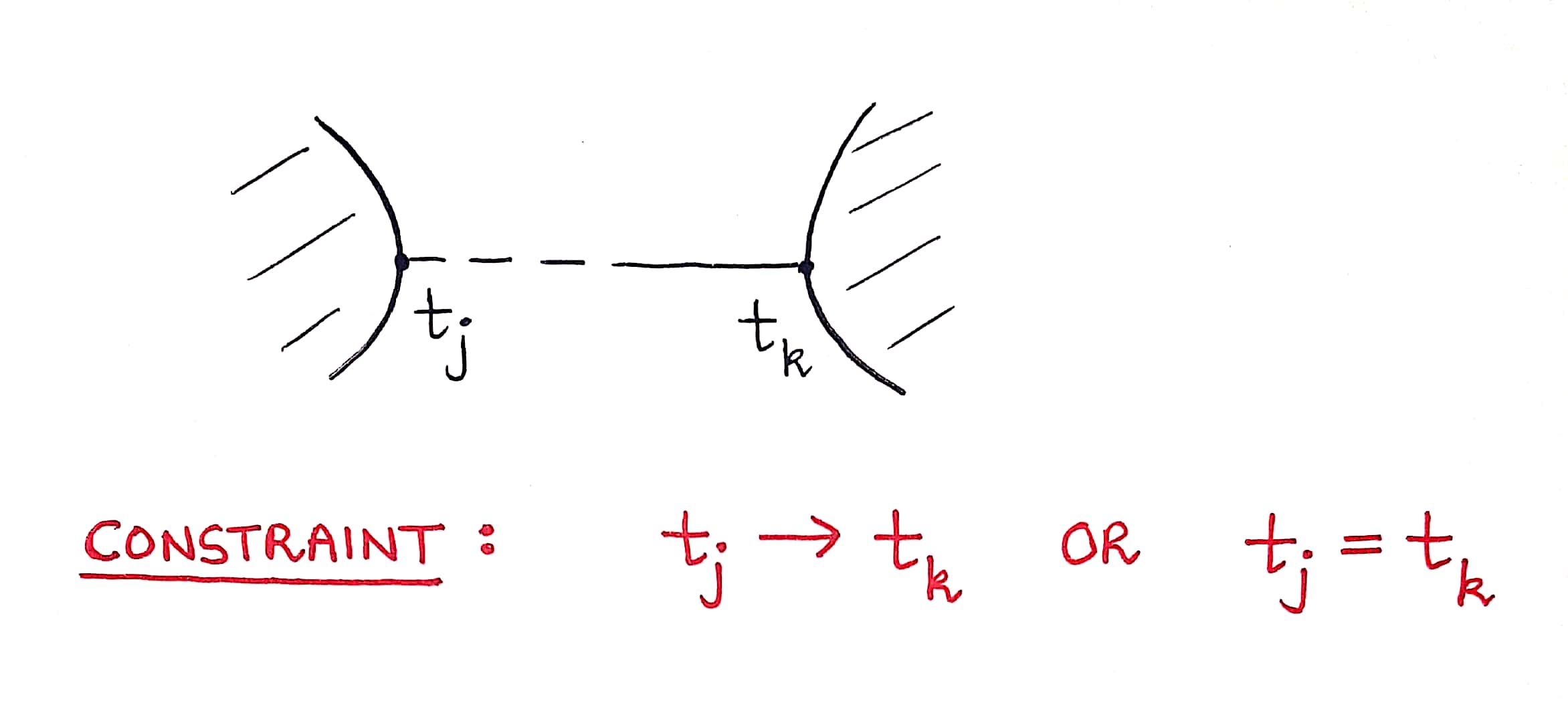}
  \caption{Constraint imposed by an internal free Wightman propagator.}
  \label{internalconstraint}
\end{figure}
\\
As before, shaded regions represent vertices. $t_j,t_k\in M_n$ refer to labels assigned to the two vertices. 
\subsubsection{\ul{Justification}}
To justify this constraint, we again take help of the same example, which is of the two point Wightman correlator:
\begin{equation}\label{anhcorrr}
    \langle x_H(t_1)x_H(t_2)\rangle.
\end{equation}
We will show that if a free Wightman propagator is coming \emph{into} a vertex from an internal point which has already been labelled $t_{1+}$, then it indeed imposes the constraint we have stated on the labels of the other vertex.

That is, we will justify:
\begin{figure}[h]
    \centering
  \includegraphics[scale=0.18]{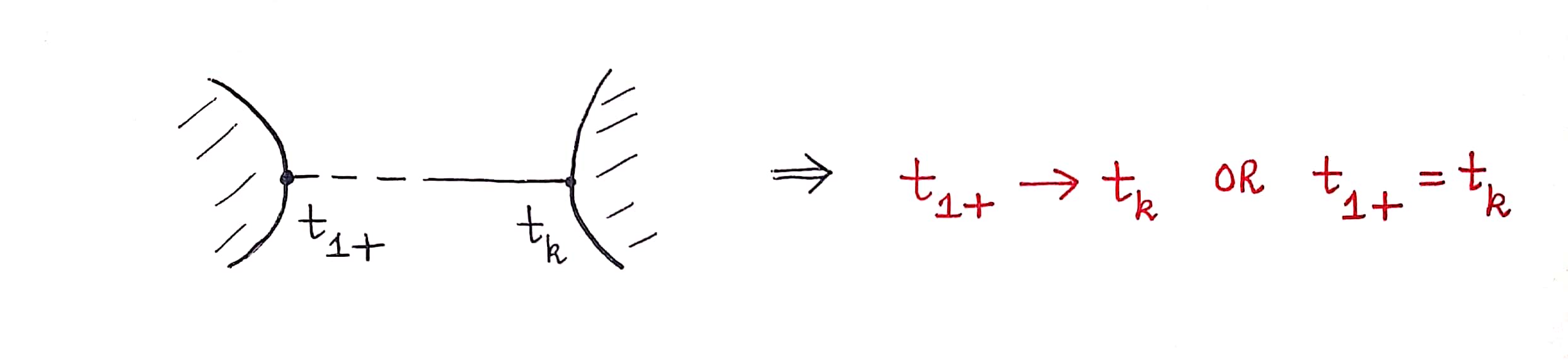}
  \caption{Justifying the constraint imposed by an internal free Wightman propagator.}
  \label{just4}
\end{figure}
\\
Since two vertices are involved in the rule we want to justify, we would have to look at the terms in the perturbation expansion of \eqref{anhcorrr} which are of second order in the coupling $\lambda$. And since one of the vertices has to be $t_{1+}$, the only terms we need to focus on are:
\begin{align}\label{freeint}
    &\Bigl(\frac{i\lambda}{4!}\Bigr)^{2}\int_{t_0}^{t_1}dt^{(1)}_{1+}\int_{t_0}^{t_1}dt^{(2)}_{1+}\langle\Bigl\{T^*x^4(t^{(1)}_{1+})x^4(t^{(2)}_{1+})\Bigr\}x(t_1)x(t_2)\rangle\notag\\+&\Bigl(\frac{i\lambda}{4!}\Bigr)\Bigl(\frac{-i\lambda}{4!}\Bigr)\int_{t_0}^{t_1}dt_{1+}\int_{t_0}^{t_1}dt_{1-}\langle\Bigl\{x^4(t_{1+})\Bigr\}x(t_1)\Bigl\{x^4(t_{1-})\Bigr\}x(t_2)\rangle\notag\\+&\Bigl(\frac{i\lambda}{4!}\Bigr)^{2}\int_{t_0}^{t_1}dt_{1+}\int_{t_0}^{t_2}dt_{2+}\langle\Bigl\{x^4(t_{1+})\Bigr\}x(t_1)\Bigl\{x^4(t_{2+})\Bigr\}x(t_2)\rangle\notag\\+&\Bigl(\frac{i\lambda}{4!}\Bigr)\Bigl(\frac{-i\lambda}{4!}\Bigr)\int_{t_0}^{t_1}dt_{1+}\int_{t_0}^{t_2}dt_{2-}\langle\Bigl\{x^4(t_{1+})\Bigr\}x(t_1)x(t_2)\Bigl\{x^4(t_{2-})\Bigr\}\rangle.
\end{align}
The first term of the above expression involves anti-time ordering. For making the further analysis clear, let us expand it using step functions of time:
\begin{align}\label{step}
    \langle\Bigl\{T^*x^4(t^{(1)}_{1+})x^4(t^{(2)}_{1+})\Bigr\}x(t_1)&x(t_2)\rangle\notag\\=&\theta(t^{(1)}_{1+}-t^{(2)}_{1+})\langle\Bigl\{x^4(t^{(2)}_{1+})x^4(t^{(1)}_{1+})\Bigr\}x(t_1)x(t_2)\rangle\notag\\+&\theta(t^{(2)}_{1+}-t^{(1)}_{1+})\langle\Bigl\{x^4(t^{(1)}_{1+})x^4(t^{(2)}_{1+})\Bigr\}x(t_1)x(t_2)\rangle.
\end{align}
The rule we want to justify, namely the situation as depicted by figure \ref{just4}, has a free Wightman propagator going \emph{into} a vertex from the internal point $t_{1+}$. The diagrammatic representation of a bi-dentate contraction as presented in \ref{contrdiag} dictates that for such a situation, the internal point $t_{1+}$ would have to contract with another internal point which is placed to its right in the interaction picture correlator. The only terms which can make this possible are any of the two terms in \eqref{step} and all the terms apart from the first term in \eqref{freeint}.

Let us first look at the terms apart from the first term in \eqref{freeint}. In all these terms, the label $t_{1+}$ already corresponds to the internal point we are talking about. Thus, the labels for the other vertex in \ref{just4} are given by $t_{1-},t_{2+}$ and $t_{2-}$ as far as these terms are concerned.

Now focus on \eqref{step}. As we have already discussed in \ref{not}, all labels of the form $t^{(i)}_{j+}$ are equivalent and would be collectively denoted as $t_{j+}$. Thus, \eqref{step} dictates that the other vertex in figure \ref{just4} also gets a label of $t_{1+}$.

Summing up, the set of possible labels for the other vertex in figure \ref{just4} is $\{t_{1+},t_{1-},t_{2+},t_{2-}\}$, which is in agreement with the constraint we wanted to justify. 

On the next page, we summarise the constraints we have discussed.
\newpage
\subsection*{A Summary of the Constraints}
Let us collate all the constraints we have established through the previous section into one place. This would facilitate future reference.
\begin{itemize}
    \item Constraint imposed by an external free Wightman propagator coming into a vertex:
    \begin{figure}[h]
    \centering
  \includegraphics[scale=0.12]{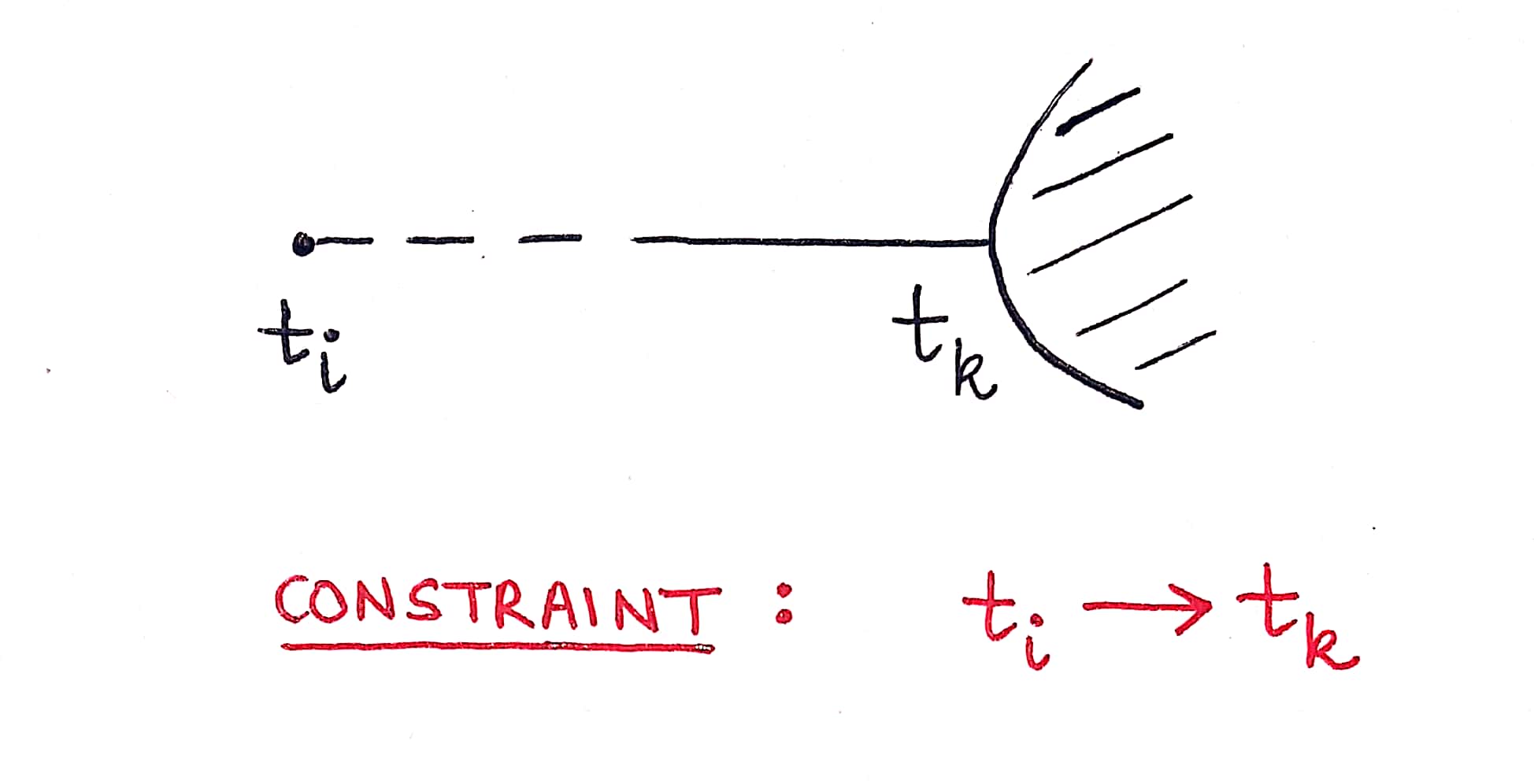}
\end{figure}
   \item Constraint imposed by an external free Wightman propagator going out of a vertex:
   \begin{figure}[h]
    \centering
  \includegraphics[scale=0.12]{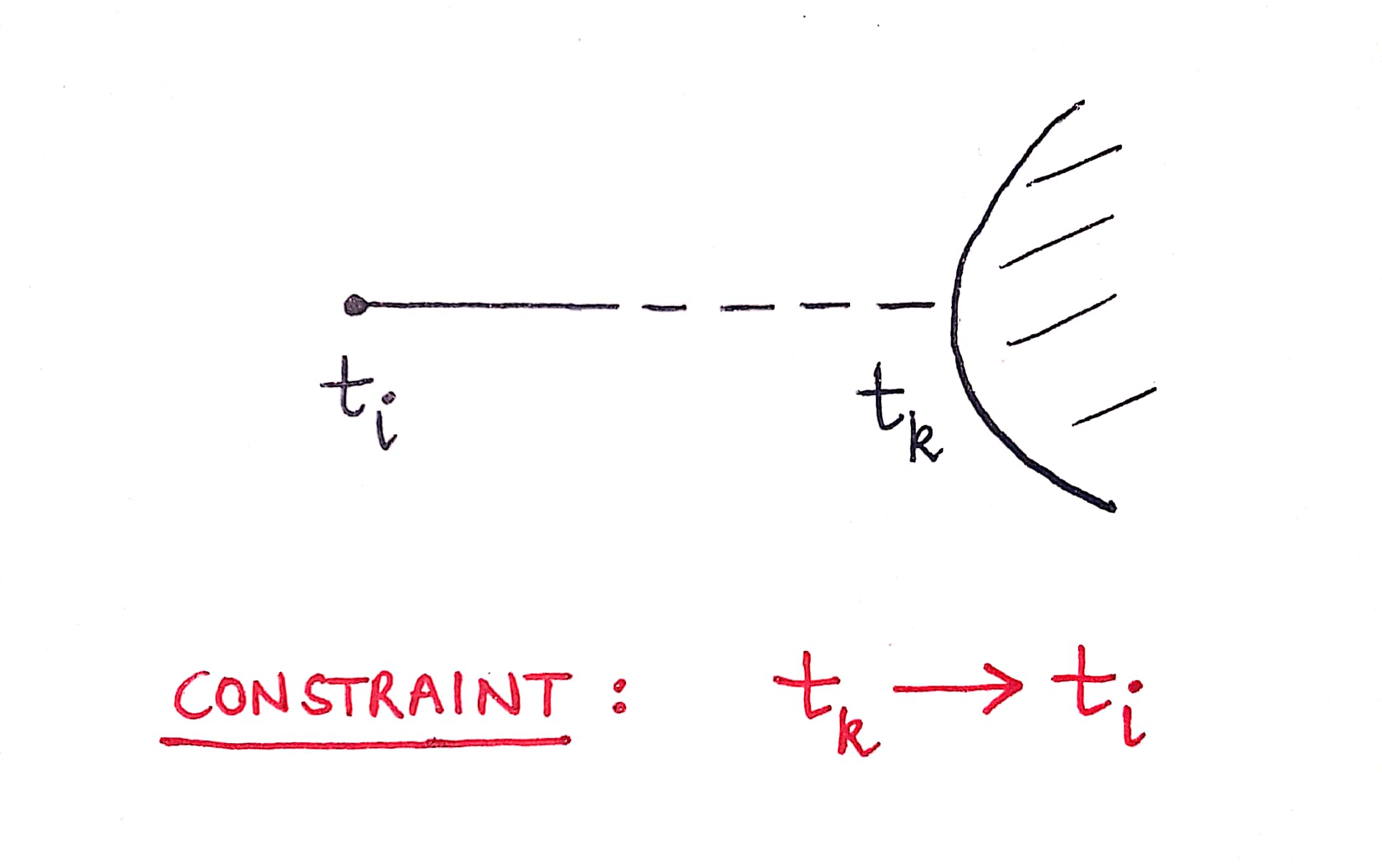}
\end{figure}
   \item Constraint imposed by an internal free Wightman propagator:
   \begin{figure}[h]
    \centering
  \includegraphics[scale=0.16]{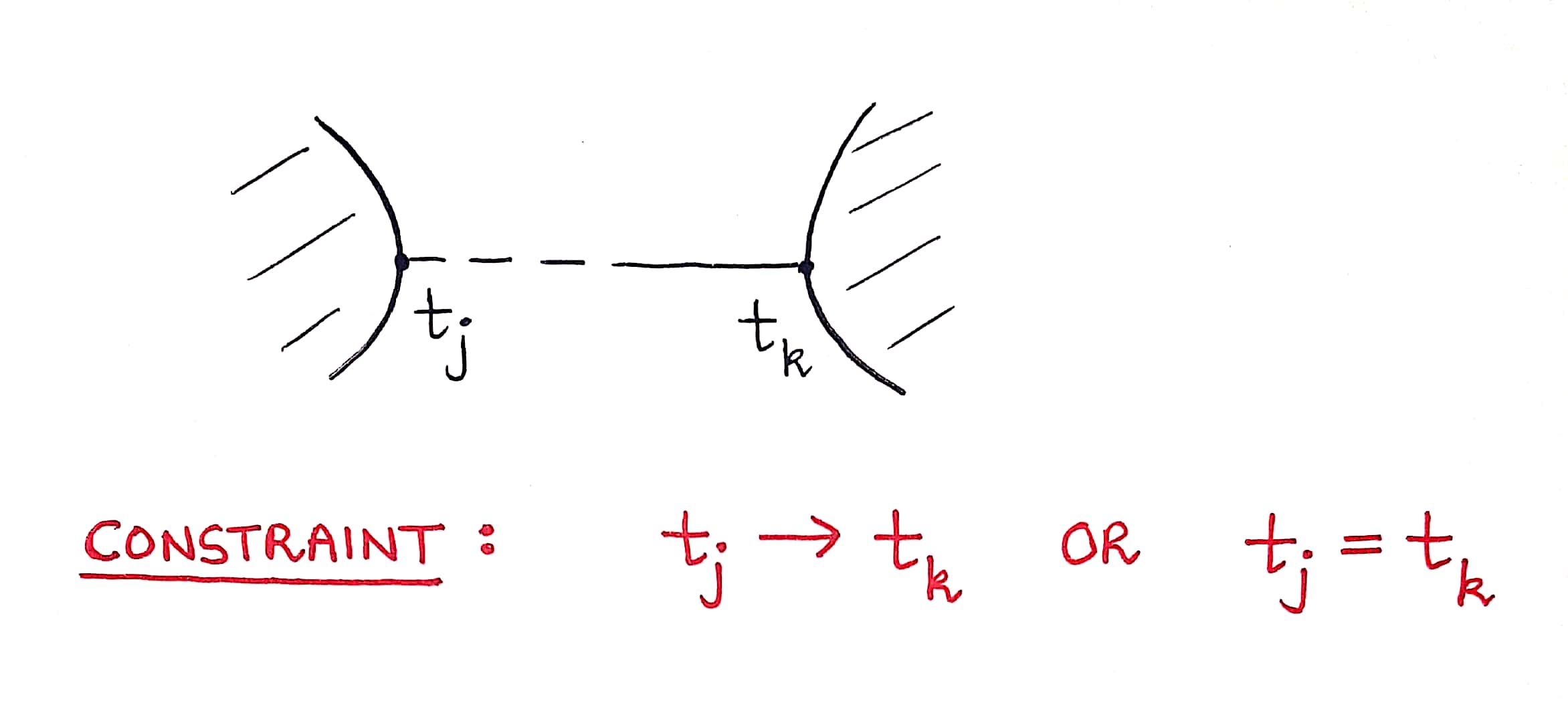}
\end{figure}
\end{itemize}
\subsection*{The Final Demand}
We have thus established precise mathematical forms of the constraints which free Wightman propagators impose on the labels of a vertex they happen to meet at. The final demand of this procedure of labelling the vertices of a diagram is the most natural one. It states:
\\[15pt]
\emph{All the vertices of a given diagram must be labelled in such a way that all the constraints imposed by the free Wightman propagators on their labels are simultaneously fulfilled}.
\\[15pt]
The examples we work out in the next section will make the above statement more clear.
\subsection*{Worked Out Examples}
We now illustrate some examples wherein we label the vertices of a diagram using the new procedure we have introduced. The free Wightman propagators pose certain constraints on the allowed labels of the vertices and finally, they are labelled in such a way that all these constraints are simultaneously fulfilled.
\subsubsection*{\underline{Example 1}}
Consider labelling the vertex of the following diagram which appears at order $\lambda$ of the perturbation expansion of $\langle x_H(t_1)x_H(t_2)\rangle$:
\begin{figure}[h]
    \centering
  \includegraphics[scale=0.1]{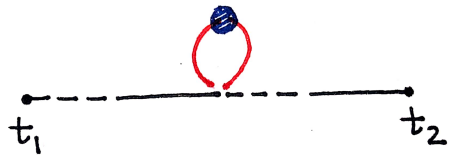}
  \caption{A diagram appearing at order $\lambda$ of the perturbation expansion of $\langle x_H(t_1)x_H(t_2)\rangle$.}
  \label{vertexlabel1}
\end{figure}
\\
Let us call the label of this vertex as $t_k$. That is:
\begin{figure}[h]
    \centering
  \includegraphics[scale=0.092]{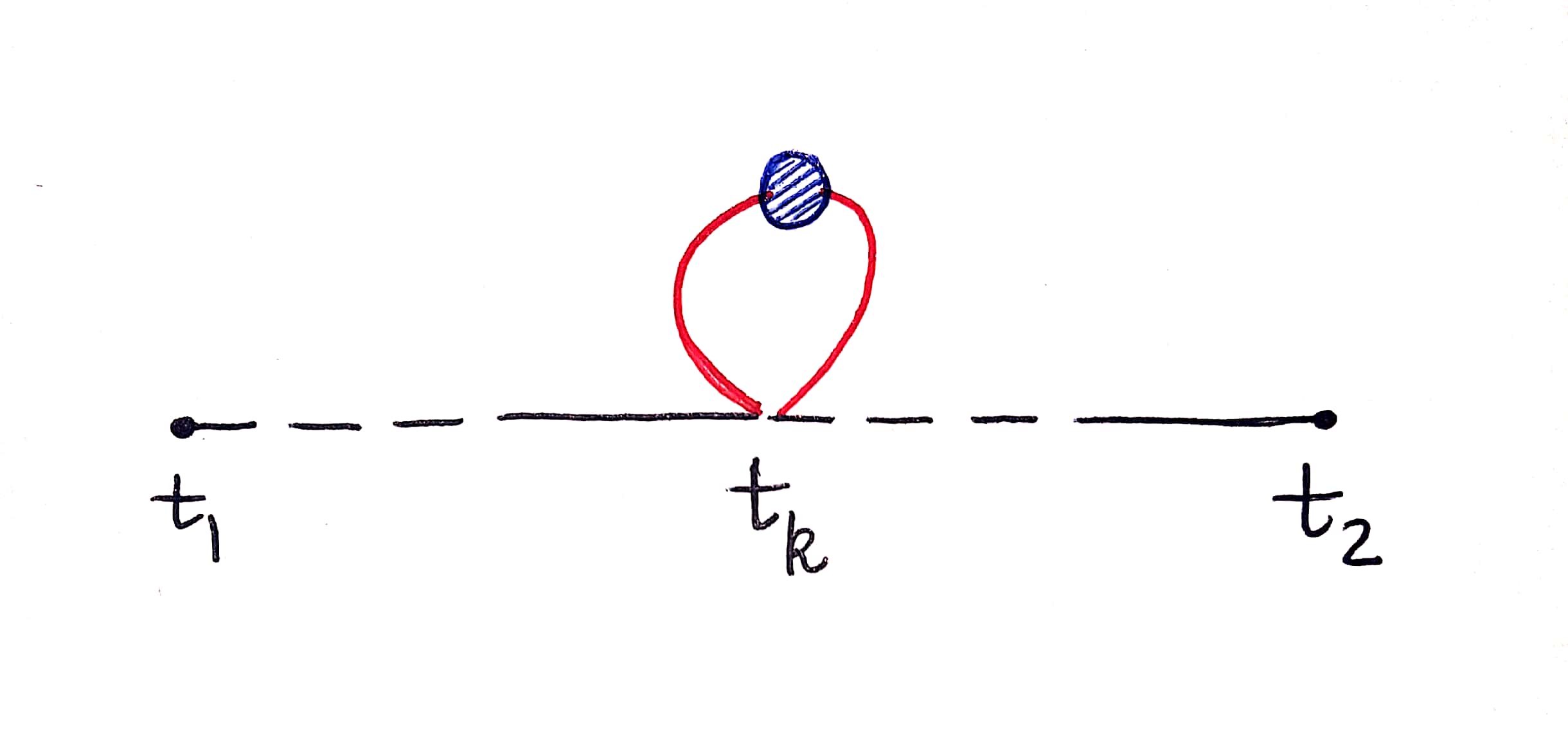}
\end{figure}
\\
First of all, we must have:
\begin{equation}
    t_k\in M_2=\{t_{1+},t_{1-},t_{2+},t_{2-}\},
\end{equation}
which is trivial to realise.

The free Wightman propagator coming from the external point $t_1$ into this vertex imposes the constraint:
\begin{equation}
    t_1\rightarrow t_k.
\end{equation}
In addition to this, the other free Wightman propagator which is going out of this vertex into the external point $t_2$ imposes the constraint:
\begin{equation}
    t_k\rightarrow t_2.
\end{equation}
The only possible labels which fulfill both of these constraints are:
\begin{equation}
    t_k=t_{1-},t_{2+}.
\end{equation}
This matches with what we had arrived at through first principles in \ref{not}.
\subsubsection*{\underline{Example 2}}
Consider labelling the vertices of the following diagram which appears at order $\lambda^2$ of the perturbation expansion of $\langle x_H(t_1)x_H(t_2)\rangle$:
\begin{figure}[h]
    \centering
  \includegraphics[scale=0.12]{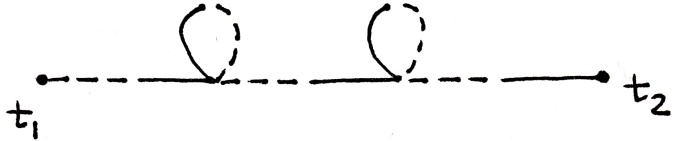}
  \caption{A diagram appearing at order $\lambda^2$ of the perturbation expansion of $\langle x_H(t_1)x_H(t_2)\rangle$.}
  \label{vertexlabel3}
\end{figure}
\\
To start with, let us label the two vertices as $t_k$ and $t_l$. That is:
\begin{figure}[h]
    \centering
  \includegraphics[scale=0.13]{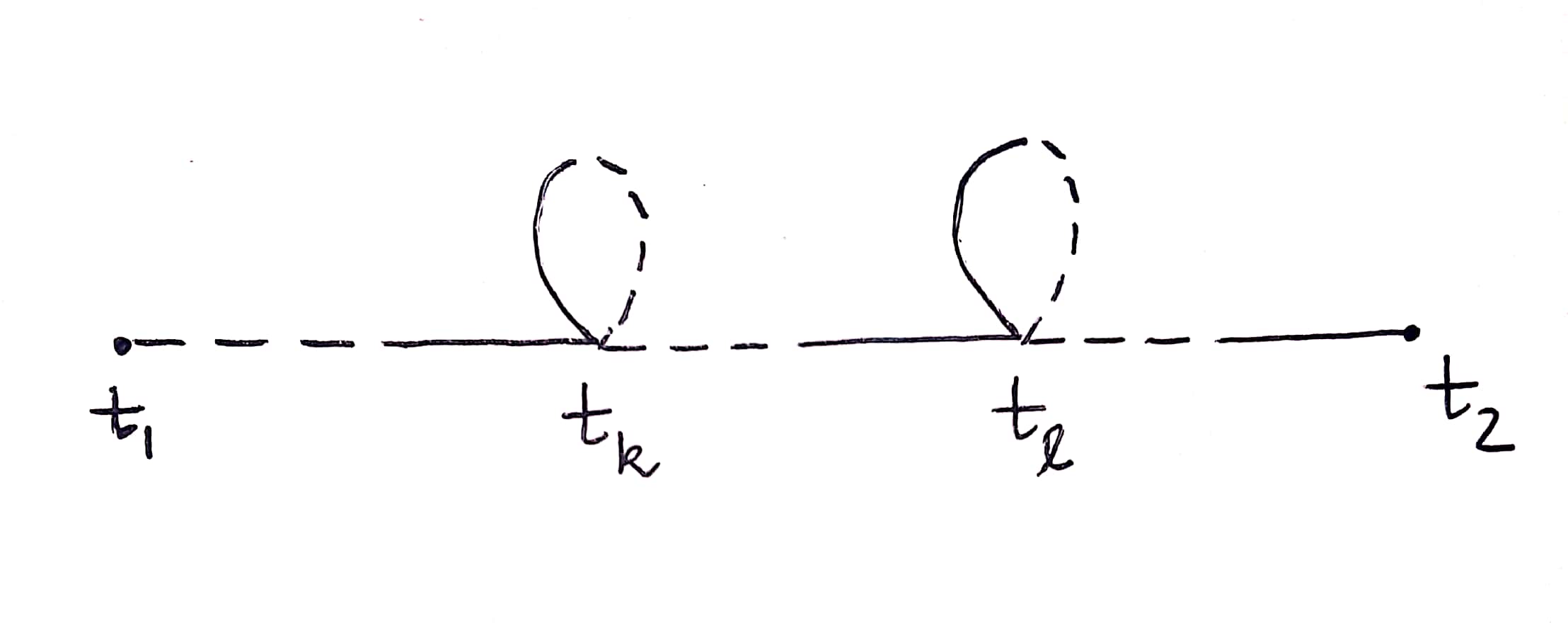}
\end{figure}
\\
First of all, since we are looking at a diagram which arises in the perturbation expansion of $\langle x_H(t_1)x_H(t_2)\rangle$, there is the trivial requirement that:
\begin{equation}\label{con1}
    t_k,t_l\in M_2=\{t_{1+},t_{1-},t_{2+},t_{2-}\}.
\end{equation}
Now look at the free Wightman propagator which is coming into the vertex labelled $t_k$ from the external point $t_1$. This imposes the constraint:
\begin{equation}\label{con2}
    t_1\rightarrow t_k.
\end{equation}
The internal free Wightman propagator imposes the constraint:
\begin{equation}\label{con3}
    t_k\rightarrow t_l\quad\text{or}\quad t_k=t_l.
\end{equation}
Finally, the free Wightman propagator which is going out from the vertex labelled $t_l$ into the external point $t_2$ imposes the constraint:
\begin{equation}\label{con4}
    t_l\rightarrow t_2.
\end{equation}
Summarising, the constraints \eqref{con1},\eqref{con2},\eqref{con3},\eqref{con4} which need to be fulfilled in this situation are:
\begin{align}
    t_k,t_l\in M_2&=\{t_{1+},t_{1-},t_{2+},t_{2-}\},\notag\\
    t_1&\rightarrow t_k,\notag\\
    t_k&\rightarrow t_l\quad\text{or}\quad t_k=t_l,\notag\\
    t_l&\rightarrow t_2.
\end{align}
The only combinations of labels which fulfill these constraints are:
\begin{equation}
    (t_k,t_l)=(t_{1-},t_{1-})\quad\text{or}\quad (t_{1-},t_{2+})\quad\text{or}\quad (t_{2+},t_{2+}),
\end{equation}
which matches with what we had found out through first principles in \ref{not}.
\subsection{Reading Off Vertex Factors from Vertex Labels}\label{reading}
The previous section has established a quick way to label the vertices of diagrams. Once this is done, reading off the vertex factors from these assigned labels is a very trivial task. Specifically:
\begin{itemize}
    \item\emph{Each label assigned to a vertex will correspond to a vertex factor of the vertex. Specifically:}
    \begin{gather}
        t_{j+}\sim +i\lambda\int_{t_0}^{t_j}dt_{j+},\notag\\
        t_{j-}\sim -i\lambda\int_{t_0}^{t_j}dt_{j-},
    \end{gather}
    \emph{where the symbol $\sim$ is to be read as} `corresponds to'.
    \item\emph{If a vertex has no possible label that can be assigned to it, then the vertex factor of this vertex is 0.}
\end{itemize}
\newpage
\section{Step Function Weights for Internal Propagators}\label{stepfn}
Recall that in the section \ref{anhov}, we had outlined three diagrammatic concepts which emerge through the analysis of Wightman correlators in general states of the anharmonic oscillator. These are absent in the corresponding analysis of the free oscillator. They are the concepts of interaction vertices, step function weights for internal propagators and symmetry factors. In the previous sections, we have studied interaction vertices in detail. We now move on to step function weights for internal propagators. Before we do so, we mention two important observations:
\begin{itemize}
    \item Given two vertices. In a given diagram, free Wightman propagators joining these two vertices will either all be going into one vertex from the other, or all be going into the other from the first. Such a situation can never happen where in a diagram, some of the free Wightman propagators joining these two vertices are going into one from the other and some are going the other way round. That is:
    \begin{figure}[h]
    \centering
  \includegraphics[scale=0.12]{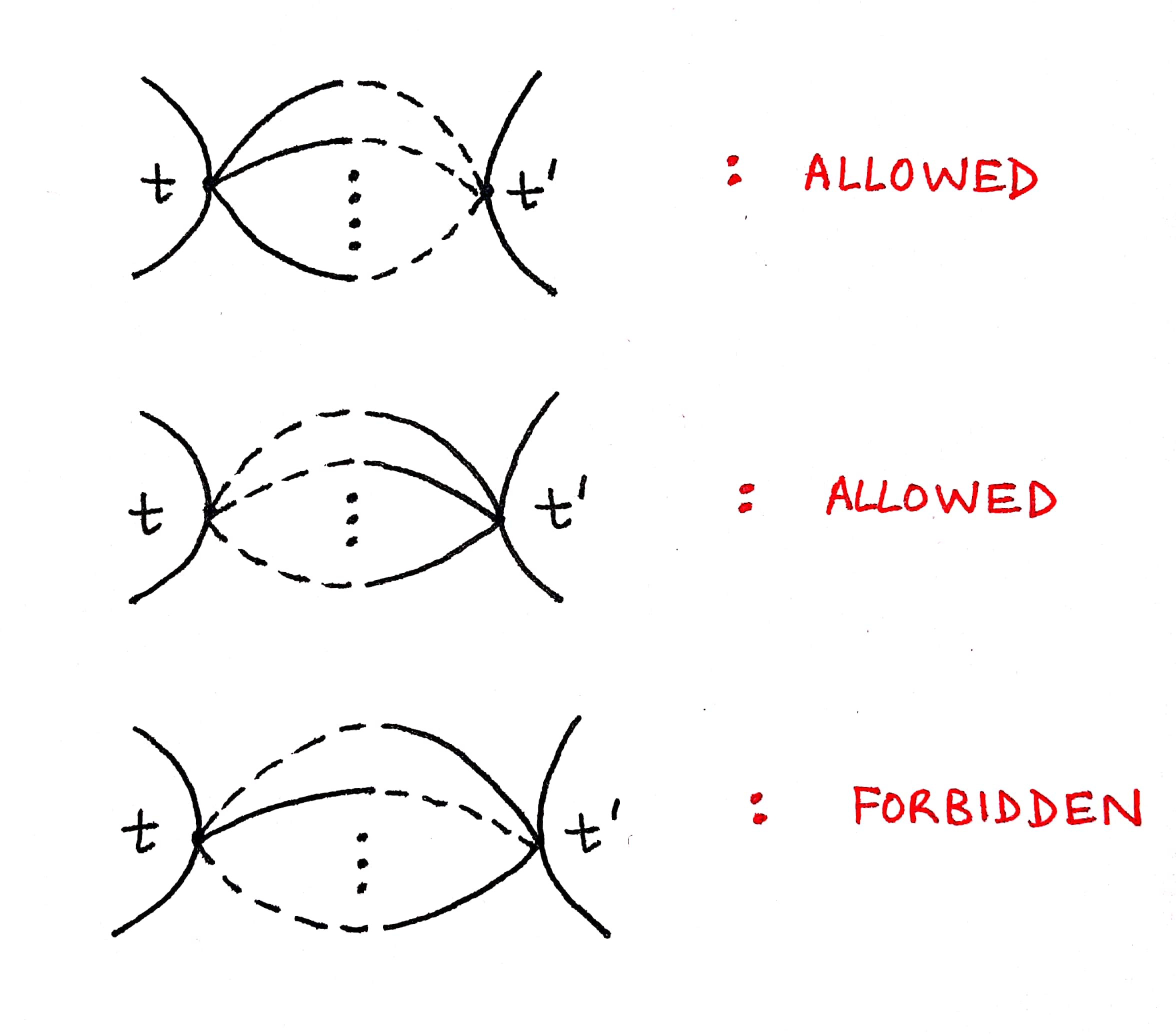}
  \caption{A crucial observation.}
  \label{allowedforbidden}
\end{figure}
\\
    \item Step function weights are only assigned to free Wightman propagators joining vertices \emph{with the same labels}.
\end{itemize}
It is easy to realise the truth of these claims once one delves into the perturbative calculation of anharmonic Wightman correlators. We will not be justifying them here. On the other hand, we will present what the step function weights mentioned in the second point above must be, and justify them instead.
\newpage
Let us study the two point anharmonic Wightman correlator:
\begin{equation}\label{obj}
    \langle x_H(t_1)x_H(t_2)\rangle.
\end{equation}
At the order $\lambda^2$ in the perturbation expansion for the object \eqref{obj}, one encounters the following `sunset' diagram:
\begin{figure}[h]
    \centering
  \includegraphics[scale=0.12]{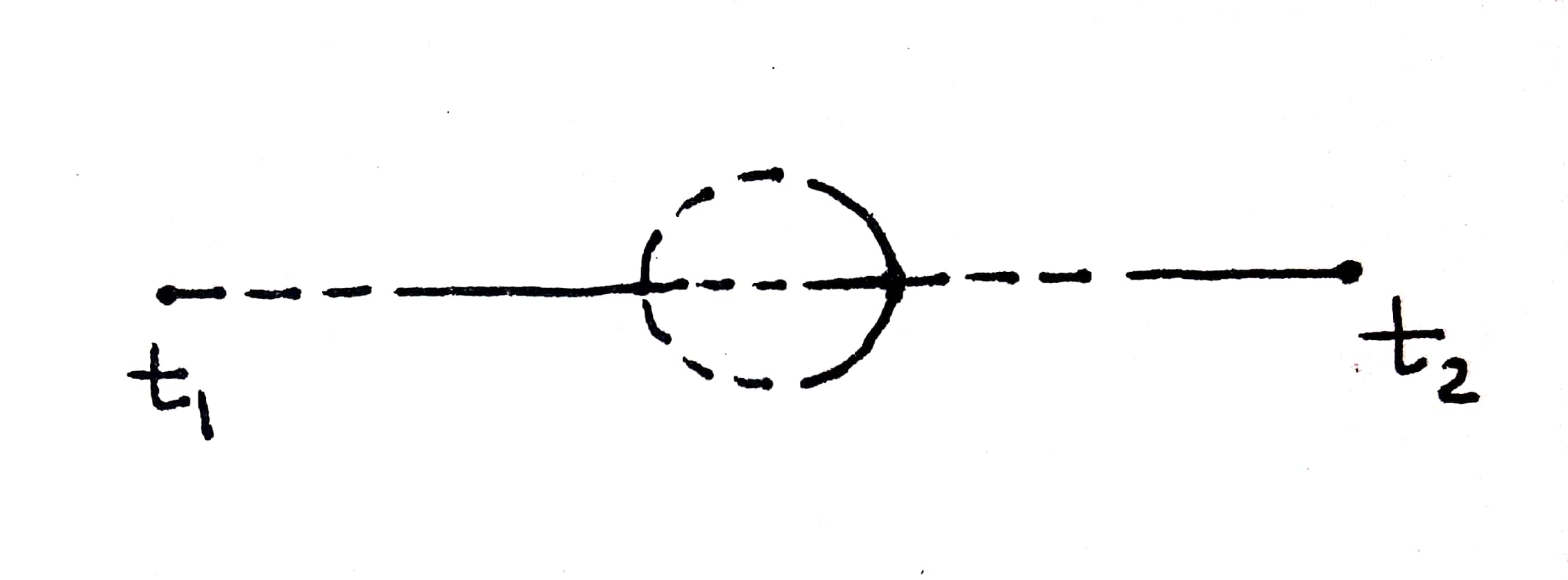}
  \caption{The sunset diagram.}
  \label{stepfn1}
\end{figure}
\\
Pause a moment to realise that this diagram agrees with the first observation pointed out at the beginning of this section. 

The labels for the two vertices in this diagram are given as:
\begin{table}[ht]
\centering
\begin{tabular}{ |p{3.5cm}|p{3.5cm}|  }
\hline
Label of Vertex 1&Label of Vertex 2\\
\hline
$t_{1-}$ & $t_{1-}$\\
$t_{1-}$ & $t_{2+}$\\
$t_{2+}$ & $t_{2+}$\\
\hline
\end{tabular}
\caption{\label{tab:table2}Labelling the vertices of the sunset diagram. \emph{Vertex 1} refers to the vertex nearer the external point $t_1$, and \emph{Vertex 2} refers to the other vertex.}
\end{table}
\\
Going by the second observation pointed out at the beginning of this section, we need only explore the first and third rows listed in the above table to know what the forms of the step function weights are. This is because it is only in these situations that both the vertices are assigned the same labels.

Consider the situation of the first row, wherein both the vertices are labelled $t_{1-}$. At the analytic level, this situation results from the term:
\begin{align}
    \Bigl(\frac{-i\lambda}{4!}\Bigr)^{2}\int_{t_0}^{t_1}dt^{(1)}_{1-}\int_{t_0}^{t_1}dt^{(2)}_{1-}\langle x(t_1)\Bigl\{Tx^4(t^{(1)}_{1-})x^4(t^{(2)}_{1-})\Bigr\}x(t_2)\rangle.
\end{align}
Expanding the time ordered product using step functions of time yields:
\begin{align}
    \langle x(t_1)\Bigl\{Tx^4(t^{(1)}_{1-})x^4(t^{(2)}_{1-})\Bigr\}x(t_2)\rangle=&\theta(t^{(1)}_{1-}-t^{(2)}_{1-})\langle x(t_1)\Bigl\{x^4(t^{(1)}_{1-})x^4(t^{(2)}_{1-})\Bigr\}x(t_2)\rangle\notag\\+&\theta(t^{(2)}_{1-}-t^{(1)}_{1-})\langle x(t_1)\Bigl\{x^4(t^{(2)}_{1-})x^4(t^{(1)}_{1-})\Bigr\}x(t_2)\rangle.
\end{align}
Let us now say that the vertex nearer $t_1$ is labelled $t^{(1)}_{1-}$ among $t^{(1)}_{1-}$ and $t^{(2)}_{1-}$. With this choice, the diagram \ref{stepfn1} results only from the contraction pattern:
\begin{align}
    \theta(t^{(1)}_{1-}-t^{(2)}_{1-})\langle\contraction{}{x}{(t_1)}{x}\contraction[2ex]{x(t_1)x(t^{(1)}_{1-})}{x}{(t^{(1)}_{1-})x(t^{(1)}_{1-})x(t^{(1)}_{1-})x(t^{(2)}_{1-})x(t^{(2)}_{1-})}{x}\contraction[1.5ex]{x(t_1)x(t^{(1)}_{1-})x(t^{(1)}_{1-})}{x}{(t^{(1)}_{1-})x(t^{(1)}_{1-})x(t^{(2)}_{1-})}{x}\contraction{x(t_1)x(t^{(1)}_{1-})x(t^{(1)}_{1-})x(t^{(1)}_{1-})}{x}{(t^{(1)}_{1-})}{x}\contraction{x(t_1)x(t^{(1)}_{1-})x(t^{(1)}_{1-})x(t^{(1)}_{1-})x(t^{(1)}_{1-})x(t^{(2)}_{1-})x(t^{(2)}_{1-})x(t^{(2)}_{1-})}{x}{(t^{(2)}_{1-})}{x}x(t_1)x(t^{(1)}_{1-})x(t^{(1)}_{1-})x(t^{(1)}_{1-})x(t^{(1)}_{1-})x(t^{(2)}_{1-})x(t^{(2)}_{1-})x(t^{(2)}_{1-})x(t^{(2)}_{1-})x(t_2)\rangle.
\end{align}
The step function factor involved above is incorporated through the following rule:
\subsubsection{\ul{Rule 1}}
\begin{itemize}
    \item\emph{A group of free Wightman propagators coming out of a vertex labelled $t^{(i)}_{j-}$ and going into another vertex labelled $t^{(k)}_{j-}$ is associated with a step function weight $\theta(t^{(i)}_{j-}-t^{(k)}_{j-})$. That is:}
\begin{figure}[h]
    \centering
  \includegraphics[scale=0.14]{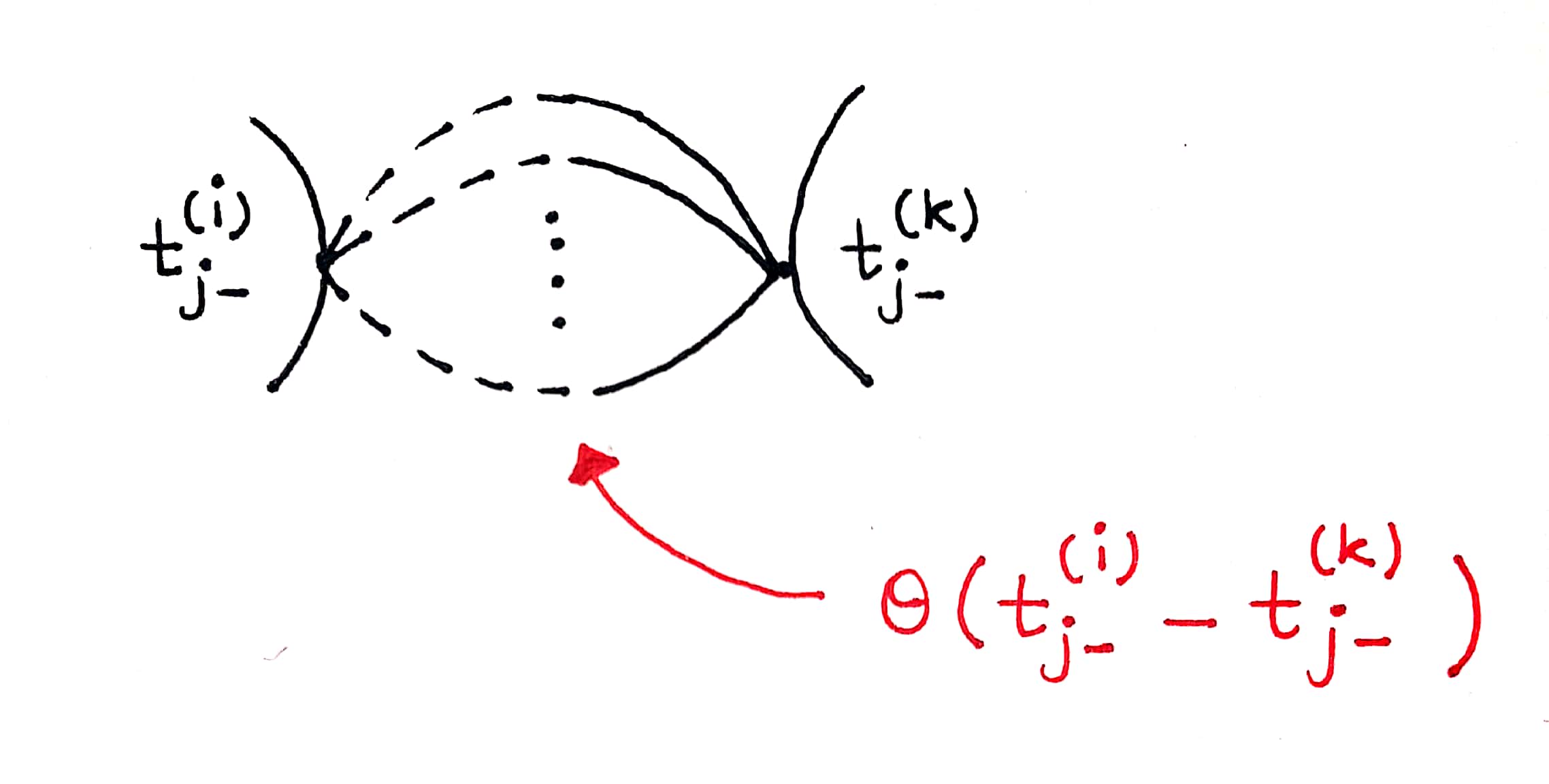}
  \caption{Rule 1.}
  \label{stepfn2}
\end{figure}
\\
\end{itemize}
Now consider the situation of the third row, wherein both the vertices are labelled $t_{2+}$. At the analytic level, this situation results from the term:
\begin{align}\label{maha}
    &\Bigl(\frac{i\lambda}{4!}\Bigr)^{2}\int_{t_0}^{t_2}dt^{(1)}_{2+}\int_{t_0}^{t_2}dt^{(2)}_{2+}\langle x(t_1)\Bigl\{T^*x^4(t^{(1)}_{2+})x^4(t^{(2)}_{2+})\Bigr\}x(t_2)\rangle.
\end{align}
Expanding the anti-time ordered product using step functions of time yields:
\begin{align}
    \langle x(t_1)\Bigl\{T^*x^4(t^{(1)}_{2+})x^4(t^{(2)}_{2+})\Bigr\}x(t_2)\rangle=&\theta(t^{(2)}_{2+}-t^{(1)}_{2+})\langle x(t_1)\Bigl\{x^4(t^{(1)}_{2+})x^4(t^{(2)}_{2+})\Bigr\}x(t_2)\rangle\notag\\+&\theta(t^{(1)}_{2+}-t^{(2)}_{2+})\langle x(t_1)\Bigl\{x^4(t^{(2)}_{2+})x^4(t^{(1)}_{2+})\Bigr\}x(t_2)\rangle.
\end{align}
Let us now say that the vertex nearer $t_1$ is labelled $t^{(1)}_{2+}$ among $t^{(1)}_{2+}$ and $t^{(2)}_{2+}$. With this choice, the diagram \ref{stepfn1} results only from the contraction pattern:
\begin{align}
    \theta(t^{(2)}_{2+}-t^{(1)}_{2+})\langle\contraction{}{x}{(t_1)}{x}\contraction[2ex]{x(t_1)x(t^{(1)}_{2+})}{x}{(t^{(1)}_{2+})x(t^{(1)}_{2+})x(t^{(1)}_{2+})x(t^{(2)}_{2+})x(t^{(2)}_{2+})}{x}\contraction[1.5ex]{x(t_1)x(t^{(1)}_{2+})x(t^{(1)}_{2+})}{x}{(t^{(1)}_{2+})x(t^{(1)}_{2+})x(t^{(2)}_{2+})}{x}\contraction{x(t_1)x(t^{(1)}_{2+})x(t^{(1)}_{2+})x(t^{(1)}_{2+})}{x}{(t^{(1)}_{2+})}{x}\contraction{x(t_1)x(t^{(1)}_{2+})x(t^{(1)}_{2+})x(t^{(1)}_{2+})x(t^{(1)}_{2+})x(t^{(2)}_{2+})x(t^{(2)}_{2+})x(t^{(2)}_{2+})}{x}{(t^{(2)}_{2+})}{x}x(t_1)x(t^{(1)}_{2+})x(t^{(1)}_{2+})x(t^{(1)}_{2+})x(t^{(1)}_{2+})x(t^{(2)}_{2+})x(t^{(2)}_{2+})x(t^{(2)}_{2+})x(t^{(2)}_{2+})x(t_2)\rangle.
\end{align}
The step function factor involved above is incorporated through the following rule:
\subsubsection{\ul{Rule 2}}
\begin{itemize}
    \item\emph{A group of free Wightman propagators coming out of a vertex labelled $t^{(i)}_{j+}$ and going into another vertex labelled $t^{(k)}_{j+}$ is associated with a step function weight $\theta(t^{(k)}_{j+}-t^{(i)}_{j+})$. That is:}
\begin{figure}[h]
    \centering
  \includegraphics[scale=0.16]{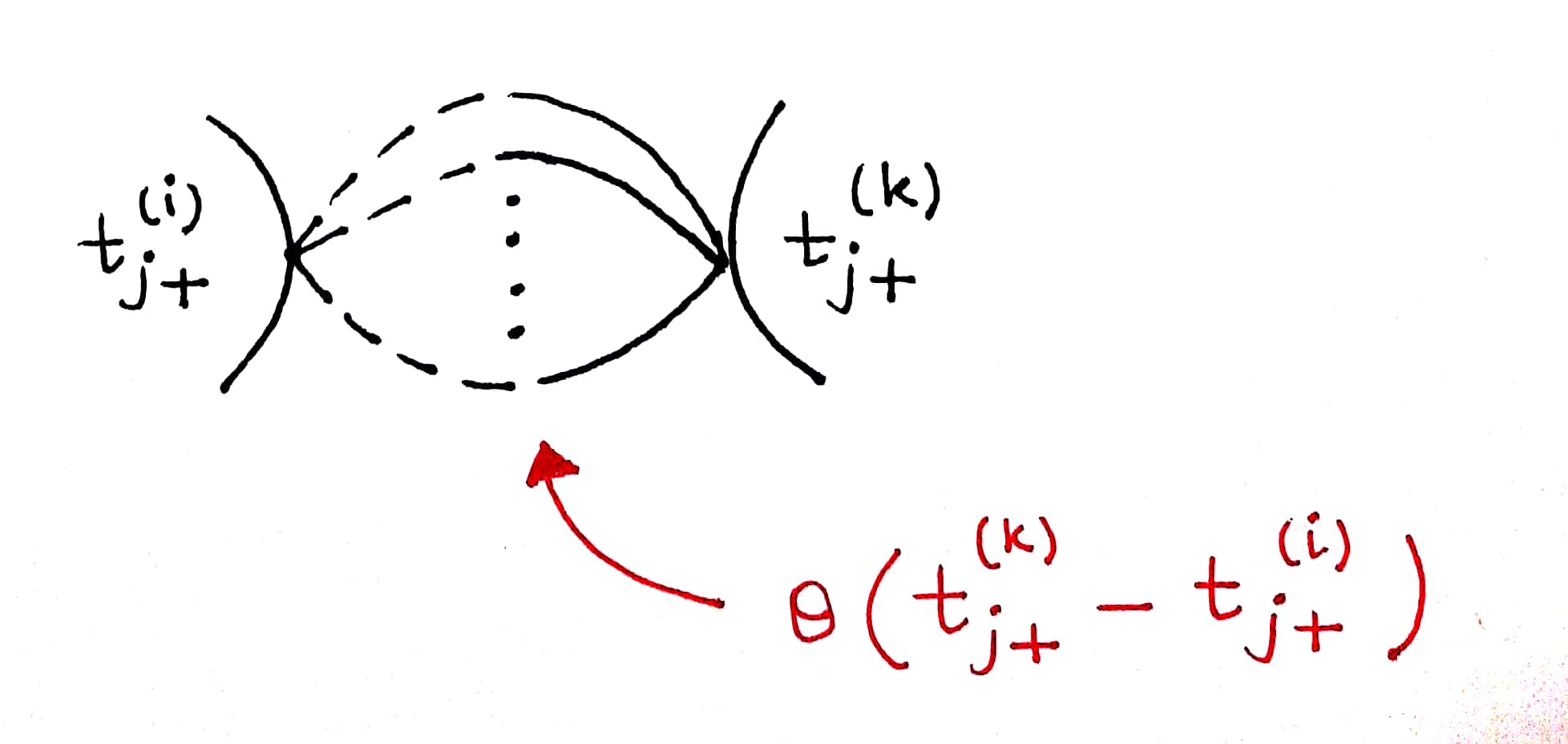}
  \caption{Rule 2.}
  \label{stepfn3}
\end{figure}
\\
\end{itemize}
Going through similar analyses, two more rules are revealed. We will not justify them. They are a simple artefact of the result that the diagrammatic representation of a bi-dentate contraction between two points involves a free Wightman propagator coming out of the point placed to the left and going into the point placed to the right in the correlator. It does not involve a free Wightman propagator coming the other way round. This was introduced in \ref{contrdiag}. Let us now list the two remaining rules.
\subsubsection{\ul{Rule 3}}
\begin{itemize}
    \item\emph{A group of free Wightman propagators coming out of a vertex labelled $t_{j\pm}$ and going into another vertex labelled $t_{k\pm}$ where $t_j\rightarrow t_k$ is associated with a step function weight of $1$. That is:}
\begin{figure}[h]
    \centering
  \includegraphics[scale=0.15]{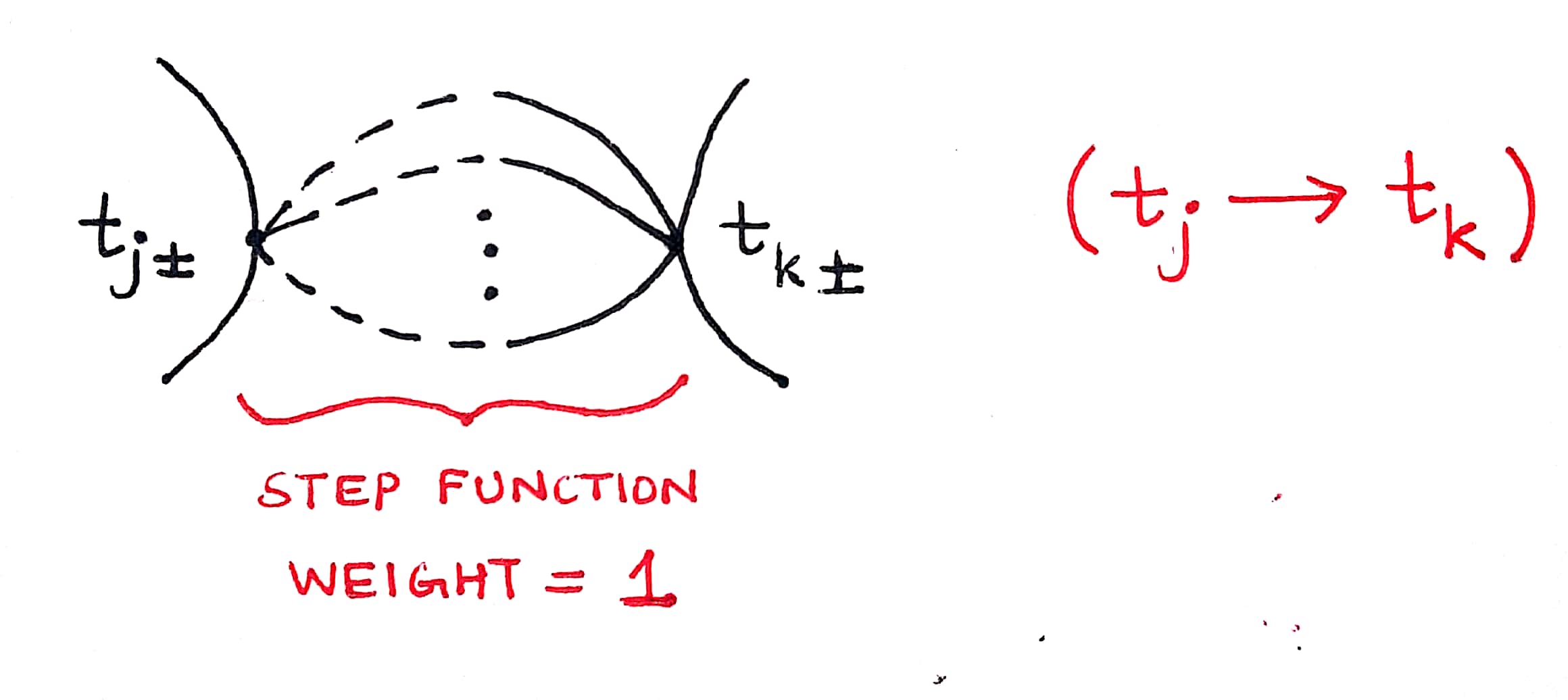}
  \caption{Rule 3.}
  \label{stepfn4}
\end{figure}
\\
\end{itemize}
\subsubsection{\ul{Rule 4}}
\begin{itemize}
    \item\emph{A group of free Wightman propagators coming out of a vertex labelled $t_{j\pm}$ and going into another vertex labelled $t_{i\pm}$ where $t_i\rightarrow t_j$ is associated with a step function weight of $0$. That is:}
\begin{figure}[h]
    \centering
  \includegraphics[scale=0.16]{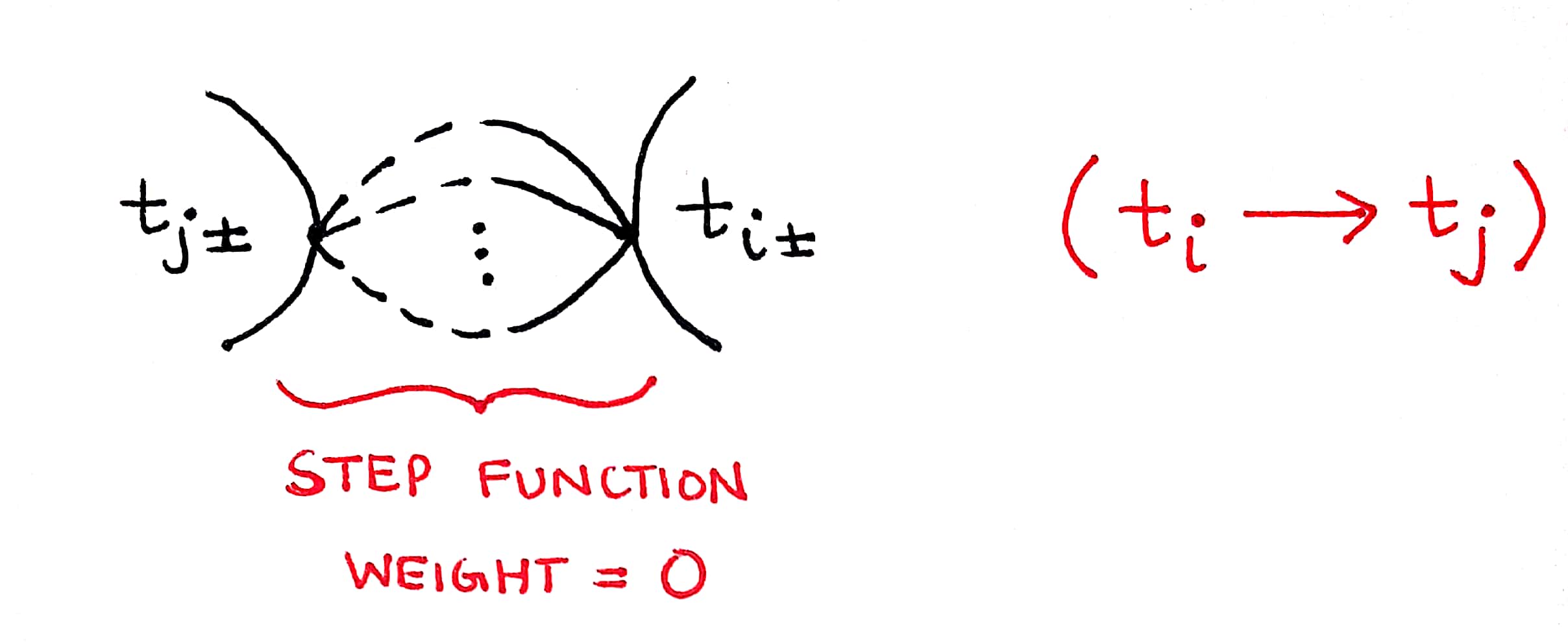}
  \caption{Rule 4.}
  \label{stepfn5}
\end{figure}
\\
\end{itemize}
In the above rules, the ordering structure `$\rightarrow$' being used is the same one which has already been introduced in \ref{notation}.
\subsection{Summary}\label{stepfnwts}
We now summarise the rules for assigning step function weights to internal propagators in diagrams.
\subsubsection{\underline{Rule 1}}
\begin{itemize}
    \item\emph{A group of free Wightman propagators coming out of a vertex labelled $t^{(i)}_{j-}$ and going into another vertex labelled $t^{(k)}_{j-}$ is associated with a step function weight $\theta(t^{(i)}_{j-}-t^{(k)}_{j-})$. That is:}
\begin{figure}[h]
    \centering
  \includegraphics[scale=0.13]{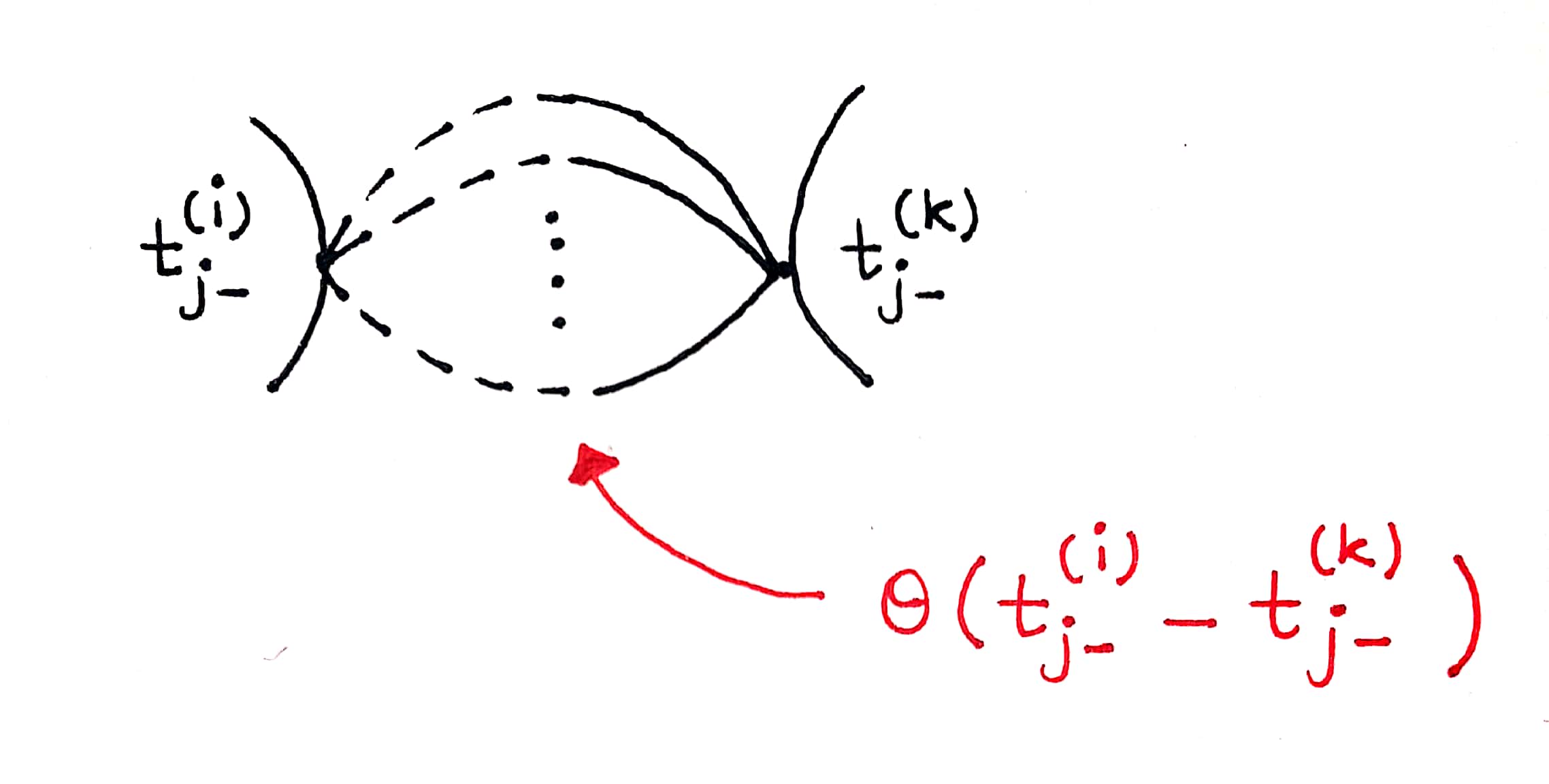}
  \caption{Rule 1.}
  \label{rule1}
\end{figure}
\end{itemize}
\subsubsection{\underline{Rule 2}}
\begin{itemize}
    \item\emph{A group of free Wightman propagators coming out of a vertex labelled $t^{(i)}_{j+}$ and going into another vertex labelled $t^{(k)}_{j+}$ is associated with a step function weight $\theta(t^{(k)}_{j+}-t^{(i)}_{j+})$. That is:}
\begin{figure}[h]
    \centering
  \includegraphics[scale=0.165]{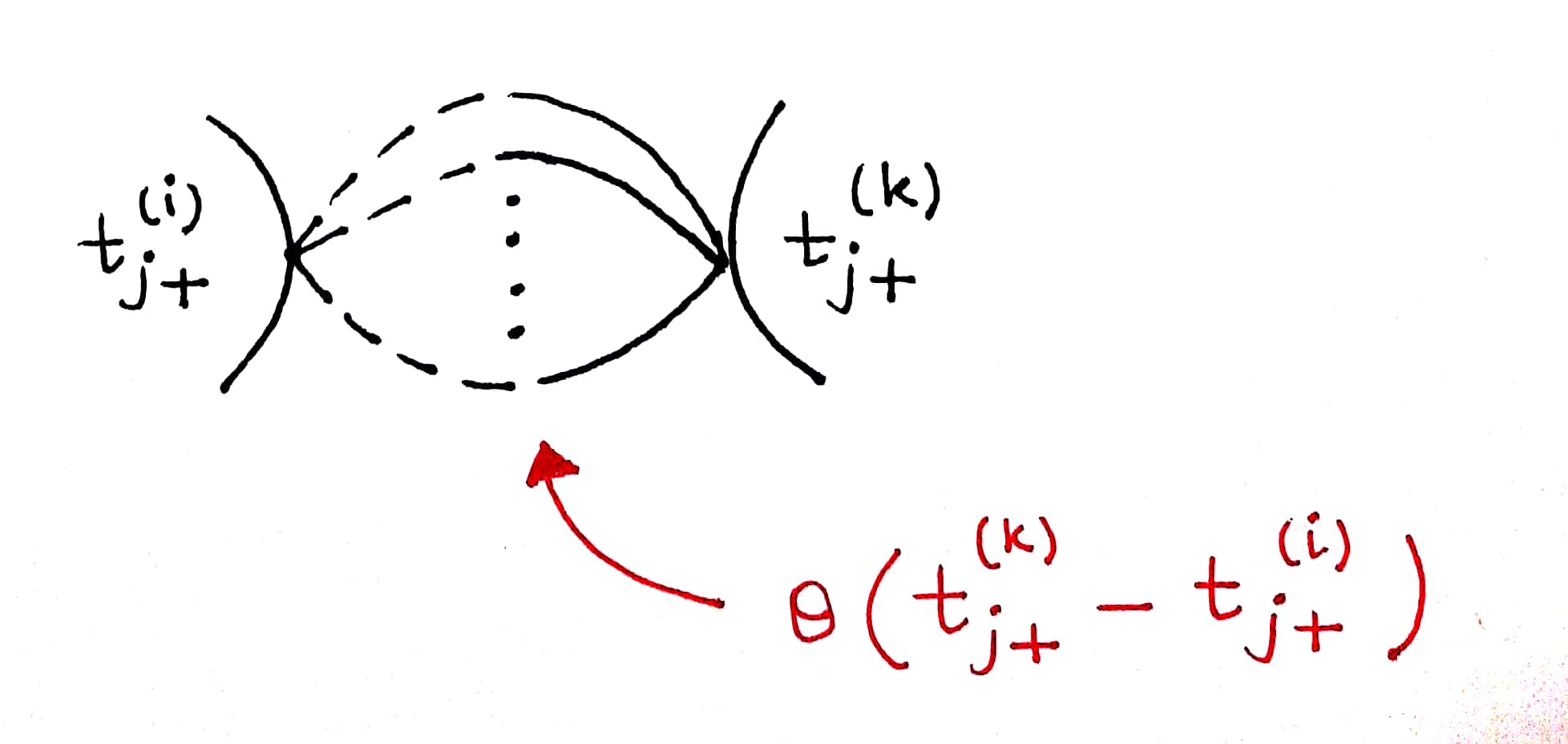}
  \caption{Rule 2.}
  \label{rule2}
\end{figure}
\end{itemize}
\subsubsection{\underline{Rule 3}}
\begin{itemize}
    \item\emph{A group of free Wightman propagators coming out of a vertex labelled $t_{j\pm}$ and going into another vertex labelled $t_{k\pm}$ where $t_j\rightarrow t_k$ is associated with a step function weight of $1$. That is:}
    \begin{figure}[h]
    \centering
  \includegraphics[scale=0.15]{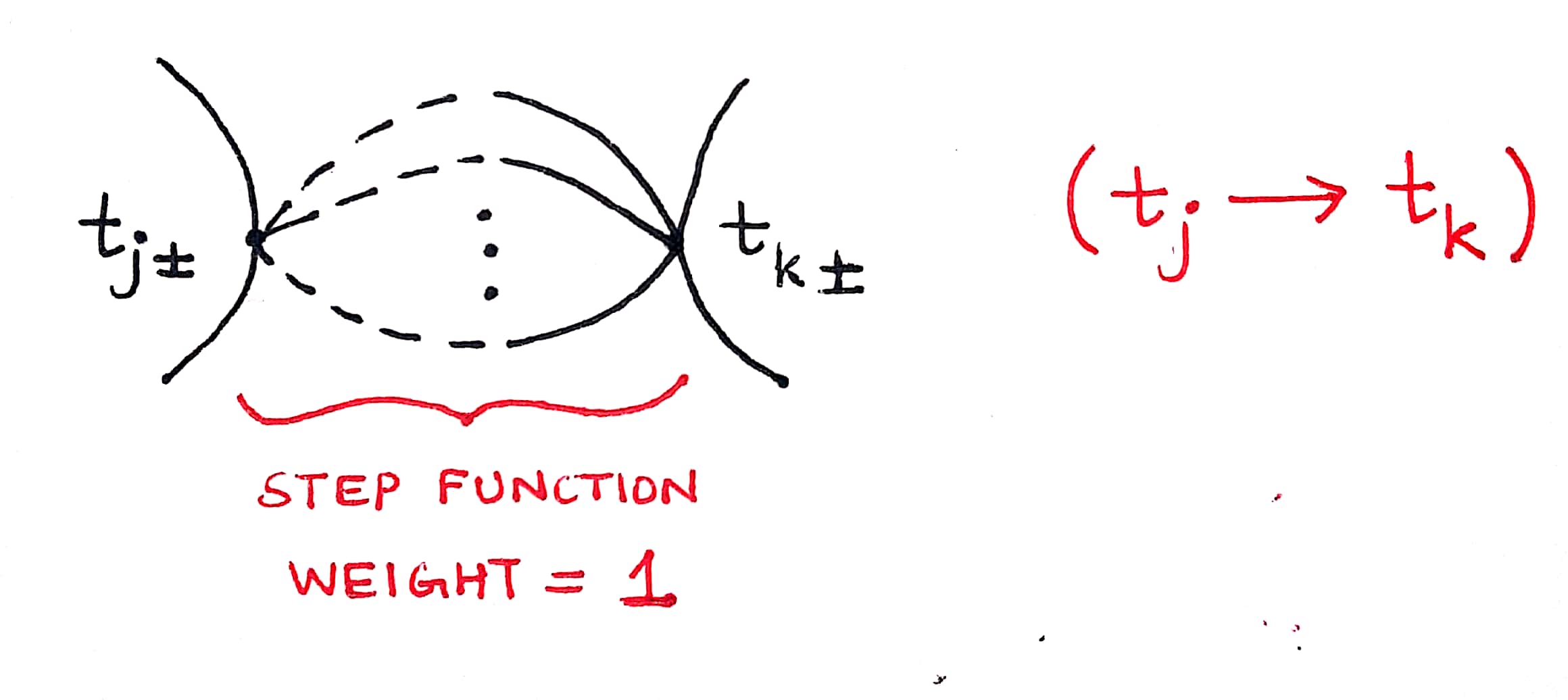}
  \caption{Rule 3.}
  \label{rule3}
\end{figure}
\end{itemize}
\newpage
\subsubsection{\underline{Rule 4}}
\begin{itemize}
    \item\emph{A group of free Wightman propagators coming out of a vertex labelled $t_{j\pm}$ and going into another vertex labelled $t_{i\pm}$ where $t_i\rightarrow t_j$ is associated with a step function weight of $0$. That is:}
    \begin{figure}[h]
    \centering
  \includegraphics[scale=0.16]{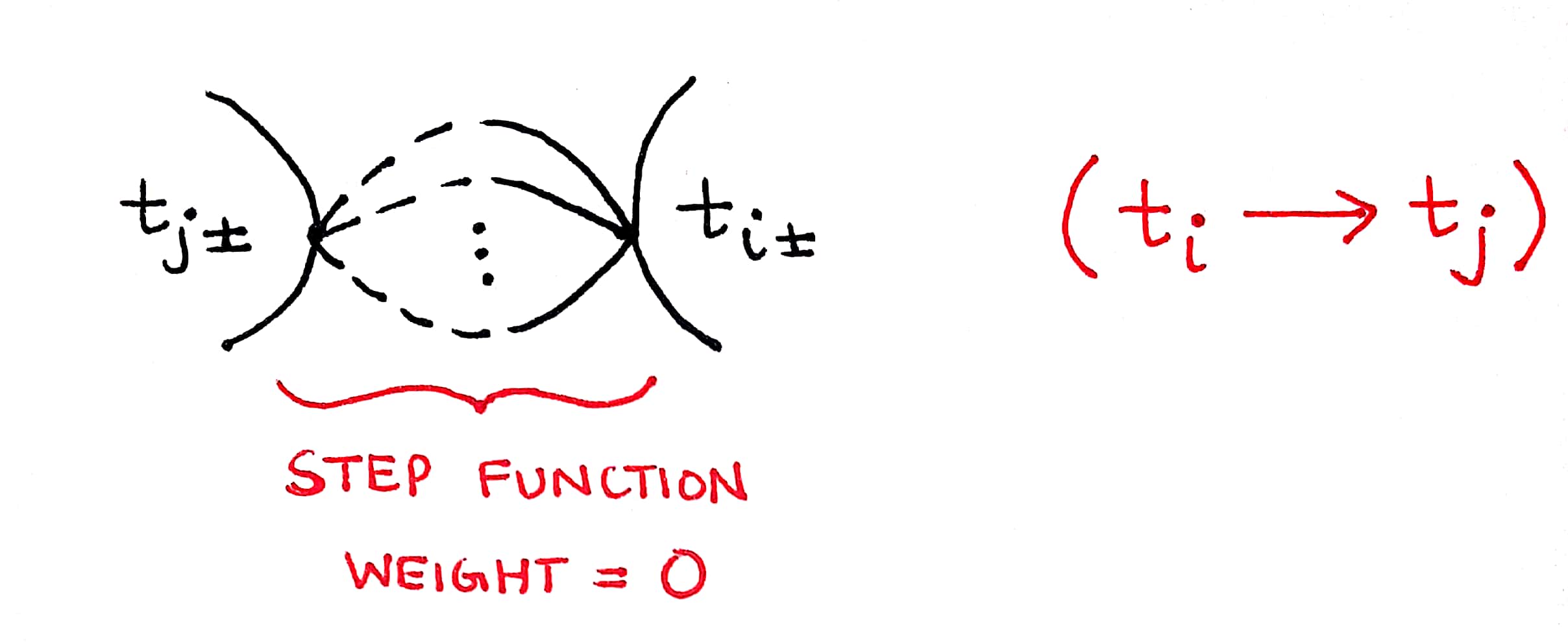}
  \caption{Rule 4.}
  \label{rule4}
\end{figure}
\end{itemize}
\section{Symmetry Factors}\label{symmetry}
We are finished discussing two of the three new diagrammatic aspects which emerge in the analysis of anharmonic Wightman correlators. The only facet remaining to be discussed is that of symmetry factors. Let us now shift our attention there.

As is also the case with traditional diagrammatic analyses of time ordered vacuum correlators, the symmetry factor of a diagram arises from the number of contraction patterns which lead to that diagram. It eventually turns out that in most cases, the symmetry factor of a diagram can be deduced by simply looking at its geometry. The need to resort to the contraction patterns which lead to it in order to find its symmetry factor is often not necessary.

In this section, we take some examples of diagrams and deduce their symmetry factors by going back to the contraction patterns which lead to them. These examples give a rough idea on how the geometry of a particular diagram may be seen as contributing to this factor. We do not lay down solid and exact rules to deduce the symmetry factor of a diagram, but just give an essence of how one may do so by looking at the geometry of the diagram. Whenever in doubt, one may always go back to the contraction patterns which lead to a diagram to compute its symmetry factor.
\newpage
As the first example, let us compute the symmetry factor of the diagram:
\begin{figure}[h]
    \centering
  \includegraphics[scale=0.09]{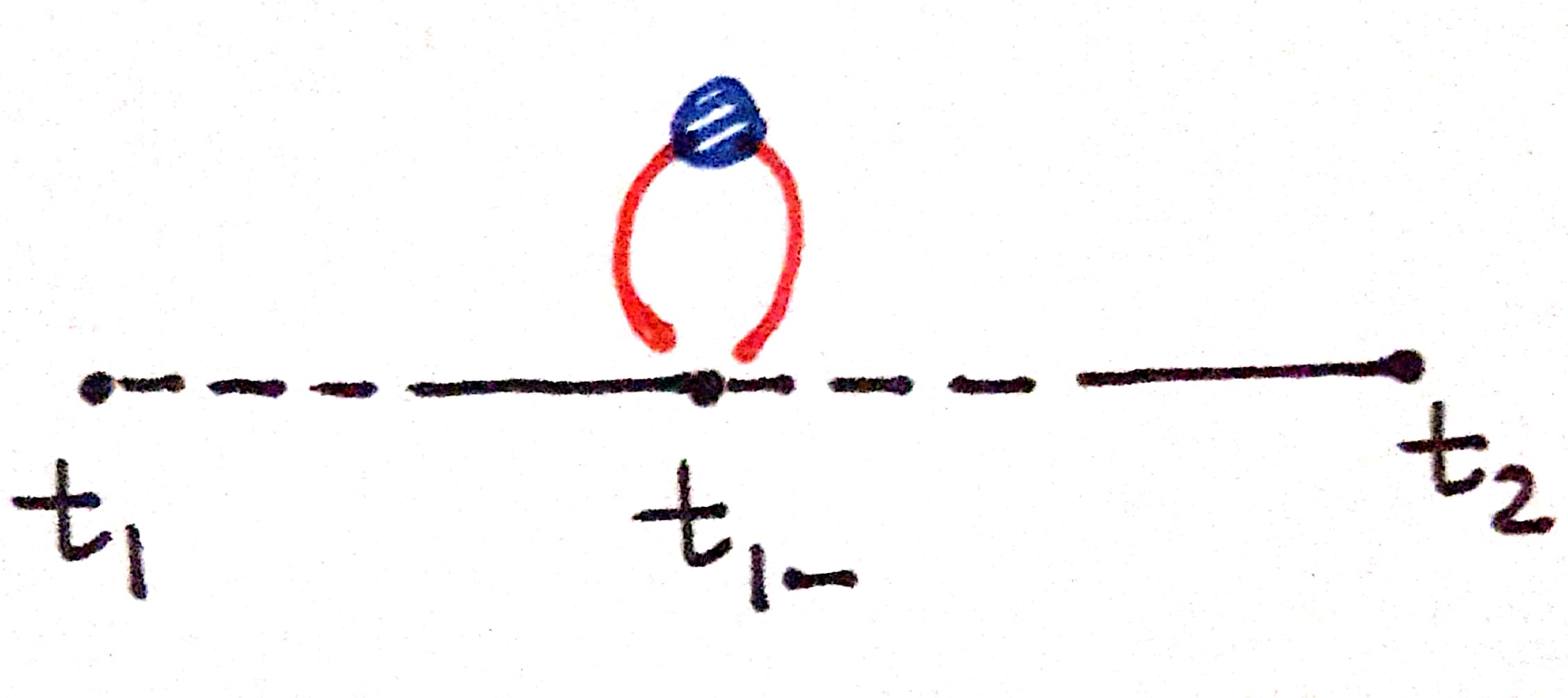}
  \caption{A diagram with symmetry factor $S=2$.}
  \label{symfac1}
\end{figure}
\\
Note that the vertex of the above diagram has already been labelled $t_{1-}$. Thus, one of the contraction patterns which leads to this diagram is:
\begin{align}\label{maha}
    \Bigl(\frac{-i\lambda}{4!}\Bigr)\int_{t_0}^{t_1}dt_{1-}\langle\contraction{}{x}{(t_1)}{x}\contraction{x(t_1)x(t_{1-})}{x}{(t_{1-})}{x}\contraction{x(t_1)x(t_{1-})x(t_{1-})x(t_{1-})}{x}{(t_{1-})}{x}x(t_1)x(t_{1-})x(t_{1-})x(t_{1-})x(t_{1-})x(t_2)\rangle.
\end{align}
If one thinks a bit more, one would realise that to get the diagram \ref{symfac1}, one only needs to contract the external point $t_1$ in \eqref{maha} with any one of the $t_{1-}$, and the external point $t_2$ with any one of the remaining $t_{1-}$. This can be done in $4\times 3=12$ ways. But there is also a factor of $4!$ in the denominator of \eqref{maha}. Combining these two, one finally gets a factor of $2$ in the denominator. This is said to be the \emph{symmetry factor} of the diagram \ref{symfac1}. Geometrically, it can be thought of as the two ways in which one can rotate the loop in diagram \ref{symfac1} by $180$ degrees about an axis which cuts the loop in half.

Similarly, the symmetry factor of the diagram:
\begin{figure}[h]
    \centering
  \includegraphics[scale=0.1]{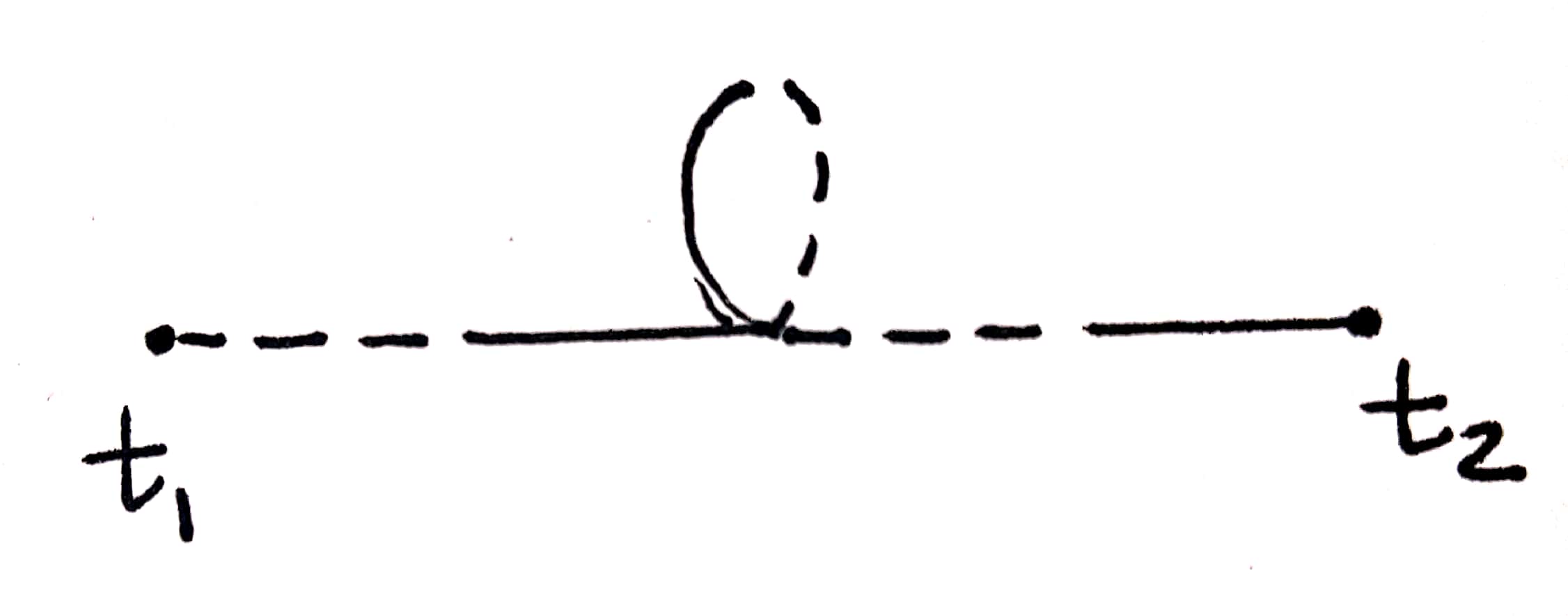}
  \caption{Another diagram with symmetry factor $S=2$.}
  \label{symfac2}
\end{figure}
\\
is also found out to be $2$. A similar geometric interpretation works for this diagram too.

\begin{figure}[h]
    \centering
  \includegraphics[scale=0.07]{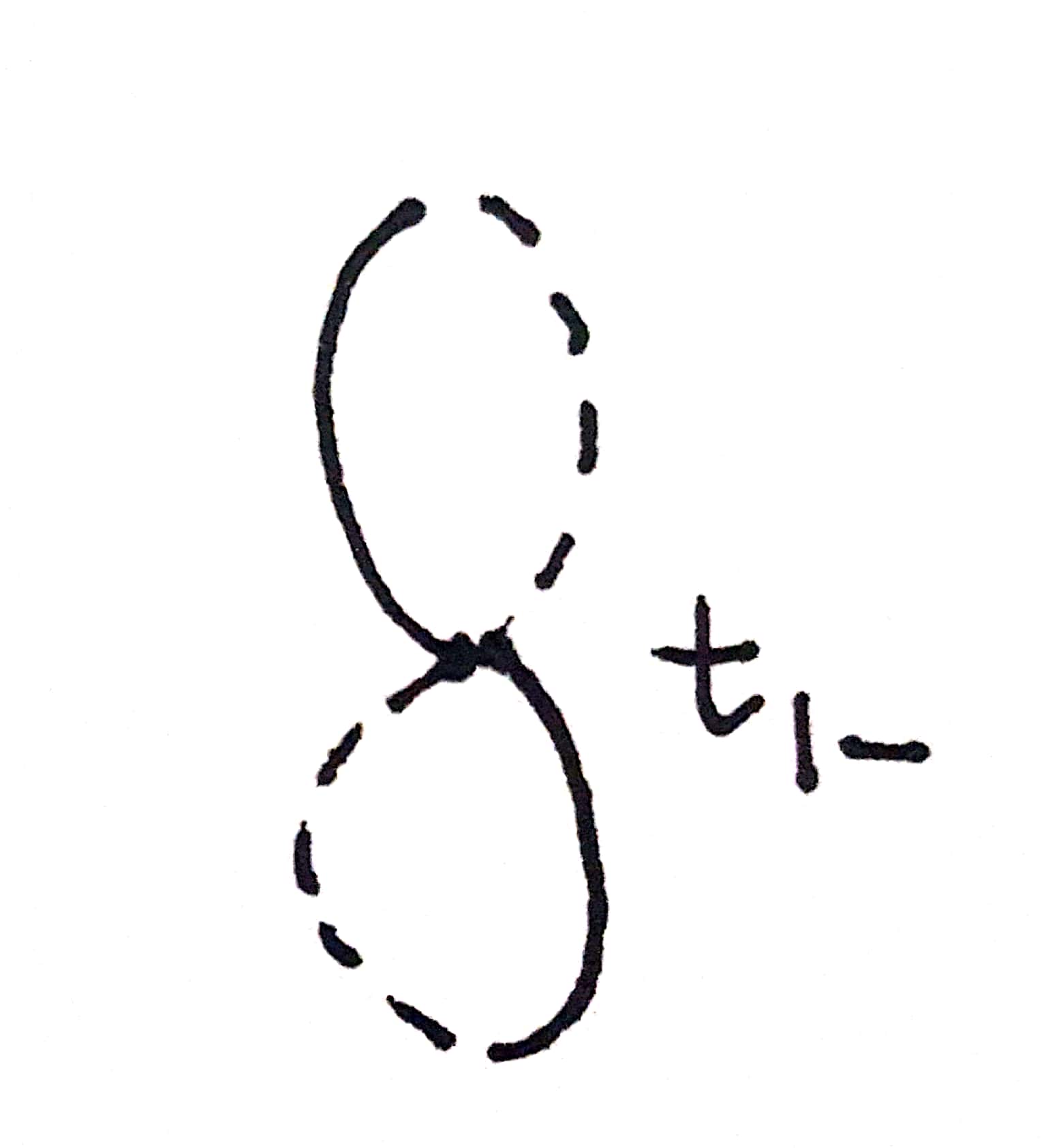}
  \caption{A vacuum bubble with symmetry factor $S=8$.}
  \label{symfac3}
\end{figure}

Consider now the vacuum bubble shown in figure \ref{symfac3} on the next page. It arises in the analysis of the two point anharmonic Wightman correlator $\langle x_H(t_1)x_H(t_2)\rangle$.
\\
Note that the vertex of the above vacuum bubble has already been labelled $t_{1-}$. Thus it arises only from the contraction patterns:
\begin{align}\label{maha2}
    &\Bigl(\frac{-i\lambda}{4!}\Bigr)\int_{t_0}^{t_1}dt_{1-}\langle\contraction{x(t_1)}{x}{(t_{1-})}{x}\contraction{x(t_1)x(t_{1-})x(t_{1-})}{x}{(t_{1-})}{x}\contraction[1.5ex]{}{x}{(t_1)x(t_{1-})x(t_{1-})x(t_{1-})x(t_{1-})}{x}x(t_1)x(t_{1-})x(t_{1-})x(t_{1-})x(t_{1-})x(t_2)\rangle\notag\\+&\Bigl(\frac{-i\lambda}{4!}\Bigr)\int_{t_0}^{t_1}dt_{1-}\langle\contraction{x(t_1)}{x}{(t_{1-})x(t_{1-})}{x}\contraction[1.5ex]{x(t_1)x(t_{1-})}{x}{(t_{1-})x(t_{1-})}{x}\contraction[2ex]{}{x}{(t_1)x(t_{1-})x(t_{1-})x(t_{1-})x(t_{1-})}{x}x(t_1)x(t_{1-})x(t_{1-})x(t_{1-})x(t_{1-})x(t_2)\rangle\notag\\+&\Bigl(\frac{-i\lambda}{4!}\Bigr)\int_{t_0}^{t_1}dt_{1-}\langle\contraction[1.5ex]{x(t_1)}{x}{(t_{1-})x(t_{1-})x(t_{1-})}{x}\contraction{x(t_1)x(t_{1-})}{x}{(t_{1-})}{x}\contraction[2ex]{}{x}{(t_1)x(t_{1-})x(t_{1-})x(t_{1-})x(t_{1-})}{x}x(t_1)x(t_{1-})x(t_{1-})x(t_{1-})x(t_{1-})x(t_2)\rangle.
\end{align}
As can be checked, each term in \eqref{maha2} gives rise to a single copy of the vacuum bubble \ref{symfac3}. Thus, the final numerical factor accompanying this vacuum bubble would be $3/4!=1/8$. This reveals its symmetry factor being $8$. Geometrically, it can be thought of as resulting from the $2$ ways in which each of the loops can be rotated. This contributes a factor of $2\times 2=4$. The additional factor of $2$ can be thought of as coming from the two ways in which the two loops can interchange their positions.

Another vacuum bubble which arises from the contraction patterns \eqref{maha2} is:
\begin{figure}[h]
    \centering
  \includegraphics[scale=0.07]{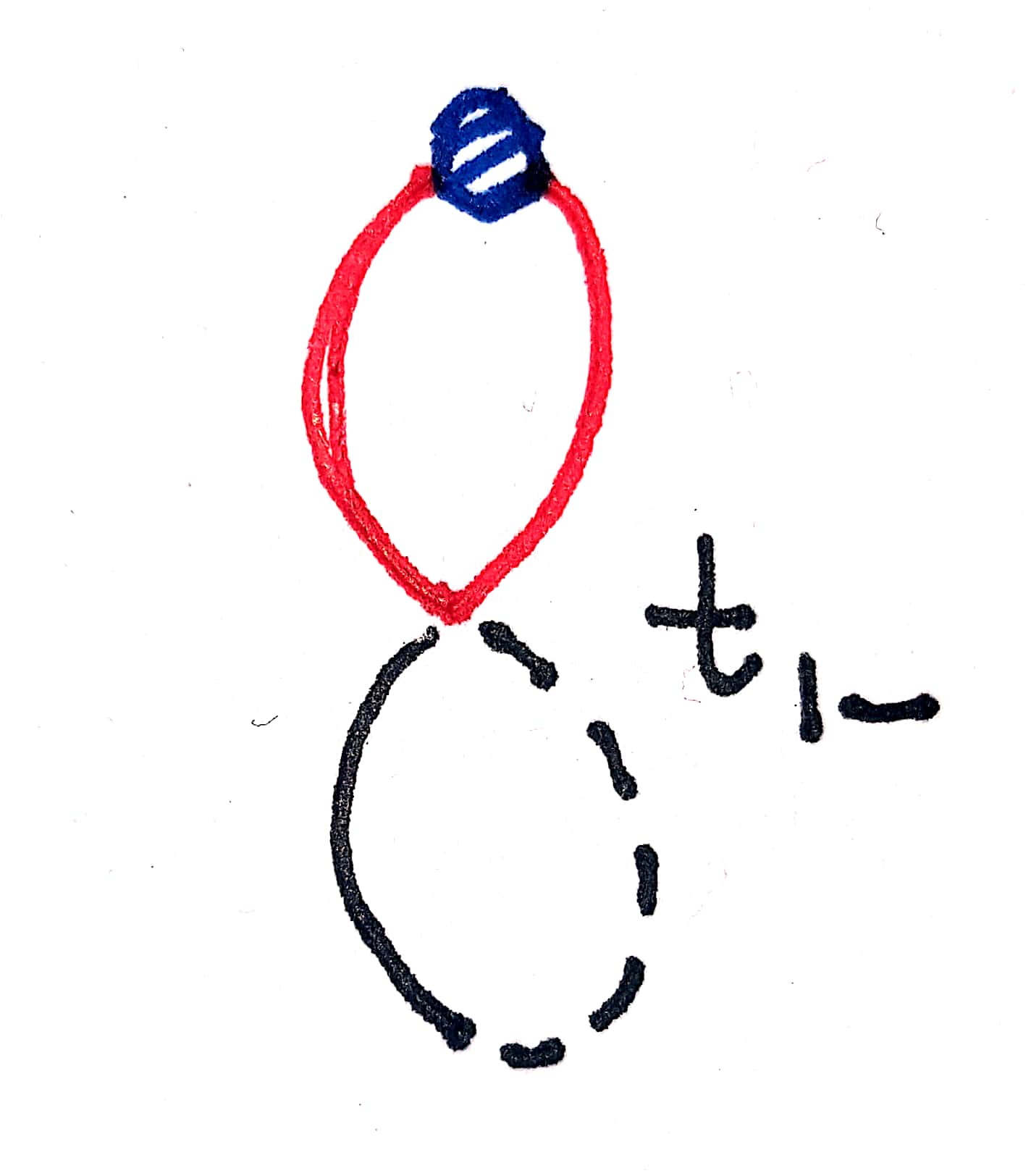}
  \caption{A vacuum bubble with symmetry factor $S=4$.}
  \label{symfac4}
\end{figure}
\\
However, as one can readily check, each term of \eqref{maha2} now contributes \emph{two} copies of this vacuum bubble. The final numerical factor accompanying this vacuum bubble would thus be $(2\times 3)/4!=1/4$. Therefore, the symmetry factor of this vacuum bubble is $4$. Geometrically, it can be thought of as resulting from the $2$ ways in which each of the loops may be rotated. The factor arising from interchanging the loops does not make an appearance here. One can justify this by stating that only similar loops may be interchanged with each other. The loops in the vacuum bubble we are analysing are not similar. One of them is a free Wightman loop, whereas the other is a loop containing a cumulant blob.
\newpage
Some other symmetry factors which may be found through similar analyses are:
\begin{figure}[h]
    \centering
  \includegraphics[scale=0.13]{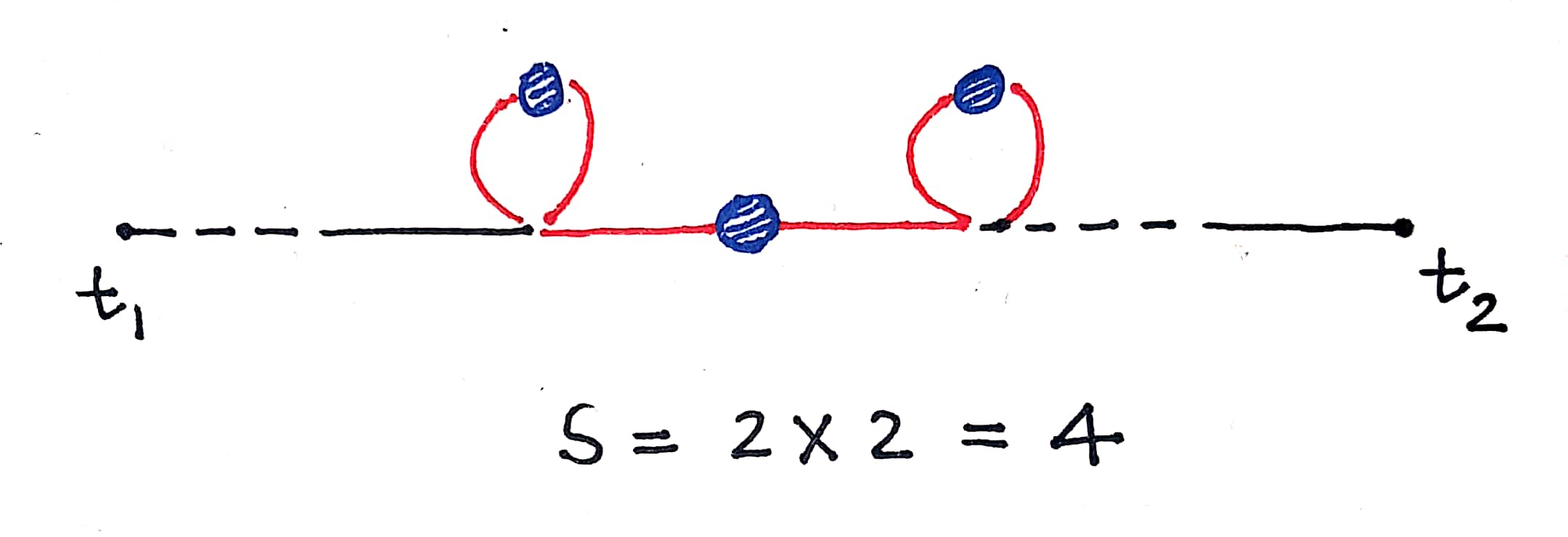}
  \caption{A diagram with symmetry factor $S=2\times 2=4$.}
  \label{symfac5}
\end{figure}
\\
because each loop can be rotated by $180$ degrees independently about axes which cut them in half.
\begin{figure}[h]
    \centering
  \includegraphics[scale=0.12]{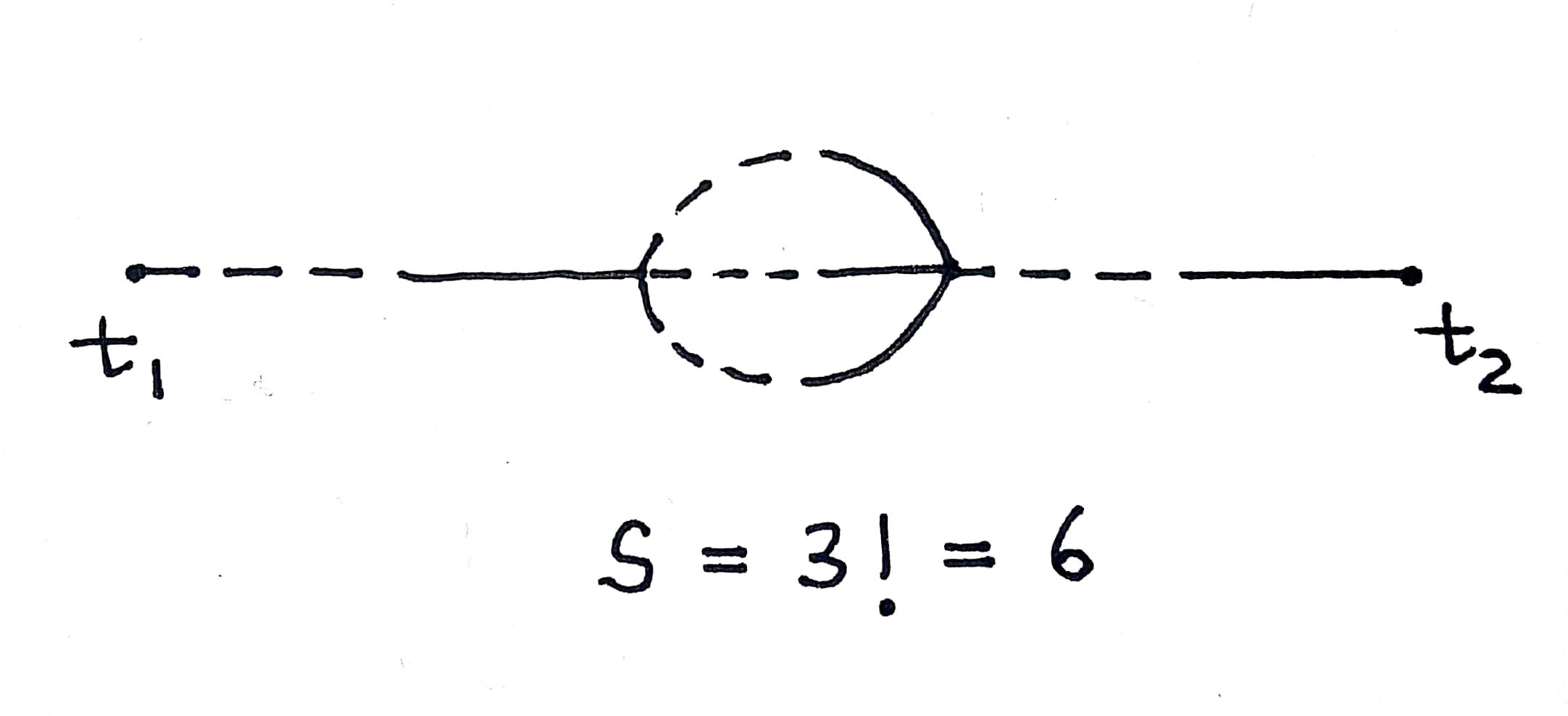}
  \caption{A diagram with symmetry factor $S=3!=6$.}
  \label{symfac6}
\end{figure}
\\
because the internal propagators can be permuted among themselves in $3!$ ways.
\begin{figure}[h]
    \centering
  \includegraphics[scale=0.07]{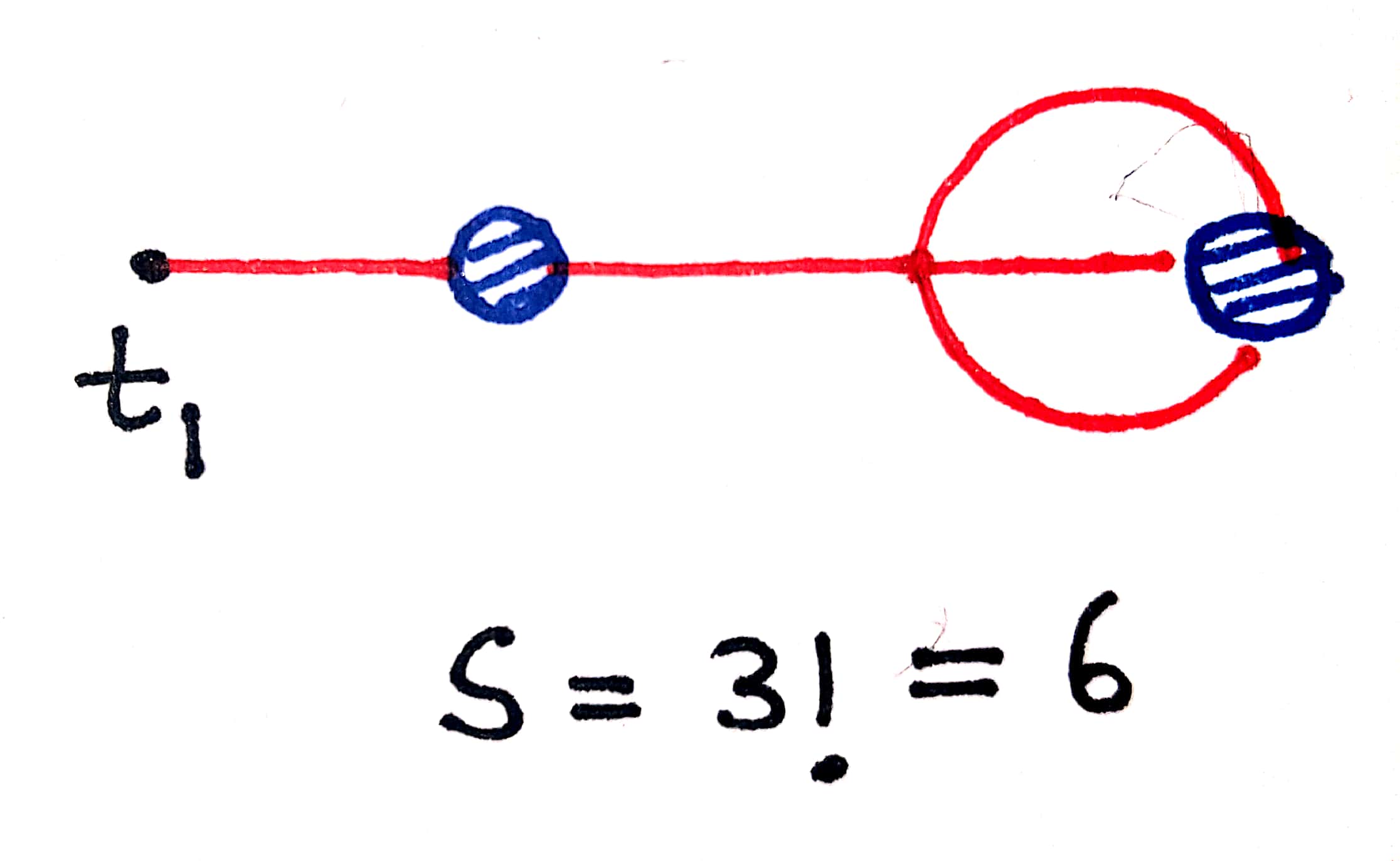}
  \caption{Another diagram with symmetry factor $S=3!=6$.}
  \label{symfac7}
\end{figure}
\\
because the three red propagators emanating from the cumulant blob at the end can be permuted among themselves in $3!$ ways.
\newpage
\section{The Feynman Rules}\label{feynrules}
We are now ready to state the `Feynman rules' for computing Wightman correlators in general states of the anharmonic oscillator. Suppose we aim to compute the object:
\begin{equation}\label{feynrul}
    \langle x_H(t_1)x_H(t_2)\dots x_H(t_n)\rangle,
\end{equation}
using diagrams. How do we proceed? 

To do so, one must follow the steps:
\begin{itemize}
    \item Draw all possible diagrams with the external points $t_1,t_2,\dots,t_n$. The components which have to be used to draw these diagrams are free Wightman propagators, red propagators and cumulant blobs. Refer to \ref{diagrams} for their description. However, the following points must be kept in mind:
    \begin{itemize}
        \item A free Wightman propagator connecting an external point with another must always be coming out of the external point which is placed to the left and going into the external point which is placed to the right in the correlator \eqref{feynrul}.
        \item Interaction vertices are intersection points of four propagators. These propagators can either be free Wightman propagators or red propagators. While applying this definition, a loop of a free Wightman propagator must be counted as two free Wightman propagators.
        \item Cumulant blobs are connected to points only through red propagators.
        \item Any number of red propagators may originate from a cumulant blob.
        \item Cumulant blobs are never connected to other cumulant blobs.
        \item A red propagator can never loop back into a cumulant blob.
        \item A group of free Wightman propagators joining two vertices must be either all going out from one of the vertices to the other or all coming in from one of the vertices to the other.
    \end{itemize}
    \item Assign all the vertices in the diagram their labels through the procedure outlined in \ref{labelling}.
    \item For a given combination of labels of the vertices:
    \begin{itemize}
        \item Associate each diagrammatic component with its corresponding analytic expression. This has been listed in \ref{diagrams}.
        \item Assign step function weights to all the groups of internal propagators in the diagram. This has been explained in \ref{stepfnwts}.
        \item Read off the vertex factors from the vertices' labels as outlined in \ref{reading}.
        \item Divide by the symmetry factor of the diagram.
    \end{itemize}
    \item Repeat the above step for all possible combinations of vertex labels. 
\end{itemize}
\section{A Practical Guide for Calculations}
Let us implement the procedure outlined in the previous section to compute some contributions to the anharmonic 2-point Wightman correlator in a general state diagrammatically. A notation we would use here is that we would denote the free Wightman propagator coming into the time $t_j$ from the time $t_i$ as $D_w(t_i-t_j)$. That is:
\begin{equation}
    \frac{\hbar}{2\omega}e^{-i\omega(t_i-t_j)}\doteq D_w(t_i-t_j).
\end{equation}
We also introduce a new function $\mathcal{C}_2(t_i,t_j)$ defined as:
\begin{equation}
    \mathcal{C}_2(t_i,t_j)\equiv C_2(t_i,t_j)-D_w(t_i-t_j).
\end{equation}
The benefit of introducing it is that it has a simple diagrammatic representation. As revealed through \ref{contrdiag}:
\begin{figure}[h]
    \centering
  \includegraphics[scale=0.13]{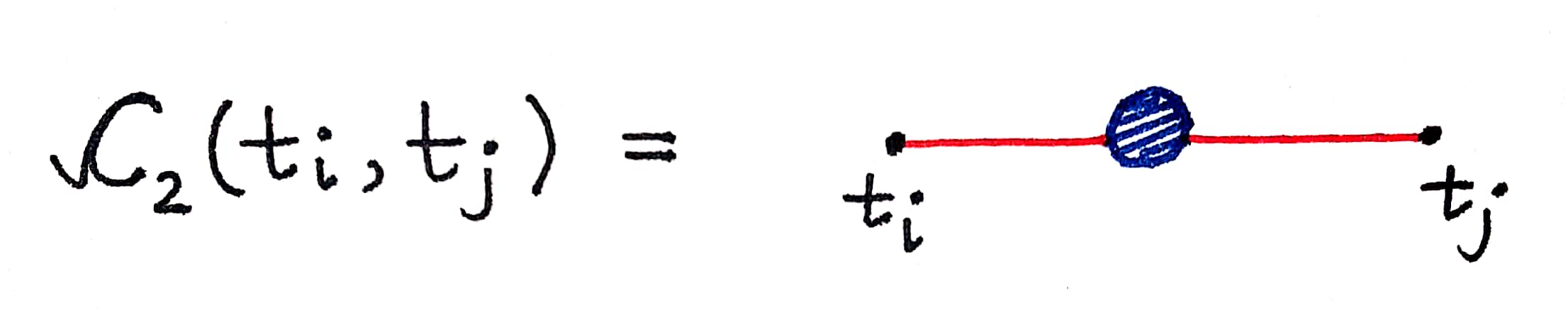}
  \caption{The diagrammatic representation of the function $\mathcal{C}_2$.}
  \label{mathcalc}
\end{figure}
\\
\iffalse\\[10pt]
\begin{center}
\textcolor{blue}{\textbf{(Continued on next page)}}
\end{center}
\newpage\fi
Let us now come to the point. A few diagrams which form a part of $\langle x_H(t_1)x_H(t_2)\rangle$ are:
\begin{figure}[h]
    \centering
  \includegraphics[scale=0.2]{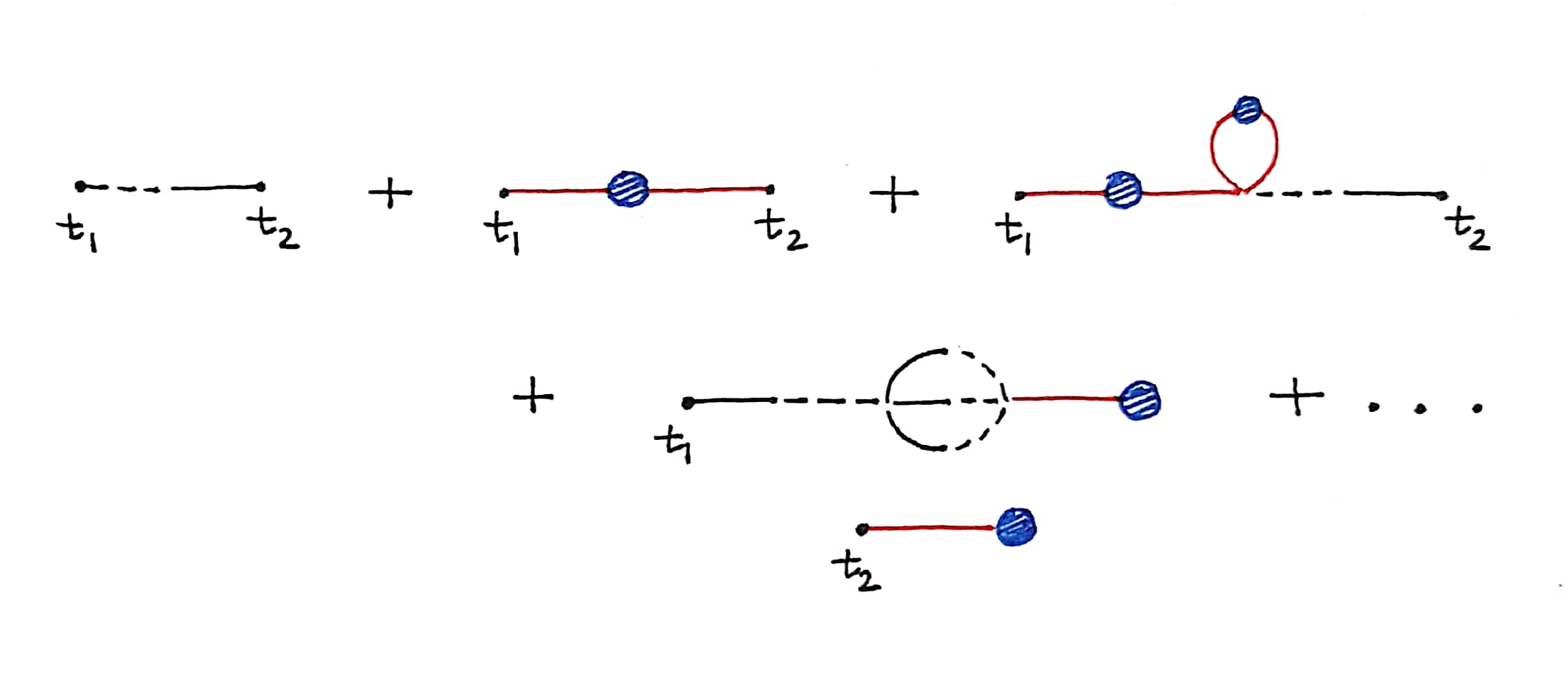}
  \caption{Some diagrams which contribute to $\langle x_H(t_1)x_H(t_2)\rangle$.}
  \label{twoptfull}
\end{figure}
\newpage
The first two diagrams of figure \ref{twoptfull} on the previous page are also obtained in the study of the free 2-point Wightman correlator. They simply sum up to the second cumulant $C_2$. That is:
\begin{figure}[h]
    \centering
  \includegraphics[scale=0.19]{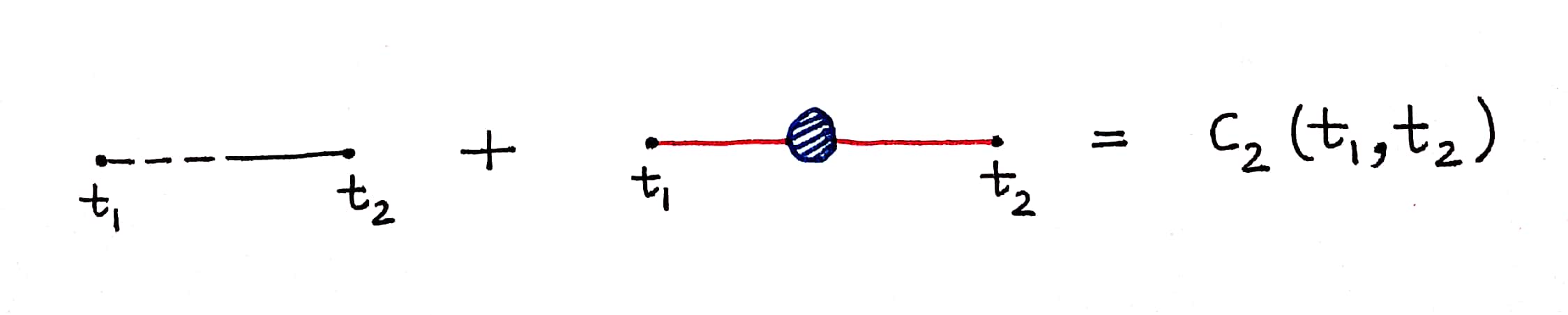}
  \caption{Sum of the first two diagrams in figure \ref{twoptfull}.}
  \label{sumc2}
\end{figure}
\\
Consider evaluating the third diagram of figure \ref{twoptfull}:
\begin{figure}[h]
    \centering
  \includegraphics[scale=0.1]{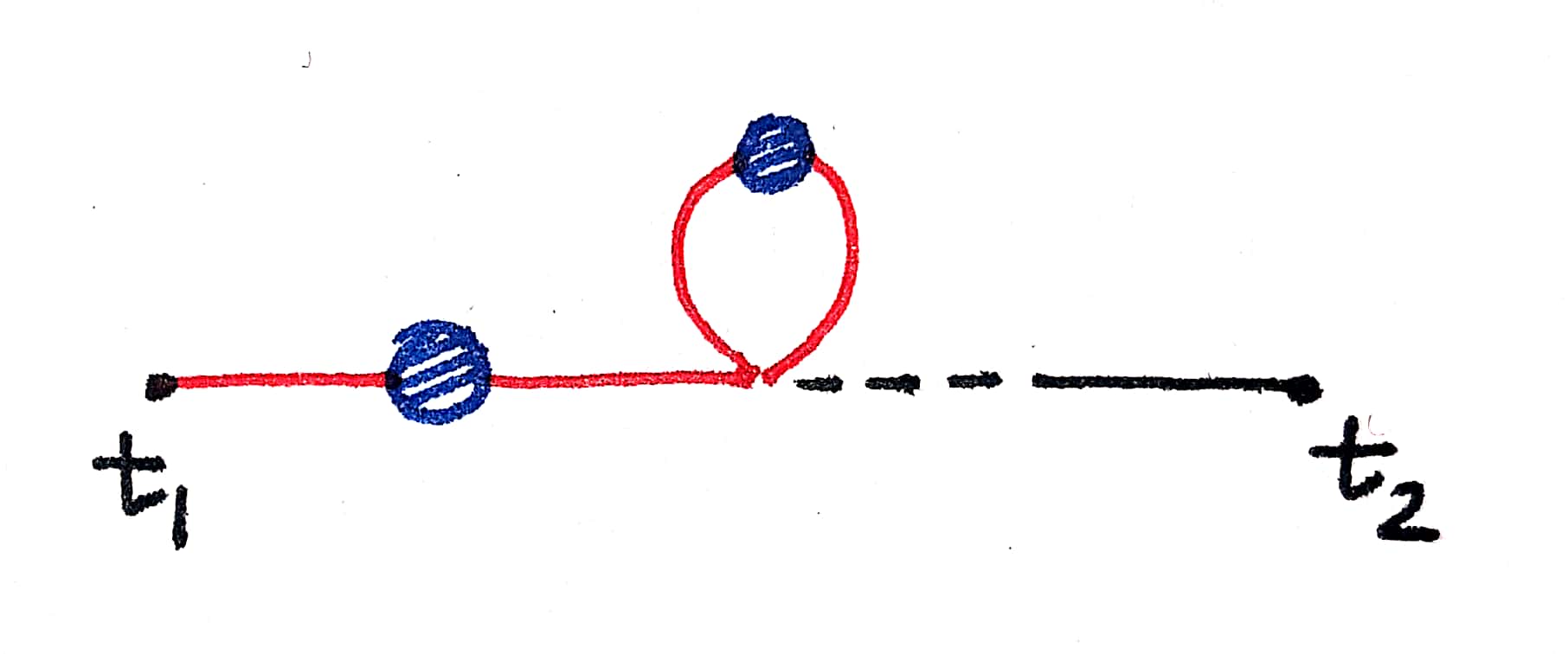}
  \caption{A diagram which contributes to $\langle x_H(t_1)x_H(t_2)\rangle$.}
  \label{thirddiag}
\end{figure}
\\
Let us now label its vertex. To start with, assume that its label is $t_k$. That is:
\begin{figure}[h]
    \centering
  \includegraphics[scale=0.09]{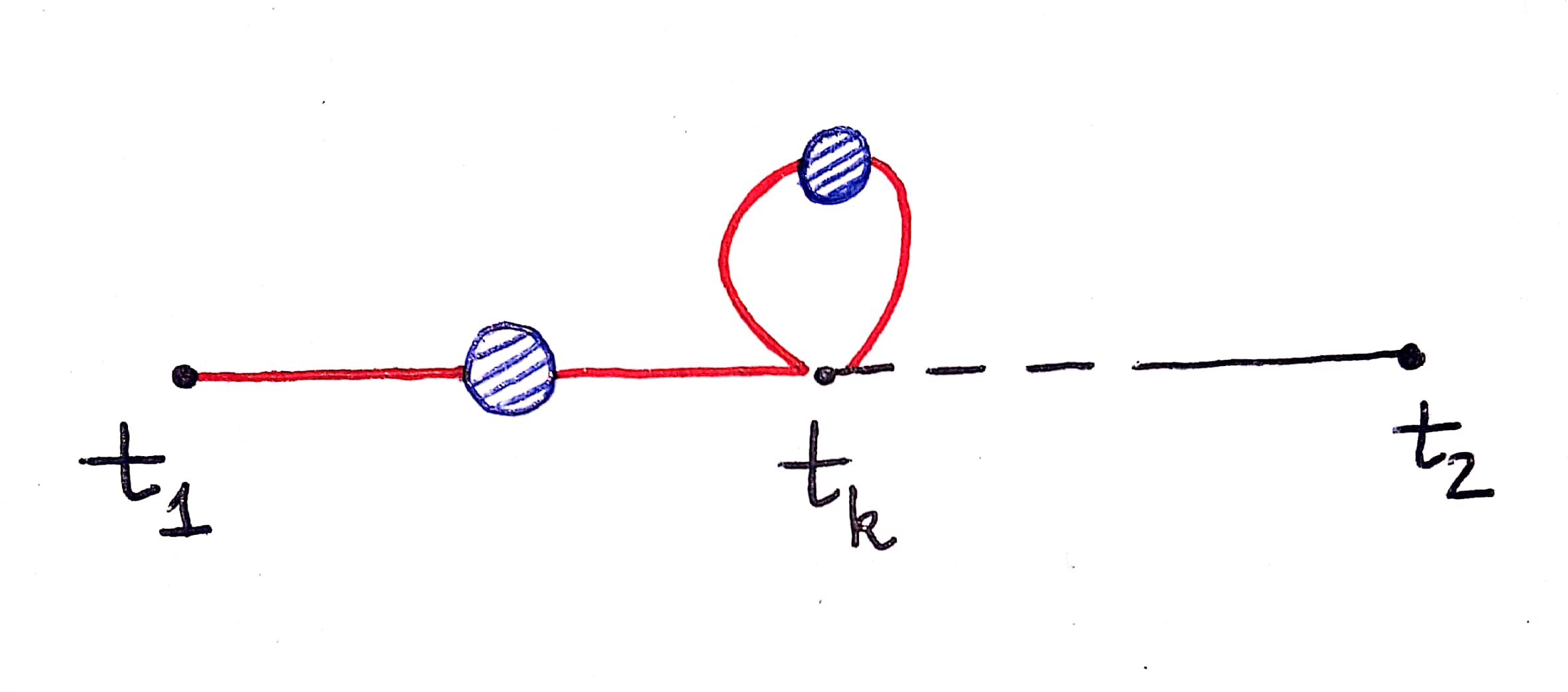}
\end{figure}
\\
Since we are studying the 2-point Wightman correlator $\langle x_H(t_1)x_H(t_2)\rangle$, there is the obvious requirement that:
\begin{equation}
    t_k\in M_2=\{t_{1+},t_{1-},t_{2+},t_{2-}\}.
\end{equation}
Moreover, the free Wightman propagator going out of the vertex into the external point $t_2$ imposes the constraint:
\begin{equation}
    t_k\rightarrow t_2.
\end{equation}
There is no other constraint apart from this one. Abiding by it, the set of permissible labels for this vertex is:
\begin{equation}
    t_k=\{t_{1+},t_{1-},t_{2+}\}.
\end{equation}
Let us first assign this vertex the label $t_{1+}$. In this case, the vertex factor reads:
\begin{equation}
    +i\lambda\int_{t_0}^{t_1}dt_{1+},
\end{equation}
and the diagram evaluates to:
\begin{equation}\label{a}
    +\frac{i\lambda}{2}\int_{t_0}^{t_1}dt_{1+}\{\mathcal{C}_2(t_1,t_{1+})\}\cdot\{\mathcal{C}_2(t_{1+},t_{1+})\}\cdot\{D_w(t_{1+}-t_2)\}.
\end{equation}
Note that the factor of $2$ in the denominator is the symmetry factor of the diagram. Similarly, when this diagram is assigned the label $t_{1-}$, the vertex factor reads:
\begin{equation}
    -i\lambda\int_{t_0}^{t_1}dt_{1-},
\end{equation}
and the diagram evaluates to:
\begin{equation}\label{b}
    -\frac{i\lambda}{2}\int_{t_0}^{t_1}dt_{1-}\{\mathcal{C}_2(t_1,t_{1-})\}\cdot\{\mathcal{C}_2(t_{1-},t_{1-})\}\cdot\{D_w(t_{1-}-t_2)\}.
\end{equation}
And when assigned the label $t_{2+}$, the vertex factor reads:
\begin{equation}
    +i\lambda\int_{t_0}^{t_2}dt_{2+},
\end{equation}
and the diagram evaluates to:
\begin{equation}\label{c}
    +\frac{i\lambda}{2}\int_{t_0}^{t_2}dt_{2+}\{\mathcal{C}_2(t_1,t_{2+})\}\cdot\{\mathcal{C}_2(t_{2+},t_{2+})\}\cdot\{D_w(t_{2+}-t_2)\}.
\end{equation}
Thus, the total value of the diagram \ref{thirddiag} is the sum of \eqref{a}, \eqref{b} and \eqref{c}. This is:
\\[10pt]
\fcolorbox{black}{gray!10}{\parbox{35em}{
\begin{equation}
    +\frac{i\lambda}{2}\int_{t_0}^{t_2}dt_{2+}\{\mathcal{C}_2(t_1,t_{2+})\}\cdot\{\mathcal{C}_2(t_{2+},t_{2+})\}\cdot\{D_w(t_{2+}-t_2)\}.
\end{equation}}}
\\[45pt]
Now consider the fourth diagram of the figure \ref{twoptfull}:
\begin{figure}[h]
    \centering
  \includegraphics[scale=0.09]{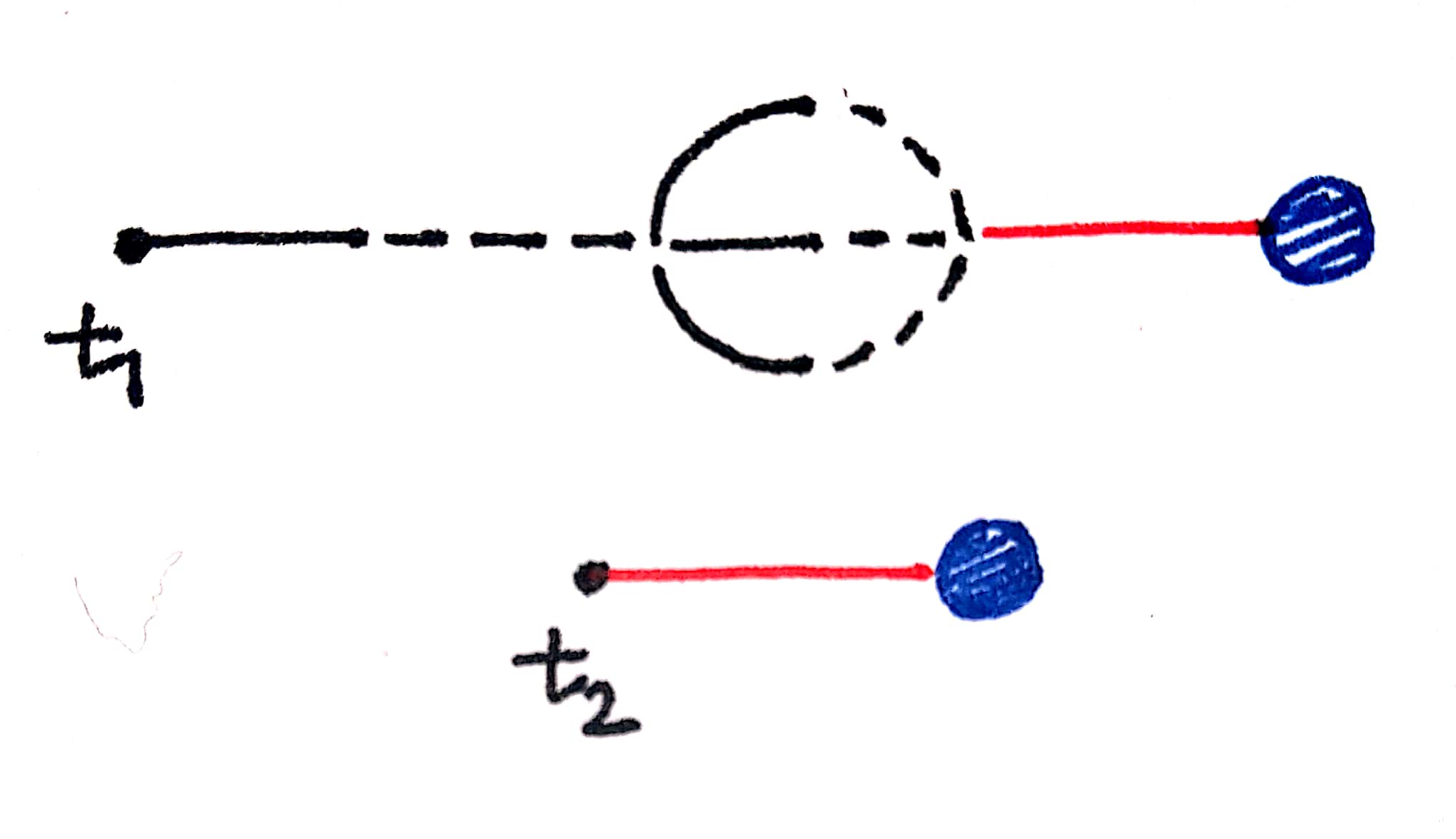}
  \caption{A diagram which contributes to $\langle x_H(t_1)x_H(t_2)\rangle$.}
  \label{fourthdiag}
\end{figure}
\newpage
Let us now label its two vertices. To start with, let the label of the vertex nearer $t_1$ be $t_k$ and that of the other vertex be $t_l$. That is:
\begin{figure}[h]
    \centering
  \includegraphics[scale=0.1]{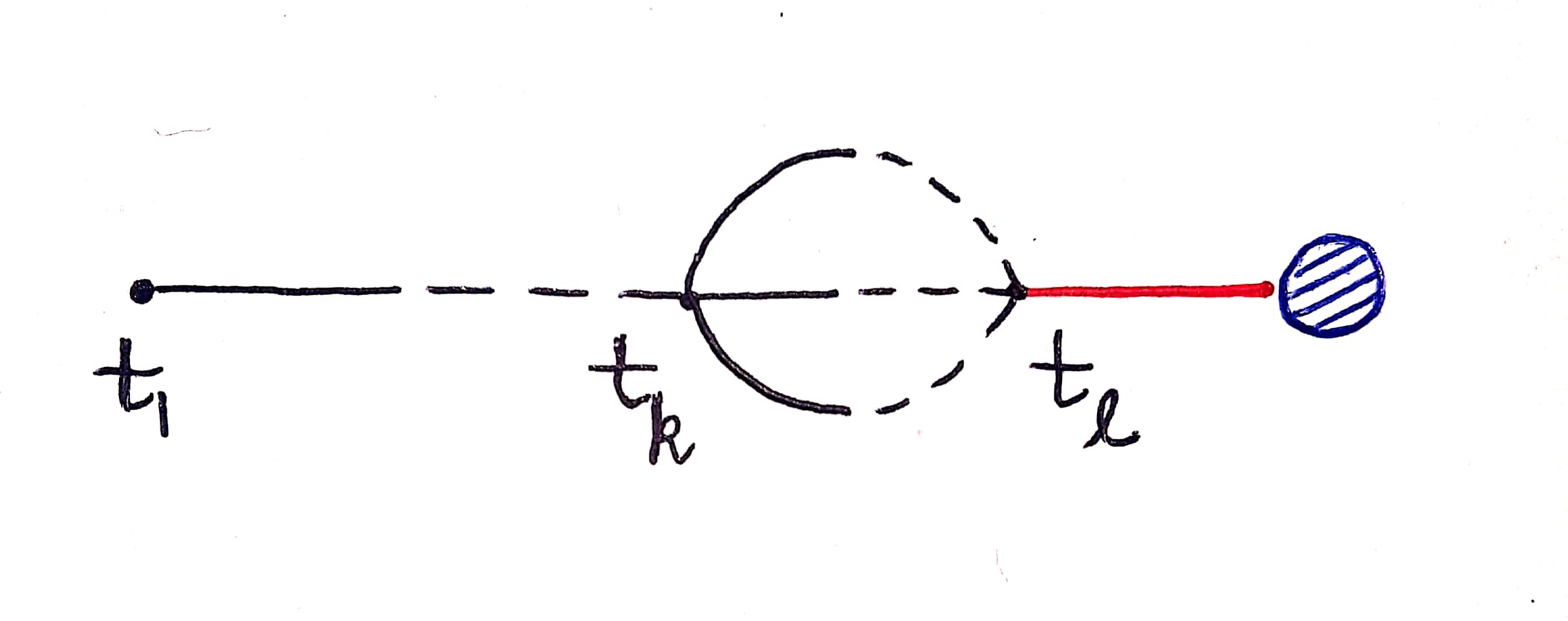}
\end{figure}
\\
Since this diagram arises in the perturbation expansion of the 2-point Wightman correlator, we must have:
\begin{equation}
    t_k,t_l\in M_2=\{t_{1+},t_{1-},t_{2+},t_{2-}\}.
\end{equation}
Now specifically look at the vertex nearer the external point $t_1$. The free Wightman propagator going out of it into the external point $t_1$ imposes the constraint:
\begin{equation}\label{constraint1}
t_k\rightarrow t_1.
\end{equation}
In addition to this, all the three internal Wightman propagators coming into this vertex from the other vertex impose the same constraint:
\begin{equation}\label{constraint2}
    t_l\rightarrow t_k\quad\text{or}\quad t_l=t_k.
\end{equation}
The constraints \eqref{constraint1} and \eqref{constraint2} are easily solved to yield:
\begin{equation}
    t_k=t_l=t_{1+}.
\end{equation}
With all this done, we label this diagram as:
\begin{figure}[h]
    \centering
  \includegraphics[scale=0.1]{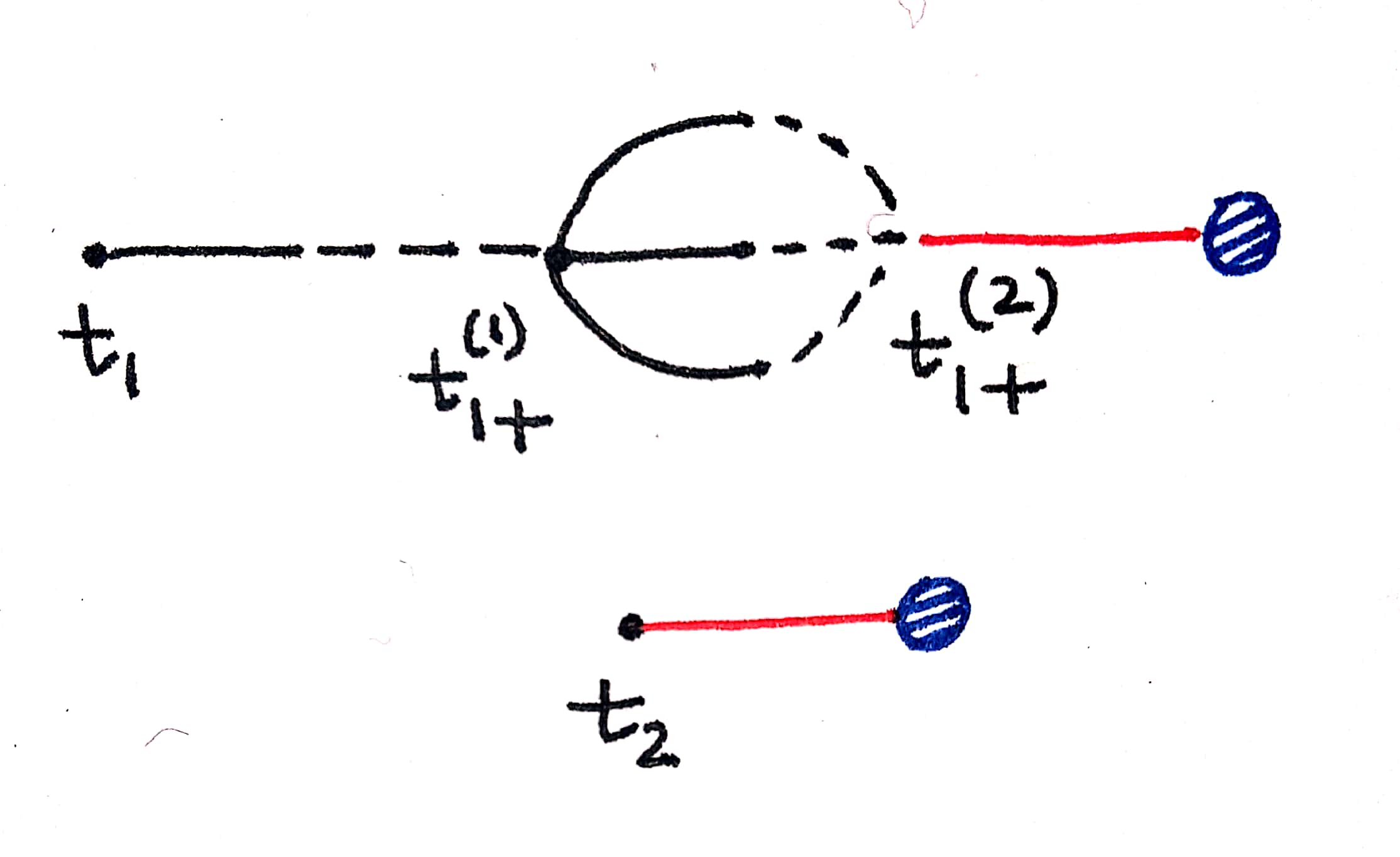}
  \caption{The diagram \ref{fourthdiag} with both its vertices labelled.}
  \label{fourthdiaglabelled}
\end{figure}
\\
Note that we have tacked on different superscripts onto the two labels here. This is a general procedure one needs to follow in cases where more than one vertex gets assigned the same label. Since the labels are essentially integration variables, one must distinguish them in some way before writing down the analytic expression corresponding to the diagram. We prefer to achieve this task using superscripts.

With the labelling done, the diagram evaluates to:
\\[10pt]
\fcolorbox{black}{gray!10}{\parbox{45em}{
\begin{equation}
    \frac{(i\lambda)^2}{6}\int_{t_0}^{t_1}dt^{(1)}_{1+}\int_{t_0}^{t_1}dt^{(2)}_{1+}\,\theta(t^{(1)}_{1+}-t^{(2)}_{1+})\cdot\{D_w(t^{(1)}_{1+}-t_1)\}\cdot\{D_w(t^{(2)}_{1+}-t^{(1)}_{1+})\}^{3}\cdot\{C_1(t^{(2)}_{1+}\}\cdot\{C_1(t_2)\}.
\end{equation}}}
\\[10pt]
Few things to note here. The factor of $6$ in the denominator is the symmetry factor of the diagram. Observe that the appropriate step function weight has been assigned to this diagram going by Rule 2 of \ref{stepfnwts}. The factor of $C_1(t_2)$ comes from the disconnected piece of the diagram.
\chapter{Conclusion and Further Aims}
We have developed a diagrammatic formalism to compute Wightman correlators in general states of the simple harmonic and anharmonic oscillators. We began this journey by realising that any such diagrammatic formalism first requires the establishment of a Wick's theorem. The traditional form of Wick's theorem, which is widely used in the computation of time ordered correlators in the vacuum state, was not found to hold once general states were brought into the picture. However, we were successful in developing a generalised Wick's theorem which allowed us to analytically compute Wightman correlators in general states of the simple harmonic oscillator. This generalised Wick's theorem was based on the concept of cumulants. Cumulants are objects which harbour all the information about the general state being considered. At the level of Wick contractions, this generalised Wick's theorem brought in concepts of contractions of a position operator with itself, as well as simultaneous contractions of more than two position operators. This was novel, for the traditional form of Wick's theorem involves only pairwise contraction structures. 

With a Wick's theorem then in our hands, a diagrammatic formalism to compute Wightman correlators in general states of the simple harmonic oscillator followed naturally. Two important aspects emerged at this juncture. Firstly, it was observed that the diagrammatics involved \emph{directional} propagators. Propagators symmetric under the exchange of their end points like the Feynman propagator were not observed. This was expected, since we were looking at Wightman correlators. These correlators, by definition, do not obey any time or anti-time ordering. As a result, no symmetry under the exchange of time arguments is expected. Secondly, a novel diagrammatic component emerged. These were cumulant blobs. These blobs were diagrammatic representations of cumulants, and hence, contained all the information about the general state in which Wightman correlators of the system were being computed.

Once a diagrammatic formalism was established for computing Wightman correlators in general states of the simple harmonic oscillator, we moved on to doing the same for a more realistic case, namely that of a weakly coupled anharmonic oscillator. While developing the diagrammatic formalism to compute Wightman correlators in general states of the anharmonic oscillator, we encountered three new aspects which were absent in similar considerations for the simple harmonic oscillator. They were those of interaction vertices, step function weights for internal propagators and symmetry factors. All these three facets were developed and finally, a protocol to compute Wightman correlators in general states of the anharmonic oscillator using diagrams was laid down.

We have accomplished all the above tasks using the operator formalism of canonical quantization. One of the next steps we would like to take would be to develop a path integral formulation of the same. What motivates us to think in this direction? There are three reasons for it which immediately strike the mind. One, it is easier to manifest symmetries which the system displays into its path integral. This is because the symmetries of a system are best incorporated through its Lagrangian and the Lagrangian is the main ingredient which goes into the path integral of the system. The second motivation for developing a path integral formulation is closely related to the first. It is that in practice, it is easier to guess the correct form of the Lagrangian of a theory, and hence its path integral too. Again, this is because the Lagrangian is the main component of the path integral. Thirdly, the path integral allows one to easily change coordinates between very different canonical descriptions of the same quantum system.

Some steps have already been taken in this direction. The Schwinger-Keldysh formalism \cite{kam} is a well established technique to compute a special type of Wightman correlators, namely contour ordered correlators. Developing on this, path integral techniques have also been set up to compute general Wightman correlators \cite{loga}. Attempts have also been made to incorporate the aspect of general initial states into the Schwinger-Keldysh formalism \cite{tifr}. Our aim in this regard would be a mix of all the above. We would like to incorporate information about the general initial state in the form of cumulants into the path integral for a general Wightman correlator.

Once we develop a path integral formalism for Wightman correlators in general states of the simple harmonic and anharmonic oscillators, we would like to extend our results to their natural generalisations, namely free and interacting quantum fields. This may prove to be fruitful since quantum fields are much more prevalent in Nature than harmonic oscillators. Other rewarding avenues which may be pursued hereon would be to develop a similar diagrammatic formalism for Wightman correlators in general states of fermionic systems and open quantum systems. In addition to this, one may also use the techniques developed here to set up a diagrammatics for computing out-of-time-order correlators (OTOCs), objects which have received a lot of attention in recent years, of all these systems in general states.      
\chapter{Appendix A}\label{appA}
\section{Overview}
In this Appendix, we explain the origin of the coefficients $\{\xi\}$ and $\{\chi\}$ used in our analyses.

We start by giving the definition of the coefficients $\{\xi\}$. We then explicitly compute cumulants till the fourth cumulant. The coefficients $\{\chi\}$ are introduced consequently. Some mathematical relationships between the $\{\xi\}$ and the $\{\chi\}$ emerge as a result. Generating functions for these two sets of coefficients are then defined, and are observed to obey a simple relationship among themselves. Finally, it is shown how Wightman correlators of the free oscillator can be expressed in terms of the $\{\chi\}$. This is demonstrated till the 4-point Wightman correlator.
\section{The Coefficients $\{\xi\}$}
Consider the free oscillator to be in a general density matrix $\rho$. The coefficients $\{\xi\}$ for this state are defined as:
\begin{equation}\label{def}
    \xi_{mn}\equiv\langle(a^{\dagger})^{m}a^{n}\rangle,
\end{equation}
where we have used the notation:
\begin{equation}
    \langle\dots\rangle\equiv\Tr[\rho\dots].
\end{equation}
Any general Wightman correlator of the system can then be expressed using these coefficients.

We now list down Wightman correlators of the system in a general state explicitly upto the 4-point function. This would serve two purposes. One, it would justify the statement made that any Wightman correlator can be expressed in terms of the coefficients $\{\xi\}$. And two, it would be used as a future reference in this Appendix. 
\newpage
\begin{align}\label{eq1}
   \langle x(t_1)\rangle&=\xi_{01}f_{+}+\xi_{10}f_{-},\notag\\
   \langle x(t_1)x(t_2)\rangle&=\xi_{02}f_{++}+\xi_{20}f_{--}\notag\\
    &\quad+\xi_{11}(f_{-+}+f_{+-})\notag\\
    &\quad+f_{+-},\notag\\
   \langle x(t_1)x(t_2)x(t_3)\rangle&=\xi_{03}f_{+++}+\xi_{30}f_{---}\notag\\
    &\quad+\xi_{12}(f_{+-+}+f_{++-}+f_{-++})+\xi_{21}(f_{--+}+f_{-+-}+f_{+--})\notag\\ 
    &\quad+\xi_{01}(f_{+-+}+2f_{++-})+\xi_{10}(f_{-+-}+2f_{+--}),\notag\\
   \langle x(t_1)x(t_2)x(t_3)x(t_4)\rangle&=\xi _{04}\;f_{++++}+\xi _{40}\;f_{----}\notag\\&\quad+\xi_{13}(f_{+-++}+f_{++-+}+f_{+++-}+f_{-+++})\notag\\&\quad+\xi _{31}(f_{---+}+f_{--+-}+f_{-+--}+f_{+---})\notag\\&\quad+\xi_{22}(f_{-+-+}+f_{-++-}+f_{+--+}+f_{+-+-}+f_{++--}+f_{--++})\notag\\&\quad+\xi _{02}(f_{+-++}+2f_{++-+}+3 f_{+++-})+\xi _{20}(f_{--+-}+2f_{-+--}+3 f_{+---})\notag\\&\quad+\xi_{11}(f_{-+-+}+2f_{-++-}+2f_{+--+}+3f_{+-+-}+4f_{++--})\notag\\&\quad+(f_{+-+-}+2 f_{++--}),
\end{align}
where:
\begin{equation}
    f_{\sigma_1\sigma_2\dots\sigma_n}=\Bigl(\frac{\hbar}{2\omega}\Bigr)^{n/2}\exp{\bigl(i\omega\;[\sigma_1t_1+\sigma_2t_2+\dots+\sigma_nt_n]\;\bigr)}.
\end{equation}
\section{Computation of Cumulants}
Recall the definitions of the cumulants as introduced in \eqref{cumu}. The first four cumulants are explicitly given by the expressions:
\begin{align}
    C_1(t_1)&\equiv\langle x(t_1)\rangle,\notag\\
    C_2(t_1,t_2)&\equiv\langle x(t_1)x(t_2)\rangle\notag\\&\quad-C_1(t_1)\cdot C_1(t_2),\notag\\
    C_3(t_1,t_2,t_3)&\equiv\langle x(t_1)x(t_2)x(t_3)\rangle\notag\\&\quad-C_1(t_1)\cdot C_2(t_2,t_3)-C_1(t_2)\cdot C_2(t_1,t_3)-C_1(t_3)\cdot C_2(t_1,t_2)\notag\\&\quad-C_1(t_1)\cdot C_1(t_2)\cdot C_1(t_3),\notag\\
    C_4(t_1,t_2,t_3,t_4)&\equiv\langle x(t_1)x(t_2)x(t_3)x(t_4)\rangle\notag\\&\quad-C_1(t_1)\cdot C_3(t_2,t_3,t_4)-C_1(t_2)\cdot C_3(t_1,t_3,t_4)-C_1(t_3)\cdot C_3(t_1,t_2,t_4)\notag\\&\quad-C_1(t_4)\cdot C_3(t_1,t_2,t_3)\notag\\&\quad-C_1(t_1)\cdot C_1(t_2)\cdot C_2(t_3,t_4)-C_1(t_1)\cdot C_1(t_3)\cdot C_2(t_2,t_4)\notag\\&\quad-C_1(t_1)\cdot C_1(t_4)\cdot C_2(t_2,t_3)-C_1(t_2)\cdot C_1(t_3)\cdot C_2(t_1,t_4)\notag\\&\quad-C_1(t_2)\cdot C_1(t_4)\cdot C_2(t_1,t_3)-C_1(t_3)\cdot C_1(t_4)\cdot C_2(t_1,t_2)\notag\\&\quad-C_1(t_1)\cdot C_1(t_2)\cdot C_1(t_3)\cdot C_1(t_4).
\end{align}
It is clear from the above that each cumulant is completely expressed in terms of Wightman correlators in the general state. These correlators are, in their turn, expressible in terms of the coefficients $\{\xi\}$. See \eqref{eq1} as an example.

Thus, we can express any cumulant in terms of the coefficients $\{\xi\}$. Doing this for the first four cumulants yields:
\begin{align}\label{eq}
    C_1&=\xi_{01}f_{+}+\xi_{10}f_{-},\notag\\
    C_2&=\bigl[-\xi_{01}^2+\xi_{02}\bigr]f_{++}+\bigl[\xi_{20}-\xi_{10}^2\bigr]f_{--}\notag\\
    &\quad+\bigl[\xi_{11}-\xi_{01}\xi_{10}\bigr](f_{-+}+f_{+-})\notag\\
    &\quad+\xi_{00}f_{+-},\notag\\
    C_3&=[2 \xi_{01}^3-3 \xi_{02} \xi_{01}+\xi_{03}]f_{+++}+[2 \xi _{10}^3-3 \xi _{20} \xi _{10}+\xi _{30}]f_{---}\notag\\&\quad+[2 \xi _{10} \xi_{01}^2-2 \xi _{11} \xi_{01}-\xi_{02} \xi _{10}+\xi _{12}](f_{-++}+f_{++-}+f_{+-+})\notag\\&\quad+[2 \xi_{01} \xi _{10}^2-2 \xi _{11} \xi _{10}-\xi_{01} \xi _{20}+\xi _{21}](f_{--+}+f_{-+-}+f_{+--}),\notag\\
    C_4&=\bigl[-6 \xi_{01}^4+12 \xi_{02} \xi_{01}^2-4 \xi_{03} \xi_{01}-3 \xi_{02}^2+\xi_{04}\bigr]f_{++++}+\bigl[-6 \xi _{10}^4+12 \xi _{20} \xi _{10}^2-4 \xi _{30} \xi _{10}-3 \xi _{20}^2+\xi _{40}\bigr]f_{----}\notag\\&\quad+\bigl[-6 \xi _{10} \xi_{01}^3+6 \xi _{11} \xi_{01}^2+6 \xi_{02} \xi _{10} \xi_{01}-3 \xi _{12} \xi_{01}-\xi_{03} \xi _{10}-3 \xi_{02} \xi _{11}+\xi _{13}\bigr](f_{+-++}+f_{++-+}+f_{+++-}\notag\\&\quad\quad\;+f_{-+++})\notag\\&\quad+\bigl[-6 \xi_{01} \xi _{10}^3+6 \xi _{11} \xi _{10}^2+6 \xi_{01} \xi _{20} \xi _{10}-3 \xi _{21} \xi _{10}-3 \xi _{11} \xi _{20}-\xi_{01} \xi _{30}+\xi _{31}\bigr](f_{---+}+f_{--+-}+f_{-+--}\notag\\&\quad\quad\;+f_{+---})\notag\\&\quad+\bigl[-6 \xi _{10}^2 \xi_{01}^2+2 \xi _{20} \xi_{01}^2+8 \xi _{10} \xi _{11} \xi_{01}-2 \xi _{21} \xi_{01}+2 \xi_{02} \xi _{10}^2-2 \xi _{11}^2-2 \xi _{10} \xi _{12}-\xi_{02} \xi _{20}+\xi _{22}\bigr](f_{-+-+}\notag\\&\quad\quad\;+f_{-++-}+f_{+--+}+f_{+-+-}+f_{++--}+f_{--++}).
\end{align}
With all these calculations now in place, we are now ready to introduce the coefficients $\{\chi\}$.
\section{The Coefficients $\{\chi\}$}
Before introducing the coefficients $\{\chi\}$, it is fruitful to introduce the functions $F_{nk}$ as already defined in \eqref{fnk}. Recall that the functions $F_{nk}$ were defined as:
\begin{equation}\label{fnk1}
    F_{nk}\equiv\sum_{\pi_k}f_{\sigma_1\sigma_2\dots\sigma_n},
\end{equation}
where $\pi_k$ denotes a permutation of the list $(\sigma_1\sigma_2\dots\sigma_n)$ such that $k$ of them are (+) and the remaining $n-k$ are (-). This definition then directs us to sum over all the $n\choose k$ such possible permutations. We are now equipped to define the coefficients $\{\chi\}$ in general states of the free oscillator.

The coefficients $\{\chi\}$ associated to a general state of the free oscillator are defined through the equation:
\begin{equation}\label{chi1}
    C_{n}\equiv\sum_{m=0}^{n}\chi_{m(n-m)}F_{n(n-m)}+\delta_{n,2}f_{+-}.
\end{equation}
where $C_n$ stands for the $n^{th}$ cumulant in the general state being considered.

What we now want to do is that we want to relate the coefficients $\{\chi\}$ thus introduced to the already established coefficients $\{\xi\}$ in a general state of the free oscillator. 

This is a very simple task. The set of equations \eqref{eq} already present us with the expressions for the cumulants upto the fourth cumulant in terms of the coefficients $\{\xi\}$. On the other hand, the definition \eqref{chi1} gives us the expression for a general cumulant in terms of the coefficients $\{\chi\}$. By writing down the expression for a particular cumulant first in terms of the coefficients $\{\xi\}$, then in terms of the coefficients $\{\chi\}$, and finally equating them, we would get the desired relationships between the two sets of coefficients $\{\xi\}$ and $\{\chi\}$.

As an example, the third cumulant $C_3$ in \eqref{eq} can be rewritten invoking the functions $F_{nk}$ as:
\begin{align}\label{shri1}
    C_3&=[2 \xi_{01}^3-3 \xi_{02} \xi_{01}+\xi_{03}]F_{33}+[2 \xi _{10}^3-3 \xi _{20} \xi _{10}+\xi _{30}]F_{30}\notag\\&\quad+[2 \xi _{10} \xi_{01}^2-2 \xi _{11} \xi_{01}-\xi_{02} \xi _{10}+\xi _{12}]F_{32}\notag\\&\quad+[2 \xi_{01} \xi _{10}^2-2 \xi _{11} \xi _{10}-\xi_{01} \xi _{20}+\xi _{21}]F_{31}.
\end{align}
On the other hand, the definition \eqref{chi1} dictates:
\begin{align}\label{shri2}
    C_3&=\chi_{03}F_{33}+\chi_{30}F_{30}+\chi_{12}F_{32}+\chi_{21}F_{31}.
\end{align}
Comparing \eqref{shri1} with \eqref{shri2} yields:
\begin{align}
    \chi_{03}&=2 \xi_{01}^3-3 \xi_{02} \xi_{01}+\xi_{03}\notag\\
    \chi_{12}&=2 \xi _{10} \xi_{01}^2-2 \xi _{11} \xi_{01}-\xi_{02} \xi _{10}+\xi _{12}.
\end{align}
We list only these $\{\chi\}$ since $\chi_{ij}=\chi^{*}_{ji}$. Repeating a similar procedure for the other cumulants $C_1, C_2$ and $C_4$ yields the additional relations:
\begin{align}
    \chi_{00}&=\xi_{00}=1\notag\\
    \chi_{01}&=\xi_{01}\notag\\
    \chi_{02}&=-\xi_{01}^2+\xi_{02}\notag\\
    \chi_{11}&=\xi_{11}-\xi_{01}\xi_{10}\notag\\
    \chi_{04}&=-6 \xi_{01}^4+12 \xi_{02} \xi_{01}^2-4 \xi_{03} \xi_{01}-3 \xi_{02}^2+\xi_{04}\notag\\
    \chi_{13}&=-6 \xi _{10} \xi_{01}^3+6 \xi _{11} \xi_{01}^2+6 \xi_{02} \xi _{10} \xi_{01}-3 \xi _{12} \xi_{01}-\xi_{03} \xi _{10}-3 \xi_{02} \xi _{11}+\xi _{13}\notag\\
    \chi_{22}&=-6 \xi _{10}^2 \xi_{01}^2+2 \xi _{20} \xi_{01}^2+8 \xi _{10} \xi _{11} \xi_{01}-2 \xi _{21} \xi_{01}+2 \xi_{02} \xi _{10}^2-2 \xi _{11}^2-2 \xi _{10} \xi _{12}-\xi_{02} \xi _{20}+\xi _{22}.
\end{align}
\newpage
\subsection{Final Results}
We now collate all the results that have been obtained through the previous sections into one place.

Considering upto the fourth cumulant, the coefficients $\{\chi\}$ in terms of the coefficients $\{\xi\}$ are given as:
\begin{align}\label{eq4}
    \chi_{00}&=\xi_{00}=1\notag\\
    \chi_{01}&=\xi_{01}\notag\\
    \chi_{02}&=-\xi_{01}^2+\xi_{02}\notag\\
    \chi_{11}&=\xi_{11}-\xi_{01}\xi_{10}\notag\\
    \chi_{03}&=2 \xi_{01}^3-3 \xi_{02} \xi_{01}+\xi_{03}\notag\\
    \chi_{12}&=2 \xi _{10} \xi_{01}^2-2 \xi _{11} \xi_{01}-\xi_{02} \xi _{10}+\xi _{12}\notag\\
    \chi_{04}&=-6 \xi_{01}^4+12 \xi_{02} \xi_{01}^2-4 \xi_{03} \xi_{01}-3 \xi_{02}^2+\xi_{04}\notag\\
    \chi_{13}&=-6 \xi _{10} \xi_{01}^3+6 \xi _{11} \xi_{01}^2+6 \xi_{02} \xi _{10} \xi_{01}-3 \xi _{12} \xi_{01}-\xi_{03} \xi _{10}-3 \xi_{02} \xi _{11}+\xi _{13}\notag\\
    \chi_{22}&=-6 \xi _{10}^2 \xi_{01}^2+2 \xi _{20} \xi_{01}^2+8 \xi _{10} \xi _{11} \xi_{01}-2 \xi _{21} \xi_{01}+2 \xi_{02} \xi _{10}^2-2 \xi _{11}^2-2 \xi _{10} \xi _{12}-\xi_{02} \xi _{20}+\xi _{22},
\end{align}
with $\xi_{ij}=\xi_{ji}^{*}$ and hence, $\chi_{ij}=\chi^{*}_{ji}$.
\section{Generating Functions for the $\{\xi\}$ and $\{\chi\}$}\label{gf}
We have now seen a concrete example of how one should proceed if one has to establish mathematical relationships between the two sets of coefficients, namely the $\{\xi\}$ and the $\{\chi\}$.

However, it would be good if we could write down a single equation which would encompass all the relationships between the $\{\xi\}$ and $\{\chi\}$ as witnessed in \eqref{eq4}, in addition to predicting similar relationships between the other $\{\xi\}$ and $\{\chi\}$ which will appear in the cumulants and correlators higher than their 4-point counterparts.

Before arriving at this prospective equation, let us first think what form it should take. The most elegant way through which it can manifest would be that it be an equality between two \emph{representative} objects. One of these objects would be a representative of all the $\{\xi\}$, and the other, of all the $\{\chi\}$. This single equation would then house all the relationships between the $\{\xi\}$ and the $\{\chi\}$. Specific relations, like the ones listed in \eqref{eq4}, may consequently be churned out from this equation through a simple mathematical process.

So what should these representative objects be? 

They are \emph{generating functions}.

One defines the generating function for the $\{\xi\}$ as:
\begin{equation}\label{def2}
    Z_P(\lambda,\bar{\lambda})\equiv\nsum_{m,n=0}^{\infty}\frac{\lambda^m}{m!}\frac{\bar{\lambda}^{n}}{n!}\;\xi_{mn}.
\end{equation}
The $\{\xi\}$ are obtained from $Z_{P}$ as:
\begin{equation}
    \xi_{mn}=\frac{\partial^{m+n}}{\partial^{m}\lambda\;\partial^{n}\bar{\lambda}}\;Z_{P}(\lambda,\bar{\lambda})\Bigl|_{\lambda,\bar{\lambda}=0}.
\end{equation}
Similarly, one defines the generating function $Z_{\chi}(\lambda,\bar{\lambda})$ for the coefficients $\{\chi\}$ as:
\begin{equation}
    Z_{\chi}(\lambda,\bar{\lambda})\equiv\nsum_{m,n=0}^{\infty}\frac{\lambda^m}{m!}\frac{\bar{\lambda}^{n}}{n!}\;\chi_{mn},
\end{equation}
and the $\{\chi\}$ are obtained from $Z_{\chi}$ as:
\begin{equation}
    \chi_{mn}=\frac{\partial^{m+n}}{\partial^{m}\lambda\;\partial^{n}\bar{\lambda}}\;Z_{\chi}(\lambda,\bar{\lambda})\Bigl|_{\lambda,\bar{\lambda}=0}.
\end{equation}
\subsection{Relation between $Z_{P}$ and $Z_{\chi}$}\label{rel}
It turns out that the single equation which contains all the relationships between the $\{\xi\}$ and $\{\chi\}$ in it is:
\begin{equation}
    \boxed{Z_{\chi}=\ln Z_{P}}
\end{equation}
\section{Wightman Correlators in terms of $\{\chi\}$}
We now come to the final aim of this Appendix. We will show how Wightman correlators in general states of the free harmonic oscillator can be re-expressed in terms of the $\{\chi\}$ rather than the $\{\xi\}$.

We do this because it opens up the scope of developing a diagrammatic formalism to represent Wightman correlators of the oscillator in general states.

Wightman correlators of the oscillator in a general state (upto the 4-point correlator) have already been expressed in terms of the $\{\xi\}$ in \eqref{eq1}. We also know the $\{\chi\}$ in terms of the $\{\xi\}$ through \eqref{eq4}. So all one needs to do in order to express the Wightman correlators in terms of the $\{\chi\}$ is to invert the relations \eqref{eq4} and plug back the result into \eqref{eq1}.

The result of inversion is:
\begin{align}
    \xi_{00}&=\chi_{00}=1\notag\\
    \xi_{01}&=\chi_{01}\notag\\
    \xi_{02}&=\chi_{02}+\chi_{01}^{2}\notag\\
    \xi_{11}&=\chi_{11}+\chi_{01}\chi_{10}\notag\\
    \xi_{03}&=\chi_{01}^3+3 \chi_{02} \chi_{01}+\chi_{03}\notag\\
    \xi_{12}&=\chi _{10} \chi_{01}^2+2 \chi _{11} \chi_{01}+\chi_{02} \chi _{10}+\chi _{12}\notag\\
    \xi_{04}&=\chi_{01}^4+6 \chi_{02} \chi_{01}^2+4 \chi_{03} \chi_{01}+3 \chi_{02}^2+\chi_{04}\notag\\
    \xi_{13}&=\chi _{10} \chi_{01}^3+3 \chi _{11} \chi_{01}^2+3 \chi_{02} \chi _{10} \chi_{01}+3 \chi _{12} \chi_{01}+\chi_{03} \chi _{10}+3 \chi_{02} \chi _{11}+\chi _{13}\notag\\
    \xi_{22}&=\chi _{10}^2 \chi_{01}^2+\chi _{20} \chi_{01}^2+4 \chi _{10} \chi _{11} \chi_{01}+2 \chi _{21} \chi_{01}+\chi_{02} \chi _{10}^2+2 \chi _{11}^2+2 \chi _{10} \chi _{12}+\chi_{02} \chi _{20}+\chi _{22}.
\end{align}
Plugging these relations into \eqref{eq1} yields our final result:
\begin{align}
    \langle x(t_1)\rangle&=\chi_{01}f_{+}+\chi_{10}f_{-}\notag\\
    \langle x(t_1)x(t_2)\rangle&=(\chi_{02}+\chi_{01}^{2})f_{++}+(\chi_{20}+\chi_{10}^{2})f_{--}\notag\\
    &\quad+(\chi_{11}+\chi_{01}\chi_{10})(f_{-+}+f_{+-})\notag\\
    &\quad+\chi_{00}f_{+-}\notag\\
    \langle x(t_1)x(t_2)x(t_3)\rangle&=(\chi_{01}^3+3 \chi_{02} \chi_{01}+\chi_{03})f_{+++}+(\chi _{10}^3+3 \chi _{20} \chi _{10}+\chi _{30})f_{---}\notag\\
    &\quad+(\chi _{10} \chi_{01}^2+2 \chi _{11} \chi_{01}+\chi_{02} \chi _{10}+\chi _{12})(f_{+-+}+f_{++-}+f_{-++})\notag\\&\quad+(\chi_{01} \chi _{10}^2+2 \chi _{11} \chi _{10}+\chi_{01} \chi _{20}+\chi _{21})(f_{--+}+f_{-+-}+f_{+--})\notag\\
    &\quad+(\chi_{01})(f_{+-+}+2f_{++-})+(\chi_{10})(f_{-+-}+2f_{+--})\notag\\
    \langle x(t_1)x(t_2)x(t_3)x(t_4)\rangle&=(\chi_{01}^4+6 \chi_{02} \chi_{01}^2+4 \chi_{03} \chi_{01}+3 \chi_{02}^2+\chi_{04})\;f_{++++}\notag\\&\quad+(\chi _{10}^4+6 \chi _{20} \chi _{10}^2+4 \chi _{30} \chi _{10}+3 \chi _{20}^2+\chi _{40})\;f_{----}\notag\\&\quad+(\chi _{10} \chi_{01}^3+3 \chi _{11} \chi_{01}^2+3 \chi_{02} \chi _{10} \chi_{01}+3 \chi _{12} \chi_{01}+\chi_{03} \chi _{10}+3 \chi_{02} \chi _{11}\notag\\&\quad\quad\;+\chi _{13})(f_{+-++}+f_{++-+}+f_{+++-}+f_{-+++})\notag\\&\quad+(\chi_{01} \chi _{10}^3+3 \chi _{11} \chi _{10}^2+3 \chi_{01} \chi _{20} \chi _{10}+3 \chi _{21} \chi _{10}+3 \chi _{11} \chi _{20}+\chi_{01} \chi _{30}\notag\\&\quad\quad\;+\chi _{31})(f_{---+}+f_{--+-}+f_{-+--}+f_{+---})\notag\\&\quad+(\chi _{10}^2 \chi_{01}^2+\chi _{20} \chi_{01}^2+4 \chi _{10} \chi _{11} \chi_{01}+2 \chi _{21} \chi_{01}+\chi_{02} \chi _{10}^2+2 \chi _{11}^2\notag\\&\quad\quad\;+2 \chi _{10} \chi _{12}+\chi_{02} \chi _{20}+\chi _{22})(f_{-+-+}+f_{-++-}+f_{+--+}+f_{+-+-}\notag\\&\quad\quad\;+f_{++--}+f_{--++})\notag\\&\quad +(\chi_{02}+\chi_{01}^{2})(f_{+-++}+2f_{++-+}+3 f_{+++-})\notag\\&\quad+(\chi _{10}^2+\chi _{20})(f_{--+-}+2f_{-+--}+3 f_{+---})\notag\\&\quad+(\chi_{11}+\chi_{01}\chi_{10})(f_{-+-+}+2f_{-++-}+2f_{+--+}+3f_{+-+-}+4f_{++--})\notag\\&\quad+\chi _{00}(f_{+-+-}+2 f_{++--}).
\end{align}
\\[70pt]
\begin{center}
\scalebox{2.5}{\adforn{18}}
\end{center}
\chapter{Appendix B}\label{appB}
In this appendix, we explicitly compute the generating function for the cumulants in some special states of the free oscillator. Consequently, we will also compute the cumulants themselves in some of these states. The states which would be considered thus would be the vacuum state, a coherent state and the thermal state. Refer to \ref{splstates} for an introduction to these states of the harmonic oscillator.
\section{Guiding Equation}
The guiding equation for computing the generating function of the cumulants in a general state of the harmonic oscillator is \eqref{s1}:
\begin{equation}\label{appgf}
    Z_{\chi}(\lambda,\bar{\lambda})=\ln\Tr[\rho e^{\lambda a^{\dagger}}e^{\bar{\lambda}a}].
\end{equation}
\section{The Vacuum State}
The generating function for the cumulants \eqref{appgf} in the vacuum state, denoted $Z^{0}_{\chi}$, reads:
\begin{equation}
   Z^{0}_{\chi}(\lambda,\bar{\lambda})=\ln\bra{0}e^{\lambda a^{\dagger}}e^{\bar{\lambda}a}\ket{0}. 
\end{equation}
Since $a\ket{0}=0=\bra{0}a^{\dagger}$, we have:
\begin{equation}
    e^{\bar{\lambda}a}\ket{0}=\ket{0}.
\end{equation}
Similarly,
\begin{equation}
    \bra{0}e^{\lambda a^{\dagger}}=\bra{0}.
\end{equation}
Thus:
\begin{equation}
    Z^{0}_{\chi}=0.
\end{equation}
This tells us that apart from $\chi_{00}$, which is not obtainable from the generating function, \emph{all the cumulants in the vacuum state vanish}.
\\
\\
The zeroth cumulant, $\chi_{00}$, equals the norm of the vacuum state, which is unity.
\section{Coherent State}
The generating function for the cumulants \eqref{appgf} in a coherent state $\ket{\phi}$, denoted $Z^{\phi}_{\chi}$, reads:
\begin{equation}
   Z^{\phi}_{\chi}(\lambda,\bar{\lambda})=\ln\bra{\phi}e^{\lambda a^{\dagger}}e^{\bar{\lambda}a}\ket{\phi}. 
\end{equation}
Since $a\ket{\phi}=\phi\ket{\phi}$, one gets:
\begin{equation}
    e^{\bar{\lambda}a}\ket{\phi}=e^{\bar{\lambda}\phi}\ket{\phi}.
\end{equation}
Similarly:
\begin{equation}
    \bra{\phi}e^{\lambda a^{\dagger}}=\bra{\phi}e^{\lambda\phi^*}.
\end{equation}
Thus:
\begin{align}
    Z^{\phi}_{\chi}(\lambda,\bar{\lambda})&=\ln\,(e^{\lambda\phi^*+\bar{\lambda}\phi})\notag\\
    &=\lambda\phi^*+\bar{\lambda}\phi.
\end{align}
Using \eqref{s2} now, the coefficients $\{\chi^{\phi}\}$ for a coherent state $\ket{\phi}$ are given as:
\begin{gather}
    \chi^{\phi}_{01}=\phi,
    \chi^{\phi}_{10}=\phi^*,
\end{gather}
and, as usual:
\begin{equation}
    \chi^{\phi}_{00}=\braket{\phi|\phi}=1.
\end{equation}
Apart from these, all the remaining $\{\chi^\phi\}$ vanish.
\section{Thermal State}
A thermal state at inverse temperature $\beta$ is one with the density matrix:
\begin{equation}
    \rho_{th}=\frac{e^{-\beta\hbar\omega(a^{\dagger}a+1/2)}}{Z},
\end{equation}
where $Z$ is the partition function given by:
\begin{equation}
    Z=\Tr[e^{-\beta\hbar\omega(a^{\dagger}a+1/2)}]=\frac{1}{2\sinh{(\frac{\beta\hbar\omega}{2})}}.
\end{equation}
In this case, the generating function for the $\{\xi\}$ reads:
\begin{align}\label{trace}
    Z^{\beta}_{\xi}&=\Tr[\rho_{th}\,e^{\lambda a^{\dagger}}e^{\bar{\lambda}a}]\notag\\
    &=\frac{\Tr[e^{-\beta\hbar\omega(a^{\dagger}a+1/2)}e^{\lambda a^{\dagger}}e^{\bar{\lambda}a}]}{Z}.
\end{align}
Evaluating the trace in the numerator of \eqref{trace} in the coherent basis yields:
\begin{equation}\label{cohtrace}
    \Tr[e^{-\beta\hbar\omega(a^{\dagger}a+1/2)}e^{\lambda a^{\dagger}}e^{\bar{\lambda}a}]=\int d[\phi_1,\bar{\phi}_1]e^{-|\phi_1|^2}\bra{\phi_1}e^{-\beta\hbar\omega(a^{\dagger}a+1/2)}e^{\lambda a^{\dagger}}e^{\bar{\lambda}a}\ket{\phi_1},
\end{equation}
where the measure of integration is:
\begin{equation}
    \int d[\phi_1,\bar{\phi}_1]=\int\frac{d(\text{Re}\,\phi_1)\,d(\text{Im}\,\phi_1)}{\pi}.
\end{equation}
First of all, the factor of $e^{-\frac{\beta\hbar\omega}{2}}$ can be taken out from \eqref{cohtrace} giving:
\begin{equation}\label{cohtrace2}
    \Tr[e^{-\beta\hbar\omega(a^{\dagger}a+1/2)}e^{\lambda a^{\dagger}}e^{\bar{\lambda}a}]=e^{-\frac{\beta\hbar\omega}{2}}\int d[\phi_1,\bar{\phi}_1]e^{-|\phi_1|^2}\bra{\phi_1}e^{-\beta\hbar\omega a^{\dagger}a}e^{\lambda a^{\dagger}}e^{\bar{\lambda}a}\ket{\phi_1}.
\end{equation}
We now insert coherent completeness relations of the form:
\begin{equation}
    \int d[\phi_k,\bar{\phi}_k]e^{-|\phi_k|^2}\ket{\phi_k}\bra{\phi_k}=\mathbb{I}
\end{equation}
between each exponential operator in \eqref{cohtrace2} yielding:
\begin{align}\label{cohtrace3}
    \Tr[e^{-\beta\hbar\omega(a^{\dagger}a+1/2)}e^{\lambda a^{\dagger}}e^{\bar{\lambda}a}]&=e^{-\frac{\beta\hbar\omega}{2}}\int\prod_{k=1}^{3}d[\phi_k,\bar{\phi}_k]e^{-|\phi_k|^2}\bra{\phi_1}e^{-\beta\hbar\omega a^{\dagger}a}\ket{\phi_2}\notag\\&\hspace{70pt}\bra{\phi_2}e^{\lambda a^{\dagger}}\ket{\phi_3}\bra{\phi_3}e^{\bar{\lambda}a}\ket{\phi_1}.
\end{align}
Using the identity:
\begin{equation}\label{identity}
    \bra{\phi_i}\mu^{a^{\dagger}a}\ket{\phi_j}=e^{\mu\bar{\phi}_i\phi_{j}},
\end{equation}
the first matrix element of \eqref{cohtrace3} becomes:
\begin{equation}\label{firstmat}
    \bra{\phi_1}e^{-\beta\hbar\omega a^{\dagger}a}\ket{\phi_2}=e^{\rho(\omega)\bar{\phi}_1\phi_2},
\end{equation}
where:
\begin{equation}
    \rho(\omega)=e^{-\beta\hbar\omega}.
\end{equation}
The second and third matrix elements of \eqref{cohtrace3} are easily evaluated as:
\begin{align}\label{secondmat}
    \bra{\phi_2}e^{\lambda a^{\dagger}}\ket{\phi_3}&=e^{\lambda\bar{\phi}_2}\braket{\phi_2|\phi_3}\notag
    \\&=e^{\lambda\bar{\phi}_2+\bar{\phi}_2\phi_3},
\end{align}
and similarly:
\begin{align}\label{thirdmat}
    \bra{\phi_3}e^{\bar{\lambda}a}\ket{\phi_1}&=e^{\bar{\lambda}\phi_1}\braket{\phi_3|\phi_1}\notag
    \\&=e^{\bar{\lambda}\phi_1+\bar{\phi}_3\phi_1}.
\end{align}
\newpage
Putting \eqref{firstmat}, \eqref{secondmat} and \eqref{thirdmat} into \eqref{cohtrace3}, one gets:
\begin{align}\label{final}
    \Tr[e^{-\beta\hbar\omega(a^{\dagger}a+1/2)}e^{\lambda a^{\dagger}}e^{\bar{\lambda}a}]&=e^{-\frac{\beta\hbar\omega}{2}}\int\prod_{k=1}^{3}d[\phi_k,\bar{\phi}_k]e^{-|\phi_k|^2}e^{\rho(\omega)\bar{\phi}_1\phi_2}e^{\lambda\bar{\phi}_2+\bar{\phi}_2\phi_3}\notag\\&\hspace{130pt}\cdot e^{\bar{\lambda}\phi_1+\bar{\phi}_3\phi_1}.
\end{align}
The integral on the RHS of \eqref{final} is a trivial multivariable Gaussian integral which is easily solved to give the answer:
\begin{equation}
    \frac{1}{1-\rho(\omega)}\exp\Bigl[\frac{\lambda\bar{\lambda}\rho(\omega)}{1-\rho(\omega)}\Bigr].
\end{equation}
Thus:
\begin{equation}
    \Tr[e^{-\beta\hbar\omega(a^{\dagger}a+1/2)}e^{\lambda a^{\dagger}}e^{\bar{\lambda}a}]=\frac{e^{-\frac{\beta\hbar\omega}{2}}}{1-\rho(\omega)}\exp\Bigl[\frac{\lambda\bar{\lambda}\rho(\omega)}{1-\rho(\omega)}\Bigr],
\end{equation}
and finally, from \eqref{trace}:
\begin{equation}\label{finale}
    Z^{\beta}_{\xi}=\exp\Bigl[\frac{\lambda\bar{\lambda}\rho(\omega)}{1-\rho(\omega)}\Bigr].
\end{equation}
It is fruitful to bring in the Bose factor $n_B$ here, which is defined as:
\begin{equation}
    n_B=\frac{1}{e^{\beta\hbar\omega}-1}=\frac{\rho(\omega)}{1-\rho(\omega)}.
\end{equation}
In terms of the Bose factor, the generating function for the $\{\xi\}$ reads:
\begin{equation}
    Z^{\beta}_{\xi}=\exp\bigl[\lambda\bar{\lambda}n_B\bigr],
\end{equation}
and the generating function of the cumulants:
\begin{equation}
    Z^{\beta}_{\chi}=\ln Z^{\beta}_{\xi}=\lambda\bar{\lambda}n_B.
\end{equation}
From this, we see that the only non-vanishing cumulant of the thermal state (apart from $\chi_{00}=1$) is $\chi_{11}$, which is given by:
\begin{equation}
    \chi_{11}=\frac{\partial^2 Z^{\beta}_{\chi}}{\partial\lambda\,\partial\bar{\lambda}}\,\biggl|_{\lambda,\bar{\lambda}=0}=n_B.
\end{equation}
\\[50pt]
\begin{center}
\scalebox{2.5}{\adforn{18}}
\end{center}
\printbibliography

@article{lsz,
  author =       "H.Lehmann, K.Symanzik, and W.Zimmerman",
  title =        "Zur Formulierung quantisierter Feldtheorien",
  journal =      "Nuovo Cimento 1(1)",
  pages =        "205",
  year =         "1955",
  DOI =          "",
  
}

@article{dan,
    author =       "P. Danielewicz",
    journal =      "Ann. Phys.",
    volume =       "152",
    pages =        "239",
    year =         "1984",
    
}

@article{bot,
    author =       "W. Botermans and R. Malfliet",
    title =        "",
    journal =      "Phys. Reports",
    volume =       "198",
    number =       "",
    pages =        "115",
    year =         "1990",
    DOI =          ""
}

@book{pes,
    title = {An Introduction to Quantum Field Theory},
    author = {M.Peskin and D.Schroeder},
    isbn = {978-020-15-0397-5},
    series = {},
    year = {1995},
    publisher = {Westview Press, Chicago}
}

@article{wck,
    author =       "G. C. Wick",
    title =        "",
    journal =      "Phys. Rev.",
    volume =       "80",
    number =       "",
    pages =        "268",
    year =         "1950",
    DOI =          ""
}

@article{caz,
    author =       "M. A. Cazalilla and M. Rigol",
    title =        "",
    journal =      "New J. Phys.",
    volume =       "12",
    number =       "",
    pages =        "",
    year =         "2010",
    DOI =          ""
}

@article{brif,
    author =       "C. Brif, R. Chakrabarti and H. Rabitz",
    title =        "",
    journal =      "New J. Phys.",
    volume =       "12",
    number =       "",
    pages =        "",
    year =         "2010",
    DOI =          ""
}

@misc{abb,
  author = {Timothy G. Abbott},
  title = {{Feynman Diagrams in Quantum Mechanics}},
  howpublished = "\url{http://web.mit.edu/tabbott/www/papers/feynman-diagrams.pdf}"
}

@book{klein,
    title = {Critical Properties of $\phi^4$-Theories},
    author = {Hagen Kleinert and Verena Schulte-Frohlinde},
    isbn = {978-981-02-4658-7},
    series = {},
    year = {2001},
    publisher = {World Scientific}
}

@book{kam,
    title = {Field Theory of Non-Equilibrium Systems},
    author = {Alex Kamenev},
    isbn = {978-052-17-6082-9},
    series = {},
    year = {2011},
    publisher = {Cambridge University Press}
}

@book{breu,
    title = {The Theory of Open Quantum Systems},
    author = {Heinz-Peter Breuer and Francesco Petruccione},
    isbn = {978-019-92-1390-0},
    series = {},
    year = {2007},
    publisher = {Oxford University Press}
}

@article{feyn,
    author =       "R.P Feynman and F.L Vernon Jr.",
    title =        "The Theory of a General Quantum System Interacting with a Linear Dissipative System",
    journal =      "Annals of Physics",
    volume =       "24",
    number =       "",
    pages =        "118-173",
    year =         "October 1963",
    DOI =          "https://doi.org/10.1016/0003-4916(63)90068-X"
}

@article{malda,
   title={A bound on chaos},
   volume={2016},
   ISSN={1029-8479},
   url={http://dx.doi.org/10.1007/JHEP08(2016)106},
   DOI={10.1007/jhep08(2016)106},
   number={8},
   journal={Journal of High Energy Physics},
   publisher={Springer Science and Business Media LLC},
   author={Maldacena, Juan and Shenker, Stephen H. and Stanford, Douglas},
   year={2016},
   month={Aug}
}

@article{von,
   title={Operator Hydrodynamics, OTOCs, and Entanglement Growth in Systems without Conservation Laws},
   volume={8},
   ISSN={2160-3308},
   url={http://dx.doi.org/10.1103/PhysRevX.8.021013},
   DOI={10.1103/physrevx.8.021013},
   number={2},
   journal={Physical Review X},
   publisher={American Physical Society (APS)},
   author={von Keyserlingk, C. W. and Rakovszky, Tibor and Pollmann, Frank and Sondhi, S. L.},
   year={2018},
   month={Apr}
}

@misc{tong,
  author = {David Tong},
  title = {{Quantum Field Theory}},
  howpublished = "\url{https://www.damtp.cam.ac.uk/user/tong/qft/qft.pdf}",
  year = {2006}
}

@article{leeu1,
   title={Equilibrium and nonequilibrium many-body perturbation theory: a unified framework based on the Martin-Schwinger hierarchy},
   volume={427},
   ISSN={1742-6596},
   url={http://dx.doi.org/10.1088/1742-6596/427/1/012001},
   DOI={10.1088/1742-6596/427/1/012001},
   journal={Journal of Physics: Conference Series},
   publisher={IOP Publishing},
   author={van Leeuwen, Robert and Stefanucci, Gianluca},
   year={2013},
   month={Mar},
   pages={012001}
}

@article{hall,
  author =       "A.G Hall",
  title =        "Non-Equilibrium Green Functions:\;Generalised Wick's Theorem \& Diagrammatic Perturbation with Initial Correlations",
  journal =      "J.Phys.A:Math.Gen.",
  volume =       "214",
  number =       "8",
  year =         "1975",
}

@article{craig,
  author =       "R.A Craig",
  title =        "Perturbation Expansion for Real Time Green's Functions",
  journal =      "J. Math. Phys.",
  volume =       "605",
  number =       "9",
  year =         "1968",
  DOI="https://doi.org/10.1063/1.1664616"
}

@article{wagner,
  author =       "Mathias Wagner",
  title =        "Expansions of Non-Equilibrium Green's Functions",
  journal =      "Phys. Rev. B",
  volume =       "44",
  number =       "12",
  year =         "1991",
  DOI="https://doi.org/10.1103/PhysRevB.44.6104"
}

@article{garny,
   title={Kadanoff-Baym equations with non-Gaussian initial conditions: The equilibrium limit},
   volume={80},
   ISSN={1550-2368},
   url={http://dx.doi.org/10.1103/PhysRevD.80.085011},
   DOI={10.1103/physrevd.80.085011},
   number={8},
   journal={Physical Review D},
   publisher={American Physical Society (APS)},
   author={Garny, Mathias and Müller, Markus Michael},
   year={2009},
   month={Oct}
}

@article{leeu2,
   title={Wick theorem for general initial states},
   volume={85},
   ISSN={1550-235X},
   url={http://dx.doi.org/10.1103/PhysRevB.85.115119},
   DOI={10.1103/physrevb.85.115119},
   number={11},
   journal={Physical Review B},
   publisher={American Physical Society (APS)},
   author={van Leeuwen, R. and Stefanucci, G.},
   year={2012},
   month={Mar}
}

@article{fauser1,
  title={Non-equilibrium quantum field theory and perturbation theory},
  author={R. Fauser and H. H. Wolter},
  journal={Nuclear Physics},
  year={1995},
  volume={584},
  pages={604-620},
  DOI="https://doi.org/10.1016/0375-9474(94)00493-7"
}

@article{fauser2,
  title={Cumulants in Perturbation Expansions for Non-Equilibrium Field Theory},
  author={R. Fauser and H. H. Wolter},
  journal={Nuclear Physics A},
  year={1996},
  volume={600},
  pages={491-508},
  DOI="https://doi.org/10.1016/0375-9474(96)00023-1"
}

@misc{brouder,
  author =       "Christian Brouder",
  title =        "The Structure of Green Functions in Quantum Field Theory with a General State ",
  journal =      "Recent Developments in Quantum Field Theory, Leipzig, Germany",
  pages =        "163--175",
  year =         "2007",
  howpublished="\url{https://hal.archives-ouvertes.fr/hal-00184086}"
}

@article{loga,
   title={Classification of out-of-time-order correlators},
   volume={6},
   url={http://dx.doi.org/10.21468/SciPostPhys.6.1.001},
   DOI={10.21468/scipostphys.6.1.001},
   number={1},
   journal={SciPost Physics},
   publisher={Stichting SciPost},
   author={Haehl, Felix and Loganayagam, R. and Narayan, Prithvi and Rangamani, Mukund},
   year={2019},
   month={Jan}
}

@article{tifr,
   %title={Nonequilibrium field theory for dynamics starting from arbitrary athermal initial conditions},
   %volume={99},
   %ISSN={2469-9969},
   %url={http://dx.doi.org/10.1103/PhysRevB.99.054306},
   %DOI={10.1103/physrevb.99.054306},
   %number={5},
   %journal={Physical Review B},
   %publisher={American Physical Society (APS)},
   %author={Chakraborty, Ahana and Gorantla, Pranay and Sensarma, Rajdeep},
   %year={2019},
   %month={Feb}
%}

@book{carm,
    title = {Quantum-Classical Correspondence for the Electromagnetic Field I: The Glauber-Sudarshan P Representation},
    author = {Howard J. Carmichael},
    isbn = {978-3-642-08133-0},
    series = {},
    year = {1999},
    publisher = {Springer, Berlin, Heidelberg}
}
\end{document}